\documentclass[fleqn,usenatbib]{mnras}

\usepackage{newtxtext,newtxmath}

\usepackage[T1]{fontenc}

\DeclareRobustCommand{\VAN}[3]{#2}
\let\VANthebibliography\thebibliography
\def\thebibliography{\DeclareRobustCommand{\VAN}[3]{##3}\VANthebibliography}

\usepackage{graphicx}	
\usepackage{amsmath}	
\usepackage{xspace}
\usepackage{comment}
\usepackage{siunitx}

\newcommand{\cm}{\text{cm}}
\newcommand{\erg}{\text{erg}}
\newcommand{\Msun}{\text{M}_\odot}

\newcommand{\K}{\text{K}}

\newcommand{\kpc}{\text{kpc}}
\newcommand{\Mpc}{\text{Mpc}}

\newcommand{\yr}{\text{yr}}
\newcommand{\eV}{\text{eV}}
\newcommand{\kms}{\text{km}\,\text{s}^{-1}}

\newcommand{\kb}{k_\text{B}}

\newcommand{\dd}{\text{d}}

\newcommand{\colibre}{COLIBRE}
\newcommand{\eagle}{EAGLE}
\newcommand{\flamingo}{FLAMINGO}
\newcommand{\herons}{\textsc{hbt-herons}\xspace}
\newcommand{\swift}{\textsc{Swift}\xspace}
\newcommand{\sphenix}{\textsc{sphenix}\xspace}

\newcommand{\panphasia}{\textsc{panphasia}\xspace}
\newcommand{\monofonIC}{\textsc{monofonic}\xspace}
\newcommand{\soap}{\textsc{soap}\xspace}
\newcommand{\chimes}{\textsc{chimes}\xspace}

\title[The COLIBRE simulation project]{The COLIBRE project: cosmological hydrodynamical simulations of galaxy formation and evolution}
\author[J. Schaye et al.]{Joop Schaye,$^{1}$\thanks{E-mail: schaye@strw.leidenuniv.nl}
Evgenii Chaikin,$^{1}$
Matthieu Schaller,$^{2,1}$ 
Sylvia Ploeckinger,$^{3}$ 
Filip Hu\v{s}ko,$^{1}$ \newauthor
Robert J. McGibbon,$^{1}$ 
James W. Trayford,$^{4}$ 
Alejandro Ben\'{i}tez-Llambay,$^{5}$ 
Camila Correa,$^{1}$ 
Carlos S. Frenk,$^{6}$ \newauthor
Alexander J. Richings,$^{7,8}$ 
Victor J. Forouhar Moreno,$^{1}$ 
Yannick M. Bah\'{e},$^{9,10}$ 
Josh Borrow,$^{11}$  \newauthor
Anna Durrant,$^{12,1}$
Andrea Gebek,$^{13}$
John C. Helly,$^{6}$ 
Adrian Jenkins,$^{6}$
Cedric G. Lacey,$^{6}$ 
Aaron Ludlow,$^{14}$ \newauthor 
and Folkert S. J. Nobels$^{1,15}$ \\
$^{1}$Leiden Observatory, Leiden University, PO Box 9513, 2300 RA Leiden, the Netherlands \\
$^{2}$Lorentz Institute for Theoretical Physics, Leiden University, PO Box 9506, 2300 RA Leiden, the Netherlands \\
$^{3}$Department of Astrophysics, University of Vienna, T\"{u}rkenschanzstrasse 17, 1180 Vienna, Austria \\
$^{4}$Institute of Cosmology and Gravitation, University of Portsmouth, Dennis Sciama Building, Burnaby Road, Portsmouth PO1 3FX, UK \\
$^{5}$Dipartimento di Fisica G. Occhialini, Universit\`{a} degli Studi di Milano Bicocca, Piazza della Scienza, 3 I-20126 Milano MI, Italy \\
$^{6}$Institute for Computational Cosmology, Department of Physics, University of Durham, South Road, Durham, DH1 3LE, UK \\ 
$^{7}$Centre for Data Science, Artificial Intelligence and Modelling, University of Hull, Cottingham Road, Hull, HU6 7RX, UK \\
$^{8}$E. A. Milne Centre for Astrophysics, University of Hull, Cottingham Road, Hull, HU6 7RX, UK \\
$^{9}$School of Physics and Astronomy, University of Nottingham, University Park, Nottingham NG7 2RD, UK\\
$^{10}$Institute of Physics, Ecole Polytechnique F\'{e}d\'{e}rale de Lausanne (EPFL), Observatoire de Sauverny, 1290 Versoix, Switzerland \\
$^{11}$Department of Physics and Astronomy, University of Pennsylvania, 209 South 33rd Street, Philadelphia, PA 19104, USA \\
$^{12}$Astrophysics Research Institute, Liverpool John Moores University, Liverpool L3 5RF, UK\\
$^{13}$Sterrenkundig Observatorium, Universiteit Gent, Krijgslaan 281 S9, B-9000, Gent, Belgium \\
$^{14}$International Centre for Radio Astronomy Research, University of Western Australia, 35 Stirling Highway, Crawley, Western Australia, 6009, Australia \\ 
$^{15}$Netherlands Organisation for Applied Scientific Research (TNO), Molengraaffsingel 8, 2629 JD Delft, the Netherlands\\
}

\date{Accepted XXX. Received YYY; in original form ZZZ}

\pubyear{\the\year{}}

\begin{document}
\label{firstpage}
\pagerange{\pageref{firstpage}--\pageref{lastpage}}
\maketitle

\begin{abstract}
We present the \colibre\ galaxy formation model and the \colibre\ suite of cosmological hydrodynamical simulations. \colibre\ includes new models for radiative cooling, dust grains, star formation, stellar mass loss, turbulent diffusion, pre-supernova stellar feedback, supernova feedback, supermassive black holes and active galactic nucleus (AGN) feedback. The multiphase interstellar medium is explicitly modelled without a pressure floor. Hydrogen and helium are tracked in non-equilibrium, with their contributions to the free electron density included in metal-line cooling calculations. The chemical network is coupled to a dust model that tracks three grain species and two grain sizes. In addition to the fiducial thermally-driven AGN feedback, a subset of simulations uses black hole spin-dependent hybrid jet/thermal AGN feedback. To suppress spurious transfer of energy from dark matter to stars, dark matter is supersampled by a factor 4, yielding similar dark matter and baryonic particle masses. The subgrid feedback model is calibrated to match the observed $z\approx0$ galaxy stellar mass function, galaxy sizes, and black hole masses in massive galaxies. The \colibre\ suite includes three resolutions, with particle masses of $\sim10^5$,  $10^6$, and $10^7\,\Msun$ in cubic volumes of up to 100, 200, and 400~cMpc on a side, respectively. The largest runs use 136 billion ($5\times3008^3$) particles.  We describe the model, assess its strengths and limitations, and present both visual impressions and quantitative results. Comparisons with various low-redshift galaxy observations generally show very good numerical convergence and excellent agreement with the data.
\end{abstract}




\section{Introduction}

Hydrodynamical simulations following the concurrent formation and evolution of cosmological structures and the galaxies that they contain have become a central part of research in extragalactic astronomy and cosmology. Starting from redshifts $z\sim 10^2$ using initial conditions inferred from the cosmic microwave background (CMB) and surveys of large-scale structure, such simulations follow the growth of density fluctuations and the formation and evolution of galaxies \citep[for recent reviews, see e.g.][]{Vogelsberger2020, Crain2023, Feldmann2025}. They serve a wide range of purposes. These include gaining insight into astrophysical processes, interpreting observations of galaxies and diffuse gas, investigating the effect of baryon physics on cosmological probes, testing data analysis techniques, and guiding the design of new observational campaigns.

Contemporary cosmological simulations of galaxy formation in representative volumes (typically $25 \lesssim L/\Mpc\lesssim 300$ on a side) include HorizonAGN \citep{Dubois2014}, \eagle\ \citep{Schaye2015,Crain2015}, IllustrisTNG \citep{Pillepich2018TNGmethod,Pillepich2018TNG,Nelson2019}, Simba \citep{Dave2019}, ASTRID \citep{Bird2022}, CROCODILE \citep{Oku2024}, and (X)FABLE \citep{Bigwood2025}. These simulations produce populations of galaxies with properties that agree well in many respects with observational data. Larger-volume, but lower-resolution simulations such as BAHAMAS \citep{McCarthy2017}, MillenniumTNG \citep{Pakmor2023}, \flamingo\ \citep{Schaye2023,Kugel2023}, and Magneticum Pathfinder \citep{Dolag2025} reproduce many (mostly spatially unresolved) galaxy properties, as well as a variety of observations of large-scale structure and clusters of galaxies. Collections of simulations that each zoom into a region centred on a single galaxy or galaxy group, such as APOSTLE \citep{Sawala2016}, Auriga \citep{Grand2017}, FIRE \citep{Hopkins2018FIRE2, Hopkins2023FIRE3}, NIHAO-UHD \citep{Buck2020}, HESTIA \citep{Libeskind2020}, and ARTEMIS \citep{Font2020}, can be used to compare with more detailed observations of galaxies. Overall, these models are much more realistic than their earlier counterparts. However, significant discrepancies remain between the data and each of the aforementioned simulations.

While many high-resolution zoom-in cosmological simulations allow gas to cool to temperatures $T\ll 10^4\,\K$ \citep[e.g.][]{Hopkins2018FIRE2,Hopkins2023FIRE3,Agertz2021,Dubois2021,Mina2021,Applebaum2021,Gutcke2022,Rey2025}, most simulations of representative volumes that have been run to redshift $z=0$ suppress this cold gas phase by imposing an effective equation of state, or a temperature or pressure floor, on gas with densities typical of the interstellar medium (ISM), i.e.\ total hydrogen number densities $n_\text{H} \gtrsim 10^{-1}\,\cm^{-3}$. A notable exception is the 22.1~comoving Mpc (cMpc) FIREbox simulation \citep{Feldmann2023}. FIREbox predicts realistic cold gas masses, and it represents an important step towards applying sophisticated physical models to statistically representative volumes. However, even though FIREbox employs the FIRE-2 galaxy formation model, which is consistent with the observationally inferred galaxy mass -- halo mass relation when used in higher-resolution zoom-in simulations \citep{Hopkins2018FIRE2}, FIREbox itself does not predict realistic galaxy stellar masses (see \S\ref{sec:gsmf}). This discrepancy underscores the importance of testing galaxy formation models in representative volumes. FIREbox does not include active galactic nucleus (AGN) feedback, has no model for dust grain evolution, and computes the abundances of atomic and molecular hydrogen using an equilibrium model instead of tracking them directly. Another exception is the 25~cMpc Romulus25 simulation \citep{Tremmel2017}. While Romulus25 models many aspects of galaxy formation, and includes a model for the dynamical friction experienced by BHs that is more physical than those used in most other simulations including those presented here, it omits some of the dominant cooling channels. In particular, Romulus25 does not account for molecular cooling or dust processes and, for $T> 10^4\,\K$, it does not include metal-line cooling. 

Simulations of representative volumes that stop at high redshift can more easily afford the computational expense associated with the direct modelling of cold gas. However, studies based on such simulations run the risk of reproducing high-$z$ observations with models that would not work at lower redshift. This risk is a real worry because models that suffer from numerical overcooling, and which thus produce present-day galaxies that are too massive, too old, and too spatially concentrated, will tend to predict high-$z$ galaxies that undergo a more intense phase of `compaction' \citep[see e.g.][for a simulation study of compaction in high-$z$ galaxies]{Zolotov2015}. Such galaxies are more clumpy, have larger masses, reach higher star formation rates but quench earlier, which may, for example, help reproduce observations of submm galaxies \citep[e.g.][]{Kumar2025} and the abundance of massive, quenched galaxies \citep[e.g.][]{Carnall2024,Baker2025}. 

The successful reproduction of observations of galaxy populations by cosmological simulations relies on the use of subgrid models for unresolved processes, in particular galactic winds driven by feedback from star formation and AGN feedback from accreting supermassive black holes (BHs). The subgrid models for feedback are designed to circumvent `numerical overcooling'. This overcooling is understood to result from limited resolution, which causes the energy/momentum injected by stars or AGN to initially be distributed over too much gas mass, leading to too low post-shock temperatures and therefore too short cooling times \citep{DallaVecchia2012}. 

Methods to mitigate numerical overcooling include decoupling the winds from the hydrodynamics in the ISM \citep{Springel2003}, which is, for example, used in IllustrisTNG, Simba, and ASTRID, or injecting the feedback energy less often but in larger amounts \citep{Booth2009,DallaVecchia2012}, as is, for example, used in \eagle\ for stellar and AGN feedback and in IllustrisTNG and Simba for AGN feedback. Although such subgrid prescriptions can suppress overcooling, they also introduce new artefacts and problems. Wind particles that are temporarily decoupled from hydrodynamical forces cannot heat or inject turbulence in the ISM and ignore the work done by winds in the ISM \citep[e.g.][]{DallaVecchia2008}. The alternative of launching particles at high velocities or heating them to high temperatures can cause the feedback to be too bursty, can create overly large bubbles in the ISM, and can result in poor sampling of the feedback processes \citep[e.g.][]{Bahe2016}. Using either method, overcooling may still occur downstream of the wind injection, or, at the other extreme, radiative cooling losses may be underestimated. 

Subgrid prescriptions for feedback necessarily introduce free parameters that are more numerical than physical. These parameters help control the effectiveness of the feedback, but their values cannot be predicted from first principles. Examples include the initial wind mass loading, velocity, or temperature increase, and, if applicable, the conditions that must be satisfied for wind particles to recouple with the hydrodynamics. The values of at least some of these parameters are calibrated, either explicitly or implicitly, to broadly reproduce a chosen set of observables. For example, \eagle\ is calibrated to reproduce the $z\approx 0$ galaxy stellar mass function, size-mass relation of disc galaxies, and BH masses for massive galaxies \citep{Crain2015}. IllustrisTNG uses these same constraints, as well as the stellar-to-halo-mass relation, the mass-metallicity relation, the mass scale above which star formation is quenched, the halo gas fraction in galaxy groups, and the cosmic star formation history \citep{Pillepich2018TNGmethod}. It is important to note that calibration is not limited to the choice of parameter values, because the results tend to be sensitive to tacit numerical choices, such as the way the feedback energy is distributed among the gas resolution elements surrounding young stars \citep[e.g.][]{Chaikin2022}. 

As discussed in \citet{Schaye2015}, there are multiple reasons why the calibration of the subgrid model for feedback should generally depend on numerical resolution. Firstly, because a higher resolution allows one to directly simulate smaller scales and therefore new phenomena. Secondly, because the implications of keeping the subgrid parameters fixed depend on arbitrary choices in the parametrization. For example, if we use the wind velocity or temperature increase as a parameter and keep it fixed, then a higher mass resolution implies a smaller amount of injected energy per feedback event. If, on the other hand, we use the energy per event as the parameter that is kept fixed, then a higher mass resolution corresponds to a higher velocity or temperature. 

The necessity of calibration weakens the ab initio predictive power of hydrodynamical simulations relative to, say, dark-matter-only (DMO) simulations. In addition, since baryonic processes and subgrid models are not scale free and simulations cannot resolve all relevant scales, a change in resolution effectively changes the galaxy formation model, regardless of whether the subgrid model is recalibrated, making tests of numerical convergence harder to interpret \citep[e.g.][]{Ludlow2020}. 

A key motivation for preventing the formation of a cold ISM phase is to prevent artificial fragmentation due to the failure to resolve the Jeans scales \citep[e.g.][]{Bate1997,Schaye2008,Robertson2008}. However, in the cold ISM turbulent velocities are expected to be supersonic, which means the effective Jeans scales are typically larger than the thermal ones. Furthermore, since gravity is typically softened on scales greater than the Jeans length in the cold phase, artificial fragmentation is generally not a bottleneck in practice, provided the gravitational and hydrodynamic force resolutions are properly adjusted \citep{Ploeckinger2024}. 

Paradoxically, imposing a pressure floor may in fact make the simulations more sensitive to resolution. This is because in the absence of a cold phase with a small volume filling factor, the ISM is too smooth. If in reality the cold phase accounts for a large fraction of the interstellar gas mass, then keeping this gas in a more diffuse, volume-filling phase results in the cooling time of the hot phase created by feedback becoming too short, thereby exacerbating (downstream) numerical overcooling. 

For Lagrangian methods, it is conventional to quote the gravitational softening length when indicating the spatial resolution. While this common practice may seem sensible for DMO simulations, it is at best questionable for hydrodynamical simulations whose physical approximations break down, or subgrid models kick in, on scales larger than the softening length. For example, the ISM in simulations that do not include a cold, molecular gas phase is less clumpy. One could therefore argue that, irrespective of their formal resolution, the actual resolution of such simulations is similar to the Jeans scale in warm ($T\sim 10^4\,\K)$ gas with density $n_\text{H} \sim 10^{-1}\,\cm^{-3}$, which is $\sim 1~\kpc$, because the physics is clearly not modelled correctly on smaller scales.

In fact, the softening length is a poor estimate of the spatial resolution even for purely gravitational processes because gravitational interactions between simulation particles and fluctuations in the potential due to particle noise tend to randomize particle angular momenta and drive the system to equipartition, thus transferring energy from high- to low-kinetic energy particles \citep[e.g.][]{Sellwood2013,Sellwood2015}. Since dark matter (DM) particles on halo orbits tend to move faster than the particles orbiting in a galaxy disc, at least in the direction perpendicular to the disc, in simulations halo particles tend to heat up the disc. The stellar halo similarly heats up the disc, but this is less important because the DM halo contains more energy than its stellar counterpart. Because gravitational softening suppresses fluctuations in the potential due to discreteness effects, a larger softening length can actually improve the effective spatial resolution, for both DMO \citep{Ludlow2019CDM} and hydrodynamical simulations \citep{Ludlow2020}.

The spurious transfer of energy is exacerbated by the fact that nearly all cosmological simulations use the same number of DM particles as baryonic particles, or fewer DM particles than baryonic particles. A counterexample is Romulus25 \citep{Tremmel2017}, which supersamples the DM by a factor 3.375. Not supersampling the DM implies that the DM particles tend to be at least a factor $(\Omega_\text{m}-\Omega_\text{b})/\Omega_\text{b}\approx 5.3$ times more massive than the baryonic particles. This difference in particle masses causes energy transfer from the DM to the baryons even if they were to follow the same phase space distribution \citep{Ludlow2019}. Conversely, using equal-mass dark matter and baryonic particles does not eliminate spurious energy transfer if the two components follow different phase space distributions, as is generally the case \citep{Ludlow2023}.

In the real Universe this type of energy transfer is negligible if the DM particle mass is microscopic, because then the particle sampling noise in the gravitational potential is negligible and the associated relaxation time is very long. Although spurious transfer of energy to gas is typically unimportant because it is radiated away, it can be problematic for the stellar component, particularly for older stars. As a result, galaxies, particularly their inner regions, tend to get puffed up, artificially increasing their size \citep{Ludlow2021} and reducing the disc-to-bulge ratio \citep{Wilkinson2023}. For the half-mass radius, the effect only becomes negligible when there are $\gtrsim 10^5$ DM particles in the halo \citep{Ludlow2023}. Because softening suppresses the spurious transfer of energy, the effective resolution can worsen when the gravitational softening length is reduced. The most direct way to suppress spurious energy transfer is to increase the number of DM particles.

Motivated by these considerations, we present here the \colibre\footnote{\colibre\ is a project of the Virgo consortium for cosmological supercomputer simulations. The acronym stands for \textbf{COL}d \textbf{I}sm and \textbf{B}etter \textbf{RE}solution, where the `better resolution' refers to the larger ratio of DM to baryonic particles compared with most previous simulations, which increases the effective resolution for a fixed baryonic particle mass and gravitational softening length. The explicit simulation of the cold ISM also increases the physical resolution relative to simulations that use the same particle mass and force resolution, but impose a pressure floor on the ISM.} suite of cosmological hydrodynamical simulations. \colibre\ is the latest in the series of simulation projects that includes OWLS \citep{Schaye2010}, Cosmo-OWLS \citep{LeBrun2014}, \eagle\ \citep{Schaye2015}, BAHAMAS \citep{McCarthy2017}, and \flamingo\ \citep{Schaye2023}. Its objectives are most comparable to those of \eagle, but it improves on that model in many ways. In terms of numerics, the main improvements include the suppression of the spurious transfer of energy discussed above by using four times more DM than baryonic particles, the use of up to 20 times more particles in total ($5\times 3008^3$ compared to $2\times 1504^3$ for \eagle), the use of a new simulation code \citep[\swift;][]{Schaller2024swift}, and a new structure finder \citep[\herons;][]{Forouhar2025Herons}. In terms of physics, all subgrid prescriptions are new or have been significantly modified. The most important new features are that cooling below $10^4\,\K$ is allowed (i.e.\ no equation of state is imposed on the ISM), the growth and composition of dust grains are simulated, self-shielding and molecules are included and coupled to the dust physics, pre-supernova stellar feedback is included, and the sampling of supernova and AGN feedback is improved. In addition, besides the fiducial model that uses thermally-driven AGN feedback, we include simulations that employ hybrid thermally-driven/kinetic jet-driven AGN feedback, where the hybrid model tracks the BH spin. The availability of simulations that achieve similar levels of agreement with observed galaxy populations, but which employ different types of feedback from accreting BHs, can provide insight into the physical mechanisms involved and the systematic uncertainty associated with the critically important but poorly understood process of AGN feedback. 

As was the case for \eagle, the subgrid feedback is calibrated to reproduce as closely as possible the observed $z\approx 0$ galaxy stellar mass function, galaxy half-mass radii, and BH masses of massive galaxies. While this calibration was done by hand for \eagle, \colibre\ uses machine learning to perform the calibration to the mass function and sizes at our lowest resolution, similar to what was done in \citet{Kugel2023} for \flamingo. We then make small adjustments by hand when we increase the resolution or switch from thermal to hybrid AGN feedback. The calibration process for the fiducial simulations is detailed in the companion paper \citet{Chaikin2025calibration}, while the adjustments made for the hybrid AGN feedback simulations are presented in \citet{Husko2025method}. In addition to the parameters associated with stellar and AGN energy feedback, the dust and chemistry modules include parameters whose values were chosen based on comparison of test simulations with observations. The effects of varying these parameters are demonstrated in \citet{Trayford2025} and \citet{Correa2025chemo}, respectively. The radiative cooling module includes parameters associated with local radiation fields and self-shielding whose effect is described in \citet{Ploeckinger2025}. Finally, the effects of the parameter choices associated with the subgrid prescriptions for star formation, pre-supernova stellar feedback, and supernova~Ia are discussed in \citet{Nobels2024}, \citet{BenitezLlambay2025}, and \citet{Nobels2025}, respectively.

\begin{figure*} \label{fig:cubes}
    \includegraphics[width=\textwidth]{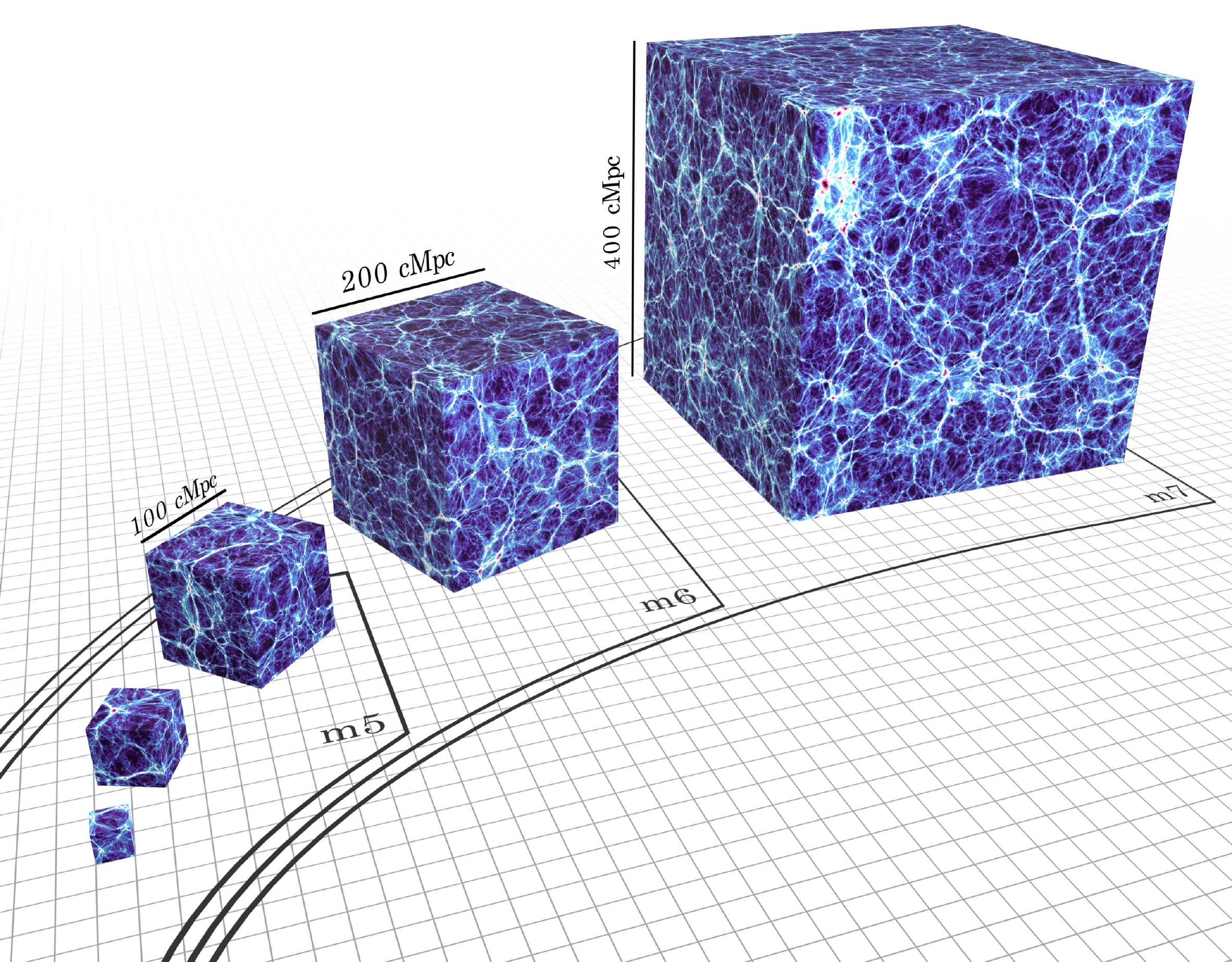}
    \caption{The five cubic \colibre\ boxes, which have side lengths ranging from 25 to 400 comoving Mpc. Colour shows total surface density (in 5~Mpc thick faces) at $z=0$. The volumes available at high (m5), intermediate (m6), and low (m7) resolution are indicated. Note, however, that at the time of writing the 50 and 100~Mpc high-resolution simulations have not yet reached redshift $z=0$.}
\end{figure*}

The \colibre\ suite presented here has a `wedding cake' design, with larger volumes using a progressively lower resolution (see Fig.~\ref{fig:cubes} and \S\ref{sec:firstlook} for visual impressions). As we find the model to converge better than \eagle, we use a wider range of resolutions. The two flagship simulations introduced here each follow $5\times 3008^3$ particles in cubic volumes of 200 and 400 cMpc on a side, yielding a particle mass, for both baryons and DM, of $\sim 10^6$ and $\sim 10^7~\Msun$, respectively. We also present a higher-resolution model with particle mass $\sim 10^5~\Msun$ in a 25~cMpc box that has reached $z=0$, while, at the time of writing, 50 and 100~cMpc versions of the same model are still running. We will refer to the high, intermediate, and low \colibre\ resolutions as m5, m6, and m7, respectively. 

We view \colibre\ as a stepping stone between very high-resolution simulations of ISM patches \citep[e.g.][]{Kim2023Tigress-ncr}, of isolated galaxies \citep[e.g.][]{Rosdahl2015,Semenov2017,Smith_Matthew2021,Richings2022,Katz2022}, or of individual (dwarf) galaxies in a cosmological setting \citep[e.g.][]{Mina2021,Applebaum2021,Gutcke2022,Rey2025}, and the previous generation of simulations of representative volumes. Although \colibre\ cannot resolve individual molecular clouds, it captures key effects of a cold interstellar phase that has a small volume filling factor while comprising a large fraction of the ISM mass. Including a cold phase also opens up an entirely new set of questions that the simulations can address, such as the evolution and distribution of atomic and molecular hydrogen, and the effect of dust on chemistry and observables on scales exceeding molecular cloud sizes. Although simulations that pressurize the ISM and do not explicitly model the evolution of dust grains have been used to investigate these questions, their analysis requires strong assumptions and relies on post-processing of gas and dust distributions that are unlikely to be particularly realistic.

This paper is organized as follows. Section~\ref{sec:methods} summarizes the simulation methods, including the gravity and hydrodynamics solver, initial conditions and cosmological parameters, identification of haloes and galaxies, and data products. Section~\ref{sec:subgrid} presents the subgrid prescriptions. Sections~\ref{sec:methods} and \ref{sec:subgrid} are rather technical sections and can be skipped by readers mostly interested in the results, although we recommend reading at least the brief overview of the subgrid prescriptions at the start of Section~\ref{sec:subgrid}. Section~\ref{sec:sims} gives an overview of the simulation runs. In order to give some intuition for what the simulations look like, we present visual impressions in Section~\ref{sec:firstlook} and, in Section~\ref{sec:phys_cond}, we characterize the physical conditions using phase diagrams and the properties of the gas from which stars are born. We compare the simulation results with the observations used to calibrate the stellar and AGN feedback in Section~\ref{sec:obs_cal}. In Section~\ref{sec:obs_other} we test predictions for other observations, including the cosmic star formation history; the $z=0$ relations between galaxy stellar mass and star formation rates, quenched fractions, \ion{H}{i} masses, molecular masses, gas and stellar metallicities, dust masses and grain sizes; and X-ray luminosities from the circumgalactic medium as a function of halo mass. In general, we find very good numerical convergence and excellent agreement with the data. In Section~\ref{sec:hybrid} we compare the predictions of the simulations using hybrid and fiducial AGN feedback, finding only small differences and similar levels of agreement with the data as shown in Section~\ref{sec:obs} for the fiducial simulations. Finally, we summarize and discuss our main results in Section~\ref{sec:conclusions}.

Images, videos, and interactive visualisations can be found on the \colibre\ website\footnote{\url{https://colibre-simulations.org/}}.

\section{Simulation methods} \label{sec:methods}
In this section we discuss the gravity and hydrodynamics solver (\S\ref{sec:swift}), the initial conditions and the cosmological parameters (\S\ref{sec:ICs}), the (sub)halo finding (\S\ref{sec:herons}), and the main available data products (\S\ref{sec:data_products}). The subgrid models for unresolved processes are discussed in the next section (\S\ref{sec:subgrid}).

\subsection{The gravity and hydrodynamics solvers} \label{sec:swift}
The simulations were performed using a version of the \swift code\footnote{\url{www.swiftsim.com}} \citep{Schaller2024swift} that combines the open-source version 2025.04 with new subgrid modules. \swift uses task-based parallelism within compute nodes and non-blocking message passing interface between compute nodes. 

The simulation contains collisionless cold dark matter (CDM) and baryonic particles. We use four times more CDM than baryonic particles, which helps suppress spurious transfer of energy from CDM to stellar particles and hence  improves the effective resolution for the baryonic component \citep{Ludlow2019,Ludlow2021,Ludlow2023}. The baryonic particles are initially all gas particles, but during the simulation they can be converted into stellar or BH particles, which are both collisionless. Stellar and gas particles can, respectively, lose and gain mass due to stellar mass loss (see \S\ref{sec:chemo}). While mass loss associated with stellar evolution will not reduce the masses of stellar particles by more than a factor of $\approx 2$, there is no limit to the mass that can be gained by gas particles. To prevent large differences between the masses of baryonic particles, gas particles whose mass exceeds $4m_\text{g}$ are split into two nearly co-spatial equal mass particles, where $m_\text{g}$ is the mean initial gas particle mass. Photons, relativistic neutrinos, and dark energy only affect the expansion history, while non-relativistic neutrinos also influence the growth of structure during the simulation (see below).

The short- and long-range gravitational forces are computed using a 4$^{\text{th}}$-order fast multipole method \citep{Greengard1987, Cheng1999, Dehnen2014} and a particle-mesh method solved in Fourier space, respectively, following the force splitting approach of \cite{Bagla2003}. The accuracy of the gravity solver is mainly controlled by an adaptive acceptance criterion for the fast multipole method similar to that proposed by \cite{Dehnen2014}.

The equations of hydrodynamics are solved with smoothed particle hydrodynamics (SPH) using the \sphenix implementation  \mbox{\citep{Borrow2022sphenix}}. \sphenix, which was specifically designed for galaxy formation simulations, uses a density-energy formulation of the equations of motion combined with artificial viscosity and conduction terms. Viscosity and conduction limiters prevent spurious energy losses in feedback-generated shocks and radiative cooling events. We use a quartic spline SPH kernel with a smoothing length of $\eta=1.2348$ in units of the local mean particle separation, $(m_\text{g}/\rho_\text{g})^{1/3}$, where $\rho_\text{g}$ is the gas particle's density. This choice corresponds to 64.90 weighted neighbours (we use a tolerance of $\pm 0.02$ neighbours). Stellar (and BH) particles also require a smoothing length because they can transfer (accrete) mass and energy to their gas neighbours. To reduce the computational cost, we reduce $\eta$ to 1.1642 for stellar (but not for BH) particles, which corresponds to 54 weighted neighbours (and we use a tolerance of $\pm 1$). 

We stress that we use the same implementation and parameter values for \sphenix\ as in \citet{Borrow2022sphenix} and refer the interested reader to that paper for the results of a variety of standard hydrodynamical test problems. The performance is satisfactory, although \citet{Braspenning2023} show that cloud-wind interactions converge very slowly if the density contrast is high ($\gg 10$). However, at least for the numerical resolution of the simulations presented here, we expect details of the subgrid prescriptions for feedback to dominate over differences due to the implementation of the hydrodynamics solver \citep[e.g.][]{Scannapieco2012,Schaller2015hydro}.

The effects of neutrinos on the expansion rate and on the growth of density fluctuations are modelled using the semi-linear method of \citet{alihaimoud2013}, which was also used in the BAHAMAS simulations \citep{McCarthy2018} and is compared with other methods in \citet{Elbers2021}. A linear perturbation integrator is run on the fly to compute the evolution of neutrino density perturbations sourced by the gravitational potential of the baryons and CDM. The gravitational force due to the neutrino perturbations is then included in the total gravitational force exerted on the baryons and CDM. This method self-consistently models the mutual dynamical response of fluctuations in the baryons + CDM and fluctuations in the neutrinos, but it does not capture the effect of neutrino self-gravity on the evolution of neutrino perturbations. Capturing this effect would require the use of neutrino particles, as was done in the large-volume \flamingo\ simulations \citep{Schaye2023}, but the effect is negligible on galaxy scales. 

Time integration is performed using a standard leapfrog scheme. The individual time steps of the particles are set by the minimum of the gravity time step ($\Delta t = \left(0.025 \epsilon/a_\text{acc}\right)^{1/2}$), where $\epsilon$ is the gravitational softening length and $a_\text{acc}$ is the acceleration, and the Courant–Friedrichs–Lewy \citep{Courant1928} condition for hydrodynamics with parameter value $0.2$. In addition, the time steps of young stellar particles (ages $< 40$~Myr) are limited to be at most 1~Myr (\S\ref{sec:chemo}), while gas particles can have their time step limited by turbulent diffusion (\S\ref{sec:diffusion}) and BH particles by their accretion rates (\S\ref{sec:agn_fb}). We activate gas particles that are directly heated/accelerated as a result of any kind of energy feedback, i.e.\ stellar winds, radiation pressure, \ion{H}{II} regions, core collapse supernovae (CCSNe) (in either thermal or kinetic form), type Ia SNe (SNIa), AGN (in both thermal and kinetic form). Following \citet{Durier2012}, the time step of inactive neighbouring gas particles is limited to at most 4 times the minimum time step of the neighbours in order to properly evolve the fluid even in extreme shocks. Finally, the time step is limited by the requirement that the SPH smoothing length does not change by more than a factor of $2^{1/3}$ in successive time steps.  

Gravity is softened using the \citet{Wendland1995} C2 kernel on scales smaller than the kernel size $3\epsilon$, where $\epsilon$ is the `Plummer-equivalent' kernel size that we refer to as the softening length. All particles use the same spatially fixed, but time-evolving gravitational softening length (see Table~\ref{tbl:simulations}). Appendix~\ref{app:softening} contains a discussion of the potential consequences of gravitational softening for the gravitational collapse of gas clouds. We impose a floor on the gas particle SPH smoothing length of $10^{-8}\epsilon$, which is sufficiently small to suppress spurious subkernel instabilities \citep{Ploeckinger2024}.

\subsection{Initial conditions and cosmological parameters} \label{sec:ICs} 
The initial conditions (ICs) are generated with \monofonIC \citep{Hahn2020,Michaux2021} using second-order Lagrangian perturbation theory. CDM and baryon particles are set up in a two-stage process. First, the combined mass-weighted CDM + baryon fluid is initialised as a single fluid, accounting for the presence of neutrinos. This single fluid is then split into separate components with distinct transfer functions. Linear power spectra and transfer functions are computed using \textsc{class} \citep{lesgourgues11,lesgourgues11b}. To limit cosmic variance without compromising the ability to carry out zoom-in simulations, we use `partially fixed ICs' \citep{Angulo2016}: the amplitudes of modes with $(kL)^2 < 1025$ are set to the mean variance, where $k$ is the wavenumber and $L$ is the side length of the cubic simulation volume. The Gaussian random fields were chosen from subregions of \panphasia\ to allow future zoom-in resimulations \citep{Jenkins2013}, see Appendix~\ref{app:panphasia} for details. The starting redshift, $z=63$, is a compromise between the desire to limit discreteness errors, which become larger if the simulation starts earlier, and the accuracy of perturbation theory, which worsens if we start later. 
  
The unperturbed particle load for the initial conditions is created as follows.  The entire periodic volume is tiled with uniform cubic cells with the cell centres forming a primitive cubic lattice.  An SPH particle is placed at the centre of each of these cubic cells. Equal-mass CDM particles are placed so that there are CDM particles at the vertices and in the centres of the sides of every cubic cell. By symmetry, the gravitational force acting on all particles is zero. Each cell has 8 vertices and 6 sides. Each vertex is shared by 8 cells and each side by 2 cells. The total number of CDM particles per SPH particle in the entire simulation volume is therefore $8/8+6/2=4$. 

Following \citet{Hahn2021}, the perturbation theory result is then imprinted in two steps. First, growing mode adiabatic perturbations are realised by perturbing the positions and velocities of the particles using 2nd order Lagrangian perturbation theory. Both particle species are treated in the same way during this phase. Second, compensated isocurvature perturbations, including leading order corrections to their evolution from decaying modes, are imprinted by altering the masses and relative velocities of baryonic and CDM particles. This procedure results in r.m.s.\ particle mass variations of $\sim 1$ per cent. \citet{Hahn2021} show that discreteness errors are suppressed by perturbing the masses of the particles rather than their positions.

Each hydrodynamic simulation is complemented by a corresponding DMO simulation that uses the same initial phases and the same total number of particles. To create the initial conditions for the DMO simulation, we converted the baryonic particles from the corresponding hydrodyamic run into CDM particles. The DMO simulations therefore also use two-fluid initial conditions. 

The values of the cosmological parameters for all runs presented here are as follows. The dimensionless Hubble constant, $h=0.681$; the density parameter for the total amount of non-relativistic matter, $\Omega_\text{m}=0.306$, with contributions from CDM, $\Omega_\text{CDM}=0.256011$, baryons, $\Omega_\text{b}=0.0486$, and non-relativistic neutrinos, $\Omega_{\nu,\text{NR}}=0.001389$; the dark energy density parameter, $\Omega_\Lambda=0.693922$; the density parameter for photons, $\Omega_\gamma = 0.000053$; the density parameter for relativistic neutrinos, $\Omega_{\nu,\text{R}}=0.000025$; the amplitude of the primordial matter power spectrum, $A_\text{s} = 2.099\times 10^{-9}$; the power-law index of the primordial matter power spectrum, $n_\text{s}=0.967$. The initial baryonic mass fractions of hydrogen and helium are $X=0.756$ and $Y=1-X=0.244$, respectively\footnote{The transfer function used for the initial conditions assumes fractions of 0.754579 and 0.245421, respectively.}. Note that the values of the Hubble and density parameters are given at $z=0$. The following derived parameters may also be useful. The amplitude of the initial power spectrum parametrized as the root-mean-square mass density fluctuation in spheres of radius $8~h^{-1}\,\Mpc$ extrapolated to $z=0$ using linear theory, $\sigma_8 = 0.807$; the amplitude of the initial power spectrum parametrized as $S_8\equiv \sigma_8\sqrt{\Omega_\text{m}/0.3} = 0.815$.  
 
These values of the cosmological parameters are identical to those used for the fiducial \flamingo\ simulations \citep{Schaye2023,Kugel2023}. They are the maximum posterior likelihood values from the Dark Energy Survey year three (DES Y3; \citealt{Abbott2022}) `3×2pt + All Ext.' $\Lambda$CDM cosmology. These values assume a spatially flat universe and are based on the combination of constraints from three DES~Y3 two-point correlation functions: cosmic shear, galaxy clustering, and galaxy-galaxy lensing, with constraints from external data from baryon acoustic oscillations (BAO), redshift-space distortions, type Ia supernovae, and Planck observations of the CMB (including CMB lensing), Big Bang nucleosynthesis, and local measurements of the Hubble constant (see \citealt{Abbott2022} for details). The neutrino density corresponds to $\sum m_\nu c^2 =0.06~\eV$, which is the minimum mass allowed by neutrino oscillation experiments \citep{Esteban2020,Salas2021}, and is consistent with the 95 per cent confidence upper limit of 0.13~eV from DES~Y3. In this model the neutrino contribution is provided by one massive and two massless species.

\subsection{Structure finding} \label{sec:herons}
To identify collapsed structures such as galaxies and clusters of galaxies, the snapshots are post-processed with three different algorithms. 

First, a friends-of-friends (FoF) halo finder is run on the DM particles\footnote{Following convention, we often refer to DM particles, in which case we actually mean CDM particles. Massive neutrinos do not contribute to the mass of CDM particles, which implies that neutrinos are implicitly assumed not to cluster on the scales of interest here. This is a good assumption for galaxies, but it is not quite correct for massive clusters of galaxies.}, which links all DM particles separated by less than the linking length of 0.2 times the mean distance between DM particles. Baryonic particles are then linked to the nearest DM particle within the linking length, if any. FoF groups whose total number of DM, stellar, gas, and BH particles is smaller than 32 are discarded. 

Second, the subhalo finder \herons\footnote{\url{https://hbt-herons.strw.leidenuniv.nl/}} \citep{Forouhar2025Herons} is run, which uses the FoF catalogue and snapshot data as inputs. \herons is a modified version of the hierarchical bound-tracing algorithm used in the \textsc{hbt+} code \citep{Han2012HBT,Han2018HBT+}. The algorithm assumes that structure forms hierarchically, i.e.\ that every satellite subhalo was once a central subhalo. \herons improves on \textsc{hbt+} in a variety of ways, some of which are critically important for hydrodynamical simulations.  \citet{Forouhar2025Herons} show that by tracking subhaloes in time, \herons outperforms structure finders that only use instantaneous phase-space information, which tend to have difficulty identifying satellites near halo centres, even if these contain very large numbers of bound particles. Furthermore, satellites found by phase-space finders often undergo unphysically large and non-monotonic changes in mass when they cross pericentre, a problem that is nearly absent for \herons \citep{Chandro-Gomez2025}. 

\herons starts by iteratively identifying self-bound\footnote{\herons defines the binding energy of a particle as the difference between its gravitational potential energy and its kinetic energy in the rest frame of the subhalo, where the former is computed from the particles making up the subhalo and the latter includes the thermal energy in the case of gas particles.} subhaloes within the highest redshift output and then sequentially processes each subsequent snapshot. Tracer particles, for which we use the 10 most bound particles of DM or stellar type, are subsequently used to track the descendants of the subhaloes and identify their host FoF haloes in the next snapshot, also allowing for the possibility that subhaloes may currently not have a FoF host. If a FoF halo contains multiple subhaloes, then the central subhalo is chosen from among the subhaloes whose masses exceed 0.8 times the mass of its most massive subhalo in the previous snapshot. If more than one subhalo fulfils this criterion, then the one with the lowest specific orbital kinetic energy in the centre-of-mass frame of the FoF halo is chosen as the central. A new subhalo is assigned to any FoF halo that does not yet contain any provided the new subhalo contains at least 20 bound particles and at least 10 tracer (i.e.\ DM or stellar) particles. Any particle that is no longer bound to the subhalo is removed from it.

Central subhaloes can accrete mass from surrounding diffuse matter as well as from their satellites, but satellite subhaloes can generally only accrete mass that has been part of their own satellites (i.e.\ satellites of satellites).
An exception to this `hierarchical' transfer of mass exists, as gas particles can be transferred between subhaloes without any prior connections under the following specific condition. Before unbinding, \herons finds the 10 nearest tracer particles for all gas particles in a FoF halo. If none of these neighbours are part of the same subhalo as the gas particle in question, then the gas particle is reassigned to the subhalo that contains its nearest tracer particle. This reassignment of gas particles relaxes the assumption of hierarchical assembly for gas, which, given its collisional nature, is not as well motivated as for dark matter and stars. 

Subhaloes that overlap strongly in phase space are merged. Subhaloes containing fewer than 20 bound particles or fewer than 10 tracer particles become `orphans', which may reappear as bound subhaloes in later snapshots. The location of a subhalo is defined by that of its most bound particle. For further details, see \citet{Forouhar2025Herons}.

Third, the Spherical Overdensity and Aperture Processor (\soap) tool \citep{McGibbon2025} is used to compute a large number of (sub)halo properties in a variety of apertures. \soap\ takes the subhalo locations and the particle IDs of the subhaloes' bound particles, which are provided by \herons, as input. All galaxy and halo properties, other than their locations and bound particle membership, are taken from \soap rather than from \herons. \soap computes properties in a variety of 3D and projected apertures, including apertures of fixed proper sizes and mean internal overdensities, and including either all or only bound particles. The spherical overdensity radius $R_{\Delta\text{c}}$ is defined as the radius within which the mean internal total matter density is $\Delta$ times the redshift-dependent critical density of the Universe. The mass inside $R_{\Delta\text{c}}$ is labelled $M_{\Delta\text{c}}$. Spherical overdensity apertures are only computed for central subhaloes, whereas properties in apertures of fixed proper size are computed for all subhaloes. In this work, galaxy properties will be computed using the bound particles within 3D apertures of radius 50~proper kpc (pkpc) centred on the most bound particle of the subhalo, a choice motivated by the finding of \citet{deGraaff2022} that this aperture reproduces masses inferred from S\'ersic fits to mock Sloan Digital Sky Survey images. 

Several caveats should be noted. First, the value of the FoF linking length that we use is conventional and is motivated by the virial overdensity predicted by analytic spherical collapse calculations \citep[for a discussion, see][]{More2011}. However, the precise value is arbitrary because collapse is not spherically symmetric. Moreover, FoF haloes cannot be determined observationally. This is problematic because even though halo mass is generally expressed as a spherical overdensity mass instead of as an FoF mass, the FoF linking length does determine the number of haloes and hence the number of central galaxies. Indeed, \citet{Forouhar2025Herons} show that varying the linking length by 20 per cent leads to $\approx 6$ per cent changes in the halo mass function at $M_\text{200c}=10^{12}\,\Msun$. Second, while \herons\ outperforms other structure finders when it comes to identifying satellite subhaloes, it cannot find galaxies that were born without DM if such objects exist. We note that algorithms that do find such objects will generally also fragment galaxies into self-bound subunits such as star clusters and molecular clouds, a problem that becomes more severe in simulations that resolve the multiphase ISM. Third, \herons does not account for the accretion of mass by satellites, except for accretion from satellites of the satellite in question and through gas reassignment. 
Fourth, \herons does not account for mergers between satellites that were independently accreted by their host halo and therefore lack a hierarchical connection. Most particles from the merged subhalo are therefore assigned to the host halo instead of the descendant satellite, resulting in the presence of a small number of residual substructures in the central subhalo. We find the incidence of such cases to be low and, when present, the substructures are only evident in a small fraction of the stellar component. Fifth, because every \herons subhalo must have been identified as a central at some point, subhaloes that were both born and became satellites in between consecutive snapshots are missed. However, we find that the results are converged for the number of snapshots that we use. Sixth, the (sub)halo catalogues extend down to particle numbers for which resolution effects are often important, so it is always necessary to use a convergence test to determine the actual resolution limit for the question of interest.

\subsection{Data products} \label{sec:data_products}
We save the state of the simulation at 128 redshifts between $z=30$ and 0. This includes 36 full snapshots and 92 reduced snapshots, also referred to as `snipshots' (see supplementary Table \texttt{output\_list.txt} for all the output times). Snapshots store many properties of the dark matter, gas, stellar and BH particles. While snipshots contain all BH properties, they only record a small subset of key gas and stellar properties and we remove the dark matter properties after completion of the structure finding (see supplementary Table~\texttt{output\_properties.pdf} for a list of the particle properties that are saved). For three small-volume simulations (L050m6, L050m6h, and L012.5m5; Table~\ref{tbl:simulations}) we save many more snipshots (2,000 for L050m6 and L050m6h, 10,000 for L012.5m5) to facilitate video creation and to enable projects requiring particle tracking at very high cadence. 

We save auto and cross power spectra at 123 redshifts between $z=30$ and 0 for matter--matter, CDM--CDM, gas--gas, stars--stars, gas--matter, CDM--gas, gas--(stars+BHs), CDM--(stars+BHs), pressure--pressure, and matter--pressure, where matter is the sum of the densities of CDM, gas, stars, and BHs, but not neutrinos, and where pressure refers to the electron pressure. 

To facilitate the production of mock quasar absorption spectra at high-cadence without having to read all the data, we frequently save the SPH particles that overlap with 35 randomly positioned lines parallel to the $x$ axis of the cubic simulation volume and similarly for the $y$ and $z$ axes. These `line-of-sight' outputs commence at expansion factor $a=0.1$ and are produced at intervals of $\Delta a = 0.01a$, yielding 232 output times in total.

As described in Section~\ref{sec:herons}, snapshots and snipshots are processed using the \herons structure finder and the \soap tool, yielding catalogues of many (sub)halo properties for multiple apertures, as well as merger trees.

\section{Subgrid prescriptions} \label{sec:subgrid}
As is the case for any simulation of a macroscopic system, prescriptions are required to model what happens below the resolution scale (`subgrid'). In this section we will summarize our subgrid models for radiative cooling and heating (\S\ref{sec:cooling}), dust grains (\S\ref{sec:dust}), star formation (\S\ref{sec:SF}), stellar mass loss and the rates of SNIa (\S\ref{sec:chemo}), turbulent diffusion (\S\ref{sec:diffusion}), stellar winds, radiation pressure from starlight, and \ion{H}{ii} regions, i.e.\ stellar feedback processes that start earlier than supernova feedback (\S\ref{sec:early_fb}), supernova (SN) feedback (\S\ref{sec:sn_fb}), BHs (\S\ref{sec:bhs}), and AGN feedback (\S\ref{sec:bhs}). 
The level of detail provided in the subsections depends on whether the subgrid module is described in a dedicated paper and on the importance of the subgrid process for the results and interpretation of the simulations. The calibration of the subgrid feedback parameters to the $z=0$ galaxy stellar mass function, galaxy sizes, and BH masses is summarized in \S\ref{sec:calibration}. 

Before describing the individual subgrid models, we provide a brief overview of the main features in the order in which they are discussed in the subsequent subsections:
\begin{itemize}
    \item Radiative cooling is allowed down to $\approx 10$~K \citep{Ploeckinger2025}. Direct modelling of the cold interstellar gas phase necessitates including more physics than is required for simulations that impose an equation of state. In particular, we allow for the formation of molecules in both the gas phase and on the surface of dust grains, self-shielding of the ionizing and photo-dissociating radiation by gas and dust, photo-electric heating by dust, photo-ionization, photo-dissociation, and photo-heating by the metagalactic radiation field, the interstellar radiation field, and interstellar cosmic rays. We compute the radiative cooling and heating rates, as well as the changes in the species fractions using our \chimes \citep{Richings2014a,Richings2014b} chemical reaction network, which includes the time-dependent evolution of 157 ions and molecules from 11 elements. However, to save computing time, we only treat the hydrogen and helium species fully in non-equilibrium. Cooling rates due to metals are computed element-by-element using tables assuming equilibrium species fractions, but the rates are corrected for the non-equilibrium free electron densities provided by the reaction network for hydrogen and helium. We use the local (thermal plus turbulent) Jeans length to estimate the shielding column density, the intensity of the interstellar radiation field, and the cosmic ray density.
    \item The formation, growth and destruction of dust is computed on the fly, using a model consisting of three grain species and two grain sizes \citep{Trayford2025}. The dust model is coupled to the model for radiative cooling. Reactions and cooling rates involving grains account for the amount and properties of the dust that is tracked in the simulation. Gas metal cooling rates account for the depletion of individual elements onto dust grains.
    \item Star formation is implemented using a Schmidt law combined with a gravitational instability criterion \citep{Nobels2024}. While the pressure-dependent model used in \eagle\ is ideal for simulations without a cold ISM, because it enables the direct implementation of the observed kpc-scale Kennicutt-Schmidt star formation surface density law, simulations like \colibre\ that model the multiphase ISM can be more ambitious and aim to predict the coarse-grained star formation law.
    \item The nucleosynthetic stellar yields from asymptotic giant branch (AGB) stars, CCSNe, and SNIa use up to date data from the literature \citep{Correa2025chemo}, and the SN Ia delay time distribution is calibrated to the observed evolution of the cosmic SN Ia rate.
    \item Besides the 11 elements that dominate the radiative cooling rates, we track the abundances of the s-process elements Ba and Sr, which are produced during the AGB phase, and the r-process element Eu, whose production by neutron star mergers, common-envelope jets SNe, and collapsars is modelled \citep{Correa2025chemo}.
    \item The turbulent diffusion of element mass fractions and the mass fractions of different types of dust grains are modelled \citep{Correa2025chemo}.
    \item Feedback processes from massive stars that begin before the first CCSNe explode, i.e.\ stellar winds, radiation pressure, and \ion{H}{ii} regions, are included \citep{BenitezLlambay2025}. 
    \item The model for CCSN feedback includes not only a stochastic thermal component \citep[as in][but modified to improve the sampling of feedback events in low-density gas]{DallaVecchia2012} that can drive galactic winds, but also a lower-energy kinetic component that helps drive turbulence \citep{Chaikin2023}.
    \item SN Ia energy feedback is implemented using the same stochastic thermal prescription used for CCSNe in order to mitigate numerical overcooling. 
    \item While our fiducial runs use purely thermally-driven AGN feedback \citep[as in][but modified to improve the sampling of feedback events for low-mass BHs]{Booth2009}, we repeat a subset of simulations using the \citet{Husko2025method} hybrid model that combines thermally-driven wind and kinetic jet feedback. These hybrid AGN feedback simulations track the spin of the BHs. 
\end{itemize}
The remainder of this section is intended to serve as a reference describing the methods that can be skipped by readers interested mostly in the results.

For the equations in this section, we assume the primordial hydrogen mass fraction $X=0.756$ \citep[e.g.][]{Aver2021} when converting gas mass densities, $\rho_\text{g}$, into total hydrogen number densities, $n_\text{H}$. Note that unless otherwise stated, we use each particle's actual element mass fractions during the simulation. For equations that depend on the gas particle mass, we assume the mean initial mass for our intermediate-resolution (i.e.\ m6 resolution; see \S\ref{sec:sims}) simulations. 

\subsection{Radiative cooling} \label{sec:cooling} 
Radiative cooling and heating rates are computed using \textsc{hybrid-chimes}, which is detailed in \citet[][we use their `weak ISRF' model]{Ploeckinger2025} and builds on the work of \citet{Ploeckinger2020}. The main differences from the subgrid model for radiative cooling used in the simulations presented in \citet{Ploeckinger2025} are that in \colibre\ the chemistry and cooling are coupled to the abundances of dust grains that are tracked by each gas particle, as described in Section~\ref{sec:dust}.

Cooling is allowed down to temperatures\footnote{To be precise, in \swift cooling is allowed down to the internal energy corresponding to neutral gas (mean molecular weight $\mu=1.224$) with a temperature of 10~K. In addition, \chimes imposes a minimum temperature of 10~K.} of 10~K. Rates are computed element by element, accounting for the individual abundances of all the elements that \citet{Wiersma2009Cooling} showed contribute significantly to the cooling: H, He, C, N, O, Ne, Mg, Si, S, Ca, and Fe. The abundances of species, i.e.\ ion and molecule mass fractions, and cooling rates are computed using an updated version of the \chimes\footnote{\url{https://richings.bitbucket.io/chimes/home.html}} reaction network \citep{Richings2014a,Richings2014b} for the radiation field, cosmic ray, and shielding prescriptions from \citet{Ploeckinger2025}. \chimes includes 157 ions and molecules. It accounts for the processes of collisional ionization, radiative and di-electronic recombination, grain-surface recombination, charge transfer, collisional excitation, Bremsstrahlung, molecule formation in gas and on dust grains, rovibrational and rotational cooling, Compton heating and cooling (using relativistic corrections, see \citealt{Ploeckinger2025}), cosmic ray ionization and heating, photoionization, photodissociation, gas and dust photo-electric heating, dust depletion, dust cooling, and self-shielding by dust and gas.

The abundances of free electrons and the nine H and He species \ion{H}{i}, \ion{H}{ii}, H$^-$, \ion{He}{i}, \ion{He}{ii}, \ion{He}{iii}, H$_2$, H$_2^+$, and H$_3^+$ are tracked individually, i.e.\ without assuming equilibrium. Reactions involving dust grains, such as the formation of molecular hydrogen on grain surfaces, account for the abundances of grain species tracked in the simulation. Although \textsc{hybrid-chimes} includes the option to boost the rates of reactions on the surface of dust grains, including the formation of H$_2$, this option is not used in \colibre. Cooling due to the HD molecule is added assuming a constant $n_\text{HD}/n_{\text{H}_2} = 1.65\times 10^{-5}$ \citep{Pettini2001}. 

The abundances of metal ions and molecules containing heavy elements (CO, H$_2$O, and OH) are tabulated using \chimes as a function of density, temperature, metallicity, and redshift, assuming chemical equilibrium (i.e.\ ionization equilibrium and steady-state chemistry). However, cooling rates due to heavy elements are computed using the non-equilibrium free electron abundance\footnote{By `non-equilibrium free electron density' we mean the sum of the non-equilibrium contributions from hydrogen and helium and the (generally much smaller) equilibrium contribution from heavy elements.}, so the equilibrium assumption is only used for the calculation of the ion and molecule fractions of heavy elements. 

Gas cooling rates account for the depletion of heavy elements on dust grains. The dust cooling rate (from recombinations on the grain surface and from grain-gas collisions), the dust photoelectric heating rate, and the formation rate of molecular hydrogen on dust grains account for the grain sizes tracked in the simulation, assuming that the rates scale in proportion with the surface area to volume ratio, i.e.\ inversely with the linear grain size for our spherical dust grains. 

\colibre\ does not include radiative transfer, but we do model some of its main effects in an approximate manner. While cooling radiation is assumed to escape freely\footnote{For molecular cooling from CO, H$_2$O and OH \chimes accounts for self-absorption of the cooling radiation.}, we do account for the attenuation of photoionization, photodissociation, and photoheating rates, which is a critical ingredient for simulations aiming to capture the multiphase ISM. Shielding by dust\footnote{For dust shielding of hydrogen and helium species we use the dust-to-gas ratio predicted by the simulation. For dust shielding of metal species we assume a dust-to-gas ratio of $6.6 \times 10^{-3}$, as in the \textsc{cloudy} ISM grain model \citep{Ferland2017}. \citet{Ploeckinger2025} assumed $5.5\times 10^{-3}$ in order to match the depletion from \citet{Jenkins2009}. This is not necessary for \colibre\ because depletion and dust-to-gas ratios are treated self-consistently in the live dust model.}, \ion{H}{i}, H$_2$, \ion{He}{i}, \ion{He}{ii}, and CO in the gas near the receiving gas particle is modelled. Surface densities are estimated using the local Jeans column density, which is the product of the density of the species of interest and the Jeans length. This approximation, which was used to model the transitions from the ionized to atomic and from the atomic to molecular phases by \citet{Schaye2001maxNHI,Schaye2004}, reproduces the ionized to atomic transition in cosmological radiative transfer simulations \citep{Rahmati2013}, which, however, lack the resolution and physics to model the atomic to molecular transition. Because the use of the Jeans column density implies the assumption that the gas is self-gravitating, it may overestimate the effect of shielding for pressure-confined clouds. On the other hand, if poorly resolved pressure-confined clouds are embedded in a resolved self-gravitating structure, such as the thick disc of warm gas, which is a situation quite typical for the ISM at the resolution of \colibre, then the Jeans column density may still provide a reasonable estimate of the shielding column.

Because we expect supersonic turbulence in the cold phase, we use the maximum of the thermal and turbulent Jeans scales, where the latter assumes a 1D turbulent velocity dispersion of $6~\kms$. 
At very low densities, corresponding to dynamical time-scales that are long compared to other relevant times-scales, local hydrostatic equilibrium is not expected, and we therefore cap the maximum Jeans length to 10 (50) kpc before (after) reionization. While the shielding of the hydrogen and helium species, whose abundances are computed in non-equilibrium, is modelled using the self-consistently predicted dust grain abundances, the shielding of heavy elements, for which cooling rates are tabulated, assumes a dust-to-metal ratio that depends on the density and temperature but saturates to a constant value in the ISM.

The calculations of species abundances and cooling rates account for the presence of the CMB, the UV/X-ray background from galaxies and quasars, an interstellar radiation field (ISRF), and interstellar cosmic rays. We adopt a modified version of the \citet{Faucher2020} model for the redshift-dependent metagalactic UV/X-ray background. Following~\citet{Ploeckinger2020}, at $z > 7.2$ ($z > 3$) we attenuate the \citet{Faucher2020} spectrum above the \ion{H}{i} (\ion{He}{ii}) ionization energies using \ion{H}{i} (\ion{He}{ii}) column densities tuned to match the effective photo-ionization and photo-heating rates that \citet{Faucher2020} finds are needed to match observations of the intergalactic medium (IGM) and the optical depth seen by the CMB. Hydrogen and helium reionization complete at $z\approx 7$ and $z\approx 3$, respectively. \citet{Ploeckinger2025} show that the model reproduces the observed thermal evolution of the IGM. Due to the non-equilibrium treatment of hydrogen and helium, no extra energy injections around the times of hydrogen and helium reionization (as for example used in \eagle) are needed \citep[e.g.][]{Puchwein2015}.

The incident ISRF has the fixed spectral shape from \citet{Black1987}, which combines the radiation field in the solar vicinity from \citet{Mathis1983} with the Galactic soft X-ray background from \citet{Bregman1986}. The intensity of the ISRF and the cosmic ray density are assumed to scale with the local star formation rate surface density, which is assumed to scale with the gas surface density to the power 1.4, as for the observed \citet{Kennicutt1998} law. The estimate for the local gas surface density is computed in the same way as the self-shielding column density. We use the `weak ISRF' model of \citet{Ploeckinger2025} because we found that for dwarf galaxies ($M_*\sim 10^9\,\Msun$) their fiducial ISRF, which is normalized to the \citet{Black1987} amplitude for solar neighbourhood conditions, results in an unnecessarily large fraction of star formation occurring in warm ($T\sim 10^4\,\K$) gas. The amplitude of the ISRF saturates at a total hydrogen column density $N_\text{H}=10^{22}\,\cm^{-2}$, because we do not expect the scaling with the Jeans column density to hold inside molecular clouds. The amplitude of the cosmic ray density saturates at $N_\text{H}=10^{21}\,\cm^{-2}$, in agreement with the observations of \citet{Indriolo2015}. The amplitudes of both the ISRF and the cosmic ray density are cut off below a proper density of $n_\text{H} = 10^{-2}\,\cm^{-3}$ and an overdensity of 100, because such low (over)densities correspond to conditions typical for the IGM rather than the ISM.

As we cannot resolve the initial phases of SN explosions and AGN winds, species abundances, i.e.\ ion and molecule fractions, are set to their equilibrium values when the particles are directly heated by thermal SN or thermal AGN feedback, i.e.\ when their temperature is increased due to the explicit injection of feedback energy (see \S\ref{sec:sn_fb} and \S\ref{sec:agn_fb}).

For each particle at each time step, we compute the internal energy that the particle is expected to reach at the end of its time step based on its current net radiative cooling rate. The time derivative of the internal energy is then adjusted so that the particle will drift to this final value over the course of its time step. Note that if a particle is drifting between time steps at the time a snapshot is written, then the recorded internal energy is not fully consistent with the recorded ion and molecule abundances, because in \chimes the abundances are immediately set to the final values at the end of the time step. Because the method of drifting the internal energy is only appropriate if the particle is on a time step that is short compared to the radiative cooling time, we adopt the same modification of this standard prescription as used in the \flamingo\ simulations: if the internal energy is expected to decrease by at least one third during the time step, then we immediately set the internal energy to the expected final value. This modification also improves consistency between the internal energy and species abundances saved in snapshots for particles that are in between time steps. A more accurate solution for the treatment of rapidly cooling gas would be to limit the time step to be small compared with the radiative cooling time, but this would be computationally prohibitively expensive.

\subsection{Dust grains} \label{sec:dust} 
The dust content is a subgrid property of gas particles. The model for the origin and evolution of dust grains is described and tested in \citet{Trayford2025}. Here we will only give a brief summary.

Gas particles track the fraction of the mass that is in dust grains of three different chemical compositions and two different grain sizes, i.e.\ six dust mass fractions in total. The three species are graphites, which are pure carbon, and two kinds of silicates, namely forsterite ($\text{Mg}_2\text{Si}\text{O}_4$) and fayalite ($\text{Fe}_2\text{Si}\text{O}_4$), which are both olivine. The grains are assumed to be spherical with radii $r_\text{g}=0.01$ or 0.1~\textmu m.

Grains are seeded by AGB stars using the metallicity-dependent dust yields of \citet{DellAgli2017}, assuming that stars with zero age main sequence mass $0.85 \le M/\Msun \le 3.5$ and $3.5 < M/\Msun \le 5$ produce graphites and silicates, respectively. CCSN dust yields are adapted by combining the dust nucleation yields from \citet{Zhukovska2008} with the total metallicity-dependent yields from \citet{Correa2025chemo} that are discussed in \S\ref{sec:chemo}. The total amount of silicates produced by AGB stars and CCSN depends on the amounts of silicon and oxygen in the ejecta. Forsterite and fayalite are assumed to each account for half of the released silicate mass. Ninety per cent of the dust produced (by mass) is assumed to consist of large (i.e.\ 0.1~\textmu m) grains. 

The metal yields used to enrich the gas are computed as the difference between the total metal yields (see \S\ref{sec:chemo}) and the dust metal yields.

The precise values used for grain seeding, including the assumed fractions of large and small grains, are unimportant because the grain properties are mainly determined by grain evolutionary processes. We note that this situation differs from that in some other simulations \citep[e.g.][]{McKinnon2016,Dave2019}, which assume the dust yields of \citet{Dwek1998} that are an order of magnitude higher than ours. See \citet{Trayford2025} for further discussion.

Following \citet{Hirashita2014}, gas accretion causes grains to grow exponentially in mass with an e-folding time of
\begin{equation}
    \tau_\text{acc} = \tau_\text{G}\left(\frac{r_\text{g}}{0.1 \text{\textmu m}}\right)\left(\frac{X_j}{X_{j,\odot}}\right)^{-1} \left(\frac{n^\prime_\text{H}}{10~\cm^{-3}}\right)^{-1} 
    \left(\frac{T}{10~\text{K}}\right)^{-1/2} ,
\end{equation}
where $\tau_\text{G}=180~$Myr and 99.3~Myr for graphites and silicates, respectively, and $X_j$ is the gas-phase mass fraction of the bottleneck element, i.e.\ the element whose abundance limits would first limit the amount of the grain species that can be formed. The fiducial values correspond to a sticking probability of 0.3 for gas accretion. The mass in each dust component is increased by an exponential factor based on the time-scale computed for that component using this equation.

Because our simulations cannot resolve the dense parts of molecular clouds where grain growth is most efficient, we expect to underestimate the rate of grain growth in dense gas. We therefore increase the density appearing in the gas accretion rate by a subgrid clumping factor, $C$, i.e.\ $n_\text{H}^\prime = C n_\text{H}$. Since we do not wish to boost the rates when the gas densities are sufficiently low that we do not expect significant resolution limitations, and since we do not wish to boost the densities without limit, we adopt the following functional form for the clumping factor
\begin{equation} \label{eq:clumping_factor}
    C = 
    \begin{cases}
        1 & n_\text{H} \le 0.1~\cm^{-3} ,\\
        \left (\frac{n_\text{H}}{10^{-1}~\cm^{-3}}\right )^{2/3} & 0.1~\cm^{-3} < n_\text{H} \le 100~\cm^{-3} ,\\
        100 & n_\text{H} > 100~\cm^{-3} .
    \end{cases}
\end{equation}
\citet{Trayford2025} show that for high densities ($n_\text{H}\gg 10~\cm^{-3}$) the results are insensitive to the precise value of the maximum clumping factor, provided it is $\gtrsim 10$, because the dust depletion fractions saturate (at a value of $2/3$ for graphite, as discussed below, and at levels determined by the abundance of the bottleneck element for silicates).

Thermal sputtering, i.e.\ destruction by collisions with hot gas particles, causes the grain mass to decline exponentially with an e-folding time of \citep{Tsai1995}
\begin{equation}
    \tau_\text{sp} = 0.85~\text{Myr} \left(\frac{r_\text{g}}{0.1 \text{\textmu m}}\right) \left(\frac{n_\text{H}}{10~\cm^{-3}}\right)^{-1} \left [1 + \left (\frac{T}{2\times 10^6\,\K}\right )^{-2.5}\right ] .
\end{equation}
As was the case for gas accretion, we compute the change in mass of each dust component using the e-folding time for that specific component, which for the case of thermal sputtering varies with the grain size but not with the grain composition. 
Note that the thermal sputtering time-scale depends on the SPH density, i.e.\ it is not reduced by the clumping factor, because thermal sputtering is due to collisions of grains with hot gas particles instead of with other dust grains.
Dust in gas that is converted into stars and in gas directly heated by thermal SN or thermal AGN feedback (see \S\ref{sec:sn_fb} and \S\ref{sec:agn_fb}) is assumed to be fully destroyed by astration and unresolved sputtering, respectively. 

Grain-grain collisions cause dust mass to be transferred from large to small grains. This grain shattering process proceeds at an exponential rate with an e-folding time of \citep{Aoyama2017,Granato2021}
\begin{align}
    \tau_\text{sh} = ~& 54.1~\text{Myr} \, \left (\frac{\mathcal{DTG}}{0.01}\right )^{-1} \left (\frac{r_\text{g}}{0.1~\text{\textmu m}}\right ) \nonumber \\ & \times
    \begin{cases}
        \left (\frac{n_\text{H}}{1~\cm^{-3}}\right )^{-1} & \text{for~} n_\text{H} < 1~\cm^{-3} ,\\ 
        \left (\frac{n_\text{H}}{1~\cm^{-3}}\right )^{-1/3} & \text{for~} n_\text{H} \ge 1~\cm^{-3} .
    \end{cases}
\end{align}
Here $\mathcal{DTG}$ is the dust-to-gas ratio, i.e.\ the fraction of the gas particle mass that is in dust.
Grain-grain collisions also lead to transfer of dust mass from small to large grains via coagulation, which proceeds at an exponential rate with an e-folding time based on \citet{Aoyama2017}:
\begin{equation} \label{eq:tau_co}
    \tau_\text{co} = 13.55~\text{Myr} \, f_\text{co} \left (\frac{\mathcal{DTG}}{0.01}\right )^{-1} \left (\frac{r_\text{g}}{0.1~\text{\textmu m}}\right ) \left (\frac{n_\text{H}^\prime}{10~\cm^{-3}}\right )^{-1}.
\end{equation}
Here $f_\text{co}$ is a resolution-dependent factor that we introduced to improve the convergence of the grain sizes with resolution. It is set to $10^{-0.5}$ for our lowest resolution (i.e.\ m7) simulations and to unity for all other simulations.
The shattering and coagulation time-scales assume a material density of $3~\text{g}\,\cm^{-3}$ and a coagulation velocity of $0.2~\kms$. Because shattering is thought to be important in the warm ISM, which we can resolve, while coagulation dominates in the cold and dense ISM \citep[e.g.][]{Hirashita2009}, which we cannot fully resolve, we boost the density appearing in the coagulation time-scale by the clumping factor (equation~\ref{eq:clumping_factor}), but not the density appearing in the time-scale for shattering. Note that coagulation always dominates over shattering, even if we do not boost the coagulation rate by the clumping factor. Grain growth by gas accretion is faster for smaller grains. Grain mass is then transferred to larger grains because coagulation dominates over shattering. 

Motivated by observations of molecular clouds \citep[e.g.][]{Fuente2019}, the fraction of carbon in graphite grains is limited to two thirds of the total carbon content of each gas particle. Without this limit, most of the carbon in the dense ISM would be depleted, whereas in reality, much of the carbon in high-density gas is observed to be in the form of CO. Given that we lack the resolution to track CO explicitly and that gaseous carbon is an important coolant and diagnostic, we opt to cap the maximum depletion factor. A cap on the dust depletion of oxygen, which is another important coolant in the gas phase, is unnecessary, because its depletion onto silicates is limited by the lower availability of other constituents.

The dust grains are coupled to the radiative cooling and chemistry modules described in Section~\ref{sec:cooling}. In particular, the cooling rates account for the depletion of heavy elements onto dust grains, for dust cooling and dust photoelectric heating, and for the formation of molecular hydrogen on dust grains, which all depend on the grain size. Finally, we note that dust grains are subject to turbulent diffusion using the prescription described in Section~\ref{sec:diffusion}.

\citet{Trayford2025} present further details of the dust modelling as well as a suite of simulations that vary the main free parameters and turn individual processes on or off.

\subsection{Star formation} \label{sec:SF}
The prescription for star formation is taken from \citet{Nobels2024}, to which we refer the reader for further details and tests. Briefly, star formation is implemented by stochastically converting gas particles that meet the criterion to be star-forming into collisionless stellar particles. Gas that is locally gravitationally unstable is eligible to form stars at a rate given by a \citet{Schmidt1959} law with power-law index $3/2$ and an efficiency per free-fall time of $\varepsilon = 0.01$.

Specifically, gas particles are only eligible for star formation, and hence labelled star-forming, if they satisfy the gravitational instability criterion
\begin{equation} \label{eq:SFcrit}
    \alpha \equiv \frac{\sigma_\text{th}^2 + \sigma_\text{turb}^2}{G \langle N_\text{ngb} \rangle^{2/3} m_\text{g}^{2/3} \rho_\text{g}^{1/3}} < \alpha_\text{crit},
\end{equation}
where we set $\alpha_\text{crit}=1$. Here $\sigma_\text{th}$ and $\sigma_\text{turb}$ are the particle's 3D thermal and turbulent velocity dispersions, which are computed from its temperature and the local interparticle velocity (peculiar plus Hubble) dispersion, respectively; $\left < N_\text{ngb}\right >$ is the weighted mean number of neighbours in the SPH kernel, which is about 65 for our quartic spline SPH kernel; $m_\text{g}$ is the mean, initial gas particle mass; $\rho_\text{g}$ is the gas density; $\alpha_\text{crit}$ is a constant of order unity whose precise value depends on the assumed geometry. The factor $\langle N_\text{ngb} \rangle m_\text{g}$ appears because we require a mass equal to that contained within the weighted mean SPH kernel to be gravitationally unstable. 

Hence, in higher-resolution simulations, i.e.\ for smaller $m_\text{g}$, star formation requires colder and/or denser gas. Indeed, to take advantage of the fact that higher-resolution simulations can model the ISM up to higher densities, we need to move the subgrid scale, i.e.\ the scale at which the subgrid model for star formation is applied, to higher densities as the resolution increases. Note that we therefore do not expect the physical conditions of gas that is labelled as star-forming to converge with the resolution. However, \citet{Nobels2024} show that coarse-grained quantities, such as the observable Kennicutt-Schmidt law on $\sim \kpc$ scales, can still converge with the resolution. 

It is instructive to consider the limiting case where the thermal velocity dispersion dominates over the turbulent dispersion. If $\sigma_\text{th} \gg \sigma_\text{turb}$, then equation~(\ref{eq:SFcrit}) yields the temperature-dependent density threshold\footnote{Equation~(29) of \citet{Nobels2024} expresses this as a density-dependent temperature ceiling, but the exponent of the density term was erroneously omitted from that equation. In addition, the solid lines of fig.~1 of \citet{Nobels2024}, which use this equation, are normalized too low by about 0.35~dex.}
\begin{equation} \label{eq:SFcrit_thermal}
    n_\text{H} > 1.9\times 10^{-1} ~\cm^{-3} \left( \frac{T}{10^4~\K} \right)^3 \left( \frac{m_\text{g}}{1.84\times 10^6~\Msun} \right)^{-2}, 
\end{equation}
where $n_\text{H}$ is the total hydrogen number density and we assumed a mean molecular mass appropriate for neutral gas, $\mu=1.3$, when converting between $\sigma_\text{th}$ and $T$. Similarly, if turbulence dominates, $\sigma_\text{th} \ll \sigma_\text{turb}$, then 
\begin{equation}
    n_\text{H} > 2.7\times 10^{-2} ~\cm^{-3} \left( \frac{\sigma_\text{turb}}{10~\kms} \right)^6 \left( \frac{m_\text{g}}{1.84\times 10^6~\Msun} \right)^{-2}. 
\end{equation}

The star formation rate (SFR) of gas particles that satisfy the star formation criterion (equation~\ref{eq:SFcrit}) is given by the Schmidt law 
\begin{equation}
    \dot{\rho}_\star = \varepsilon \frac{\rho_\text{g}}{t_\text{ff}} = \varepsilon \left (\frac{32G}{3\pi}\right )^{1/2} \rho_\text{g}^{3/2}, \label{eq:schmidt-law}
\end{equation}
where we set the efficiency per free-fall time $\varepsilon=0.01$. \citet{Nobels2024} and \citet{Lagos2025} show that this value of the efficiency, which is motivated by observations of giant molecular clouds in the Milky Way \citep[e.g.][]{Pokhrel2021,Hu2022SF}, the Large Magellanic Cloud \citep[e.g.][]{Ochsendorf2017}, and other nearby galaxies \citep[e.g.][]{Utomo2018,Leroy2025}, yields good agreement with the observed Kennicutt-Schmidt law. 

It is instructive to rewrite this as the gas consumption time-scale
\begin{equation}
\frac{\rho_\text{g}}{\dot{\rho}_\star} \approx 1.4\times 10^{10}~\yr \left (\frac{\varepsilon}{0.01}\right )^{-1}   \left( \frac{n_\text{H}}{10^{-1}~\cm^{-3}} \right)^{-1/2}, 
\end{equation}
which indicates that densities $n_\text{H} > 10^{-1}~\cm^{-3}$ are required for a star-forming gas particle to be expected to be converted to a star particle within a Hubble time at $z=0$.
The Schmidt law corresponds to an SFR per gas particle of
\begin{align}
    \dot{m}_\star &= m_\text{g} \frac{\dot{\rho}_\star}{\rho_\text{g}}, \\
    &\approx 1.3\times 10^{-4}~\Msun\,\yr^{-1} \left (\frac{\varepsilon}{0.01}\right )  \left( \frac{m_\text{g}}{1.84\times 10^6~\Msun} \right)  \left( \frac{n_\text{H}}{10^{-1}~\cm^{-3}} \right)^{1/2}, 
\end{align}
which, combined with the star formation criterion, gives an estimate of the minimum possible SFR.

\subsection{Stellar mass loss and chemical enrichment} \label{sec:chemo}
Our model for stellar mass loss and chemical enrichment is based on the prescription developed by \citet{Wiersma2009Chemo} for OWLS, a slightly modified version of which was used in EAGLE. It includes mass loss by winds from AGB and massive stars, CCSNe, and SNIa. As in \citet{Wiersma2009Chemo}, we track the 11 elements\footnote{S and Ca are assumed to track Si, i.e.\ their abundance relative to that of Si is assumed to be solar \citep{Asplund2009}.} that \citet{Wiersma2009Cooling} found to be important for the radiative cooling rates: H, He, C, N, O, Ne, Mg, Si, S, Ca, Fe. 

For \colibre\ we introduce several improvements:
\begin{itemize}
    \item Updated nucleosynthetic yields for all enrichment channels.
    \item An updated SNIa delay time distribution.
    \item Tracking of the s-process elements Sr and Ba, which are produced by AGB stars.
    \item Tracking of the r-process element Eu, with separate contributions from neutron star mergers, common envelope jets SNe, and collapsars.
    \item A model for unresolved small-scale mixing by turbulent diffusion (see \S\ref{sec:diffusion}). 
\end{itemize}
The model is motivated, detailed, tested, and calibrated in \citet{Correa2025chemo}. Here we only provide a brief summary.

Each stellar particle constitutes a Simple Stellar Population (SSP), i.e.\ a coeval population of stars of uniform chemical composition, whose zero age main sequence (ZAMS) mass distribution follows the \citet{Chabrier2003} initial mass function (IMF) in the range 0.1--100~$\Msun$. Stellar particles inherit their chemical composition from their parent gas particles, where heavy elements in gaseous and solid (i.e.\ dust) phases are summed. 

At each enrichment time step, which, as we discuss below, can be longer than the dynamical time step for stellar particles older than 100~Myr, we determine which ZAMS masses transition to the AGB or CCSN stage using the metallicity-dependent stellar lifetimes of \citet{Portinari1998}. Elemental mass loss, including both individually tracked elements and a total metal mass tracer (which accounts for untracked elements), is computed by interpolating nucleosynthetic yield tables as a function of the ZAMS mass and metallicity. We distinguish between the mass of an element produced or destroyed within the star, and the mass of the element that the star was born with and that simply passed through the star. While the latter depends on individual element abundances, the former depends only on the total metallicity, as the yield tables taken from the literature assume that the abundances of heavy elements relative to each other are solar.

Intermediate-mass stars, i.e.\ stars with ZAMS masses of 1--8~$\Msun$, lose mass during the AGB phase. The AGB yields are taken from \citet{Karakas2010,Fishlock2014,Doherty2014,Karakas2016,Cinquegrana2022}. However, based on the findings of \citet{Kobayashi2020}, we decrease the \citet{Karakas2016} Ba yields by a factor of 2. 
Massive stars, i.e.\ stars with ZAMS masses of 8--40~$\Msun$, lose mass as a result of both stellar winds and CCSNe. CCSN yields are taken from \citet{Nomoto2013}, while yields representing pre-SN mass loss are derived from \citet{Kobayashi2006}. The massive star yields of C and Mg were multiplied by a factor of 1.5 to improve agreement with their abundance ratios relative to Fe in Milky Way stars from the APOGEE survey \citep{Jonsson2018} in the metallicity range $[\text{Fe}/\text{H}] \ge -1$. Such adjustments are commonly made, motivated by the fact that yields computed by different studies often differ at the factor of 2 level, even for a fixed IMF and ignoring complications due to stellar rotation and binary evolution \citep[e.g.][]{Wiersma2009Chemo,Romano2010,Buck2021,Liang2023}.
Very massive stars, i.e.\ stars with ZAMS masses $>40~\Msun$, are assumed to not enrich the ISM because their ejecta, including pre-SN mass loss, fall back into a BH. Note that part of the metal yields of AGB and CCSNe are released in the form of dust grains (see~\S\ref{sec:dust} and \citealt{Trayford2025}).

\begin{figure}
    \centering
    \includegraphics[width=0.95\linewidth]{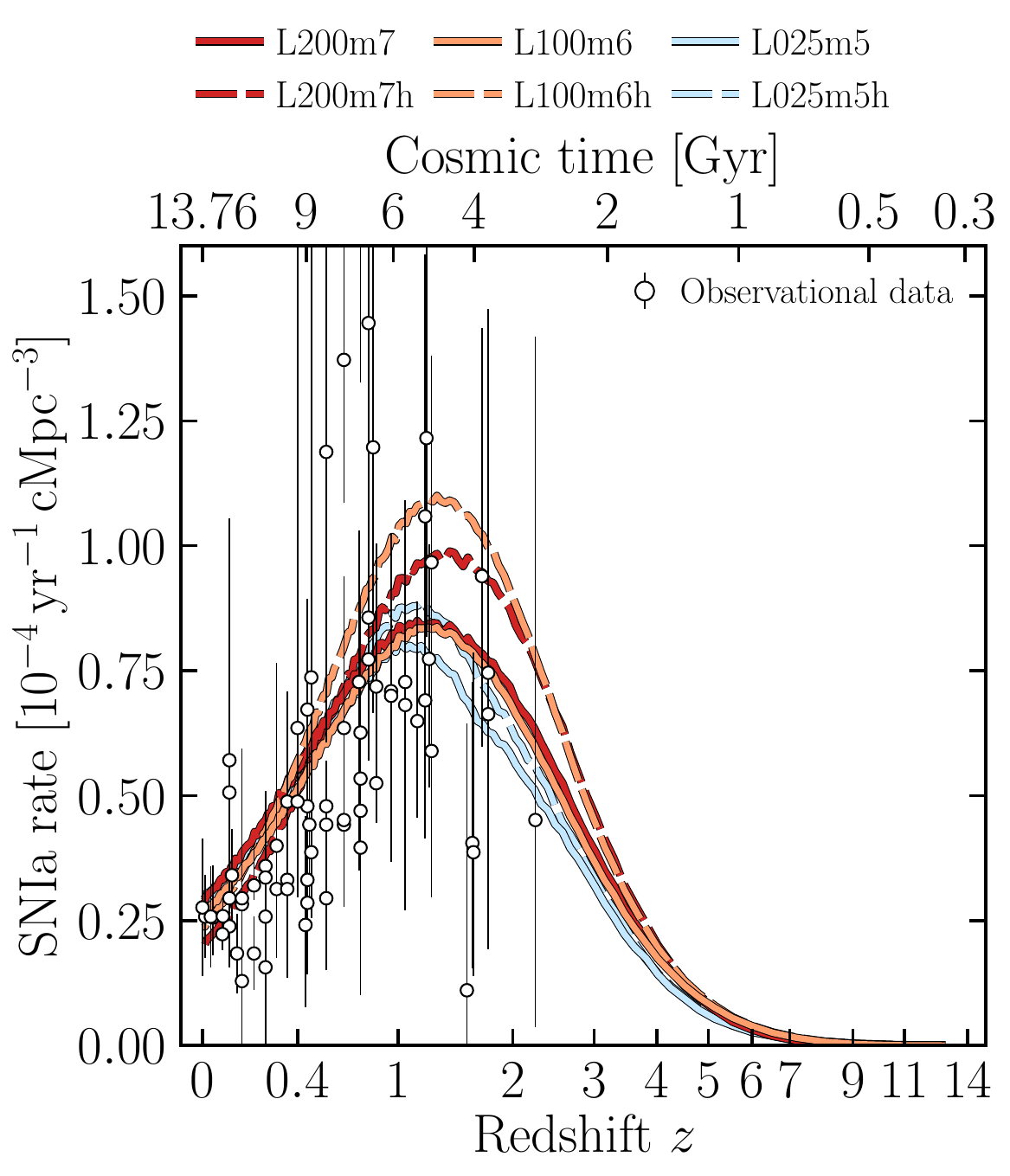}
    \caption{The evolution of the cosmic rate of SNIa per unit time and comoving volume for the fiducial (solid) and hybrid AGN feedback (dashed) simulations of different resolutions (different colours). The data points correspond to observed rates at $z<0.2$ \citep{Cappellaro1999, Hardin2000, Madgwick2003, Strolger2003, Blanc2004, Mannucci2005, Horesh2008, Dilday2010, Li2011, Quimby2012, Frohmaier2019}, $0.2<z<0.75$ \citep{Pain2002, Barris2006, Neill2006, Botticella2008, Melinder2012, Cappellaro2015} and at $z>0.75$ \citep{Tonry2003, Dahlen2004, Dahlen2008, Poznanski2007, Kuznetsova2008, Rodney2010, Graur2011, Perrett2012, Okumura2014, Rodney2014}. All simulations are in reasonable agreement with the compilation of observations.}
    \label{fig:SNIa}
\end{figure}

The rate of SNIa per unit ZAMS stellar mass for an SSP of age $t$ is given by the empirical delay time distribution function 
\begin{equation} \label{eq:SNIaDTD}
    \dot{\mathcal{N}}_\text{SNIa}(t) = \frac{\nu}{\tau} e^{-(t-t_\text{delay})/\tau} \, \Theta(t-t_\text{delay}), 
\end{equation}
where $\nu = 1.54\times 10^{-3} ~\Msun^{-1}$, 
$\tau = 2$~Gyr, $t_\text{delay} = 40$~Myr, 
and $\Theta(x)$ is the Heaviside step function. While the value of $t_\text{delay}$ is motivated by the lifetime of stars that end their lives as CCSNe, the values of $\nu$ and $\tau$ have been calibrated by \citet{Nobels2025} by comparing small-volume test simulations with observations. Fig.~\ref{fig:SNIa} shows that the model is in reasonable agreement with the observed evolution of the cosmic SNIa rate density compiled from (often discrepant) measurements taken from the literature. \citet{Nobels2025} show that the model also reproduces observations of the evolution of the specific SNIa rate as a function of galaxy stellar mass.

We adopt the SNIa yields from the W7 model of \citet{Leung2018} for near-Chandrasekhar-mass white dwarfs undergoing turbulent deflagration modelled with a deflagration-to-detonation transition. This is an updated version of the W7 model of \citet{Iwamoto1999}. 

We consider three production channels for the r-process element Eu: neutron star mergers, common-envelope jets SNe, and collapsars. The rates of these events are modelled using calibrated delay-time distributions, each with functional forms that differ from the one adopted for SNIa. To more accurately capture the rarity and stochastic nature of the r-process events, a stochastic approach is implemented. We refer to \citet{Correa2025chemo} for a detailed description.

While the yields for intermediate-mass and massive stars depend on ZAMS mass and metallicity, the yields for SNIa and the r-process production channels are assumed to be constant.
Because the contributions of the s- and r-process elements to the cooling rates are neglected, and because their contributions to the overall metallicity (on which stellar lifetimes and yields depend) are negligible, their yields can be re-scaled in post-processing without breaking the self-consistency of the simulations. For Eu, this can be done separately for the different production channels, because we track the contribution of each channel. 

The stellar mass loss of an SSP is a steeply declining function of its age. About half of the total mass lost in a Hubble time, which is a little over 40 per cent of the initial mass, is due to massive stars and hence occurs during the first $\approx 40$~Myr \citep[see e.g.\ fig.~1 of][]{Segers2016}. To ensure that this early phase is well sampled, we limit the time step (not just for enrichment, but also for the dynamics) of stellar particles younger than 40~Myr to be no longer than 1~Myr.

Because enrichment (specifically the neighbour finding) is computationally expensive, stellar particles older than 100~Myr only transfer mass after time intervals of 0.05 times their current age. All the mass lost during such a time interval is then transferred in a single time step. This enrichment time step is also applied to the SNIa energy feedback. Note that the dynamical time steps, which for collisionless particles such as stars are set by the gravity solver, are not changed for these older stellar particles. We have verified that this restricted sampling of the enrichment by older stars does not have a significant effect on the particle metallicity distribution functions. 

\subsection{Turbulent diffusion} \label{sec:diffusion}
In SPH simulations gas particles do not exchange mass, which implies that there is no mixing on mass scales smaller than the particle mass. This is not unreasonable because mixing on sub-particle mass scales is by definition unresolved. However, there are situations in which this lack of small-scale mixing can have undesirable consequences. An example, illustrated and discussed in \citet{Wiersma2009Chemo}, is the dispersal of heavy elements by galactic winds. A young star particle may enrich $\sim 10^2$ neighbouring gas particles with CCSN ejecta. If those particles are then moved to large distances by a galactic wind, the number of particles inside the radius reached by the outflow may be orders of magnitude larger, yet only the initial $\sim 10^2$ particles are enriched by the star particle. Such inhomogeneous enrichment would be unrealistic if unresolved small-scale mixing processes are in reality efficient, though in practice it may not be important if there are many enrichment events by many star particles. Since radiative cooling rates depend on the abundances at the particle/cell scale, the treatment of small-scale mixing may also affect the dynamics. 

We note that grid codes tend to have the opposite problem as Lagrangian codes. Because in grid codes gas flows are fully mixed within a cell, such simulations may suffer from numerical overmixing. 

We currently have a poor understanding of small-scale mixing processes \citep[for a review, see][]{FaucherGiguere2023}, partly because such processes are difficult to simulate, even for idealized cloud-wind interactions at very high resolution \citep[e.g.][]{Braspenning2023}. Nevertheless, it is reasonable to assume that unresolved turbulence would enhance small-scale mixing. Following early work on chemodynamics \citep[e.g.][]{MartinezSerrano2008,Greif2009,Shen2010}, we therefore implement turbulent mixing as a diffusion process,
\begin{equation} \label{eq:diffusion}
    \frac{\dd X}{\dd t} = 
    \frac{1}{\rho_\text{g}} \nabla\cdot (D \nabla X ),
\end{equation}
where $X$ is an abundance, e.g.\ the mass fraction of a particular element, $\dd/\dd t$ is the Lagrangian time derivative, and the diffusion coefficient $D$ is taken to be
proportional to the product of the Frobenius norm of the velocity shear tensor, the gas density, the SPH smoothing kernel squared and a dimensionless free parameter, $C_\text{diff}$. The diffusion coefficient therefore has dimensions $\text{mass}\,\text{length}^{-1} \text{time}^{-1}$. Note that this is equivalent to using a diffusion coefficient $D' = D/\rho_\text{g}$ that does not depend on the gas density and has the more familiar dimensions of area over time, if we replace equation~\ref{eq:diffusion} by $\frac{\dd X}{\dd t} = 
\frac{1}{\rho_\text{g}} \nabla \cdot (D'\rho_\text{g} \nabla X )$. Our treatment of diffusion is motivated by the Kolmogorov turbulence cascade. We refer the reader to \citet{Correa2025chemo} for further details.

The value of $C_\text{diff}$ varies in the literature. Based on the \citet{Correa2025chemo} comparison of $[\text{Mg}/\text{Fe}]$ as a function of $[\text{Fe}/\text{H}]$ in small test simulations and for observations of Milky Way stars, we adopt $C_\text{diff}=0.01$. \citet{Correa2025chemo} show that for $[\text{Fe}/\text{H}] > -1$, diffusion has only a limited impact on the element abundances of star particles within the disc of Milky Way-type galaxies. However, for lower metallicities, larger values of $C_\text{diff}$ boost the alpha enhancement. \citet{Correa2025chemo} also show that the effect of diffusion becomes more important at lower resolution. Including diffusion leads to better convergence of chemical abundances with numerical resolution, particularly on sub-galactic scales, but it does not have a significant effect on other galaxy properties. The fact that diffusion is more important at lower metallicities and at lower resolution is expected, as both tend to correspond to fewer enrichment events.

Turbulent diffusion is implemented for the elemental mass fractions, the mass fractions of dust grains of different chemical compositions and grain sizes, and for diagnostic tracers that track the mass fractions produced by different mass loss channels (AGB stars, massive stars and CCSNe, SNIa, different r-process channels). 

Finally, we note that turbulent diffusion can limit the time steps of gas particles, 
$(\Delta t)_i \le 0.2 h_i^2 / D_i$, where $h_i$ is the SPH smoothing length of particle $i$. 

\subsection{Stellar winds, radiation pressure, and \ion{H}{ii} regions} \label{sec:early_fb} 
Besides the radiative feedback from the local ISRF described in Section~\ref{sec:cooling}, we include three types of stellar feedback that are active before CCSNe start exploding: stellar winds, radiation pressure and \ion{H}{ii} regions. The implementation of these processes is detailed in \citet{BenitezLlambay2025}, here we will only give a brief summary. We refer to the collective effects of these processes as `pre-CCSN' or `early' stellar feedback, but note that they continue to operate after the first CCSN of an SSP explodes. 

Early stellar feedback not only provides feedback channels additional to CCSNe, but can also change the effectiveness of CCSN feedback. On the one hand, early stellar feedback can disperse the high-density stellar birth cloud \citep[e.g.][]{Chevance2022}, which can make CCSN feedback more effective by reducing radiative cooling losses for CCSNe produced by SSPs that have not yet moved out of their birth clouds. On the other hand, by smoothing the dense star-forming gas, the clustering of CCSNe is reduced, which can weaken the efficiency of the associated feedback \citep[e.g.][]{Rosdahl2015,Smith_Matthew2021}. 

Stellar winds and radiation are assumed to inject momentum at a rate taken from the BPASS stellar population synthesis models \citep[version 2.2.1;][]{Eldridge2017,Stanway2018} for the \citet{Chabrier2003} IMF with mass limits 0.1 and $100~\Msun$ that is used throughout. The rate depends on the metallicity and age of the SSP. The momentum flux due to radiation pressure is determined by convolving the emitted energy as a function of wavelength, which depends on the SSP metallicity and age, and is taken from BPASS, with the wavelength-dependent fraction that is absorbed locally by gas or dust. The column densities of gas and dust of the gas surrounding the SSP are computed in the same manner as the column densities used in the radiative cooling module (see \S\ref{sec:cooling}), i.e., they are based on the local Jeans length, and the fraction of the radiation that is absorbed is computed using \chimes. Feedback due to stellar winds and radiation pressure is implemented kinetically, using the stochastic method of \citet{DallaVecchia2008} and a kick velocity of $50~\kms$.

While the spatially varying ISRF accounts for the diffuse UV radiation field that is, for example, important for photoelectric dust heating, the ionizing radiation emitted by massive stars has a much shorter mean free path. In fact, the \ion{H}{ii} regions created by young SSPs are unresolved in our simulations, unless the density is much lower than typical of the ISM. The ionization of gas by young SSPs is implemented by stochastically labelling gas particle neighbours of young stellar particles as \ion{H}{ii} regions. Hydrogen and helium in \ion{H}{ii} regions are assumed to be singly ionized. The same applies to heavy elements in case of full non-equilibrium cooling, but in the fiducial \colibre\ setup heavy element ion fractions are assumed to be in equilibrium (see \S\ref{sec:cooling}). Gas particles in \ion{H}{ii} regions are prevented from cooling below $T=10^4\,\K$ and from forming stars. 

Every 2 Myr we compute the expected gas mass inside the Str\"omgren sphere of the \ion{H}{ii} region centred on the stellar particle powered by the ionizing luminosity of any SSP younger than 50~Myr. The mass in the \ion{H}{ii} region depends on the ionizing luminosity, which is a function of the age and metallicity of the SSP and is taken from BPASS, and the gas density, which we take to be the current density at the location of the stellar particle. The ratio of the mass in the \ion{H}{ii} region and the mass in the SPH kernel then yields the probability that a given gas neighbour is tagged as part of an \ion{H}{ii} region for the duration of the 2~Myr \ion{H}{ii} region time step. Overlap between different \ion{H}{ii} regions is ignored. After the 2~Myr have elapsed the particle's temperature and ion fractions can evolve freely, unless the particle is again tagged as part of an \ion{H}{ii} region. 

\subsection{Supernova feedback} \label{sec:sn_fb}
In \colibre\ galactic winds driven by stellar feedback are powered mainly by core collapse SNe (CCSNe) rather than by the early stellar feedback discussed in the previous section. The pre-CCSN feedback can make the CCSN feedback more efficient by reducing the gas density in the environment of young stars, which in turn reduces energy losses due to radiative cooling.

It has long been known that simulations that distribute the SN energy released by stellar particles at every time step among their gas neighbours suffer from overcooling, with inefficient feedback resulting in excessively large amounts of stellar mass \citep[e.g.][]{Katz1996}. As discussed in \citet{DallaVecchia2012}, if the sound-crossing time across the resolution element in which SN energy is injected is longer than the (post-shock) radiative cooling time, then the energy-driven expansion phase will be cut short, making feedback inefficient. There are two reasons why this problem is more severe at lower resolution. First, a lower numerical resolution implies a longer sound-crossing time because the resolution elements are larger. Second, and more importantly, if we decrease the resolution, then the radiative cooling time of the heated gas tends to become shorter due to the larger amount of mass over which the SN energy is distributed, which leads to lower peak temperatures. Hence, at the resolutions achievable for cosmological simulations of representative volumes, catastrophic overcooling will occur unless excessive radiative losses are prevented. 

Commonly used approaches to suppress numerical overcooling include using multiphase particles and injecting the energy in a hot and dilute subgrid phase \citep{Marri2003}, injecting the energy in kinetic form and temporarily decoupling wind particles from the hydrodynamics \citep{Springel2003}, temporarily turning off radiative cooling \citep[e.g.][]{Thacker2000}, or combining the energy from many SNe in order to inject it in amounts sufficient to raise the temperature to values large enough for radiative cooling to initially be inefficient \citep{DallaVecchia2012}. \colibre\ employs the last option, variants of which are sometimes referred to as `superbubble feedback' \citep{Keller2014}. 

As in \eagle, we follow \citet{DallaVecchia2012} and inject CCSN energy in thermal form using a stochastic treatment that has as parameters the injected CCSN energy in units of $10^{51}\,\erg$, $f_\text{E}$, and the desired temperature increase of the heated gas, $\Delta T_\text{SN}$. 

The amount of energy from CCSNe that becomes available for a stellar particle of initial (i.e.\ before stellar mass loss) mass $m_*$ in the time step from $t$ to $t +\Delta t$ is
\begin{equation}  \label{eq:CCSN_energy_per_dt}
  \Delta E_{\text{CCSN}} = 10^{51} \, \erg \, f_\text{E} \, \, m_*  \int_{m_\dd (t+\Delta t)}^{m_\text{d}(t)}  \, \Phi(m) \, \dd m,
\end{equation}
where $\Phi(m)$ is the IMF, which we assume to be \citet{Chabrier2003} and to cover the mass range 0.1 -- 100$~\Msun$, and $m_\text{d}(t)$ denotes the mass of the stars that explode as CCSNe at age $t$. We use the metallicity-dependent stellar lifetime tables from \citet{Portinari1998} to compute $m_\text{d}(t)$. The function $m_\text{d}(t)$ is non-zero only for zero-age main sequence masses between $m_\text{min,CCSN} = 8$ and $m_\text{max,CCSN} = 100~\Msun$, which correspond to stellar ages of $\approx 40$ and $\approx 3$~Myr, respectively. Note that \eagle\ uses $m_\text{min,CCSN} = 6~\Msun$, which yields $\approx 50$\% more CCSN energy in total.

For each SPH neighbour of a given stellar particle and for each time step, the probability that the gas particle is heated is equal to the ratio of the energy available for thermal feedback, $(1-f_\text{kin})\Delta E_{\text{CCSN}}$, and the energy required to increase the temperature of all the stellar particle's SPH neighbours by $\Delta T_\text{SN}$. Here $f_\text{kin}$ is the fraction of the CCSN energy injected in kinetic form (see below).

Equation~(18) of \citet{DallaVecchia2012} gives the density above which thermal feedback is expected to be ineffective due to catastrophic radiative losses,
\begin{equation} \label{eq:ncrit}
    n_{\text{H},t_\text{c}} = 2\times 10^2 ~\text{cm$^{-3}$} f_\text{t}^{-3/2} \left(\frac{m_\text{g}}{1.84 \times 10^{6}~\Msun}\right)^{-1/2} \left(\frac{\Delta T_\text{SN}}{10^{7.5}~\K}\right)^{3/2} ,
\end{equation}
where $f_\text{t}$ is the ratio of the radiative cooling time of the heated gas element to the sound-crossing time across the particle's SPH kernel\footnote{Note that \citet{DallaVecchia2012} adopt $f_\text{t}=10$, which is more conservative than the value of unity to which the coefficient in equation \ref{eq:ncrit} corresponds.} and we assumed the temperature after heat injection to be large compared to the temperature before injection, the gas to be ionized, and the hydrogen mass fraction to be the primordial one.

To avoid numerical overcooling in dense gas, one may wish to use very large values of $\Delta T_\text{SN}$, but this would lead to poor sampling of feedback events, and it may lead to excessively powerful explosions when feedback does occur.
The expectation value for the number of gas particles that are heated over the lifetime of a stellar particle is (equation~4 of \citealt{Chaikin2023})
\begin{equation}
\langle N_\text{heat,tot} \rangle \approx  0.91 \, (1-f_\text{kin}) \, f_\text{E} \left(\frac{m_\text{*}}{m_\text{g}}\right) \left(\frac{\Delta T_\text{SN}}{10^{7.5}~\K}\right)^{-1},
\label{eq:Nheat}
\end{equation}
where $m_\text{g}$ is the (average) mass of gas particles in the simulation. Assuming that $m_* \approx m_\text{g}$, $f_\text{E} \sim 1$, $f_\text{kin} \ll 1$ and requiring that each star particle on average deposits at least one energy injection in its lifetime (i.e.\ $\langle N_\text{heat,tot} \rangle \gtrsim 1$), gives a constraint on the heating temperature $\Delta T_\text{SN} \lesssim 10^{7.5}~\K$. This is why \eagle, which used $f_\text{kin}=0$ and $f_\text{E}\sim 1$, adopted $\Delta T_\text{SN} = 10^{7.5}~\K$.

However, heating just one gas particle per stellar particle does not sample the feedback well, particularly in less well-resolved galaxies that do not have many young stellar particles. Furthermore, releasing the energy in very energetic, sporadic explosions may create overly large bubbles in the ISM \citep[e.g.][]{Bahe2016}. To reduce these potential problems, we make two modifications relative to \eagle. 

First, following \citet{Chaikin2023}, we inject a fraction $f_\text{kin} = 0.1$ of the energy in the form of low-velocity kicks. As detailed in \citet{Chaikin2023}, the kinetic feedback is implemented by kicking particles in pairs in random but opposite directions in a manner that conserves energy, linear momentum, and angular momentum. In this scheme, the actual post-kick velocity of the wind particles relative to the star particle will differ somewhat from the target kick velocity if the particles are moving with respect to the star particles or if the two target particles have different masses. We adopt a target velocity of $50~\kms$, which energy-wise is equivalent to a temperature increment of only $\Delta T \sim 10^5~\K$, guaranteeing good sampling even for $f_\text{kin}=0.1$. \citet{Chaikin2023} show that while the thermal feedback generates a hot phase of the ISM that modulates the SFR by driving galactic outflows, the low-energy kicks increase the turbulent velocity dispersion in the neutral ISM, which in turn helps stabilize the gas against star formation (see equation~\ref{eq:SFcrit}). 

Second, instead of using a constant $\Delta T_\text{SN}$, we make it an increasing function of the gas density,
\begin{equation} \label{eq:dT_SN}
   \Delta T_\text{SN}(n_\text{H,SN}) = 10^{6.5}~\K \left(\frac{n_\text{H,SN}}{n_\text{H,pivot}}\right)^{2/3} ,
\end{equation}
Here $n_\text{H,pivot}$ is a free parameter with resolution-dependent values in the range 0.5 -- $1.0~\cm^{-3}$ for the thermal AGN feedback and 0.75 -- $1.5~\cm^{-3}$ for the hybrid AGN feedback model, respectively (see \S\ref{sec:calibration}). The logarithmic slope of $2/3$ is motivated by the cooling argument from \citet{DallaVecchia2012}, our equation~(\ref{eq:ncrit}). The gas density is computed using the SPH formalism at the location of the stellar particle each time the stellar particle can inject CCSN energy,
\begin{equation} \label{eq:density_estimate_SPH}
    n_\text{H,SN} = \frac{X}{m_\text{p}} \sum_{i=1}^{N_\text{ngb}} \, m_{\text{g},i} \,  W(|\boldsymbol{r}_* - \boldsymbol{r}_i|, h_*) ,
\end{equation}
where $X$ is the primordial hydrogen mass fraction, $m_\text{p}$ the proton mass, $m_{\text{g},i}$ is the mass of gas particle $i$, $\boldsymbol{r}_*$ and $\boldsymbol{r}_i$ are the coordinates of the stellar particle and gas particle $i$, respectively, and $W$ is the SPH kernel function with the stellar particle's smoothing length $h_*$, and the sum runs over all gas particles within the stellar kernel. 

The use of a variable $\Delta T_\text{SN}$ enables better sampling and prevents overly large bubbles in low-density gas, where overcooling is less likely to be an issue. However, to prevent very poor sampling in high-density gas and catastrophic cooling in low-density gas, we limit the allowed temperature increment to the range $\Delta T_\text{SN,min} < \Delta T_\text{SN} < \Delta T_\text{SN,max}$. The extrema were chosen on the basis of a few small test runs. Their values increase with the resolution (see Table~\ref{tbl:subgrid_pars}), but are always in the ranges $10^{6.5}~\K \le \Delta T_\text{SN,min} \le 10^7~\K$ and $10^{7.5}~\K \le \Delta T_\text{SN,max} \le 10^8~\K$. 

A third improvement relative to \eagle\ is that we do not use particle weighting, which is similar to mass weighting if the particles have similar masses, to distribute the energy among the SPH neighbours of a stellar particle. Instead, we use the isotropic implementation of \citet{Chaikin2022}, who demonstrated that isotropic weighting results in significantly reduced cooling losses compared to mass weighting because the mass tends to be concentrated in denser gas. We consider isotropic weighting to be more physical than mass weighting, since there is no reason why SN energy would be preferentially directed towards denser gas.

Any simulation that uses a subgrid prescription to suppress excessive radiative losses will need to calibrate its parameters (or functional forms) on a chosen set of observables. This is because numerical overcooling can in practice usually not be completely overcome, because radiative losses will occur in nature, and because the most appropriate values of the subgrid parameters can typically not be determined directly from observations or from models. In \eagle\ the parameter $f_\text{E}$ increases with decreasing metallicity and with increasing density. The density dependence was critical to overcome overcooling in massive galaxies \citep{Crain2015}, because ISM densities much higher than $n_{\text{H},t_\text{c}}$ are sampled in such galaxies at the resolution of \eagle. Because \colibre\ includes a multiphase ISM, a dependence on density could introduce a sudden sharp change in energy at constant pressure if a particle transitions from the warm to the cold interstellar gas phase. 

We therefore opt to make $f_\text{E}$ a function of the thermal gas pressure. Because we do not wish to change the energy budget of a stellar particle as it ages, we make $f_\text{E}$ a function of the stellar birth pressure, $P_\text{birth}$, i.e.\ the thermal pressure of its parent gas particle at the time it was converted into the stellar particle\footnote{For this purpose we define $P_\text{birth} = \rho_\text{birth} \kb T_\text{birth} / m_\text{p}$.}. The functional form for $f_\text{E}(P_\text{birth})$ is the same as that for  $f_\text{E}(n_\text{H,birth})$ in \eagle\footnote{For \eagle\ this function was expressed in the mathematically equivalent form $f_\text{E} = f_\text{E,min} + (f_\text{E,max} - f_\text{E,min})/[1 + (P_\text{birth}/ P_\text{E,pivot})^{n_\text{P}}]$ with $n_\text{P} = -1/ (\sigma_\text{P} \, \ln 10)$.},
\begin{equation} \label{eq:f_E}
    f_\text{E}(P_\text{birth}) = f_\text{E,min} +  \frac{\displaystyle  f_\text{E,max} - f_\text{E,min}}{1 +  \exp\left(-\frac{\log_{10} P_\text{birth} - \log_{10} P_\text{E,pivot}}{\sigma_\text{P}}\right)} \, ,
\end{equation}
where $f_\text{E,min}$ and $f_\text{E,max}$ are, respectively, the minimum and maximum energies that can be injected by a single SN, in units of $10^{51}~\erg$, $P_\text{E,pivot}$ is the birth pressure for which $f_\text{E} = (f_\text{E,min}+f_\text{E,max})/2$, and the parameter $\sigma_\text{P}$ determines the width of the transition from $f_\text{E,min}$ to $f_\text{E,max}$. Note that, different from \eagle, $f_\text{E}$ is independent of the metallicity. Note also that because \eagle\ assumes stars with mass $6-8~\Msun$ explode as (electron capture) SNe, the \eagle\ values of $f_\text{E}$ need to be multiplied by a factor 1.5 for a fair comparison with \colibre.

Equation~(\ref{eq:f_E}) implies that stellar particles born in higher gas-pressure environments will inject more CCSN energy. This can be interpreted as an ingredient necessary to overcome residual numerical overcooling, which can occur despite our use of stochastic thermal feedback with a variable $\Delta T_\text{SN}$ because densities can exceed $n_{\text{H},t_\text{c}}$ by large factors. It can also be interpreted as a reflection of the expectation that physical cooling losses will be smaller in superbubbles resulting from the clustering of many SNe \citep[e.g.][]{Sharma2014}. Values $f_\text{E} > 1$ can be justified on numerical grounds, i.e.\ because such values are needed to overcome numerical overcooling, or on physical grounds, e.g.\ because some SNe inject kinetic energies greater than $10^{51}~\erg$ \citep[e.g.][]{Mazzali2014} or because the IMF may become more top-heavy in starbursts \citep[e.g.][]{Gunawardhana2011}, where gas pressures are expected to be high. Values $f_\text{E} < 1$ can be justified because the use of stochastic thermal feedback can result in an underestimate of the physical radiative losses.  

As discussed in Section~\ref{sec:calibration}, we set $f_\text{E,max}=4$ and $\sigma_\text{P}=0.3$. The parameter $f_\text{E,min}$ increases with increasing resolution, but is limited to the range 0.1 -- 0.8 for the simulations used here. The value of $P_\text{E,pivot}$ also increases with the simulation resolution, with $8\times 10^3 ~\K\,\cm^{-3} < P_\text{E,pivot}/\kb < 1.5\times 10^4 ~\K\,\cm^{-3}$ for the simulations presented here (see Table~\ref{tbl:subgrid_pars}), where $\kb$ is the Boltzmann constant.  

Finally, we note that energy feedback from SNIa is implemented as for CCSNe, but with $f_\text{E}=1$, $f_\text{kin}=0$, and with equation~(\ref{eq:CCSN_energy_per_dt}) replaced by 
\begin{equation}  \label{eq:SNIa_energy_per_dt}
  \Delta E_{\text{SNIa}} = 10^{51}~\erg ~ f_\text{E} \, m_*  \int_{t}^{t+\Delta t} \dot{\mathcal{N}}_\text{SNIa}(t-t_*) \dd t,
\end{equation}
where $\dot{\mathcal{N}}_\text{SNIa}(t-t_*)$ is the rate of SNIa per unit initial stellar mass for a simple stellar population born at time $t_* < t$, see equation~(\ref{eq:SNIaDTD}). Given that compared to CCSN feedback, SNIa feedback is unimportant for the observables that we calibrate the subgrid feedback to, we set $f_\text{E}=1$ and $f_\text{kin}=0$ for simplicity. Another reason for setting $f_\text{E}=1$ is that residual numerical overcooling is less likely for SNIa than for CCSN feedback, since SNIa tend to explode long after the stars have moved out of the dense environments in which they were born. Our treatment of SNIa feedback differs from \eagle, where stochastic thermal feedback was used for CCSNe, but the SNIa energy was instead dumped thermally at each time step, causing the SNIa feedback to be ineffective due to numerical overcooling. Nevertheless, in \colibre\ SNIa feedback remains much less important than CCSNe \citep[for comparisons, see][]{Nobels2025}.

\subsection{Supermassive black holes} \label{sec:bhs}
The prescription for supermassive black holes (BHs) for the thermal AGN model is a modified version of that presented in \citet{Nobels2022} and \citet{Bahe2022}, which is itself based on \citet{Springel2005BHs}, \citet{DiMatteo2008}, and \citet{Booth2009}. We will first describe the model ingredients, i.e.\ BH seeding, gas accretion, mergers, and repositioning, and then discuss the modifications specific to the hybrid AGN feedback simulations.

\subsubsection{Seeding}

Starting from scale factor $a=1/(1+z) = 0.05$, after every\footnote{The cadence of the on-the-fly FoF finder used for seeding is the same as in the \flamingo\ simulation, but the value of $\Delta a$ was misquoted in \citet{Schaye2023}.} $\Delta a = 0.00751a$, a friends-of-friends (FoF) halo finder is run on the fly as described in Section~\ref{sec:herons}. Haloes more massive than $M_\text{FoF,seed}$ that do not yet contain a BH are seeded with a BH of mass $m_\text{BH,seed}$. The densest gas particle is then converted into a collisionless BH particle. 
Because the desired BH seed mass is generally small compared to the mass of the BH particle, $m_\text{BH,part}$, BH particles store a subgrid BH mass, $m_\text{BH} \le m_\text{BH,part}$. The subgrid BH mass is the mass that controls the physical processes related to BHs, while the BH particle mass is used in the gravity solver.  

The BH seed mass varies between $m_\text{BH,seed}=2\times 10^4~\Msun$ for m5 and $\approx 3\times 10^5~\Msun$ for m7 resolution. Such values correspond to the `heavy seed' scenario, where BHs are formed by the direct collapse of a massive gas cloud or by the rapid merging of stars and/or low-mass BHs \citep[e.g.][]{Regan2024}. The value of $m_\text{BH,seed}$ is set during calibration (see Section~\ref{sec:calibration}). It has some effect on the onset of the rapid BH growth phase through gas accretion, which can lead to quenching of star formation. BHs are seeded in haloes with $M_\text{FoF,seed} = 5\times 10^{10}~\Msun$ for m7 resolution and $M_\text{FoF,seed} = 1\times 10^{10}~\Msun$ for the m6 and m5 resolutions. The minimum value of $M_\text{FoF,seed} = 1\times 10^{10}~\Msun$ guarantees that the haloes are well resolved, i.e.\ they contain $\gg 10^2$ particles for all resolutions. At m7 resolution we use a five times higher value to ensure that the subgrid BH mass converges to the BH particle mass at large masses ($m_\text{BH}\gg m_\text{g}$). A lower value of $M_\text{FoF,seed}$ yields more seed BHs and at m7 resolution the increase in BH particle mass during mergers would then be too large for the subgrid BH mass to be able to catch up to BH particle mass via gas accretion. 

\subsubsection{Gas accretion}
BHs can grow through gas accretion and mergers with other BHs, but growth by the accretion of stars or DM is neglected. Following \citet{Krumholz2006}, the gas accretion rate is given by the Bondi-Hoyle-Lyttleton (BHL) accretion rate \citep{Bondi1952,Hoyle1939}
\begin{equation}
    \dot{m}_\text{BHL} = 4 \pi G^2 \frac{\rho_\text{g}}{c_\text{s}^3} m_\text{BH}^2,
\end{equation}
modified to account for turbulence and vorticity,
\begin{equation}
    \dot{m}_\text{accr} = \dot{m}_\text{BHL}\frac{f_\text{turb} f_\text{ang}}{(f_\text{turb}^2 + f_\text{ang}^2)^{1/2}},
\end{equation}
which interpolates smoothly between the asymptotic regimes where the accretion rate is limited by turbulence ($\dot{m}_\text{accr} \rightarrow f_\text{turb} \dot{m}_\text{BHL}$ if $f_\text{turb} \ll f_\text{ang}$) or by vorticity ($\dot{m}_\text{accr} \rightarrow f_\text{ang} \dot{m}_\text{BHL}$ if $f_\text{turb} \gg f_\text{ang}$). 

The turbulence limiter is \citep{Ruffert1994}
\begin{equation}
    f_\text{turb} = \left[ \frac{\lambda^2 + \mathscr{M}^2}{(1+\mathscr{M}^2)^4}\right]^{1/2},
\end{equation}
with $\lambda=1.1$ \citep{Krumholz2006} and where $\mathscr{M} = v/c_\text{s}$ is the Mach number, with $c_\text{s}$ the sound speed of the ambient gas. We take $v^2 = v_\text{BH}^2 + \sigma_\text{turb}^2$, where $v_\text{BH}$ and $\sigma_\text{turb}$ are, respectively, the bulk velocity relative to the BH and the velocity dispersion of the ambient gas in the frame of the BH. If $\mathscr{M} \ll 1$ then the turbulence-limited accretion rate asymptotes to $\lambda$ times the BHL rate.

The vorticity limiter is \citep{Krumholz2005}
\begin{equation}
    f_\text{ang} = 0.34 f(\omega_\star),
\end{equation}
where $\omega_\star = \omega r_\text{B} / c_\text{s}$, $\omega = |\nabla \times \mathbf{v}|$ is the vorticity of the ambient gas, $r_\text{B} = Gm_\text{BH}/c_\text{s}^2 \approx 4~\kpc\ (m_\text{BH}/10^8\,\Msun) (c_\text{s}/10~\kms)^{-2}$ is the Bondi radius, and $f(\omega_\star) = (1+\omega_\star^{0.9})^{-1}$. If the vorticity is small, i.e.\ $\omega \ll c_\text{s}/r_\text{B}$, then the vorticity-limited accretion rate asymptotes to 0.34 times the BHL rate. The reason why the accretion rate asymptotes to a value slightly smaller than the BHL rate is that we assume that even when vorticity is negligible at the Bondi radius, the accreting gas circularizes before it reaches the Schwarzschild radius \citep{Krumholz2005}.

This prescription for gas accretion differs from that of \eagle, which also used BHL accretion, but followed a different approach to account for the effect of relative motions between the BH and the gas and for the angular momentum of the gas \citep{Rosas-Guevara2015}. 

The Eddington accretion rate is the accretion rate for which the AGN luminosity equals the Eddington luminosity, i.e.\ the luminosity for which the radiation pressure due to electron scattering balances the gravity of the BH. 
Previous simulations such as \eagle\ did not allow super-Eddington accretion. However, enforcing the Eddington limit may be inappropriate, e.g.\ because accretion may be aspherical, photons may be trapped, or accretion originating from larger scales may result in a large inward momentum flux \citep[e.g.][]{Izquierdo-Villalba2023}. \citet{Husko2025SuperEddington} found that allowing super-Eddington accretion can have a large impact on galaxy evolution at high redshift. In \colibre\ the gas accretion rate is therefore not limited to the Eddington rate. We do limit it to 100 times the Eddington rate to prevent very large increases in BH mass on possibly poorly resolved time-scales. 

The Eddington rate is given by
\begin{equation} \label{eq:Edd_rate}
    \dot{m}_\text{Edd} = \frac{4 \pi G m_\text{p} m_\text{BH}}{\epsilon_\text{r} \sigma_\text{T} c} \approx 2.2 ~\text{M}_\odot \; \text{yr}^{-1} \left (\frac{\epsilon_\text{r}}{0.1}\right )^{-1} \left( \frac{m_\text{BH}}{10^8 ~\rm M_\odot} \right),
\end{equation}
where $\sigma_\text{T}$ is the Thomson cross-section, $\epsilon_\text{r}=0.1$ is the assumed radiative efficiency of the subgrid accretion disc \citep[e.g.\ similar to][]{shakura1973} and $c$ is the speed of light. Note that the BHL rate exceeds the Eddington rate for densities greater than
\begin{align}
    & n_\text{H}(\dot{m}_\text{BHL} = \dot{m}_\text{Edd})
    = \frac{X c_\text{s}^3}{\sigma_\text{T} G c \epsilon_\text{r} m_\text{BH}} ,\\
    &\approx 2.8\times 10^{-2} ~\cm^{-3} \left (\frac{c_\text{s}}{10~\kms}\right )^3 \left (\frac{\epsilon_\text{r}}{0.1}\right )^{-1} \left (\frac{m_\text{BH}}{10^8~\Msun}\right )^{-1}, 
\end{align}
To reach accretion rates 100 times the Eddington rate, 100 times higher densities are required.

Accounting for the energy that is radiated away as the gas spirals in through a subgrid accretion disc, gas accretion causes the BH to gain mass at the rate 
\begin{equation} \label{eq:bh_growth_rate}
    \dot{m}_\text{BH} = (1 - \epsilon_\text{r}) \dot{m}_\text{accr},
\end{equation}
and the BH particle mass to decrease by 
\begin{equation}
    \dot{m}_\text{BH,part} = - \epsilon_\text{r} \dot{m}_\text{accr}.
\end{equation}
If gas accretion results in $m_\text{BH} > m_\text{BH,part}$ then, in order to conserve mass-energy, mass is transferred from the surrounding gas particles to the BH particle mass. The mass and momentum taken from each SPH neighbour are proportional to that neighbour's contribution to the gas density at the location of the BH. Neighbours with mass less than half the initial baryonic particle mass are excluded. \citet{Bahe2022} find that this method of allowing the BH to `nibble' on its neighbours lets the BH subgrid mass track the BH particle mass more closely than if, as in \eagle, entire gas particles are stochastically swallowed by the BH. Nibbling also prevents the artificially large velocity kicks that occur due to momentum conservation when entire particles are swallowed. 

\subsubsection{Mergers}
BHs merge if three conditions are met. First, they must be separated by less than 3 gravitational softening lengths, $r< 3\epsilon$, where $r$ is their spatial separation. Second, their relative velocity must satisfy $\Delta v < [2G(m_\text{BH,part,1}+m_\text{BH,part,2})/r]^{1/2}$, where $m_{\text{BH,part,}i}$ is the particle mass of BH $i$. This second condition differs from \eagle, which used the \citet{Booth2009} criterion $\Delta v < \sqrt{Gm_\text{BH}/h}$, where $h$ is the SPH smoothing length of the more massive BH. As discussed in \citet{Bahe2022}, using $r$ instead of $h$ is more appropriate because merging is a gravitational process and should therefore not depend on the gas density. We use the particle BH mass rather than the subgrid BH mass used in \eagle\ and \citet{Bahe2022} because in the simulation the particle BH masses determine the particle orbits. Third, the less massive BH must be within the SPH smoothing length of the more massive BH, $r < h$. This last condition enables the code to find neighbouring BHs that are candidates for merging. 

The energy lost in gravitational waves during the merging process is assumed to be that for two non-spinning BHs with masses $m_\text{BH,1}$ and $m_\text{BH,2}$ and is subtracted from the rest mass energy of the descendant BH, i.e.\ from the sum of the rest mass energies of the merging BHs, for both the subgrid and the particle masses. It is taken from \citet{Barausse2012},
\begin{equation} \label{eq:GWloss}
    \frac{E_\text{GW}}{(m_\text{BH,1}+m_\text{BH,2})c^2} \approx 0.057191\nu + (0.037763 + 0.506)\nu^2
\end{equation}
where $\nu = m_\text{BH,1} m_\text{BH,2} / (m_\text{BH,1}+m_\text{BH,2})^2$ and terms $\mathcal{O}(\nu^3)$ are neglected. Here the $0.506\nu^2$ term accounts for the plunge, merger, and ringdown phases, while the other terms account for the inspiral phase. The energy loss is maximum for equal mass BHs, in which case it is $\approx 4.8$~per cent of the total BH rest mass energy. Note that this energy loss was not accounted for in \eagle.

When two BHs merge, we conserve linear momentum. Gravitational wave recoil is not accounted for. We remove the particle hosting the lower-mass BH from the simulation.

\subsubsection{Repositioning}
BHs are affected by dynamical friction, which causes their orbits to shrink towards the galaxy nucleus. There are several reasons why cosmological simulations will generally underestimate the effects of dynamical friction. First, except for very massive BHs in relatively high-resolution simulations, the BH particle masses are equal to or similar to the masses of other particles, whereas in reality they are many orders of magnitude higher than the masses of DM and atomic particles. Second, the contribution of small-impact parameter scatterings is underestimated due to gravitational softening \citep[e.g.][]{Tremmel2015}. Third, nuclear star clusters will dominate the dynamical friction if they are more massive than the BH that they host, as is expected for low-mass galaxies \citep[e.g.][]{Neumayer2020}, but such clusters are unresolved \citep[e.g.][]{Ma2021BHs_not_sinking}. 

To compensate for the inability to numerically resolve dynamical friction, most simulations either move the BH down the potential gradient or pin it to the centre of its halo. \citet{Bahe2022} demonstrated that implementing a BH repositioning scheme is crucial, because without any repositioning, BH growth is dramatically reduced, which in turn causes AGN feedback to become ineffective. Our treatment of BH repositioning follows the recommendations of \citet{Bahe2022}. At each time step, the BH is moved instantaneously to the location of the SPH neighbour within three gravitational softening lengths that has the lowest gravitational potential, provided it is lower than the potential at the current location of the BH. For the purpose of repositioning, we subtract the contribution of the BH to the potential in order to prevent the BH from becoming trapped in its own local potential well. The velocity of the BH is not altered when it is repositioned. However, in practice the BH will lose kinetic energy in the subhalo's rest frame as it repeatedly moves up the gravitational potential before being repositioned.    

Compared to \eagle, the most important differences in the repositioning scheme are that \eagle\ did not correct the potential for the contribution of the BH and that \eagle\ also allowed repositioning on to stellar and DM particles. The choice to only allow BHs to move to the locations of gas particles was motivated by the considerable computational expense associated with finding collisionless neighbours. However, this restriction can have undesired consequences. For example, in gas-poor galaxies the BH may not track the minimum of the gravitational potential as well as in gas-rich galaxies. BHs hosted by satellites that are being stripped of their gas may move out of their galaxy, though the fact that the BH velocities are unaltered during repositioning will generally cause them to return as long as the satellites retain some of their gas \citep{Bahe2022}. Nevertheless, satellites losing their BHs as a result of repositioning is a concern and for future simulations the computational expense associated with improved and more physical prescriptions for dynamical friction \citep[e.g.][]{Tremmel2015,Genina2024} may be considered worthwhile. However, a fully satisfactory solution to the important problem of modelling dynamical friction in cosmological simulations does not exist, particularly for BHs with $m_\text{BH}\lesssim m_\text{CDM}$.

\subsubsection{BH growth in the hybrid thermal/jet model}
As described in Section~\ref{sec:agn_fb}, a subset of our runs implement AGN feedback as a combination of thermal feedback and kinetic jet feedback. These runs include a subgrid accretion disc model that is used to track BH spins and accretion disc states, which together determine the nature and efficiency of the AGN feedback. The model, aspects of which are described in \citet{Husko2022,Husko2024,Husko2025SuperEddington} is motivated and detailed in \citet{Husko2025method}. Here we will only give a brief overview. 

The subgrid accretion disc model for the evolution of the BH spin vector accounts for gas accretion, BH mergers, jet-induced spindown, radiation torques, and Lense-Thirring torques. Equation~(\ref{eq:GWloss}) for the energy lost in gravitational waves during BH mergers is replaced by a spin-dependent version. The rate of growth of the BH through gas accretion, equation~(\ref{eq:bh_growth_rate}) for the thermal model, is replaced by
\begin{equation} \label{eq:dmbh/dt_hybrid}
    \dot{m}_\text{BH} = (1-\epsilon_\text{r}-\epsilon_\text{jet}-\epsilon_\text{wind})\epsilon_\text{accr} \dot{m}_\text{accr},
\end{equation}
where $\epsilon_\text{jet}$ is the jet efficiency, $\epsilon_\text{wind}$ is the accretion disc wind efficiency, and $\epsilon_\text{accr}$ is the accretion efficiency. The latter is the fraction of the accretion rate onto the disc, $\dot{m}_\text{accr}$, that is not ejected and actually ends up in the BH (if we ignore the loss of rest mass energy through feedback, i.e.\ the term in parentheses in equation~\ref{eq:dmbh/dt_hybrid}), which is assumed to be unity in the thermal model.

The type of accretion disc depends on the Eddington ratio (at the outer accretion disc scale, i.e.\ not reduced by the factor $\epsilon_\text{accr}$), 
\begin{equation}
    f_\text{Edd} \equiv \frac{\dot{m}_\text{accr}}{\dot{m}_\text{Edd}} .
\end{equation}
The radiative efficiency, $\epsilon_\text{r}$, that appears in the denominator of the Eddington accretion rate (equation~\ref{eq:Edd_rate}), is replaced by the expression from \citet{Novikov1973} for a relativistic Keplerian accretion disc. In this model the radiative efficiency increases with the BH spin to account for the decrease in the radius of the innermost stable circular orbit with the spin. Hence, the Eddington ratio decreases with BH spin. 

The accretion disc can be in one of three different states:
\begin{itemize}
    \item Thick disc, $f_\text{Edd}<0.01$. The disc is geometrically thick and optically thin. The accretion efficiency increases with the Eddington ratio and depends on the size of the thick component of the disc. The relation for this size includes a free parameter, $r_\text{tr,0}$, which corresponds to the dimensionless size, i.e.\ the size relative to the gravitational radius of the BH, for $f_\text{Edd}=0.01$ and is calibrated to the $z=0$ AGN bolometric luminosity function.\footnote{Although the thick disc tends to produce too low bolometric luminosities to affect the comparison to the data, the accretion efficiency affects the efficiency of feedback and hence the accretion rates and thus the fraction of BHs in the thin disc state that dominates the higher luminosities.} 
    \item Thin disc, $0.01<f_\text{Edd}<1$. The disc is geometrically thin and optically thick. The accretion efficiency $\epsilon_\text{accr} = 1$.
    \item Slim disc, $f_\text{Edd}>1$. The disc is slim, a state intermediate between the thick and thin discs. The accretion efficiency is set to $\epsilon_\text{accr} = 0.01$.
\end{itemize}
The fraction of BHs in the slim disc state is small for all BH masses and redshifts and is only non-negligible for intermediate-mass ($10^6 \lesssim m_\text{BH}/\Msun \lesssim 10^8$) BHs at high redshift ($z>3$). The thick (thin) disc fraction decreases (increases) with redshift. At a fixed redshift it is minimum (maximum) for $m_\text{BH}\sim 10^7~\Msun$. At $z=0$ the thick disc state is dominant at all masses, while for $z>2$ BHs with $m_\text{BH} > 10^6~\Msun$ are mostly in the thin disc state.

The seed BH mass is $m_\text{BH,seed}= 4\times 10^4\,\Msun$, $10^{4.75}\,\Msun \approx 5.6\times 10^4\,\Msun$, and $5\times 10^5\,\Msun$ for m5, m6, and m7 resolution, respectively. Together with the value for the pivot density for SN feedback (equation~\ref{eq:dT_SN}), its value is calibrated to reproduce the observed $z=0$ galaxy stellar mass function and size -- stellar mass relation (see~\S\ref{sec:calibration}).

\subsection{AGN feedback} \label{sec:agn_fb}

In the fiducial simulations, AGN feedback is implemented using a purely thermally-driven mode. Thermal AGN feedback is implemented using a method based on \citet{Booth2009}. To prevent numerical overcooling, the energy is not injected at every time step but in quantities sufficient to increase the temperature of a gas particle by a pre-defined amount specified by the parameter $\Delta T_\text{AGN}$. This is similar to the subgrid recipe for SN feedback (see the discussion in Section~\ref{sec:sn_fb}) and to the treatment of AGN feedback in \eagle, except that the prescription is not stochastic. Instead, as in \citet{Booth2009}, each BH saves its AGN feedback energy in a reservoir until there is sufficient energy for a single feedback event, i.e.\ to increase the temperature of at least one gas particle by\footnote{More than one particle can be heated if the BH accretes a sufficient amount of gas during a single time step. BHs are allowed to heat a maximum of 50 gas neighbours in a given time step; if this number is insufficient to use up the available AGN feedback energy, then $\Delta T_\text{AGN}$ is increased.} $\Delta T_\text{AGN}$. The energy is then injected into the SPH neighbour(s) nearest to the BH. This differs from the isotropic scheme used to distribute the SN energy (see \S\ref{sec:sn_fb}), but \citet{Chaikin2022} show that the use of the nearest particle results in a similar feedback efficiency and a similar degree of isotropy as the isotropic scheme. For AGN feedback we prefer to use the closest particle because the distance to SPH neighbours can become large in gas-poor galaxies.

The rate at which AGN feedback energy becomes available is given by
\begin{equation}
    \dot{E}_\text{AGN} = \epsilon_\text{f} \epsilon_\text{r} \dot{m}_\text{accr} c^2 ,
\end{equation}
while the observable AGN luminosity is 
\begin{equation} \label{eq:L_AGN_thermal}
    L_\text{AGN} = (1-\epsilon_\text{f})\epsilon_\text{r} \dot{m}_\text{accr} c^2,
 \end{equation}
where $\epsilon_\text{f}$ is the fraction of the intrinsic luminosity, $\epsilon_\text{r} \dot{m}_\text{accr} c^2$, that is assumed to couple to the ambient gas. Due to self-regulation, the value of $\epsilon_\text{f}$ strongly affects the masses of BHs in galaxies that are sufficiently massive for their growth to be regulated by AGN feedback \citep{Booth2009,Booth2010}. The value of $\epsilon_\text{f}$ is difficult to predict from first principles and was chosen to yield reasonable BH masses in high-mass galaxies at $z=0$. We use $\epsilon_\text{f}=0.1$ for our m7-resolution runs, but 0.05 for the m6 and m5 resolutions. 

The energy required for a single feedback event is
\begin{equation} \label{eq:E_AGN}
    \Delta E_\text{AGN,thermal} = \frac{m_\text{g} \kb \Delta T_\text{AGN}}{(\gamma-1) \mu m_\text{p}},
\end{equation}
where $\gamma$ is the ratio of specific heats, which we assume to be the value for a monatomic gas, $\gamma = 5/3$, and $\mu$ is the mean molecular weight, which we will assume to be 0.6, corresponding to an ionized gas of primordial composition.
The time interval between feedback events is 
\begin{align} \label{eq:Dt_AGN_Edd}
    \Delta t_\text{AGN} &= \Delta E / \dot{E}_\text{AGN} , \nonumber \\
    &= 3.8\times 10^4~\yr \, \left (\frac{\epsilon_\text{f}}{0.05}\right )^{-1} \left( \frac{m_\text{g}}{1.84\times 10^6~\Msun} \right) \nonumber \\
    &~~~~ \times \left (\frac{m_\text{BH}}{10^8~\Msun}\right )^{-1} \left (\frac{\Delta T_\text{AGN}}{10^9~\K}\right ) \left (\frac{\dot{m}_\text{accr}}{\dot{m}_\text{Edd}}\right )^{-1} . 
\end{align}

Unlike \eagle, we opt to scale $\Delta T_\text{AGN}$ with the BH mass,
\begin{equation} \label{eq:DeltaT_AGN}
    \Delta T_\text{AGN} = 10^9~\K \, \left ( \frac{m_\text{BH}}{10^8~\Msun}\right ). 
\end{equation}
This functional form makes $\Delta t_\text{AGN}$, and hence the time sampling of AGN feedback, independent of BH mass if the accretion rate is proportional to the Eddington rate, i.e.\ for a fixed Eddington ratio.
The use of a variable heating temperature improves the sampling of AGN feedback for low BH masses, which would otherwise have very long $\Delta t_\text{AGN}$, particularly given that the accretion rate is typically sub-Eddington. Because the change to a variable $\Delta T_\text{AGN}$ was made after the calibration of the subgrid model using a constant $\Delta T_\text{AGN} = 10^9~\K$ \citep{Chaikin2025calibration}, the normalization of equation~(\ref{eq:DeltaT_AGN}) was chosen to coincide with the value used in the calibration for the mass of BHs typical of galaxies transitioning from a state of active star formation to passive evolution. 

\begin{table*}
\centering
\caption{Fiducial values of subgrid parameters that vary between models. The columns list the simulation resolution; the type of AGN feedback; the pivot pressure for the pressure-dependent CCSN energy fraction, $P_\text{E,pivot}/\kb$ (equation~\ref{eq:f_E}); the pivot density for the density-dependent SN temperature increment, $n_\text{H,pivot}$ (equation~\ref{eq:dT_SN}); the BH seed mass, $m_\text{BH,seed}$; the minimum CCSN energy fraction, $f_\text{E,min}$; the  minimum SN temperature increment, $\Delta T_\text{SN,min}$; the maximum SN temperature increment, $\Delta T_\text{SN,max}$; the maximum AGN temperature increment, $\Delta T_\text{AGN,max}$; the FoF halo mass for BH seeding, $M_\text{FoF,seed}$; the efficiency for thermal AGN feedback, $\epsilon_\text{f}$; the pre-factor for the dust coagulation time (equation~\ref{eq:tau_co}), $f_\text{co}$.} 
\label{tbl:subgrid_pars}
    \begin{tabular}{llcccccccccc}
	\hline
	 Resolution & AGN & $P_\text{E,pivot}/\kb$ & $n_\text{H,pivot}$ & $m_\text{BH,seed}$ & $f_\text{E,min}$ & $\Delta T_\text{SN,min}$ & $\Delta T_\text{SN,max}$ & $\Delta T_\text{AGN,max}$ & $M_\text{FoF,seed}$ & $\epsilon_\text{f}$ & $f_\text{co}$\\ 
     && ($\K\,\cm^{-3}$) & ($\cm^{-3}$) & ($\Msun$) & & (K) & (K) & (K) & ($\Msun$) \\ 
	\hline
    m5 & Thermal & $1.5\times 10^4$ & 1.00 & $2\times 10^4$ & 0.8 & $10^{7.00}$ & $10^{8.0}$ & $10^{10}$ & $1\times 10^{10}$ & 0.05 & 1 \\
    m6 & Thermal & $1.0\times 10^4$ & 0.50 & $3\times 10^4$ & 0.3 & $10^{6.75}$ & $10^{8.0}$ & $10^{10}$ & $1\times 10^{10}$ & 0.05 & 1 \\
    m7 & Thermal & $8.0\times 10^3$ & 0.60 & $10^{5.5}$ & 0.1 & $10^{6.50}$ & $10^{7.5}$ & $10^{9.5}$ & $5\times 10^{10}$ & 0.10 & $10^{-0.5}$ \\
    m5 & Hybrid & $1.5\times 10^4$ & 1.20 & $4\times 10^4$ & 0.8 & $10^{7.00}$ & $10^{8.0}$ & $10^{10}$ & $1\times 10^{10}$ & 0.02 & 1 \\
    m6 & Hybrid & $1.0\times 10^4$ & 0.75 & $10^{4.75}$ & 0.3 & $10^{6.75}$ & $10^{8.0}$ & $10^{10}$ & $1\times 10^{10}$ & 0.02 & 1 \\
    m7 & Hybrid & $8.0\times 10^3$ & 1.50 & $5\times 10^{5}$ & 0.1 & $10^{6.50}$ & $10^{7.5}$ & $10^{9.5}$ & $5\times 10^{10}$ & 0.03 & $10^{-0.5}$ \\
	\hline
    \end{tabular}
\end{table*}

To prevent catastrophic overcooling at low BH masses, as well as poor sampling of AGN feedback for lower accretion rates at high BH masses, we limit $10^{6.5}~\K \le \Delta T_\text{AGN} \le \Delta T_\text{AGN,max}$, where $\Delta T_\text{AGN,max} = 10^{9.5}~\K$ in m7-resolution simulations and $10^{10}~\K$ for the m6 and m5 resolutions. We opted for a higher maximum value for the higher resolutions after some experimentation revealed that this was necessary to keep the AGN feedback efficient, because higher gas densities, which would otherwise imply greater radiative losses, are reached in the higher-resolution simulations. Note that the energy injected per feedback event scales as $m_\text{g}\Delta T_\text{AGN}$ and therefore decreases from m7 to m6 and from m6 to m5 resolution. Hence, the sampling of AGN feedback improves if the resolution increases.

The efficiency of AGN feedback on resolved scales, i.e.\ its impact on the resolved properties of galaxies and the gas around them, is mainly determined by the value of $\Delta T_\text{AGN}$. Higher values of $\Delta T_\text{AGN}$ generally imply stronger feedback, provided the value is still low enough for the feedback to be well sampled. While the value of $\epsilon_\text{f}$ had little effect on the emerging strength of AGN feedback in the \eagle\ simulations, which can be understood in terms of self-regulated AGN feedback\footnote{Self-regulated feedback results in outflows balancing inflows when averaged over suitable length and time-scales. A higher (lower) value of $\epsilon_\text{f}$ means that less (more) mass needs to be accreted by the BH to power the outflow rate that balances the galaxy-scale inflow rate. This then results in smaller (larger) BH masses, but a similar outflow rate.} \citep{Booth2010}, in \colibre\ it has a more significant effect. This is because $\Delta T_\text{AGN}$ now depends on the mass of the BH, which, in turn, is sensitive to $\epsilon_\text{f}$.

Because BHs are collisionless particles, their time steps are determined by the gravity solver and can be long compared to $\Delta t_\text{AGN}$. To ensure that the sampling of AGN feedback is not compromised, which could lead to runaway BH growth, we use $\Delta t_\text{AGN}$ as a time step limiter for BHs, but with a floor of 100~yr. In addition, we limit the BH time step to at most 4 times the minimum time step of the BH's SPH neighbours. Note that this does not prevent time-stepping issues altogether because within its time step, the BH cannot respond to changes in the physical conditions of the ambient gas. An improvement for future simulations would be to also use the hydro time step of the gas around the BH to wake up drifting BH particles. 

\subsubsection{AGN feedback in the hybrid thermal/jet model}
The hybrid model for AGN feedback is detailed and tested in \citet{Husko2025method}. Here we will only give a brief summary.

In the hybrid AGN feedback model kinetic jets are assumed to be present in each of the three accretion disc regimes discussed in the previous section (thick, thin, and slim). In the thick and slim disc states there is in addition an accretion disc wind, while for thin discs there is instead the radiation pressure driven wind that may originate from the accretion disc and/or from larger (but still unresolved) scales than the accretion disc wind. The radiatively driven and accretion disc winds are both injected as thermal energy using the method described for the thermal model but with different efficiencies. The energy from jets is injected in kinetic form in directions that are nearly parallel and anti-parallel to the BH spin axis. There are separate energy reservoirs for thermal and jet feedback. The dominant feedback mechanisms in the hybrid model are thermal feedback from the thin disc (important at all redshifts), jets from the thin disc (important at $z\gtrsim 1$) and jets from the thick disc (important at $z\lesssim 1$). 

The total AGN feedback power is 
\begin{equation}
    \dot{E}_\text{AGN} = (\epsilon_\text{f}\epsilon_\text{r} + \epsilon_\text{jet} + \epsilon_\text{wind}) \epsilon_\text{accr} \dot{m}_\text{accr} c^2 .
\end{equation}
The model assumes a magnetically-arrested disc (MAD; \citealt{Narayan2003}) with jets powered by the \citet{Blandford1977} mechanism. The efficiencies $\epsilon_\text{r}$, $\epsilon_\text{jet}$, and $\epsilon_\text{wind}$ are not free parameters. They are functions of the BH spin, which is tracked by the spin evolution model, and, except for $\epsilon_\text{r}$, the magnetization of the disc, where the magnetic fields are determined by the accretion rate and BH spin under the MAD assumption. For the thin disc the accretion disc wind efficiency $\epsilon_\text{wind}=0$, while for the thick and slim discs the coupling efficiency for radiative feedback $\epsilon_\text{f}=0$. For the thin disc $\epsilon_\text{f}$ is a free parameter that is calibrated to fit the observed $z=0$ relation between BH mass and galaxy stellar mass. We use $\epsilon_\text{f} = 0.02$ for the m5 and m6 resolutions, and $\epsilon_\text{f} = 0.03$ for the m7 resolution.

The observable AGN luminosity is 
\begin{equation}
    L_\text{AGN} = (1-\epsilon_\text{f}) \epsilon_\text{r}\epsilon_\text{accr} \dot{m}_\text{accr} c^2.
\end{equation}
For the thick and slim discs $\epsilon_\text{f}=0$ and this becomes $L_\text{AGN} = \epsilon_\text{r}\epsilon_\text{accr} \dot{m}_\text{accr} c^2$. For the thin disc, which dominates the AGN luminosity function for detectable luminosities, we have $\epsilon_\text{accr}=1$ and thus $L_\text{AGN} = (1-\epsilon_\text{f})\epsilon_\text{r} \dot{m}_\text{accr} c^2$. While the expression for the thin disc luminosity is the same as for the thermal AGN feedback model (equation~\ref{eq:L_AGN_thermal}), we emphasize that in the hybrid model the radiative efficiency depends on spin, while it is constant in the thermal model.

Jets are implemented by kicking two particles in opposite directions. The jet direction is limited to be within 7.5 degrees from the BH spin axis, but is otherwise random. Equations~(\ref{eq:E_AGN}) and (\ref{eq:DeltaT_AGN}) are, respectively, replaced by 
\begin{equation}
    \Delta E_\text{AGN,jet} = 2 \times \frac{1}{2} m_\text{g} v_\text{jet}^2,
\end{equation}
and
\begin{equation}
    v_\text{jet} = v_\text{jet,0} \left (\frac{m_\text{BH}}{10^9\,\Msun}\right )^{1/2},
\end{equation}
where $v_\text{jet,0} = 10^{4.5}\,\kms$ and the jet velocity is limited to the range $10^{2.5}~\kms \le v_\text{jet} \le v_\text{jet,0}$.
To ensure conservation of energy, the actual change in the velocity of a particle kicked into a jet differs from $v_\text{jet}$ if the particle is not at rest in the frame of the BH. 

There are three BH time-step limiters in addition to the one for thermal AGN feedback. The first is analogous to the one for thermal feedback, but applied to jet feedback. The second and third limiters ensure that the relative changes in both the magnitude and the direction of the spin are small. 

\begin{table*}
\centering
\caption{\colibre\ hydrodynamical simulations for which results are presented here. The simulations are grouped by type of AGN feedback and sorted by decreasing resolution and then by decreasing box size. The columns list the simulation identifier (where `L' is followed by the side length of the cubic simulation volume in comoving Mpc; `m' is followed by the rounded $\log_{10}$ of the mean initial particle mass in solar masses (for both baryons and DM); `h' indicates hybrid AGN feedback); the comoving box side length, $L$; the number of baryonic particles, $N_\text{b}$; the number of CDM particles, $N_\text{CDM}$; the initial mean baryonic particle mass, $m_\text{g}$; the mean CDM particle mass, $m_\text{CDM}$; the comoving gravitational softening length, $\epsilon_\text{com}$; the maximum proper gravitational softening length, $\epsilon_\text{prop}$; the redshift which the simulation had reached at the time of writing. Most combinations of resolution and AGN feedback type is also available in a L012.5 volume.}
\label{tbl:simulations}
    \begin{tabular}{lrrrlllll}
	\hline
	Identifier & $L$ & $N_\text{b}$ & $N_\text{CDM}$ & $m_\text{g}$ & $m_\text{CDM}$ & $\epsilon_\text{com}$ & $\epsilon_\text{prop}$ & $z$\\ 
	& (cMpc) &&& ($\Msun$) & ($\Msun$)  & (ckpc) & (pkpc) \\
	\hline
    \\
\multicolumn{9}{c}{Fiducial AGN feedback}\\
\hline
     	L100m5  & 100 & $3008^3$ & $4\times 3008^3$ & $2.30\times 10^5$ & $3.03\times 10^5$ & 0.9  & 0.35 & 6.0 \\
       	L050m5  & 50 & $1504^3$ & $4\times 1504^3$ & $2.30\times 10^5$ & $3.03\times 10^5$ & 0.9  & 0.35 & 1.1 \\
        L025m5  & 25 & $752^3$ & $4\times 752^3$ & $2.30\times 10^5$ & $3.03\times 10^5$ & 0.9  & 0.35 & 0\\
\hline
        L200m6  & 200 & $3008^3$ & $4\times 3008^3$ & $1.84\times 10^6$ & $2.42\times 10^6$ & 1.8  & 0.7 & 0\\
        L100m6  & 100 & $1504^3$ & $4\times 1504^3$ & $1.84\times 10^6$ & $2.42\times 10^6$ & 1.8  & 0.7 & 0\\
        L050m6  & 50 & $752^3$ & $4\times 752^3$ & $1.84\times 10^6$ & $2.42\times 10^6$ & 1.8  & 0.7 & 0\\
        L025m6  & 25 & $376^3$ & $4\times 376^3$ & $1.84\times 10^6$ & $2.42\times 10^6$ & 1.8  & 0.7 & 0\\
\hline
        L400m7  & 400 & $3008^3$ & $4\times 3008^3$ & $1.47\times 10^7$ & $1.94\times 10^7$ & 3.6  & 1.4 & 0\\
        L200m7  & 200 & $1504^3$ & $4\times 1504^3$ & $1.47\times 10^7$ & $1.94\times 10^7$ & 3.6  & 1.4 & 0\\
        L100m7  & 100 & $752^3$ & $4\times 752^3$ & $1.47\times 10^7$ & $1.94\times 10^7$ & 3.6  & 1.4 & 0\\
        L050m7  & 50 & $376^3$ & $4\times 376^3$ & $1.47\times 10^7$ & $1.94\times 10^7$ & 3.6  & 1.4 & 0\\
        L025m7  & 25 & $188^3$ & $4\times 188^3$ & $1.47\times 10^7$ & $1.94\times 10^7$ & 3.6  & 1.4 & 0\\
\hline
\\
\multicolumn{9}{c}{Hybrid AGN feedback}\\
\hline
        L025m5h & 25 & $752^3$ & $4\times 752^3$ & $2.30\times 10^5$ & $3.03\times 10^5$ & 0.9  & 0.35 & 0 \\
        L050m5h & 50 & $1504^3$ & $4\times 1504^3$ & $2.30\times 10^5$ & $3.03\times 10^5$ & 0.9  & 0.35 & 2.5 \\
\hline
        L100m6h & 100 & $1504^3$ & $4\times 1504^3$ & $1.84\times 10^6$ & $2.42\times 10^6$ & 1.8  & 0.7 & 0\\
        L050m6h & 50 & $752^3$ & $4\times 752^3$ & $1.84\times 10^6$ & $2.42\times 10^6$ & 1.8  & 0.7 & 0\\
        L025m6h & 25 & $376^3$ & $4\times 376^3$ & $1.84\times 10^6$ & $2.42\times 10^6$ & 1.8  & 0.7 & 0\\
\hline
        L200m7h & 200 & $1504^3$ & $4\times 1504^3$ & $1.47\times 10^7$ & $1.94\times 10^7$ & 3.6  & 1.4 & 0\\
        L100m7h & 100 & $752^3$ & $4\times 752^3$ & $1.47\times 10^7$ & $1.94\times 10^7$ & 3.6  & 1.4 & 0\\
        L050m7h & 50 & $376^3$ & $4\times 376^3$ & $1.47\times 10^7$ & $1.94\times 10^7$ & 3.6  & 1.4 & 0\\
        L025m7h & 25 & $188^3$ & $4\times 188^3$ & $1.47\times 10^7$ & $1.94\times 10^7$ & 3.6  & 1.4 & 0\\
	\hline
    \end{tabular}
\end{table*}

\subsection{Calibration of the feedback} \label{sec:calibration}

The calibration of the feedback is detailed in \citet{Chaikin2025calibration}. Briefly, we first created a Gaussian process emulator using a 2-level Latin hypercube of 48 L050m7 simulations, varying four subgrid parameters. The four subgrid parameters that are sampled in the hypercube are the BH seed mass, $m_\text{BH,seed}$, which affects the onset of AGN feedback, and the following three stellar feedback parameters: the fraction of energy injected as low-velocity ($50~\kms$) kinetic feedback, $f_\text{kin}$; the gas density for which the temperature of gas that is directly heated by SNe is increased by $\Delta T_\text{SN} = 10^{6.5}~\K$ (see equation~\ref{eq:dT_SN}), $n_\text{H,pivot}$; and the thermal stellar birth pressure for which the injected CCSN energy is halfway between the minimum and maximum allowed values (see equation~\ref{eq:f_E}), $P_\text{E,pivot}$. 

The emulator was then used to fit to the observed \citet{Driver2022} $z=0$ galaxy stellar mass function (SMF) from the GAMA survey and the $z < 0.05$ galaxy size -- stellar mass relation (SMR) from \citet{Hardwick2022}, where size refers to the projected half-mass radius. We chose these particular data sets because they are recent and span the required mass range. An advantage of the \citet{Driver2022} mass function, which they show is in good agreement with other measurements, is that it is provided precisely at $z=0$. While most observational studies report half-light radii, \citet{Hardwick2022} present half-mass radii, which allow for a direct comparison with the simulations without having to produce virtual observations accounting for dust attenuation. 
These observables were both fit at $z=0$ over the galaxy stellar mass range $9.0 < \log_{10} M_*/\Msun < 11.3$, where the lower and upper limits are motivated by the resolution and box size, respectively. These mass limits are well matched to the mass range spanned by the size measurements of \citet{Hardwick2022}, $9.11 < \log_{10} M_*/\Msun < 11.26$, while the \citet{Driver2022} SMF extends to both lower and higher masses. Equal weight was assigned to the SMF and SMR. The rounded best-fitting subgrid parameter values were adopted for the fiducial m7 model. 

For the higher resolutions, the parameters $n_\text{H,pivot}$, $P_\text{E,pivot}$, and $m_\text{BH,seed}$, but not $f_\text{kin}$, were then slightly adjusted by trial and error to achieve convergence for these observables in small simulation volumes (mainly L025m6 and L012.5m5). Some adjustments were also made to $\Delta T_\text{SN,min}$, $\Delta T_\text{SN,max}$, $\Delta T_\text{AGN,max}$ and $f_\text{E,min}$, which tend to increase with increasing resolution. In addition, the m5 and m6 resolutions use a five-times smaller value of $M_\text{FoF,seed}$ than m7. Finally, for m7 resolution we reduced the dust coagulation time to improve convergence for the dust grain sizes. Although the higher resolution simulations can be used to probe lower galaxy masses, we did not consider observations of galaxy sizes below the lowest mass covered by the \citet{Hardwick2022} data, $\log_{10} M_*/\Msun = 9.11$.
Note that because we recalibrate after changing the resolution, we are aiming for `weak convergence' in the language of \citet{Schaye2015}. 

See Table~\ref{tbl:subgrid_pars} for the values of the parameters that change with resolution. Only three of these parameters, $P_\text{E,pivot}$, $n_\text{H,pivot}$, and $m_\text{BH,seed}$, are critically important for the calibration to the $z=0$ SMF and SMR. Because there are degeneracies between some of the parameters in the table, this choice of critical parameters is not unique. Because the degeneracies are only partial, adjustments in some of the remaining parameters can further improve the fits. \citet{Chaikin2025calibration} shows the effects that changing the values of individual parameters has on the calibration observables for m7 resolution. 

\citet{Chaikin2025calibration} demonstrate that simpler versions of the model for CCSN feedback than our fiducial prescription cannot simultaneously fit the SMF and the SMR. A constant energy (i.e.\ constant $f_\text{E}$), purely thermal model with fixed $\Delta T_\text{SN} = 10^{7.5}\,\K$ (the fiducial value of $\Delta T_\text{SN}$ from \citet{DallaVecchia2012} used in \eagle) cannot fit the SMF for any value of $f_\text{E}$. Injecting a fraction of the energy in kinetic form, with $f_\text{kin}$ a free parameter and using a target velocity of $50~\kms$, can fit either the SMF or the SMR, but not both simultaneously. Adding a density-dependent $\Delta T_\text{SN}$ improves the fits, but good fits are only obtained when the available energy, i.e.\ $f_\text{E}$, is made to vary, which we choose to do as a function of the thermal pressure (see \S\ref{sec:sn_fb}).

The efficiency of the thermal AGN feedback, $\epsilon_\text{f}$, which is the fraction of the energy radiated by the BH accretion disc that is used to heat the ambient gas, is calibrated to give a reasonable match to the observed relation between BH and stellar masses for massive galaxies in the local Universe (see \S\ref{sec:bh_masses}). Previous simulations using the same prescription and similar resolution found that values of order 10 per cent yield acceptable results \citep[e.g.][]{Booth2009,Schaye2015}. We therefore set $\epsilon_\text{f}=0.1$ and found no reason to change it for m7 resolution. Its value was halved for the higher-resolution simulation to improve convergence. 

Similarly to when we change the resolution, when we switch to the hybrid AGN feedback model we use the best-fitting values of the thermal models as the starting point. Two of these parameters, $n_\text{H,pivot}$ and $m_\text{BH,seed}$, as well as some parameters specific to the hybrid AGN model, were adjusted by trial and error using small-volume test simulations. The calibration of the simulations using hybrid AGN feedback is discussed in detail in \citet{Husko2025method}.

\begin{figure*}
    \centering
    \includegraphics[width=\linewidth]{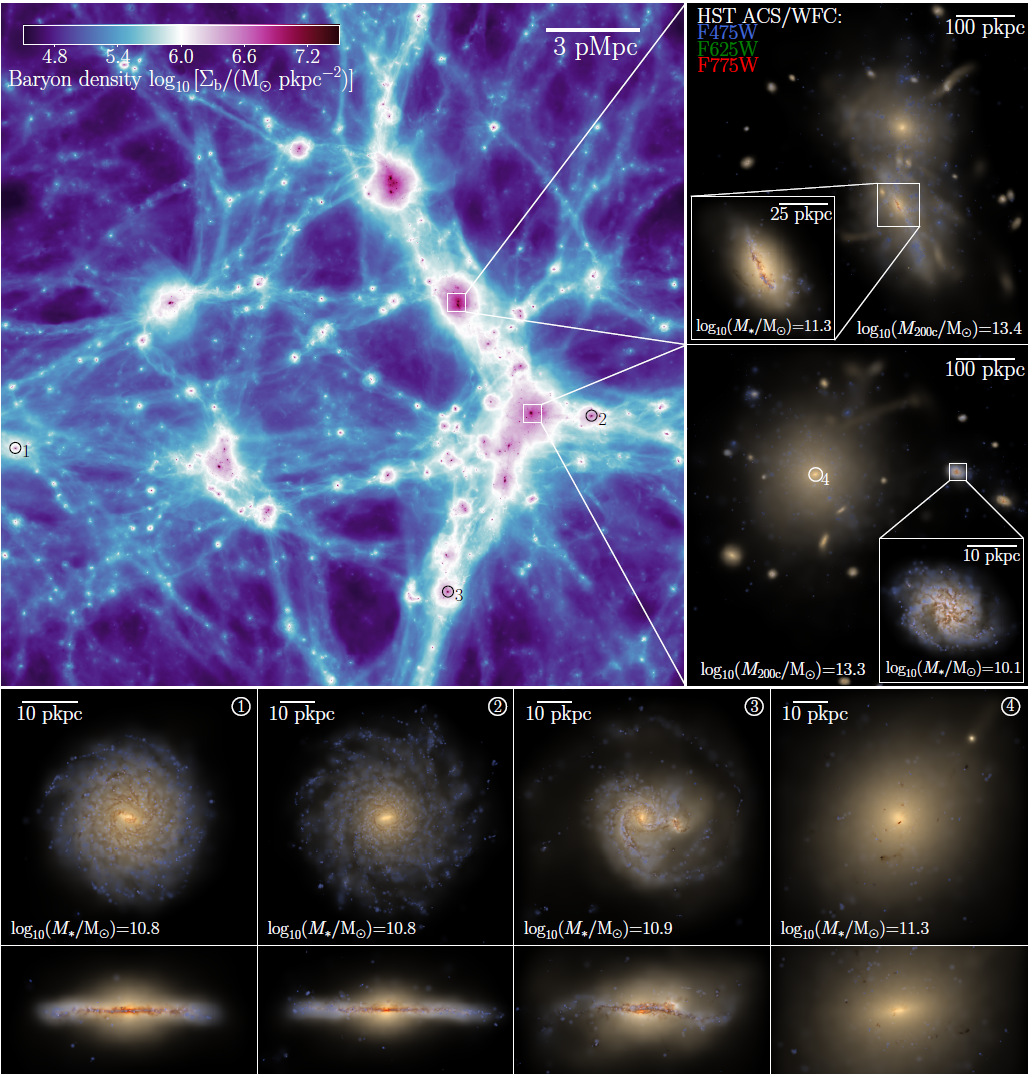}
    \caption{Visual impression of the dynamic range in the high-resolution \colibre\ simulation L025m5 at $z=0.1$. The top left panel shows a projection of the entire simulation with the colour encoding baryon surface density. The other panels zoom into different regions and show the stellar light in \textit{HST} colours  accounting for attenuation by dust. The top right and middle right panels zoom into 800~pkpc cubic regions of groups of mass $M_\text{200c} = 10^{13.4}\,\Msun$ and $10^{13.3}\,\Msun$, respectively. The insets in these two panels show 75~pkpc $\times$ 75~pkpc and 30~pkpc $\times$ 30~pkpc zoom-in images of galaxies with stellar mass $M_* = 10^{11.3}\,\Msun$ and $10^{10.1}\,\Msun$, respectively. The orientations of these two galaxies are the same as in the other images of the same regions. The two bottom rows show face- and edge-on views of four galaxies with stellar masses in the range $10^{10.8} \le M_*/\Msun \le 10^{11.3}$. These galaxies are numbered and their locations are indicated in the top left or middle right panels. Galaxy zooms show regions of 75~pkpc $\times$ 75~pkpc for face-on orientations and 75~pkpc $\times$ 30~pkpc for edge-on orientations, except for the zoom of galaxy 1 on the bottom left, which is 50~pkpc $\times$ 50~pkpc and 50~pkpc $\times$ 20~pkpc for the face- and edge-on orientations, respectively. \colibre\ simultaneously models the large-scale cosmic web and the internal structure of galaxies.}
    \label{fig:slice_L025m5}
\end{figure*}

\begin{figure*} 
    \centering
    \includegraphics[width=\linewidth]{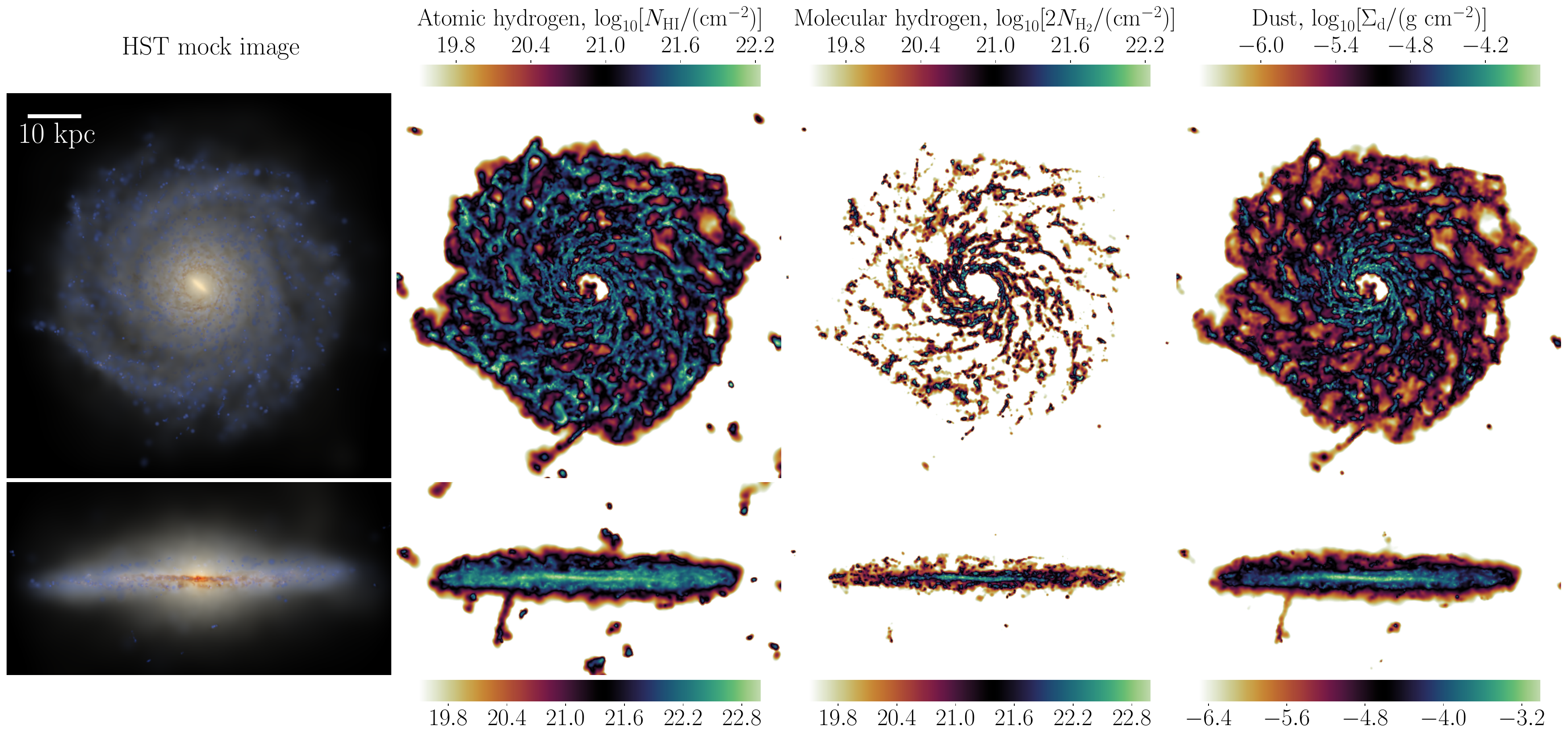}
    \caption{Face-on (top) and edge-on (bottom) view of a $z=0$ disc galaxy of mass $M_* = 8\times 10^{10}\,\Msun$ and $\text{SFR} = 2.7~\Msun\,\yr^{-1}$ in the L025m5 simulation. From left to right the columns show stellar light, \ion{H}{i} column density, H$_2$ column density times 2, and dust surface density. The image sizes are 75~kpc $\times$ 75~kpc for face-on and 75~kpc $\times$ 30~kpc for edge-on orientations. Beyond the gas-poor centre, molecular hydrogen is more concentrated towards the centre and in the spiral arms than atomic hydrogen, while the dust distribution is intermediate between that of atomic and molecular hydrogen (compare, for example, the green in the H$_2$ map with the red in the dust map).}
    \label{fig:disc}
\end{figure*}

\section{The simulations} \label{sec:sims}

The \colibre\ simulation suite includes two types of models, simulations with thermally-driven AGN feedback and simulations with hybrid, i.e.\ thermal plus jet, AGN feedback. There are three resolutions, corresponding to mean particle masses (both for baryons and CDM) of $\sim 10^5$, $10^6$, and $10^7~\Msun$. Comoving box sizes start at 25~cMpc and increase by powers of 2. The maximum box sizes are 100, 200, and 400~cMpc for the high-, intermediate-, and low-resolution runs, respectively. All but the 50 and 100~cMpc high-resolution simulations have already reached $z=0$.

Table~\ref{tbl:simulations} lists all hydrodynamic simulation runs. The run identifier indicates the box size in cMpc (prefix `L'), the rounded log base 10 of the initial mean particle mass in solar masses (prefix `m') and the type of AGN feedback (default is thermal, `h' indicates hybrid). For example, run L200m6 is a box of 200~cMpc on a side with a particle mass $\sim 10^6\,\Msun$ and using purely thermally-driven AGN feedback.

Due to their computational expense\footnote{Runs L200m6 and L400m7 took, respectively, 72 and 34 million core hours on the Cosma8 computer in Durham, UK. Both simulations were run on 160 nodes that each have 128 cores.}, the largest runs, which use $5\times 3008^3 \approx 1.4\times 10^{11}$ particles, are run only with thermal AGN feedback. The largest hybrid AGN feedback runs use $5\times 1504^3$ particles, i.e.\ 8 times fewer. All runs start with 4 times more CDM than baryonic particles. 

The gravitational softening length is the same for baryons and CDM. It is initially held constant in comoving units at $\epsilon_\text{com} = 0.9$, 1.8, and 3.6~ckpc for m5, m6, and m7, respectively. This corresponds to $\approx 1/37$ and $\approx 1/23$ times the mean interparticle spacing of baryons and CDM, respectively. At redshift $z<1.57$ the softening length is kept constant in proper units, which yields a maximum physical softening length of $\epsilon_\text{prop} = 0.35$, 0.7, and 1.4~pkpc for m5, m6, and m7, respectively. 

For each hydrodynamic simulation, there is a corresponding DMO simulation for which the initial conditions were created by converting the baryonic particles into CDM particles. Hence, the DMO and hydrodynamic simulations use the same total number of particles and identical initial conditions. This allows for more accurate comparisons between hydrodynamic and DMO simulations than the usual approach of simply omitting the baryonic component and scaling up the CDM particle masses accordingly because we maintain the separate transfer functions and do not change the particle masses. 

\begin{figure*}
    \centering
    \includegraphics[width=\linewidth]{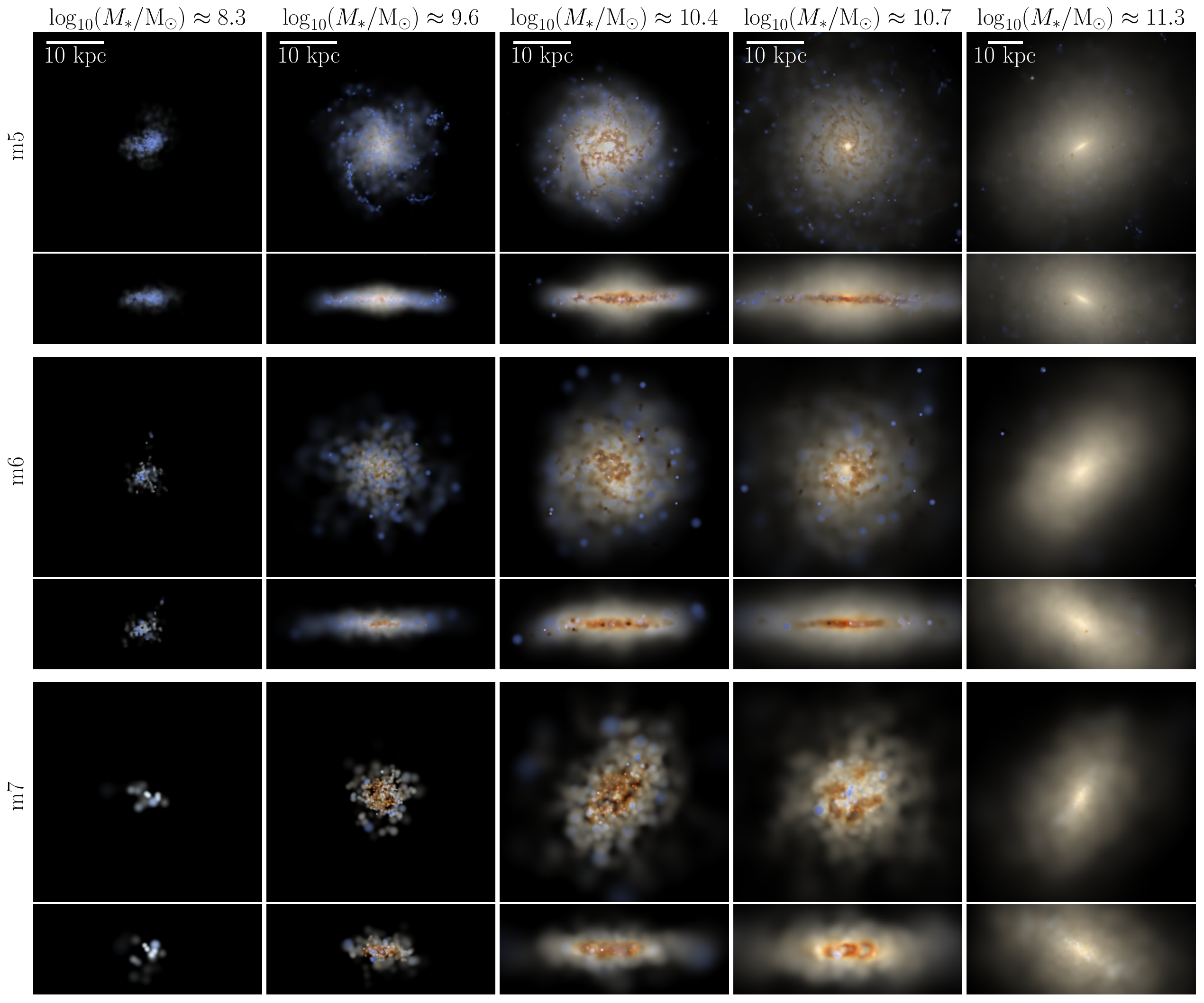}
    \caption{Comparison of galaxy morphologies across different \colibre\ resolutions. Each column shows face-on and edge-on images of the same galaxy in the L025m5 (top), L025m6 (middle), and L025m7 (bottom) simulations at $z=0$. The different columns show different galaxies with stellar mass increasing from left to right. The stellar masses for m6 resolution are indicated above each column. These can differ by $\pm 0.1$~dex for the other resolutions. The images show stellar light in \textit{HST} colours and account for attenuation by dust. They are 50~kpc across except for the last column, which is 75~kpc across. Higher-resolution simulations reveal more detail and predict thinner discs. Except for the lowest mass galaxy, the galaxy shape can be inferred at all resolutions.}
    \label{fig:res}
\end{figure*}

\begin{figure*}
    \centering
    \includegraphics[width=\linewidth]{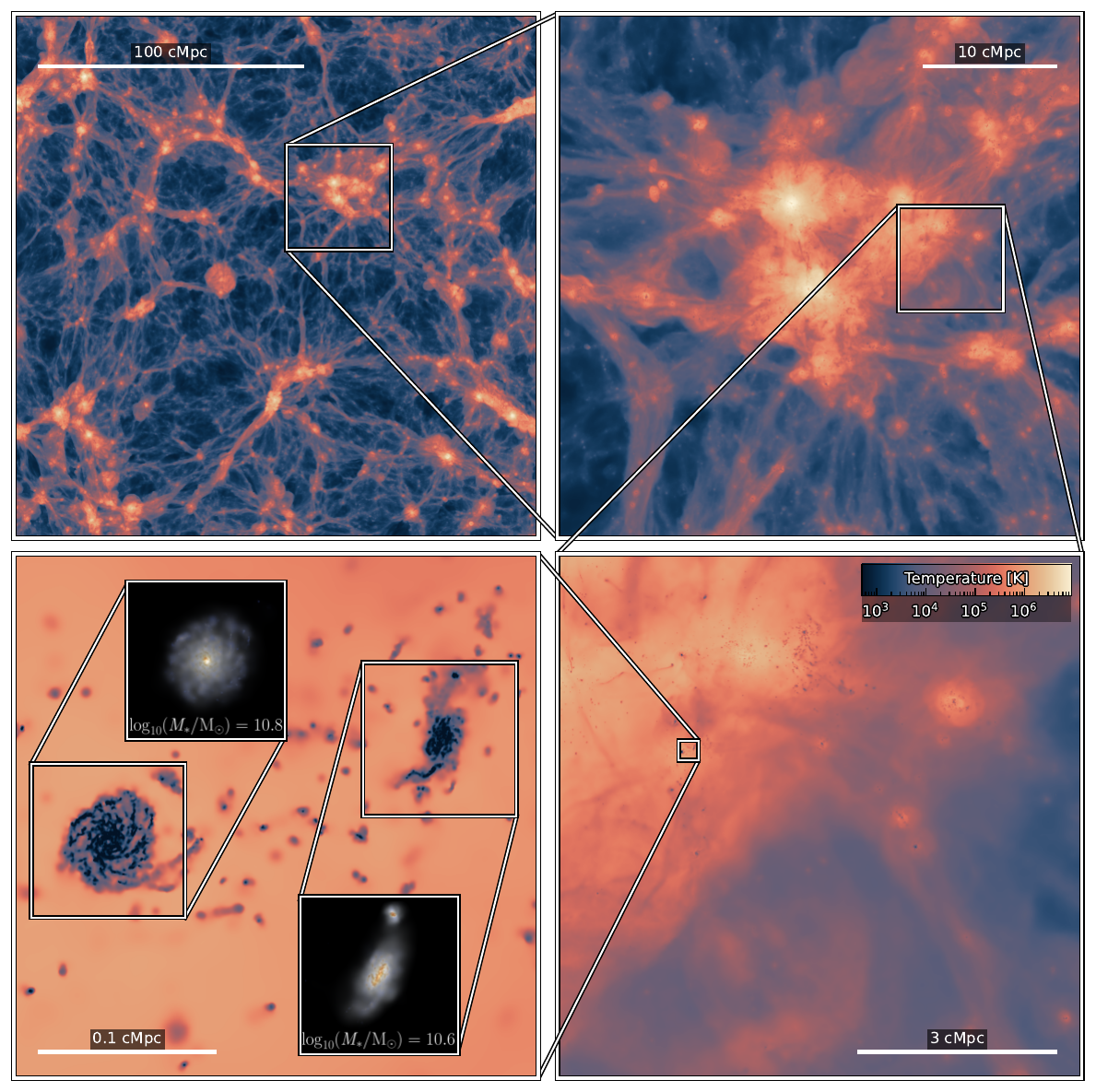}
    \caption{Visual impression of the dynamic range in the COLIBRE simulation L200m6 at $z=0.1$. The top left panel shows a 25~cMpc thick slice of the temperature (mass-weighted mean of $\log_{10}T$) of the gas. The three other main panels show consecutive zooms into the outskirts of a galaxy cluster, keeping the same 25~cMpc depth. The two smaller images in the bottom left panel show the stellar light in \textit{Euclid} colours and account for attenuation by dust. \colibre\ simultaneously models the large-scale cosmic web, the multiphase gas in the intracluster and interstellar media, and the internal structure of galaxies.}
    \label{fig:slice_L200m6}
\end{figure*}

\begin{figure}
    \centering
    \includegraphics[width=\linewidth]{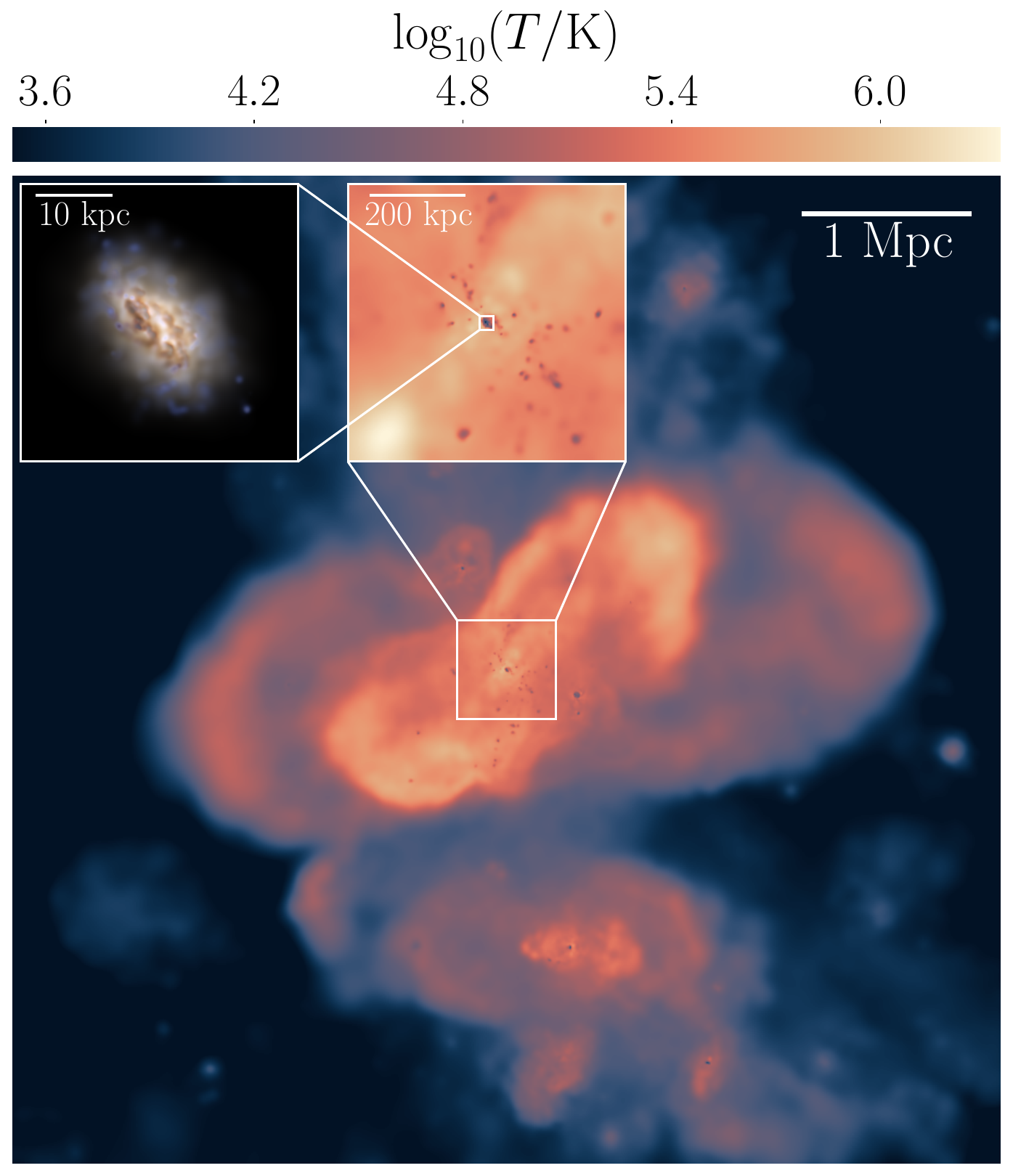}
    \caption{Visual impression of the effect of AGN jets on the gas around two star forming galaxies. The main image shows the gas temperature (mass-weighted mean of $\log_{10}T$) in a 6~Mpc cubic region in the hybrid AGN feedback L100m6h simulation at $z=0$. The image is centred on a halo of mass $M_\text{200c} \approx 10^{11.9}\,\Msun$ with a central galaxy of stellar mass $10^{10.4}\,\Msun$, star formation rate $1.6~\Msun\,\yr^{-1}$, and nuclear BH mass $10^{7.0}\,\Msun$. The other panels show consecutive zooms of the temperature in a 0.6~Mpc cubic region and of stellar light in \textit{HST} colours accounting for attenuation by dust in a 30~kpc cubic region. The galaxy below the one in the main panel has a smaller mass ($M_\text{200c} \approx 10^{11.4}\,\Msun$, $M_*\approx 10^{9.9}\,\Msun$, $\text{SFR}\approx 0.5~\Msun\,\yr^{-1}$, $m_\text{BH}\approx 10^{6.0}\,\Msun$). In 3D the two galaxies are separated by 2.8~Mpc. Shock fronts driven by AGN jet activity are clearly visible in the main panel, while the jets themselves are also visible in the first zoom-in panel.}
    \label{fig:jet_image}
\end{figure}

\section{Visual impressions} \label{sec:firstlook}
In this section we present images that give visual impressions of the \colibre\ simulations.

The top left panel of Fig.~\ref{fig:slice_L025m5} shows a projection of the L025m5 simulation at $z=0.1$. The colour coding indicates the baryon surface density. The image shows the cosmic web with haloes distributed mostly along filaments.
All other panels show stellar light accounting for attenuation by dust, based on the predicted spatial distribution, chemical composition, and sizes of the dust grains.
The insets on the right show successive zooms into two galaxy groups. 
The bottom two rows show face- and edge-on views of four galaxies with mass $M_*\sim 10^{11}\,\Msun$. These galaxies are numbered and their locations are indicated in the main panel. Galaxies 1--3 are spirals with mass $M_*\approx 10^{10.8}\,\Msun$. Galaxies 1 and 2 are relatively isolated, but galaxy 3 is in the process of merging with a companion galaxy. Galaxy 4 is the most massive object in the group shown in the middle right panel. It is a gas-poor galaxy with mass $M_*\approx 10^{11.3}\,\Msun$ and an ellipsoidal shape. For the disc galaxies dust lanes are clearly visible in the edge-on views as well as in the spiral arms for face-on views. Fig.~\ref{fig:slice_L025m5} illustrates that the L025m5 simulation resolves the detailed internal structure of realistic-looking galaxies in a variety of cosmic environments. 

The procedure for making stellar-light images is detailed in \citet{Husko2025coligal}. Briefly, luminosities are generated using the package Flexible Stellar Population Synthesis \citep[FSPS;][]{Conroy2009, Conroy2010} based on the stellar particle ages, metallicities, and redshift. The sampling of stars with ages smaller than 10~Myr is improved by using star-forming gas particles. A 3D luminosity grid is then combined with a 3D grid of the dust distribution. Dust attenuation accounts for absorption and scattering, and for the size and chemical composition of the dust grains. An RGB image is created following \citet{Lupton2004} for the \textit{HST} ACS/WFC F475W (blue), F625W (green) and F775W (red) bands. 

Fig.~\ref{fig:disc} compares the spatial distributions of different components of the ISM in face- and edge-on views of a spiral galaxy at $z=0$ in the L025m5 simulation. The galaxy has mass $M_* = 8\times 10^{10}\,\Msun$ and $\text{SFR} = 2.7~\Msun\,\yr^{-1}$. From left to right, the different panels show stellar light, \ion{H}{i} column density, H$_2$ column density times 2, and dust surface density. The central few kpc host a stellar bar and are poor in gas and dust. This nuclear region is surrounded by a ring of dusty, largely molecular gas. Compared with the atomic gas, the molecular gas and, to a lesser extent, the dust are more concentrated toward the centre and the spiral arms. 

Fig.~\ref{fig:res} illustrates the effect of simulation resolution on galaxy images. We compare \textit{HST} colour images of the stellar light of the corresponding $z=0$ galaxies in the L025m5 (top), L025m6 (middle) and L025m7 (bottom) simulations. From left to right, the different columns show galaxies with stellar mass $\log_{10} M_*/\Msun\approx 8.5$, 9.5, 10.3, 10.8, and 11.3 in L025m6, with each column showing the same galaxy matched across three resolutions. These values vary by $\approx 0.1$~dex between the different resolutions. 
Such small differences are consistent with the typical amplitudes of random deviations found when repeating cosmological simulations that, like \colibre, include stochastic subgrid models \citep{Borrow2023}. Except for the lowest-mass galaxy (the two lowest-mass galaxies), resolution m6 (m7) suffices to classify the galaxy morphology, determine its orientation, and measure spatially integrated properties such as the mass and SFR. However, higher-resolution simulations reveal increasingly detailed substructure and produce thinner discs. 

Fig.~\ref{fig:slice_L200m6} gives a visual impression of the L200m6 simulation. The four main panels show 10~Mpc thick slices, with the hue (brightness) indicating the temperature (gas surface density). Clockwise from top left the different same-size panels show images of 200, 30, 3, and 0.3~Mpc on a side, successively zooming into the outskirts of a galaxy cluster. The two small images in the bottom left panel show stellar light in \textit{Euclid} bands, accounting for dust attenuation. As the figure shows, L200m6 has a volume that is sufficiently large to include many galaxy clusters and to capture the large-scale distribution of galaxies and clusters, while having sufficient resolution to spatially resolve the structure of galaxies.

Fig.~\ref{fig:jet_image} provides a visual impression of the effect of AGN jet activity on the gas around a star-forming galaxy of stellar mass $10^{10.4}\,\Msun$ in the hybrid AGN feedback simulation L100m6h at $z=0$. The main image and the first zoom-in show the gas temperature, while the final zoom-in shows stellar light in \textit{HST} bands, accounting for dust attenuation. We can clearly see the shock fronts resulting from two different episodes of AGN activity. We have verified, using jet tracers, that these bubbles are driven mainly by jet rather than thermal AGN feedback. Below the image centre, we can see a lower-mass galaxy ($M_*\approx 10^{9.9}\,\Msun$) whose circumgalactic medium has also been shaped by past jet activity. We refrain from comparing with the same region in the thermal AGN feedback simulation because the stochastic nature of AGN feedback episodes implies that such comparisons are of limited use for individual galaxies. However, we note that the isotropic thermal AGN feedback used in the fiducial simulations also frequently results in bipolar outflows because galactic winds tend to escape perpendicular to the galaxy disc along the path of least resistance, though the outflows tend to be less extended and less collimated than for jet AGN feedback (see fig.~12 of \citealt{Husko2025method}).   

More images, also for L400m7, as well as videos and interactive visualisations, can be found on the \colibre\ website\footnote{\url{https://colibre-simulations.org/}}.

\section{Phase diagrams and stellar birth properties} \label{sec:phys_cond}
In this section we provide some physical intuition for the behaviour, outcome, and limitations of the simulations. In Section~\ref{sec:phase_diagrams} we discuss the distribution and properties of gas at $z=0$ in temperature/pressure -- density space, limited to the L200m6 simulation for brevity. In Section~\ref{sec:birth_properties} we explore the physical properties of the gas from which stars are born and in which CCSNe explode, and their dependence on numerical resolution.

\begin{figure*}
    \centering
    \includegraphics[width=0.45\linewidth]{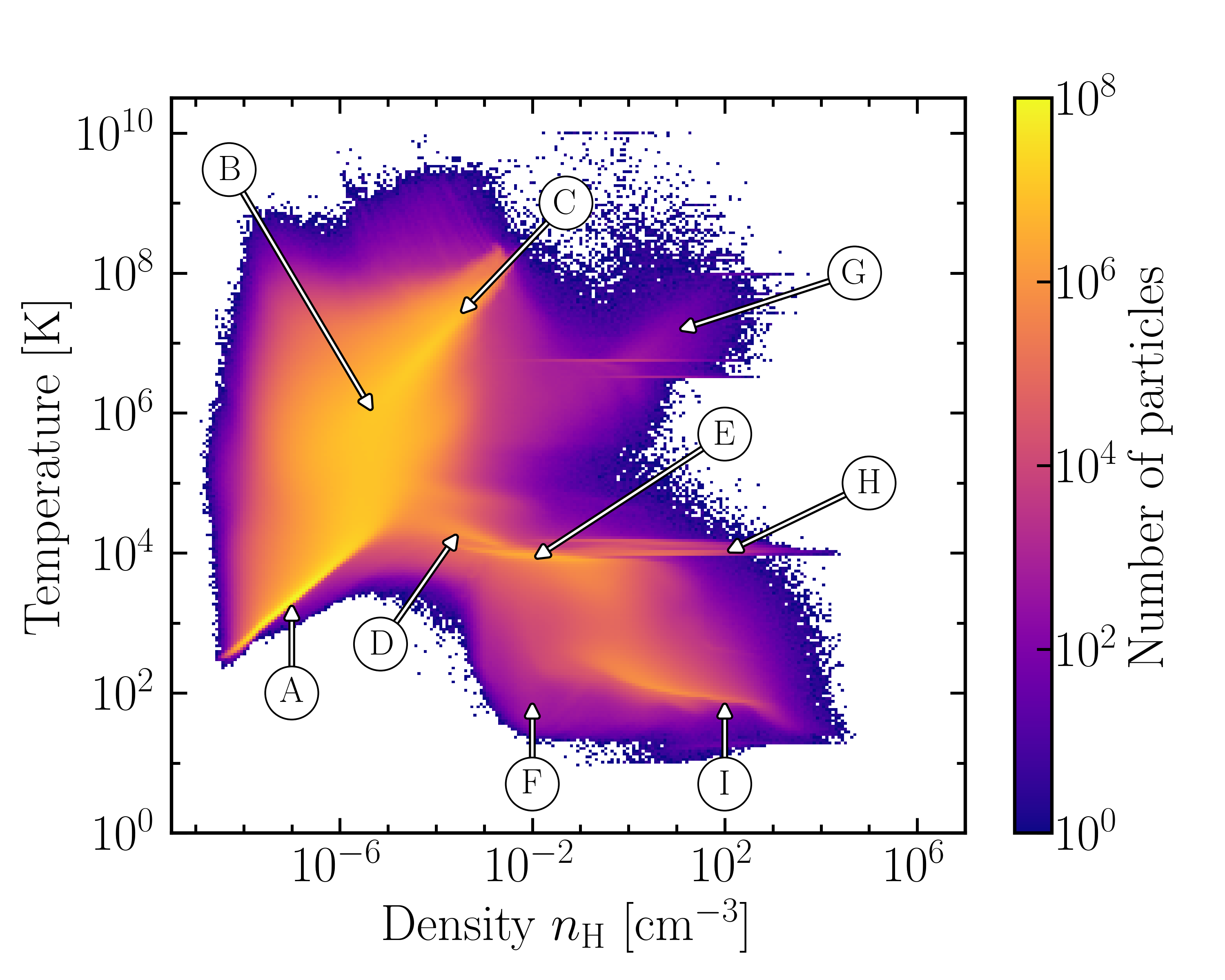}
    \includegraphics[width=0.45\linewidth]{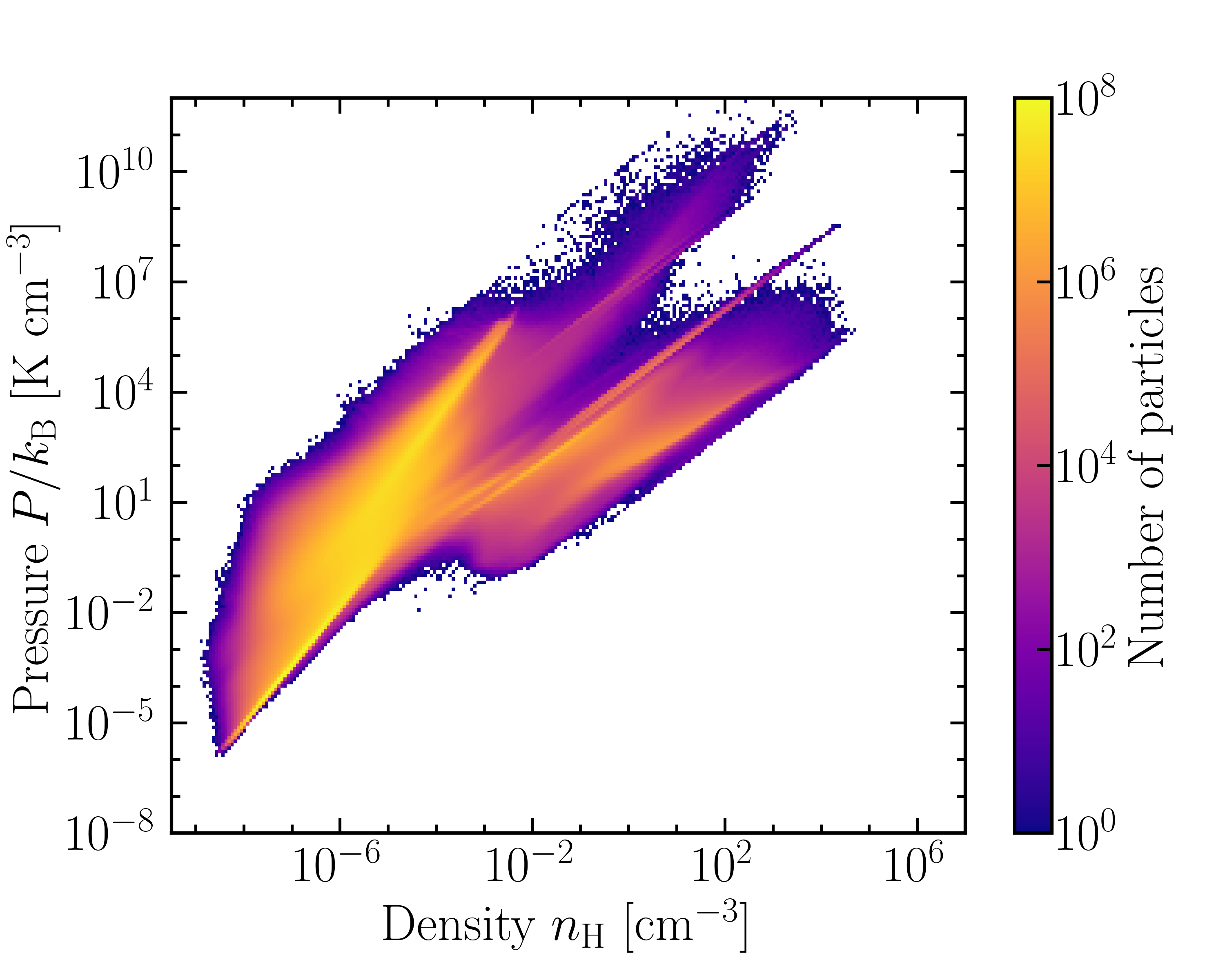}
    \caption{Distribution of gas particles in the L200m6  simulation at $z=0$ in the temperature -- density (left panel) and the thermal pressure -- density (right panel) planes. The colour coding shows the fraction of the gas particles in the pixel on a logarithmic scale. The labels in the left panel are used to facilitate identification of particular regions discussed in the text. The gas in \colibre\ spans a wide range of physical conditions, with multiple gas phases overlapping in pressure.}
    \label{fig:gas_distr_phys_props}
\end{figure*}

\subsection{Phase diagrams} \label{sec:phase_diagrams}

Fig.~\ref{fig:gas_distr_phys_props} shows the distribution of gas particles in temperature -- density (left panel) and pressure -- density (right panel) space. For densities $n_\text{H}\gtrsim 10^{-1}~\cm^{-3}$, the left panel shows concentrations of gas corresponding to the three main phases of the ISM, i.e.\ cold ($T\lesssim 10^2~\K$), warm ($T\sim 10^4~\K$), and hot ($T\gg 10^4~\K$) gas. The right panel shows that these different phases of the ISM can have the same pressures. Note that in the previous generation of simulations of representative volumes, such as \eagle, there is no cold phase and the temperature of the warm phase becomes unrealistically high because a (density-dependent) entropy floor is imposed on the ISM.

\begin{figure*}
    \centering
    \includegraphics[width=0.45\linewidth]{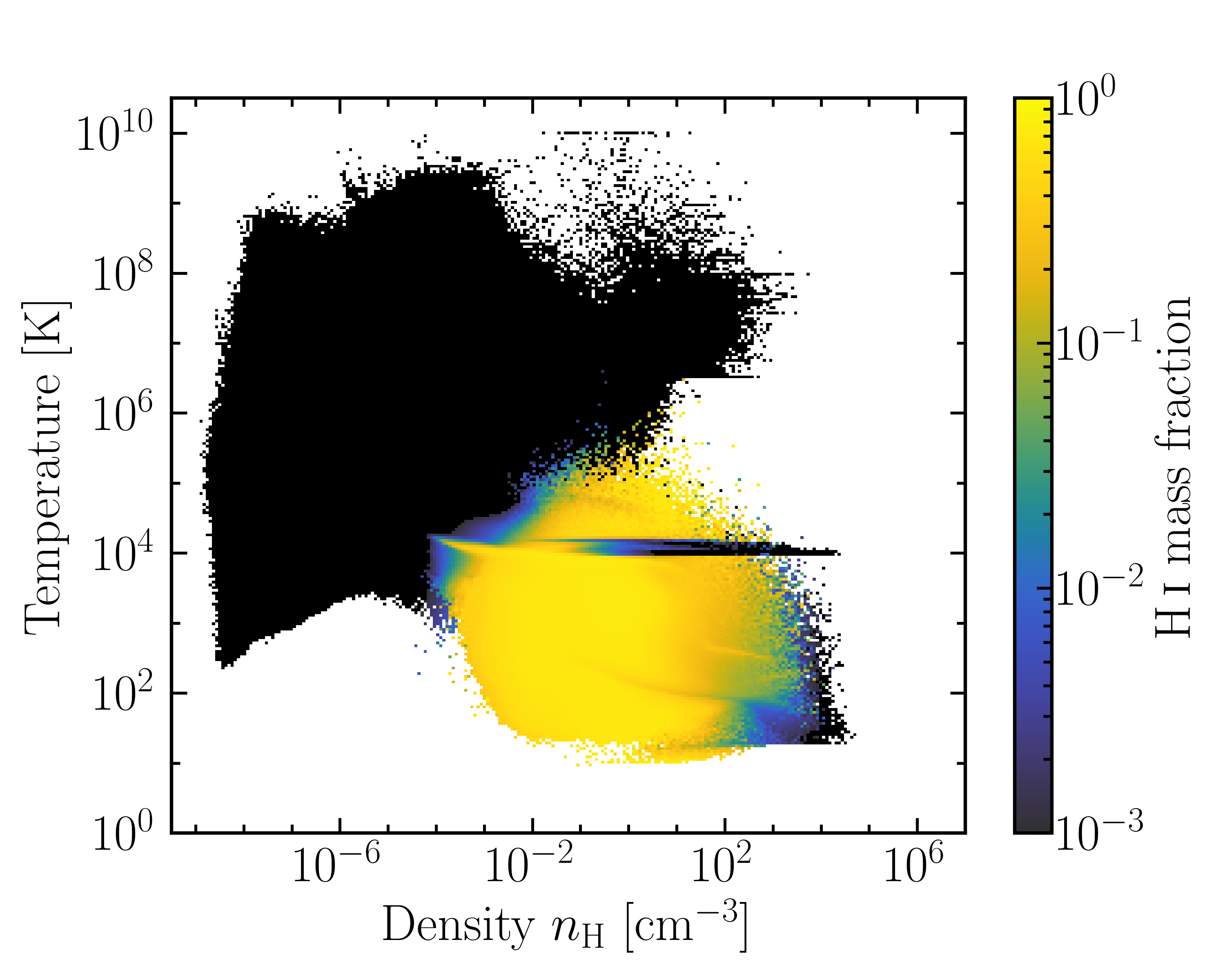} \hspace{0.01\linewidth}
    \includegraphics[width=0.45\linewidth]{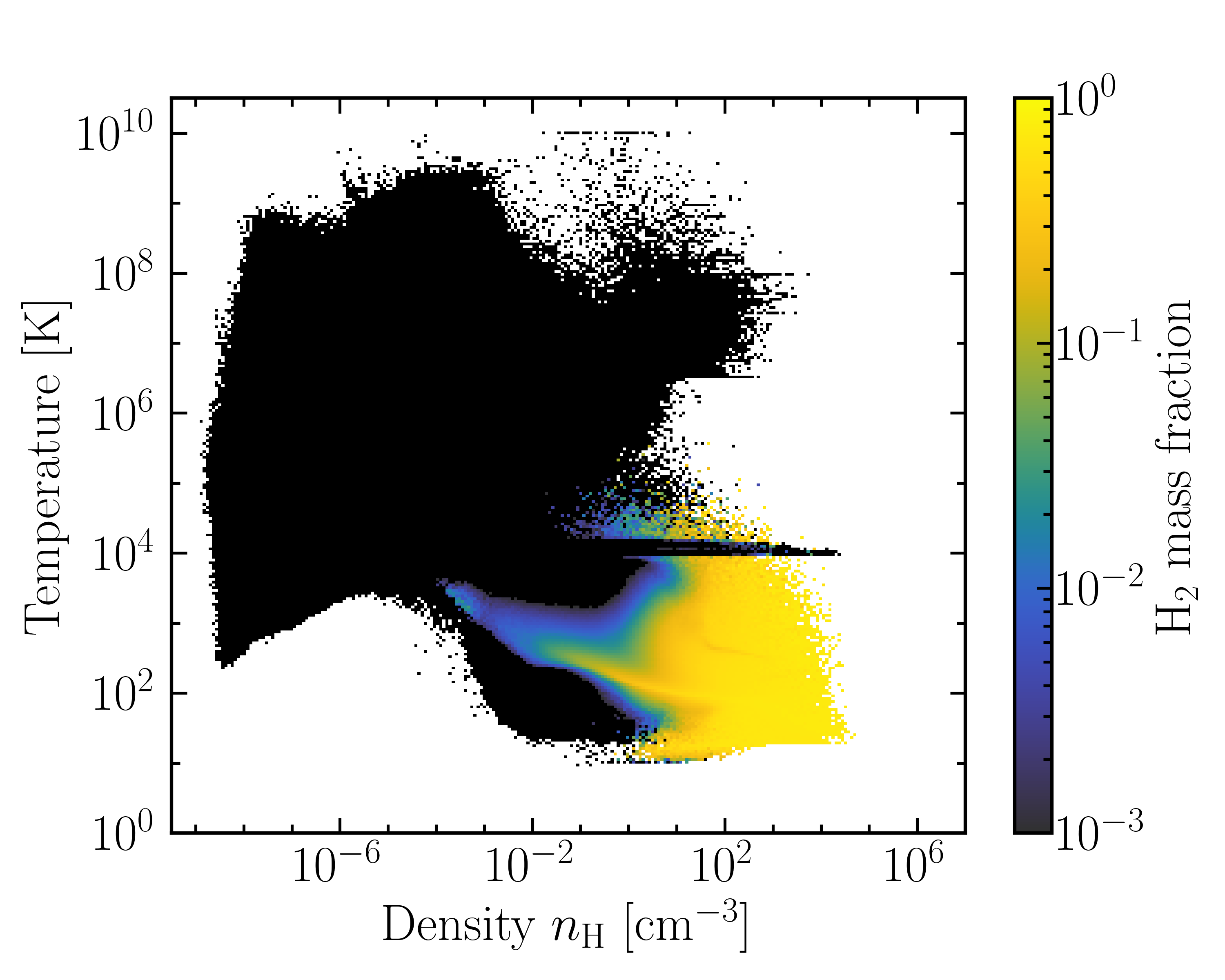} \\
    \includegraphics[width=0.45\linewidth]{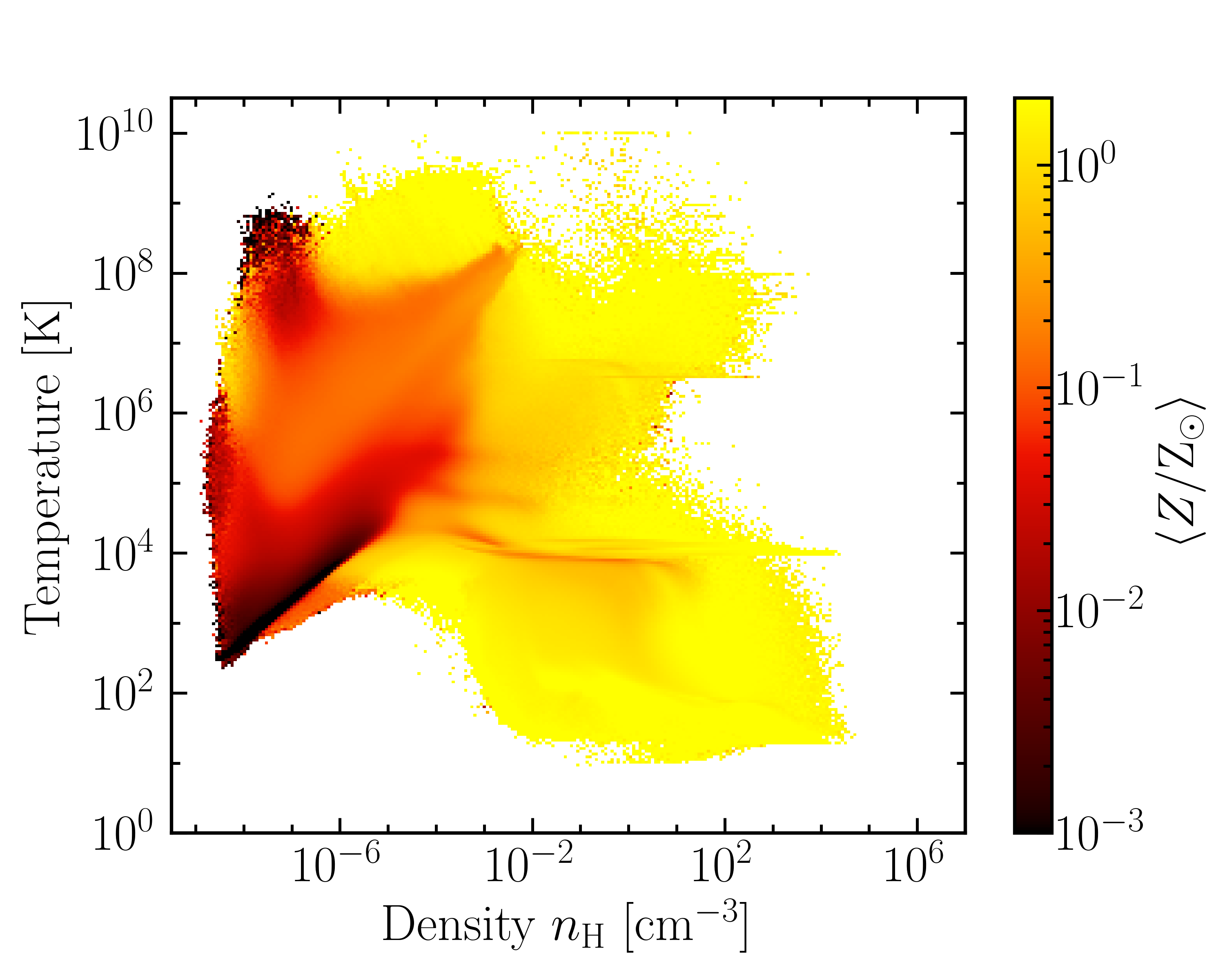}\hspace{0.01\linewidth}
    \includegraphics[width=0.45\linewidth]{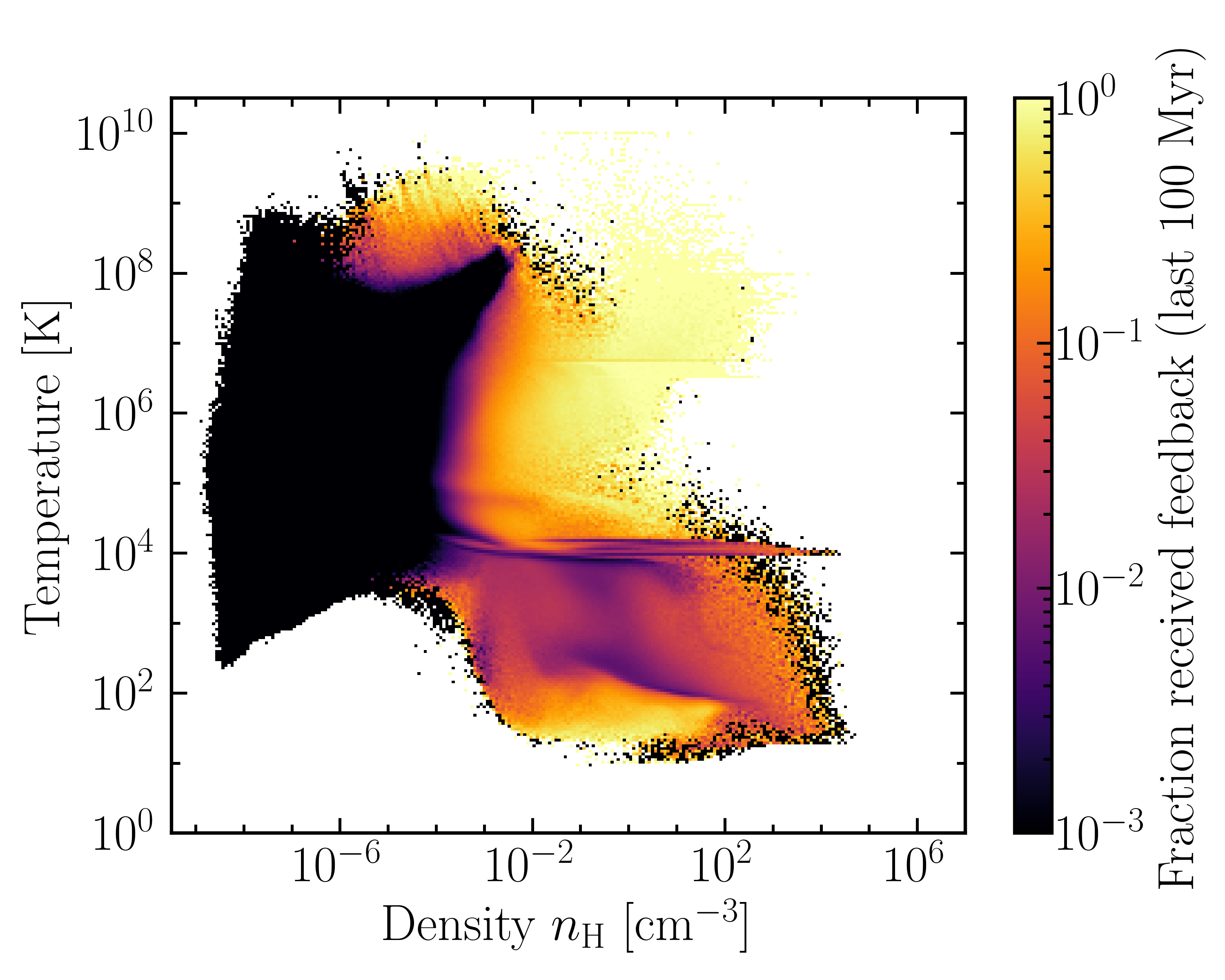} \\    \includegraphics[width=0.45\linewidth]{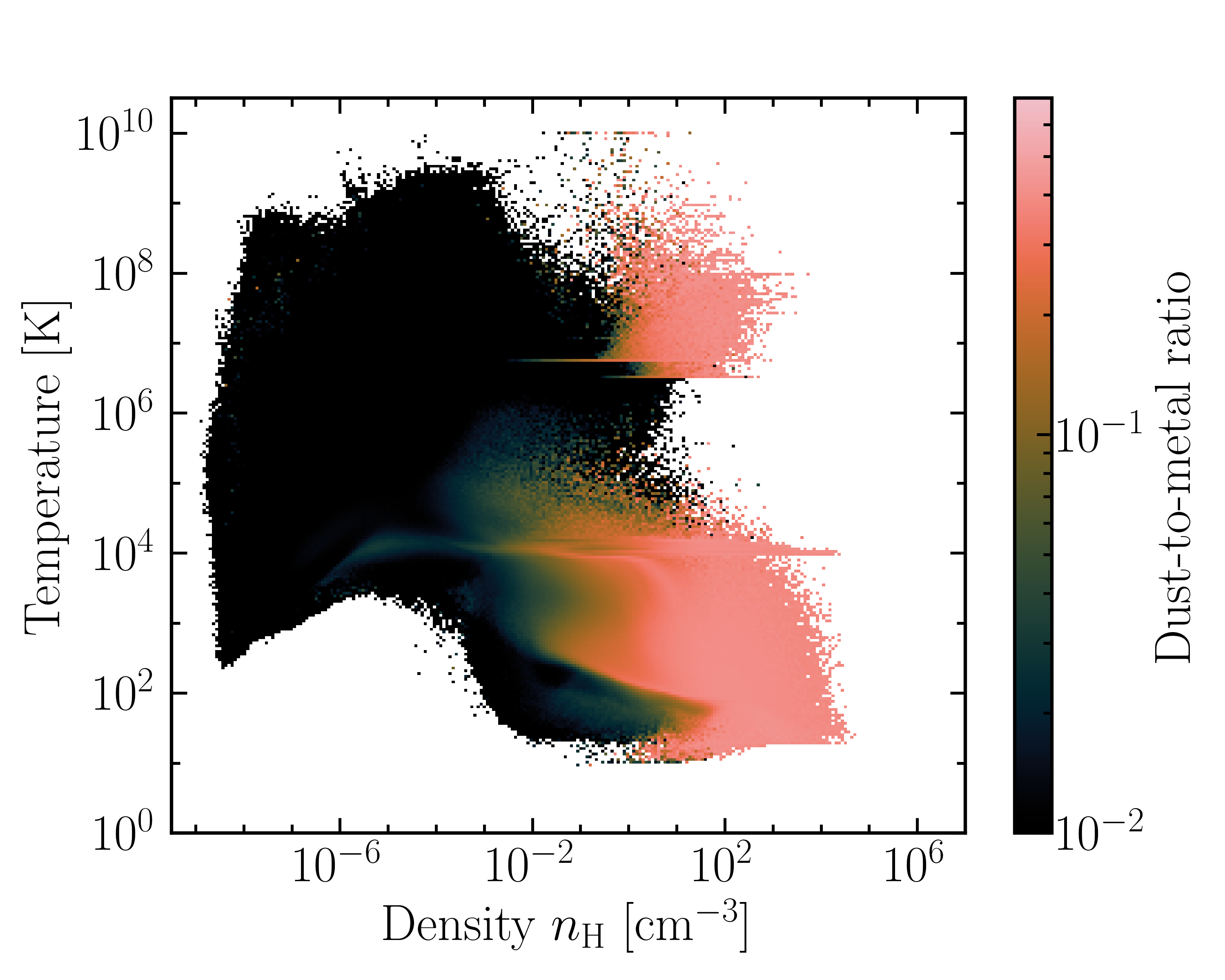} \hspace{0.01\linewidth}
    \includegraphics[width=0.45\linewidth]{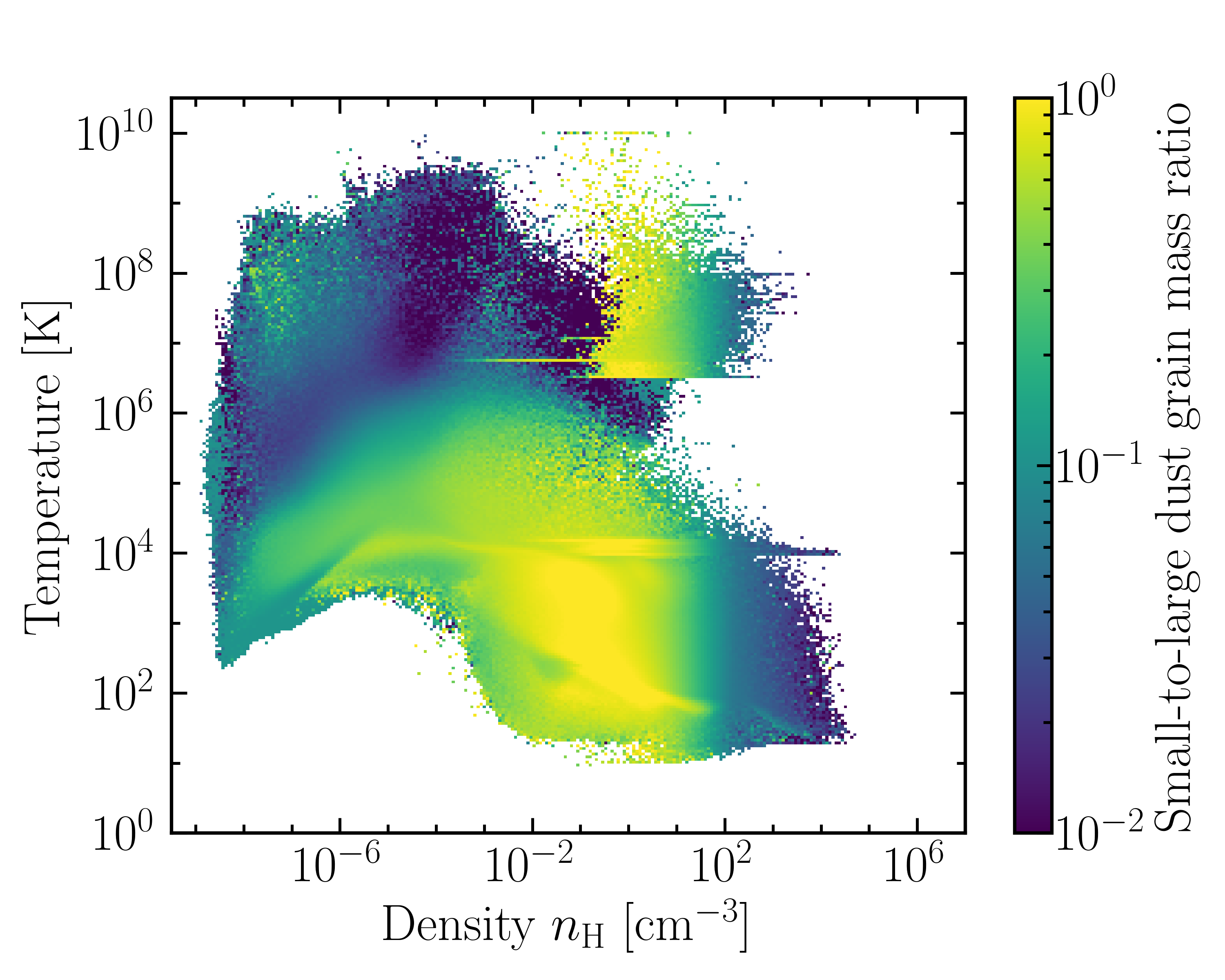}
    \caption{Various gas properties in temperature -- density space for the L200m6 simulation at $z=0$. The colour of each pixel encodes, on a logarithmic scale, the mean atomic hydrogen mass fraction (top left), the mean molecular hydrogen mass fraction (top right), the mean metallicity $Z/Z_\odot$ (using the solar abundance $Z_\odot=0.0134$; \citealt{Asplund2009}) (middle left), the fraction of particles in which thermal CCSN or AGN feedback was injected during the last 100~Myr (middle right), the mean dust-to-metal mass ratio (bottom left), and the mean mass ratio of small-to-large dust grains (bottom right). All means are computed as the ratio of the sum of the quantity in the numerator and the sum of the quantity in the denominator, e.g.\ $\Sigma_i m_{\text{HI},i}/\Sigma_i m_{\text{H},i}$ for the top left panel, where the sum is over all particles $i$ with temperatures and densities that fall into the pixel. Note that the colour scales do not span the full ranges of values.}  
    \label{fig:phase_diagrams}
\end{figure*}

Fig.~\ref{fig:phase_diagrams} shows six different phase diagrams for the L200m6 simulation at $z=0$. The six panels show the mean atomic hydrogen mass fraction (top left), the mean molecular hydrogen mass fraction (top right), the mean metallicity (middle left), the fraction of particles in which thermal CCSN or AGN feedback was injected in the last 100~Myr (middle right), the mean dust-to-metal mass ratio (bottom left), and the mean mass ratio of small-to-large dust grains (bottom right). 

The top panels of Fig.~\ref{fig:phase_diagrams} show that the cold interstellar phase is neutral (i.e.\ \ion{H}{i} or H$_2$), with the denser parts ($n_\text{H} \gtrsim 10^2~\cm^{-3}$) being substantially molecular. Comparing with the bottom left panel, we see that the temperature -- density relation of the bulk of the cold ISM, i.e.\ track I in the left panel of Fig.~\ref{fig:gas_distr_phys_props}, corresponds to features traced by both the molecular hydrogen fraction and the dust-to-metal ratio. This is expected because molecules primarily form on dust grains, and both dust and molecules play a key role in regulating the thermal balance of the ISM. The bottom right panel shows that the higher gas densities favour larger dust grain sizes, which is due to accretion of metals on to dust grains in the cold and dense ISM. 

Region~E in the left panel of Fig.~\ref{fig:gas_distr_phys_props}, which is the track with the high concentration of particles that emerges above $n_\text{H}\sim 10^{-3}\,\cm^{-3}$ and approximately follows the $T= 10^4\,\K$ isotherm, traces the thermal equilibrium temperature of low-metallicity gas (see the middle left panel of Fig.~\ref{fig:phase_diagrams}). Above $n_\text{H}\sim 10^{-0.5}\,\cm^{-3}$, the track rapidly transfers to another track at $T\sim 10^2\,\K$, after which its temperature declines gradually with increasing density (region I). The middle left panel of Fig.~\ref{fig:phase_diagrams} also shows that the transition to the cold ISM occurs at higher densities in gas with lower metallicity. 

At densities $10^{-3} \lesssim n_\text{H}/\cm^{-3} \lesssim 10^2$ there is a small fraction of gas at temperatures below the track corresponding to the thermal equilibrium curve of the cold ISM (region~F in the left panel of Fig.~\ref{fig:gas_distr_phys_props}). From Fig.~\ref{fig:phase_diagrams} we see that this gas has a much lower molecular fraction and dust-to-metal ratio than the main cold ISM track. This is likely gas that overshot the equilibrium temperature as it cooled rapidly from high ISM temperatures ($T\gg 10^4\,\K$). This is a non-equilibrium effect resulting from the temporary availability of residual free electrons due to the recombination lag captured by our time-dependent treatment of hydrogen and helium \citep[see][for a demonstration of this effect]{Richings2014a}. The middle right panel shows that a large fraction of this gas recently received thermal feedback energy through direct injection. This recently cooled gas has not yet had the time to (re)grow dust grains and (re)form molecules that, if they were present in the past, would have been rapidly destroyed when the gas was hot. 

The nearly isothermal track just above $T= 10^4\,\K$, region~H in the left panel of Fig.~\ref{fig:gas_distr_phys_props} and visible in all temperature -- density plots, which emerges above $n_\text{H}\sim 10^{-2}\,\cm^{-3}$ and extends to very high densities, consists of ionized gas, i.e.\ \ion{H}{ii} regions, as can be inferred from the two top panels of Fig.~\ref{fig:phase_diagrams}. The right panel of Fig.~\ref{fig:gas_distr_phys_props} shows that the pressures in \ion{H}{ii} regions extend to values higher than those of the cold ISM, suggesting that these regions may expand. 

The hot phase of the ISM , region~G in the left panel of Fig.~\ref{fig:gas_distr_phys_props}, is produced by stellar and AGN feedback. This can be seen from the middle right panel of Fig.~\ref{fig:phase_diagrams}, though one should keep in mind that gas that was recently shock-heated by winds, but in which thermal feedback energy was not directly injected, is not included in the fraction used for the colour coding. The horizontal features at $T\approx 10^{6.5}~\K$ and $\approx 10^{6.75}~\K$, which are most clearly visible in the left panel of Fig.~\ref{fig:gas_distr_phys_props}, reflect the minimum heating temperatures for AGN and SN feedback, respectively, while the features at $T\approx 10^{10}\,\K$ and $\approx 10^{8}~\K$ reflect the corresponding maximum heating temperatures. There are additional but weaker features at integer multiples of the maximum SN heating temperature caused by gas particles being hit by multiple SN events. Gas that is (shock-)heated by feedback will either cool radiatively, in which case it will rapidly join the warm or even the cold ISM, or, if the temperature is sufficiently high, cool adiabatically to lower densities and temperatures until it reaches pressure equilibrium with the ambient gas or until radiative cooling takes over. 

The bottom left panel of Fig.~\ref{fig:phase_diagrams} shows that the hot ISM can be dusty, though only for dense gas above the minimum feedback heating temperatures, which is an artefact caused by grain destruction lagging one time step behind the feedback: because the injection of energy by feedback takes place at the end of the time step, there has not yet been an opportunity for the dust to be destroyed. The dust in these particles will generally be destroyed by sputtering (or by hand for directly heated particles, see \S\ref{sec:dust}) in the next time step. The fact that most of the particles that are both hot ($T \gtrsim 10^7\,\K$) and dense ($n_\text{H}\gtrsim 1~\cm^{-3}$) are dusty indicates that these particles have only just been heated, as can also be inferred from the middle right panel. This implies that this hot and dense phase is very short-lived, as expected given its very high pressures (see the right panel of Fig.~\ref{fig:gas_distr_phys_props}).  

The left panel of Fig.~\ref{fig:gas_distr_phys_props} shows that the highest concentration of gas mass (region~A) follows a power-law temperature -- density relation at low densities ($n_\text{H} < 10^{-5}\,\cm^{-3}$) and low temperatures ($T\lesssim 10^4~\K$). This is the diffuse IGM, which has low overdensities (for reference, the cosmic mean density $\bar{n}_\text{H} \approx 1.9\times 10^{-7}~\cm^{-3}$) and which follows a track corresponding to the quasi-equilibrium between photoheating by the metagalactic ionizing background radiation and adiabatic cooling due to the Hubble expansion \citep{Hui1997}. Fig.~\ref{fig:phase_diagrams} shows that this gas is highly ionized (top row), nearly free of both metals (middle left) and dust (bottom left), and has not recently been injected with thermal feedback energy (middle right). 

The concentration of gas in Fig.~\ref{fig:gas_distr_phys_props} at very high temperatures, $T\gtrsim 10^7~\K$, comprises the intragroup and intracluster media, where adiabatic compression and shock heating result in a positive correlation between temperature and density (region~C). A large amount of gas also resides at intermediate temperatures ($10^5~\lesssim T/\K \lesssim 10^7$) and densities ($10^{-6}\lesssim n_\text{H}/\cm^{-3} \lesssim 10^{-4}$), i.e.\ region~B, which corresponds to the shock-heated warm-hot intergalactic medium (WHIM) and circumgalactic medium (CGM). The shock-heated gas is highly ionized, largely dust-free\footnote{The preference for large dust grain sizes in the nearly dust-free hot gas (bottom right panel of Fig.~\ref{fig:phase_diagrams}) reflects the stellar yields of small and large dust grains, since grain growth is inefficient in this hot gas in which the dust is quickly destroyed by thermal sputtering.}, and substantially enriched with metals. At densities $10^{-4} \lesssim n_\text{H}/\cm^{-3} \lesssim 10^{-1}$ a significant fraction of the gas is at temperatures $\sim 10^4~\K$, which corresponds to the cooler parts of the CGM, where photoheating is balanced by radiative cooling (region~D). Fig.~\ref{fig:phase_diagrams} shows that much of this gas has a very low metallicity and a high atomic hydrogen fraction. 

\subsection{The properties of the gas from which the stars are born and in which CCSNe explode} \label{sec:birth_properties}

In this section, we explore some of the physical properties of the gas particles at the time they were converted into stellar particles. We combine the birth properties of all stars, regardless of the redshift at which they formed and of the mass of the galaxies in which they currently reside. We thus consider the (zero age main sequence) stellar mass weighted mean properties of star-forming gas averaged over all of space and time. 

Before showing results, we caution that `star-forming gas' is a simulation concept. In \colibre\ gas becomes star-forming if it is gravitationally unstable at the resolution limit (see Section~\ref{sec:SF}). As discussed in \citet{Nobels2024}, we do not expect stellar birth properties to converge with increasing resolution because the star formation criterion is applied at the resolution limit as opposed to at a fixed mass scale. If the resolution is higher, then the mass scale at which instability is tested is smaller and higher densities, lower temperatures and/or smaller turbulent velocity dispersions are required to form stars. In higher resolution simulations there is generally less gas that is star-forming. However, because the star-forming gas has higher densities, it has a smaller gas consumption time scale, which means that observables such as the total SFR or the SFR at a fixed resolved physical scale can still be converged. Checking for convergence for observables is sensible, but checking for convergence of the amount or physical properties of star-forming gas is not. This lack of convergence of the stellar birth properties is intentional because we wish to take advantage of the ability to follow the collapse of gas clouds to higher densities in higher resolution simulations.  

We note that `star-forming gas' is also an ambiguous concept when analysing observations. If we observe a galaxy at a resolution similar to the size of the galaxy, we may find the atomic gas content to correlate well with star formation, but this association may break down on molecular cloud scales. Similarly, observations that resolve molecular cloud cores will find tighter correlations between star formation and the densest molecular gas than with the total amount of molecular gas \citep[e.g.][]{Kennicutt2012}. Continuing this trend, observations resolving protostars would indicate that only gas with near-stellar densities is `star-forming'.  

Because of the explicit resolution dependence, the properties of star-forming gas and of the environmental densities of CCSNe can also not be compared directly to those in much higher resolution simulations of individual $z=0$ low-mass dwarf galaxies \citep[e.g.][]{Hu2017, Smith_Matthew2021, Gutcke2021LYRA, Hislop2022, Steinwandel2023}. Another reason is that we combine all stars that formed in any galaxy during the history of the Universe. At higher redshift and in the higher mass galaxies that dominate the cosmic star formation history \citep[for a decomposition, see fig.~6 of][]{Chaikin2025smf_evol}, stars will likely form (and CCSNe explode) in gas with higher densities. 

\begin{figure}
    \centering
    \includegraphics[width=0.95\linewidth]{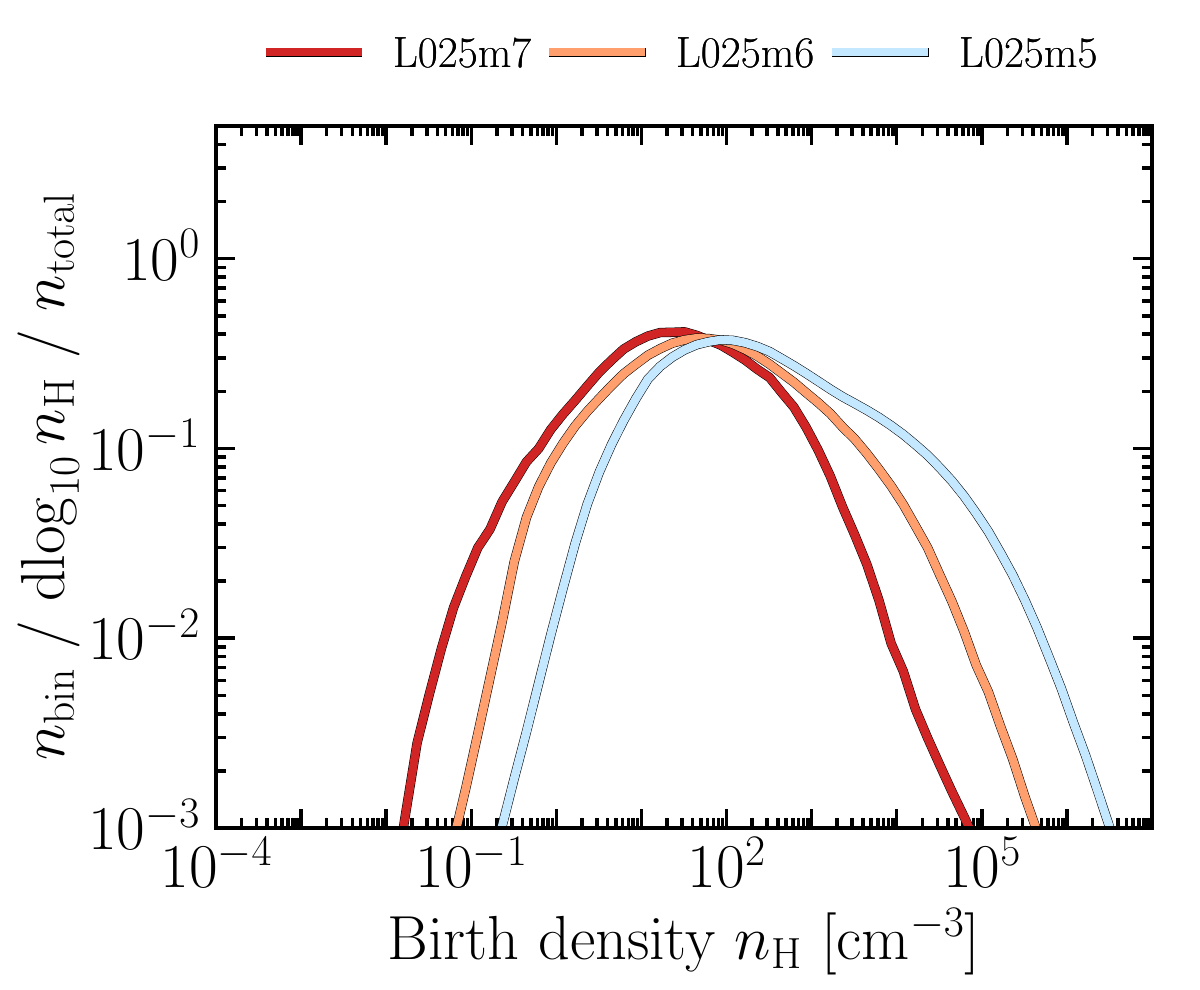}\\
    \includegraphics[width=0.95\linewidth]{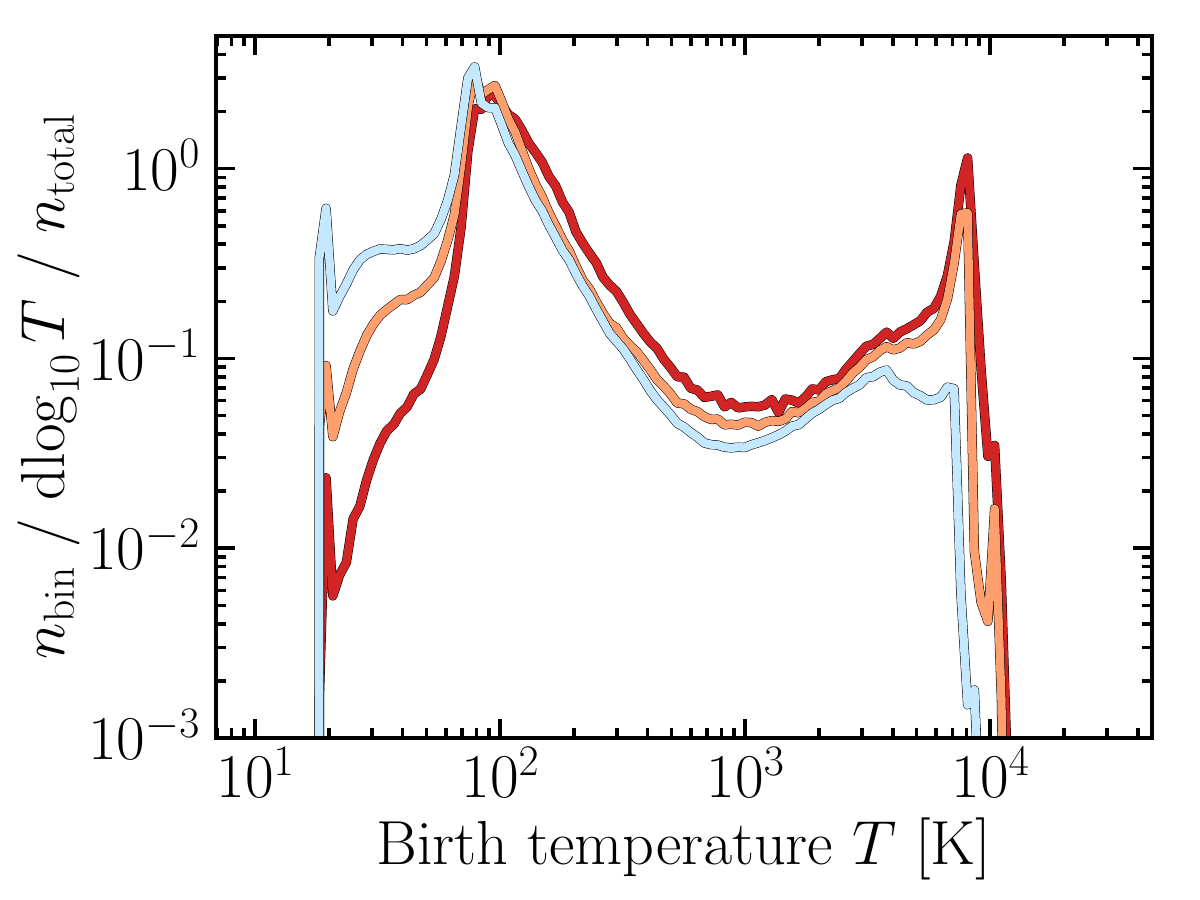}\\
    \includegraphics[width=0.95\linewidth]{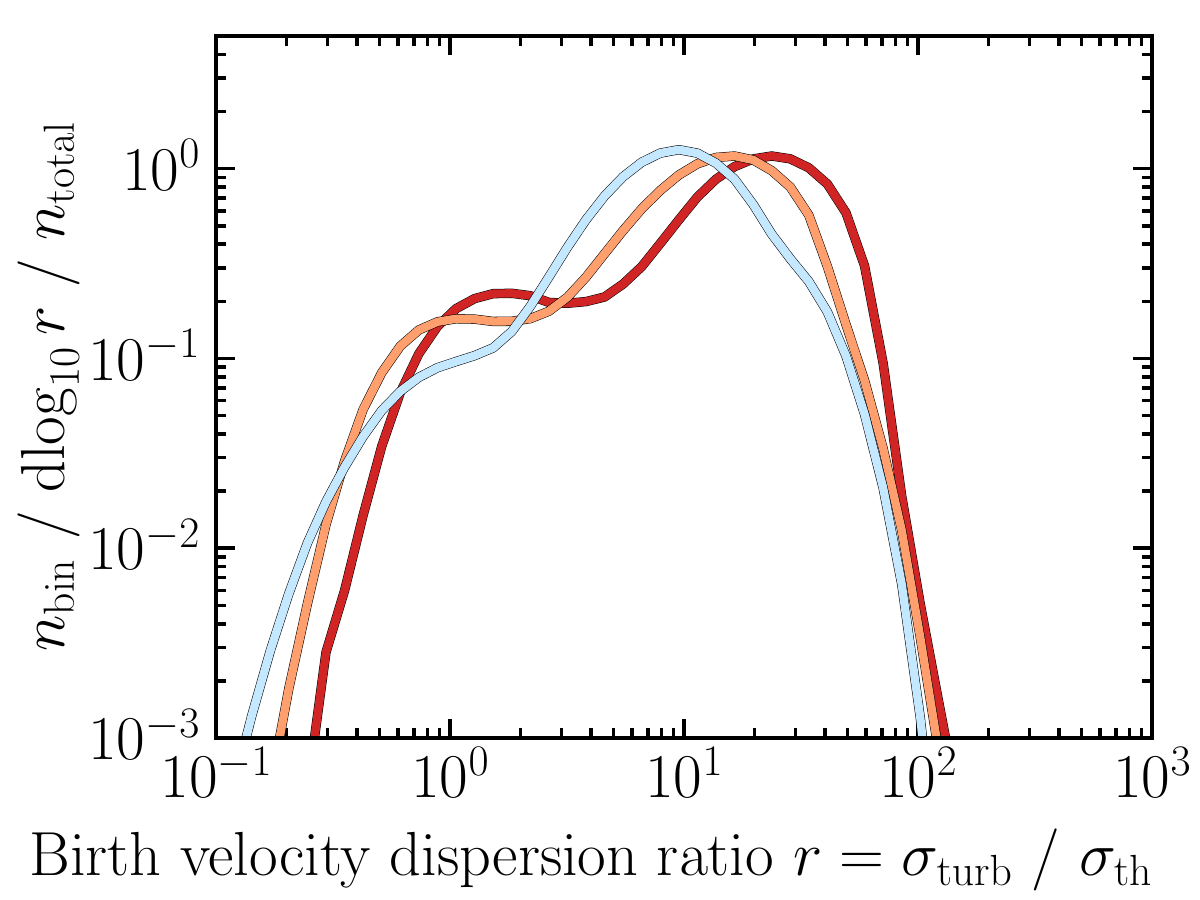}
    \caption{PDF of the density (top panel), temperature (middle panel), and the ratio ($r=\sigma_\text{turb}/\sigma_\text{th}$) of the turbulent-to-thermal velocity dispersion (bottom panel) of the gas at the time it was converted into stellar particles (at any $z\ge 0$) in simulations using the same box size (L025) but different resolutions (different colours). The turbulent dispersion, $\sigma_\text{turb}$, is the local 3D velocity dispersion between gas particles. The thermal dispersion is estimated from the birth temperature using $\sigma_\text{th} = 13.8~\kms (T/10^4\,\K)^{1/2}$, which assumes a mean molecular weight of 1.3 (which is appropriate for atomic gas but leads to an overestimate of $\sigma_\text{th}$ in molecular gas). The spike at $T\approx 20~\K$ reflects the minimum allowed internal energy. The temperature PDF is multimodal, with most stars forming in the cold ($T\ll 10^4\,\K$) interstellar gas phase. Turbulence in star-forming gas is typically supersonic ($r > 1$). As the resolution increases, star formation shifts to higher gas densities and lower temperatures.}
    \label{fig:birth_props}
\end{figure}

The top and middle panels of Fig.~\ref{fig:birth_props} show the probability density distributions (PDFs) of, respectively, the density and temperature of the gas from which the stellar particles formed. The bottom panel shows the PDF of the ratio of turbulent and thermal velocity dispersions of the gas particles that were converted into stellar particles, where the former is the local 3D velocity dispersion between particles and the latter is estimated from the gas temperature. The PDFs include all stellar particles in the simulation, i.e.\ stars formed at any redshift. The red, orange, and blue curves correspond to simulations L025m5, L025m6, and L025m7, respectively. 

The temperature PDF shows the most pronounced features. There are peaks at $T\approx 20~\K$, $\sim 10^2\,\K$, and $\sim 10^4\,\K$. The first two peaks correspond to the cold interstellar gas phase, while the last corresponds to the warm phase. The spike at $\approx 20$~K is due to the lower limit that we impose on the internal energy (corresponding to 10~K for atomic gas, but the gas is molecular at these temperatures). Most stars form in the cold phase ($T\lesssim 10^2\,\K$), but for the lower resolutions the contributions from the warm phase are significant. The density PDF is very wide, with the low-density tail corresponding to star formation in the warm phase. The bottom panel shows that turbulence is typically supersonic, with the secondary peak at subsonic/mildly supersonic values corresponding to birth temperatures $T\sim 10^4\,\K$. In Appendix~\ref{app:birth_by_mass} we show that the physical conditions correlate strongly with galaxy stellar mass, where stars in more massive galaxies tend to have formed from gas with higher densities, lower temperatures, and more supersonic turbulence. 

As resolution increases, the birth densities increase, the temperatures decrease, and the turbulence becomes less supersonic. This is expected because gas is only allowed to form stars if the kernel mass is gravitationally unstable, i.e.\ if the kernel mass exceeds the local Jeans mass. A higher resolution corresponds to a smaller kernel mass, which means that higher densities and lower total velocity dispersions are required for star formation (see equation~\ref{eq:SFcrit}). A higher density typically corresponds to a lower temperature (see Fig.~\ref{fig:phase_diagrams}). Because the turbulence in the low-temperature gas is typically supersonic, stars tend to form in regions where the turbulent velocity is relatively low. 

We note that in simulations such as \eagle, which have a baryonic resolution similar to \colibre\ m5 and m6, but impose an equation of state on the ISM, stellar birth properties are similar to those in simulations using the \colibre\ model, but with a resolution much lower than m7. In such simulations, birth temperatures are $\sim 10^4\,\K$, median birth densities are orders of magnitude lower than for similar resolution \colibre\ models, and turbulence tends to be at most mildly supersonic. Direct simulation of the cold ISM therefore results in dramatic changes to the physical conditions of the gas from which stars are born for a fixed resolution. 

For all \colibre\ resolutions, most of the stars form in the cold ISM. Although not shown, we note that a larger fraction of stars form in the warm ISM if the metallicity is lower because in that case the transition from the warm to the cold ISM occurs at higher densities. Lower metallicities tend to correspond to higher formation redshifts and lower galaxy masses.

\begin{figure}
    \centering
    \includegraphics[width=0.95\linewidth]{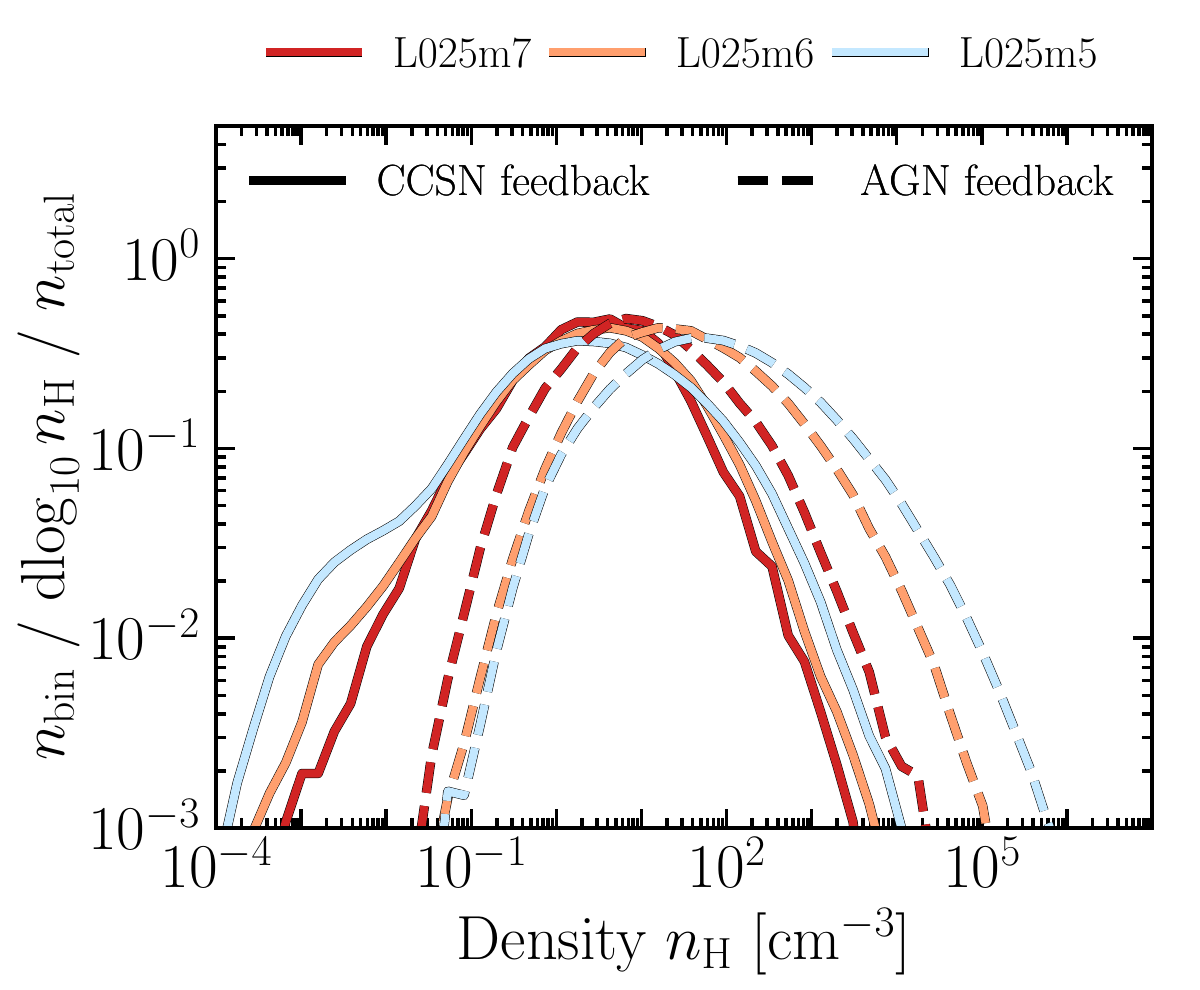}
    \caption{PDF of the density of the gas in which CCSN (solid lines) and AGN (dashed lines) thermal feedback energy is injected in simulations using the same box size (L025) but different resolutions (different colours). This is based on the last, if any, CCSN and AGN feedback events that resulted in the direct injection of thermal energy in the gas particle, which can be at any redshift $z\ge 0$. While the typical feedback gas densities for CCSNe are nearly converged with the resolution, the density of the gas in which AGN feedback is injected increases with the resolution and tends to be much higher than for CCSNe. The density of the gas in which CCSNe explode is typically much lower than the density of the gas from which the stars form, particularly for higher resolutions (compare to the top panel of Fig.~\ref{fig:birth_props}; the median birth densities are factors of $\sim 10-10^2$ higher).}
    \label{fig:feedback_props}
\end{figure}

Fig.~\ref{fig:feedback_props} shows the PDFs of the density of the gas in which thermal energy was injected by CCSNe (solid lines) and AGN feedback (dashed lines). Comparing the solid lines with the PDF of the stellar birth densities shown in the top panel of Fig.~\ref{fig:birth_props}, we note two things. First, compared to the stellar birth densities, the gas densities in which CCSNe explode are typically much lower than the stellar birth densities. Second, the density of the gas in which CCSNe explode converges much better with the resolution. The difference between the median birth and CCSN feedback densities increases from 1.0~dex at m7 to 1.9~dex at m5 resolution. Because dense cold clouds contain a large fraction of the ISM mass, the density of the intercloud medium is relatively low. While stars form in dense clouds, CCSNe explode mostly in the intercloud medium because stellar winds, photoheating, and earlier CCSNe destroy birth clouds and because the relative motion of stars and gas during the pre-CCSN phase allows many stars to move out of their birth clouds. The relatively low density of the gas in which the feedback events occur helps to reduce numerical overcooling and hence facilitates better convergence. In simulations with a multiphase ISM, such as \colibre, we therefore expect numerical overcooling to be reduced, and we can achieve better convergence for observables, compared to simulations that prevent the formation of a cold interstellar phase, such as \eagle. 

The density of the gas in which AGN feedback energy is injected tends to be much higher than that for CCSN feedback, and the difference increases with numerical resolution. This is expected because simulations with higher resolution can sample the gas closer to the BH, which tends to have higher densities, and the AGN feedback energy is given to the gas particle closest to the BH. Numerical convergence is therefore harder to achieve for AGN feedback than for CCSN feedback, although the higher maximum heating temperatures used for AGN feedback help suppress numerical overcooling. 


\section{Comparison with observations} \label{sec:obs}
In this section we compare the fiducial \colibre\ models, i.e.\ the simulations that use thermally-driven AGN feedback, with observations of a variety of fundamental galaxy properties. We will separately consider the observables used to calibrate the subgrid models for CCSN and AGN feedback (\S\ref{sec:obs_cal}) and other observables (\S\ref{sec:obs_other}). Apart from the cosmic star formation history, all the observations with which we compare are at $z\approx 0$. Comparisons with higher-redshift observations and additional observables of galaxies, including spatially resolved properties, and the IGM will be presented in future work. The simulations using the hybrid AGN feedback model are compared with the data (and the fiducial model) in Section~\ref{sec:hybrid}. Where necessary, observational data have been converted to the \citet{Chabrier2003} IMF and to the cosmology assumed in \colibre.

As discussed in Section~\ref{sec:herons}, galaxy properties are measured using \soap\ from gravitationally bound particles within a spherical aperture of radius 50~pkpc centred on the most bound particle in the \herons\ subhalo. The choice of aperture size was motivated by the results of \citet{deGraaff2022}, but as discussed by \citet{Chaikin2025smf_evol}, using a larger aperture only has a non-negligible effect for very massive galaxies, $M_* \gg 10^{11}\,\Msun$, where it gives higher masses. Because of the somewhat arbitrary and analysis-specific distinction between the brightest cluster galaxy (BCG) and the intracluster light (ICL), a dedicated analysis is required for a quantitative comparison of such massive galaxies to observations.

Nearly all plots that we show in this section have galaxy stellar mass along the $x$-axis. Observational measurements of stellar masses include both random and systematic errors. Stellar masses are typically measured by fitting stellar population synthesis models to the spatially integrated spectral energy distribution (SED). Even if the IMF is fixed, the results depend on the assumed star formation history, chemical composition and its evolution, dust obscuration, stellar rotation, stellar multiplicity, and the model/library for stellar spectra. To account for the Eddington bias resulting from the random component of these and other uncertainties, i.e.\ the fact that, for declining mass functions, random errors in the stellar mass cause more objects to scatter up to higher mass bins than to scatter down to lower mass bins, we add random errors to the predicted stellar masses. Based on a review of the literature, \citet{Chaikin2025smf_evol} adopt a lognormal distribution with mean zero and standard deviation of $\log_{10}M_*$ of
\begin{equation}\label{eq:random_scatter}
    \sigma_{\rm random}(z) = \min(0.1 + 0.1 \, z, 0.3)~\text{dex}.
\end{equation}
In Appendix~\ref{app:edd_bias} we show that at $z=0$ this amount of scatter results in only a slight smoothing of trends with mass, but that larger errors can have a strong impact for $M_* \gtrsim 10^{11}\,\Msun$. Even though they are likely significant, we do not attempt to simulate the effects of random errors in galaxy properties other than mass because such errors are less well understood than errors in stellar mass and because they do not affect the binning for the plots presented here. 

We expect the uncertainties to result not only in random errors but also in systematic errors. Accounting for systematic errors, which we do not do, would shift the entire curves along the mass axis and would change the shape of the curves if the systematic error depends on mass. \citet{Chaikin2025smf_evol} estimate the systematic error on the $z=0$ stellar mass for a fixed IMF to be $\approx 0.15$~dex in the mass range used for calibration. 

We focus on the largest simulations currently available at $z=0$ for each of the three resolutions: L400m7, L200m6 and L025m5, for which we will use red, orange, and blue line colours, respectively. Comparing these simulations tests not just for convergence with resolution, but also with volume. The L400m7 and L200m6 volumes are sufficiently large that the differences in the median trends with mass due to box size are limited to the most massive objects. However, the small volume of the L025m5 can play a significant role, mainly because the satellite fraction and the host masses of satellites are lower than for the other simulations due to the limited range of scales on which primordial fluctuations can be sampled in a volume of this size. Although we will not show them here, L025m6 and L025m7, as well as L200m7, are also available to facilitate comparing different resolutions at fixed box size. 

In all figures with stellar mass on the $x$-axis, we will switch from a solid to a dashed line style where the mass becomes smaller than 100 times the mean initial baryonic particle mass\footnote{Because of stellar mass loss, the number of stellar particles corresponding to this limit depends somewhat on the stellar age and metallicity. For m7 resolution the change of line style occurs at $\approx 140$ particles, while for m6 and m5 it corresponds to $\approx 160$ particles.} and where there are fewer than 10 objects per stellar mass bin of width 0.2~dex. We stress that the 100 stellar particle limit is just for reference and does not necessarily correspond to the resolution limit, which can and should be estimated separately for each question of interest by comparing the different resolutions, preferably using simulations of the same volume. However, it is important to note that the physical processes captured in hydrodynamical simulations using subgrid models, such as \colibre, effectively change with resolution. Hence, even if they are recalibrated to match a limited set of observations, some differences between different resolution simulations may only be indirectly caused by the numerical resolution. Instead, the differences may largely reflect the implicit freedom in the subgrid physics that remains after calibration.

\subsection{Observations used for calibration of CCSN and AGN feedback} \label{sec:obs_cal}

In this section we will compare the fiducial \colibre\ simulations with the observations used to calibrate the CCSN and AGN feedback, namely the $z=0$ galaxy stellar mass function (SMF) (\S\ref{sec:gsmf}) and the $z\approx 0$ galaxy size -- stellar mass relation (SMR) (\S\ref{sec:sizes}), as well as the $z=0$ relation between BH and galaxy stellar mass (\S\ref{sec:bh_masses}). 

As summarized in Section~\ref{sec:calibration} and discussed in detail in \citet{Chaikin2025calibration}, the calibration was carried out by first using machine learning to emulate the effect of varying four subgrid parameters, three for CCSN feedback plus the BH seed mass, at m7 resolution using volumes of 50~cMpc on a side (L050m7). The emulator predictions were then used to find the parameter values that provide the best fit to the observed SMF and SMR for the galaxy stellar mass range $9.0 < \log_{10} M_*/\Msun < 11.3$. We also consider the high-mass end of the $z=0$ relation between BH and galaxy stellar mass to belong to the calibration target observations because it motivated the choice of the AGN feedback efficiency, $\epsilon_\text{f}$. While the parameter values were kept fixed when increasing the box size, they were slightly adjusted when increasing the resolution, guided by smaller volume test simulations (L025m6 and L012.5m5).

\subsubsection{The galaxy stellar mass function} \label{sec:gsmf}
\begin{figure}
    \centering
    \includegraphics[width=0.95\linewidth]{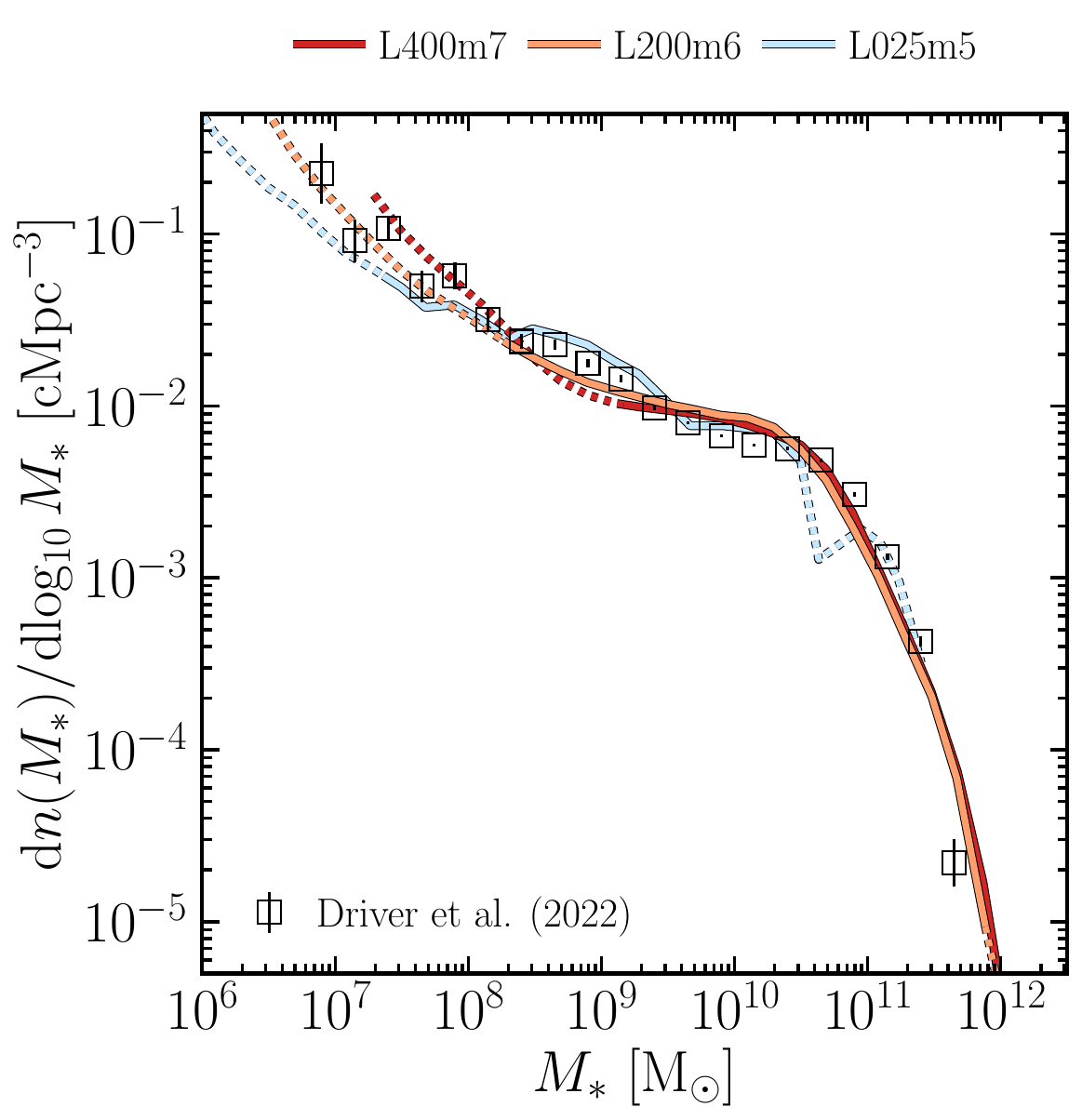}
    \caption{The $z=0$ galaxy stellar mass function in the L400m7 (red), L200m6 (orange), and L025m5 (blue) simulations. The line style switches from solid to dotted where the mass becomes smaller than 100 times the mean initial baryonic particle mass and at high mass where there are fewer than 10 objects per 0.2~dex mass bin. 
    The data points with error bars show the observed mass function from GAMA DR4 \citep{Driver2022}. 
    The different resolution simulations are in very good agreement for galaxy masses corresponding to at least a few tens of stellar particles. The agreement between the simulations and the data is excellent.}
    \label{fig:gsmf}
\end{figure}

Fig.~\ref{fig:gsmf} compares the \colibre\ SMFs to the observations of \citet{Driver2022} from the GAMA DR4 survey. We applied the corrections suggested by the authors for evolution from the actual redshift range of $z<0.1$ to $z=0$, based on the evolution measured by the DEVILS survey \citep{Thorne2021}, and for cosmic variance, obtained by normalizing to the larger-area Sloan Digital Sky Survey. Together, these corrections increase the observed number densities by 0.0807~dex.  

The agreement with the data is excellent. Taking into account the estimated systematic error on the observed stellar masses of 0.15~dex, the simulations are fully consistent with the observations from the highest masses that are sampled with at least 10 objects per mass bin down to masses corresponding to $100$ particles (the solid parts of the curves) and for m7 and m6 even down to $<10$ particles. When combined, the three simulations reproduce the observations over 5 orders of magnitude in stellar mass. Notably, for m6 and m5 the agreement with the data extends to far lower masses than the lowest mass used in the calibration ($10^9\,\Msun$). The agreement at each resolution with the same data implies excellent convergence with the numerical resolution and volume of the simulations. 

In Appendix~\ref{app:edd_bias} we show that the predictions for $M_* > 10^{11}\,\Msun$ become somewhat sensitive to the assumed random errors on the stellar masses (because of Eddington bias). \citet{Chaikin2025smf_evol} show that the same holds for the chosen aperture. Due to the steepness of the mass function, the expected systematic error on the observed stellar masses, 0.15~dex for a fixed IMF, becomes also important. This suggests that comparisons with the high-mass data would particularly benefit from applying observational techniques to virtual observations, which is, however, beyond the scope of this work. Finally, we note that \citet{Chaikin2025smf_evol} find that the agreement with the observed SMF is not limited to the redshift used for calibration (i.e.\ $z=0$), but extends up to the highest redshifts for which constraints are available, $z \lesssim 12$. 

\begin{figure}
    \centering
    \includegraphics[width=0.95\linewidth]{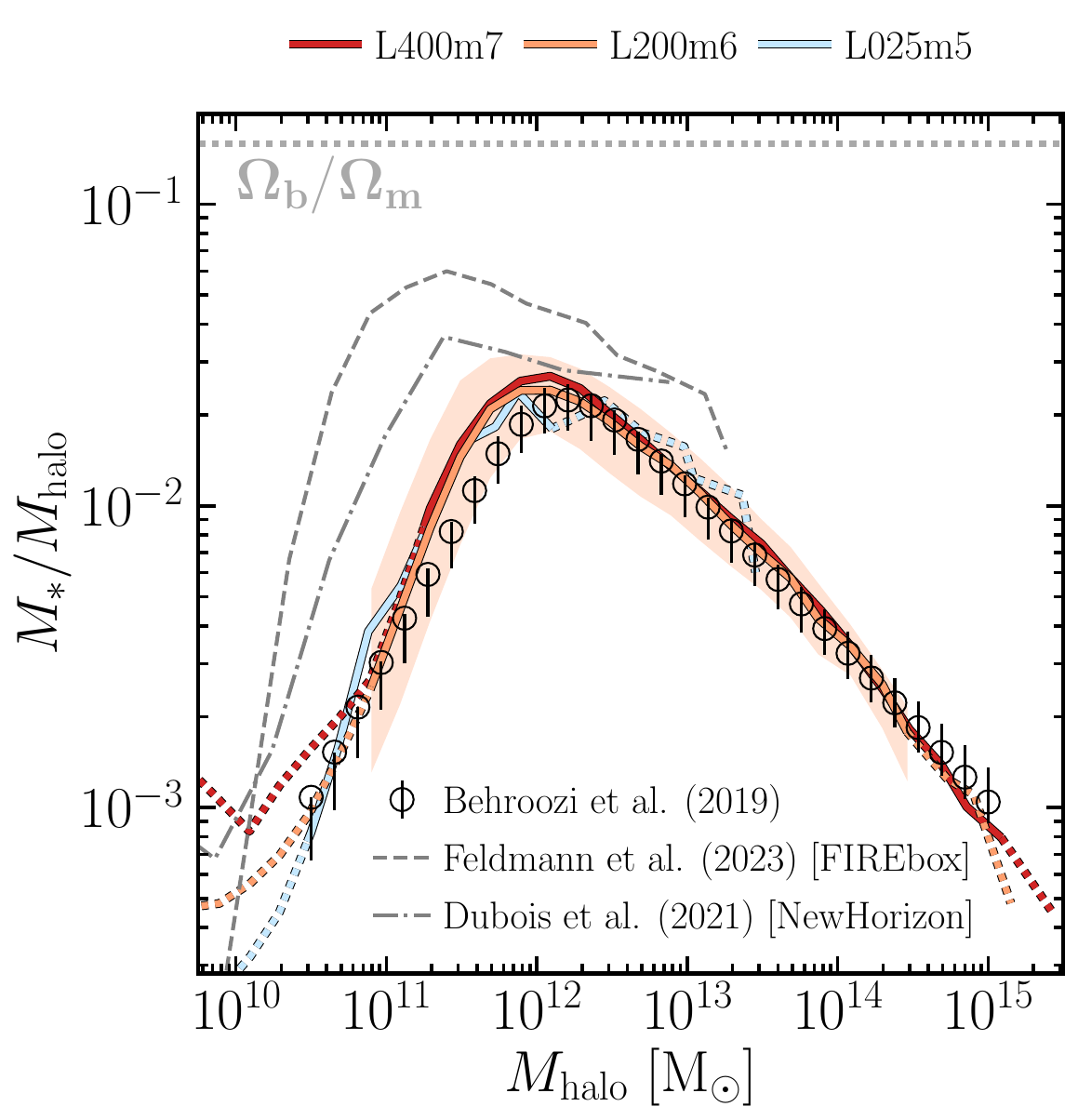}
    \caption{Median $z=0$ stellar-to-halo mass ratio for central galaxies as a function of halo mass for the L400m7 (red), L200m6 (orange), and L025m5 (blue) simulations. The orange shaded region indicates the 16th to 84th percentile range for L200m6. The line style switches from solid to dotted where the stellar mass becomes smaller than 100 times the mean initial baryonic particle mass and at high mass where there are fewer than 10 objects per 0.2~dex mass bin. The horizontal grey, dotted line indicates the universal baryon fraction. Data points with error bars show the best-fit median and the 16th-84th percentile confidence level on the median inferred from observations using the semi-empirical model UniverseMachine \citep{Behroozi2019}.  
    The thin black lines show the results from two other cosmological simulations that, like \colibre, allow cooling to cold ISM temperatures: FIREbox \citep{Feldmann2023} and NewHorizon \citep{Dubois2021}. Different resolution \colibre\ simulations are in excellent agreement with each other and with the relation inferred from observations using UniverseMachine, but there are large discrepancies with the other simulations.}
    \label{fig:smhm}
\end{figure}

Fig.~\ref{fig:smhm} shows the median stellar-to-halo mass ratio for central galaxies as a function of halo mass, which is closely related to the SMF. The simulations are compared with results inferred by \citet{Behroozi2019} by fitting the semi-empirical model UniverseMachine to a compilation of diverse observations. For consistency with \citet{Behroozi2019}, we define the halo mass as the mass within the radius for which the mean internal density is equal to the critical overdensity provided by \citet{Bryan1998}, which is based on a fit to the simulations of \citet{Eke1996} and at $z=0$ has the value $\approx 101.8$ times the critical density of the universe. However, because UniverseMachine is built on DMO simulations, it does not account for the reduction in halo mass resulting from baryonic effects, which is expected to be small at cluster masses, but $\approx 0.1$~dex for $M_\text{halo} \lesssim 10^{12}\,\Msun$ \citep[e.g.][]{Schaller2015}. While \citet{Behroozi2019} used the peak halo mass, i.e.\ the maximum mass the halo has ever reached, we use the current halo mass, but for central galaxies the difference is negligible for our purposes. To facilitate easy conversion between the stellar and halo masses of central galaxies, Fig.~\ref{fig:smhm_2panels} in Appendix~\ref{app:smhm} shows the median stellar mass as a function of halo mass and vice versa. As can be clearly seen from Fig.~\ref{fig:smhm}, the agreement with the data inferred using the semi-empirical model is very good, as expected given that both \colibre\ and UniverseMachine agree with the observed SMF. The small difference at $M_\text{halo} \sim 10^{11.5}\,\Msun$ is similar in magnitude to the effect expected from UniverseMachine’s inability to capture baryonic effects on the halo mass. 

In Fig.~\ref{fig:smhm} we also compare to two cosmological simulations that have been run to low redshift and that, like \colibre, allow cooling to cold ISM temperatures\footnote{We do not compare to the Romulus25 simulation \citep{Tremmel2017} because that simulation uses a much more basic model for radiative cooling that ignores molecules, dust grains and, for $T>10^4\,\K$, metal-line cooling. Fig.~3 of \citet{Tremmel2017}, in which the simulated stellar masses were reduced by 40 per cent to correct for observational biases, shows that the stellar masses in Romulus25 are similar to those in NewHorizon for $M_\text{halo} \gtrsim 10^{12}\,\Msun$, but closer to the semi-empirical data for lower halo masses.}. NewHorizon \citep{Dubois2021} is a re-simulation of a $\approx (16~\text{cMpc})^3$ volume taken from the HorizonAGN simulation \citep{Dubois2014} that was run to $z=0.25$. Because NewHorizon does not simulate a volume with a mean density equal to the cosmic mean, we cannot directly compare its SMF to observations, but the zoom-in initial conditions should not affect the relation between the stellar and halo masses of central galaxies, which is expected to be insensitive to the large-scale environment \citep[e.g.][]{Crain2009}. NewHorizon uses the adaptive mesh refinement code \textsc{Ramses} \citep{Teyssier2002}. The CDM particle mass is $1.2\times 10^6\,\Msun$, which is close to our intermediate resolution m6. The smallest gas cell size is 34~pc, which is larger than the minimum distance between baryonic particles in \colibre, but much smaller than our maximum physical gravitational softening length. NewHorizon includes CCSN and AGN feedback, but not feedback from stellar winds, \ion{H}{ii} regions, and type Ia SNe. Radiative cooling rates assume ionization and chemical equilibrium and there is no model for dust grains. FIREbox \citep{Feldmann2023} simulates a $(22.1~\text{cMpc})^3$ volume with the meshless finite mass \textsc{gizmo} code \citep{Hopkins2015} down to $z=0$. The CDM and gas particle masses are $3.3\times 10^5$ and $6.3\times 10^4\,\Msun$, respectively, which are comparable to our m5 resolution. Radiative cooling rates, ion and molecular fractions assume ionization and chemical equilibrium, and dust grain evolution is not modelled. Stellar feedback processes are included, but AGN feedback is not. Like \colibre, both NewHorizon and FIREbox use the current rather than the peak halo mass and define halo mass using the virial overdensity of \citet{Bryan1998} that was also used by \citet{Behroozi2019}. However, each of the four models shown in Fig.~\ref{fig:smhm} (\colibre, UniverseMachine, NewHorizon and FIREbox) uses a different structure finder and a somewhat different cosmology. The results for NewHorizon are for $z=0.25$ instead of $z=0$, but for \colibre\ the relation hardly evolves from that redshift to the present \citep[see][]{Chaikin2025smf_evol}. 

While \colibre\ is in close agreement with the stellar mass -- halo mass relation inferred from observations, there are large discrepancies with the other two simulations, which also do not agree with each other. NewHorizon and FIREbox both predict much higher galaxy masses than inferred from observations by \citet{Behroozi2019}. For NewHorizon the difference is small near the peak of galaxy formation efficiency at $M_\text{halo}\approx 1\times 10^{12}\,\Msun$, but increases to a factor of 2 at the highest halo mass sampled by that simulation, $M_\text{halo}\sim 10^{13}\,\Msun$. The difference increases sharply toward lower masses, reaching almost an order of magnitude for $M_\text{halo}< 10^{11.5}\,\Msun$. The difference with FIREbox is large at all masses. The factor of 2 difference for $M_\text{halo} \gtrsim 10^{12}\,\Msun$ could be due to the lack of AGN feedback in FIREbox, but the difference increases rapidly toward lower masses, even exceeding a factor of 10 for $M_\text{halo}\sim 10^{11}\,\Msun$. 

Finally, we refer to Appendix~\ref{app:eagle_tng} for a comparison to \eagle\ and IllustrisTNG, which, like \colibre, span multiple resolutions, but, unlike \colibre, impose an effective equation of state onto the ISM. The TNG subgrid physics was calibrated for the resolution of the TNG100 simulation, which is similar to our m6 resolution. Unlike \eagle\ and \colibre, for TNG the subgrid parameters are held fixed when the resolution is changed, such as for the higher (m5) resolution TNG50 or the lower (m7) resolution TNG300 simulations. For \eagle\ the convergence is similar to that of \colibre, but it is worse for TNG. The level of agreement with the data is slightly worse for \eagle, similarly good for TNG100, and substantially worse for TNG50 and TNG300.

\begin{figure}
    \centering
    \includegraphics[width=0.95\linewidth]{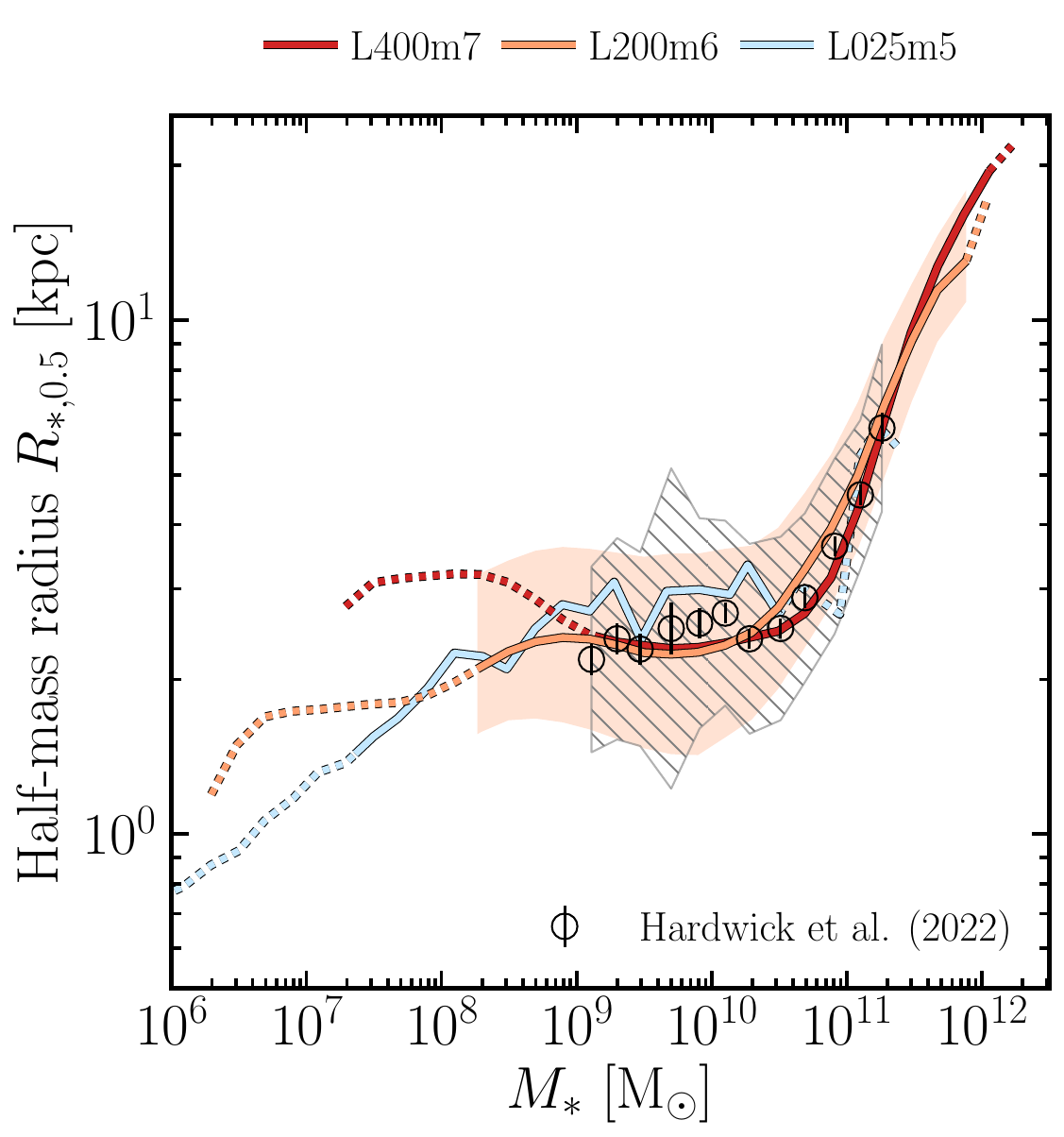}
    \caption{Median $z=0$ projected galaxy half-stellar-mass radius as a function of galaxy stellar mass for the L400m7 (red), L200m6 (orange), and L025m5 (blue) simulations. The orange shaded region indicates the 16th to 84th percentile range for L200m6. The line style switches from solid to dotted where the mass becomes smaller than 100 times the mean initial baryonic particle mass and at high mass where there are fewer than 10 objects per 0.2~dex mass bin. The data points show the observed median half-mass radii of \citet{Hardwick2022} and the error bars the $1\sigma$ uncertainty on the median. The grey hatched region shows the $1\sigma$ scatter in the observed relation. 
    For galaxies resolved with $\gtrsim 10^2$ stellar particles the agreement of the m6 and m7 resolutions with each other and with the data is excellent. At intermediate mass the sizes in m5 are slightly larger than for the lower resolutions, but the difference is small compared with the scatter and the deviation from the data is $\lesssim 0.1$~dex.}
    \label{fig:sizes}
\end{figure}

\subsubsection{The galaxy size -- mass relation} \label{sec:sizes}
The other key observable, besides the SMF, used to calibrate the subgrid feedback is the galaxy size -- mass relation (SMR), specifically the relation between galaxy stellar mass and the projected galaxy half-stellar mass radius at $z=0$. Fitting both observables simultaneously is challenging because there is tension between the two requirements \citep{Chaikin2025calibration}. Feedback not only suppresses star formation, it also tends to increase galaxy sizes at fixed mass \citep[e.g.][]{Sales2010}. The strong feedback needed to suppress the efficiency of galaxy formation to levels consistent with observation tends to result in galaxies becoming too large. The degree of complexity of the fiducial \colibre\ model for CCSN feedback, specifically the inclusion of a dependence of the injected energy on the pressure of the stellar birth cloud, is driven by the requirement to reproduce both the SMF and the SMR (\citealt{Chaikin2025calibration}, see \citealt{Crain2015} for a similar demonstration for the \eagle\ model). 

Fig.~\ref{fig:sizes} shows the median SMR for the three fiducial \colibre\ simulations L400m7 (red), L200m6
(orange), and L025m5 (blue). The orange shaded region indicates the scatter, i.e.\ the 16th to 84th percentile range, for L200m6. For clarity, we do not show the scatter for the other simulations, which is similar to that in L200m6. Galaxy sizes increase gradually with mass up to $M_* \sim 10^9\,\Msun$, but remain nearly constant between $10^9$ and $10^{10.5}\,\Msun$. The SMR steepens dramatically at $M_* \gtrsim 10^{11}\,\Msun$, i.e.\ when star formation is quenched (see \S\ref{sec:sfrs}) and mass growth is therefore expected to transition from being driven by in situ star formation to mergers. 

The convergence between the different resolutions is very good. While the m5 simulation predicts larger sizes for intermediate masses, the difference is $\lesssim 0.1$~dex. This discrepancy is probably smaller than the systematic effects due to differences in the way the sizes and masses are measured (possibly implying that we have overfitted the data) and is also small compared to the scatter. 

The simulations are compared with the observations from \citet{Hardwick2022}, which were used in the calibration of the model. The \citet{Hardwick2022} data points are based on S\'ersic fits in the $g$, $r$, and $i$ bands, converted to mass profiles based on the $r$-band magnitude and $g-i$ colours, for $\approx 1200$ stellar mass-selected galaxies at $0.01 < z < 0.05$ from the xGASS survey \citep{catinella2018}. The volumes of the L200m6 and L400m7 simulations are large enough to cover both the relatively flat part of the SMR below $M_* \approx 10^{10.5}\,\Msun$ and the relatively steep part at higher masses. The resolutions of L025m5 and L200m6 are sufficiently high to probe the relation below $10^9\,\Msun$, where they predict the SMR to steepen. The agreement of the median trends with the data is excellent over the full mass range spanned by the observations. The excellent agreement also applies to the scatter, which is indicated by the grey hatched region for the \citet{Hardwick2022} data. \citet{Ludlow2025} will show that the agreement with the observed sizes extends to other data sets, different redshifts, individual morphological classes, and different definitions of galaxy size. 

Finally, we note that the agreement with the data is slightly better than for \eagle, and much better than for IllustrisTNG (see~Appendix~\ref{app:eagle_tng}).

\subsubsection{The BH mass -- stellar mass relation} \label{sec:bh_masses}

\begin{figure}
    \centering
    \includegraphics[width=0.95\linewidth]{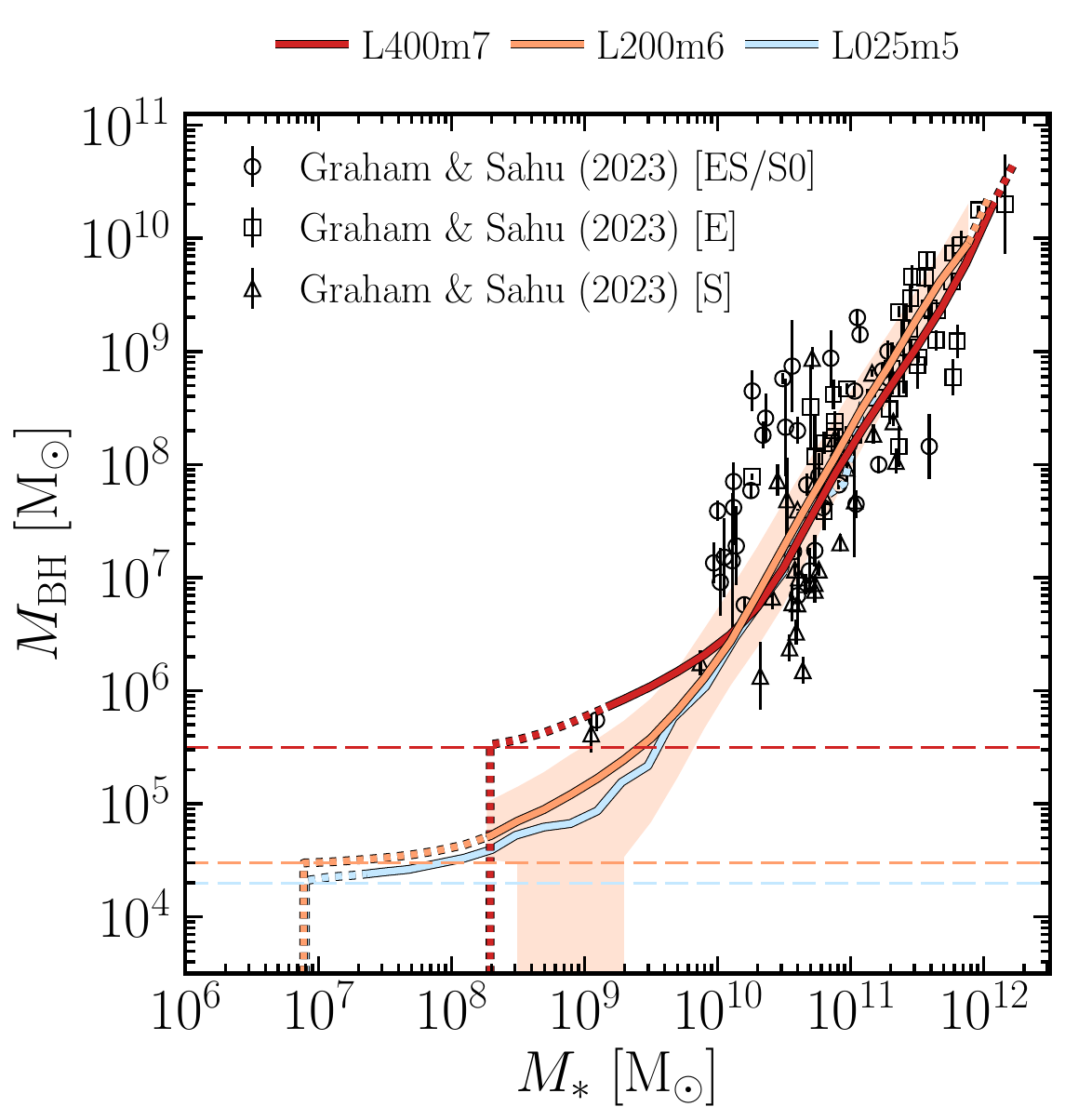}
    \caption{Median $z=0$ BH mass (i.e.\ the median mass of the most massive BH in the subhalo at $z=0$) as a function of galaxy stellar mass. The horizontal dashed lines indicate the BH seed mass, $m_\text{BH,seed}$, for the three simulation resolutions. The meanings of the line colours, line styles, and shading are as in Fig.~\ref{fig:sizes}. Data points show the compilation of direct BH mass measurements and stellar masses of local galaxies from \citet{Graham2023,Graham2024erratum}, with different symbols indicating different morphologies. For BH masses $\gg m_\text{BH,seed}$ the convergence is very good, as is the agreement between the median trend and the data.}
    \label{fig:bhs}
\end{figure}

The BH mass scales inversely with the assumed subgrid efficiency of AGN feedback, $\epsilon_\text{f}$, i.e.\ the fraction of the AGN luminosity that is coupled to the ISM \citep[e.g.][]{Booth2009,Booth2010}. Because cosmological simulations lack the resolution and physics to predict that efficiency, they cannot predict BH masses from first principles. The normalization of the observed relation between stellar mass and the mass of the supermassive BH in the galaxy nucleus, which is only determined robustly at high galaxy masses, $M_* \gg 10^{10}\,\Msun$, is therefore used to calibrate $\epsilon_\text{f}$. Because previous simulations using efficiencies of order 10 per cent gave reasonable agreement with the data \citep[e.g.][]{Booth2009,Schaye2015}, we assumed $\epsilon_\text{f}=0.1$ during the calibration of the m7 resolution using emulators described in \citet{Chaikin2025calibration}.  Although we did not fit to the observed BH masses, we would have changed the value of $\epsilon_\text{f}$ if there had been a strong disagreement with the data, and we halved the value to $\epsilon_\text{f}=0.05$ for the m6 and m5 resolutions to improve convergence. 

We note that in \colibre\ the subgrid AGN feedback efficiency can have a significant, indirect effect on galaxy properties. This is because the AGN feedback heating temperature, which is the main parameter influencing the actual efficiency of the AGN feedback on resolved scales, is proportional to the BH mass (see \S\ref{sec:agn_fb}). This situation differs from simulations that use a fixed AGN heating temperature, such as OWLS, \eagle,  and \flamingo, in which case galaxy properties other than the BH mass are nearly independent of $\epsilon_\text{f}$ \citep{Booth2010}.  

Fig.~\ref{fig:bhs} shows the median BH mass as a function of galaxy stellar mass at $z=0$. The median BH mass jumps from zero to the BH seed mass, which is indicated by a horizontal dashed line, at the stellar mass corresponding to the halo mass at which BHs are seeded. The BH seed mass depends on the resolution (see Table~\ref{tbl:subgrid_pars}) and is one of the four subgrid parameters included in the calibration of the m7 resolution using emulators. As the galaxy mass increases, the BHs lose memory of their seed mass, and the different resolutions converge. The BH seed masses hardly differ between m5 and m6, and the BH mass -- galaxy mass relations for these two resolutions are very similar over the full galaxy mass range. The low-resolution m7 simulation uses significantly larger values for the halo mass for BH seeding and the BH seed mass, but by $M_* = 10^{10}\,\Msun$ its BH masses agree very well with those in the higher-resolution simulations.

The simulations are compared with the compilation of dynamical BH mass measurements of nearby galaxies from\footnote{We corrected the measurements of \cite{Graham2023} following the erratum \cite{Graham2024erratum}.} \citet{Graham2023}, where data points  with different symbols indicate different galaxy morphologies. To facilitate a fair comparison, the stellar masses measured by \citet{Graham2023} under the assumption of a \citet{Kroupa2002} IMF have been converted to our \citet{Chabrier2003} IMF by increasing them by 0.05~dex \citep{Bernardi2010}. The agreement between the median trends and the data is excellent over the full mass range covered by the observations. However, because the scatter in the data is large and the number of measurements modest, the observations are not yet very constraining. At high mass the scatter in the simulations is smaller than for the data, but measuring BH masses is notoriously difficult, and the spread in the data may be affected by measurement errors including unrecognized systematics. 

In Appendix~\ref{app:eagle_tng} we show that the BH masses are in good agreement with those in the \eagle\ simulations. For $M_* > 10^{11}\,\Msun$ this is also the case for TNG, but at lower galaxy masses TNG predicts much higher BH masses.

\subsection{Comparison with selected other observations} \label{sec:obs_other}
We now proceed to compare the simulation predictions to observations that were not used to calibrate the subgrid models for CCSN and AGN feedback. Except for the evolution of the cosmic star formation rate density (\S\ref{sec:csfh}), we will restrict the discussion to the dependence of a range of fundamental, spatially integrated properties of $z\approx 0$ galaxies on stellar mass, leaving their evolution and spatial trends for future work. The galaxy properties considered here are specific star formation rates and quenched fractions (\S\ref{sec:sfrs}), atomic hydrogen and molecular gas masses (\S\ref{sec:cool_gas}), gas and stellar metallicites (\S\ref{sec:mzr}), dust masses and grain sizes (\S\ref{sec:dust_masses}), and CGM X-ray luminosities (\S\ref{sec:xray}).  

\subsubsection{The cosmic star formation history} \label{sec:csfh}
 
\begin{figure}
    \centering
    \includegraphics[width=0.95\linewidth]{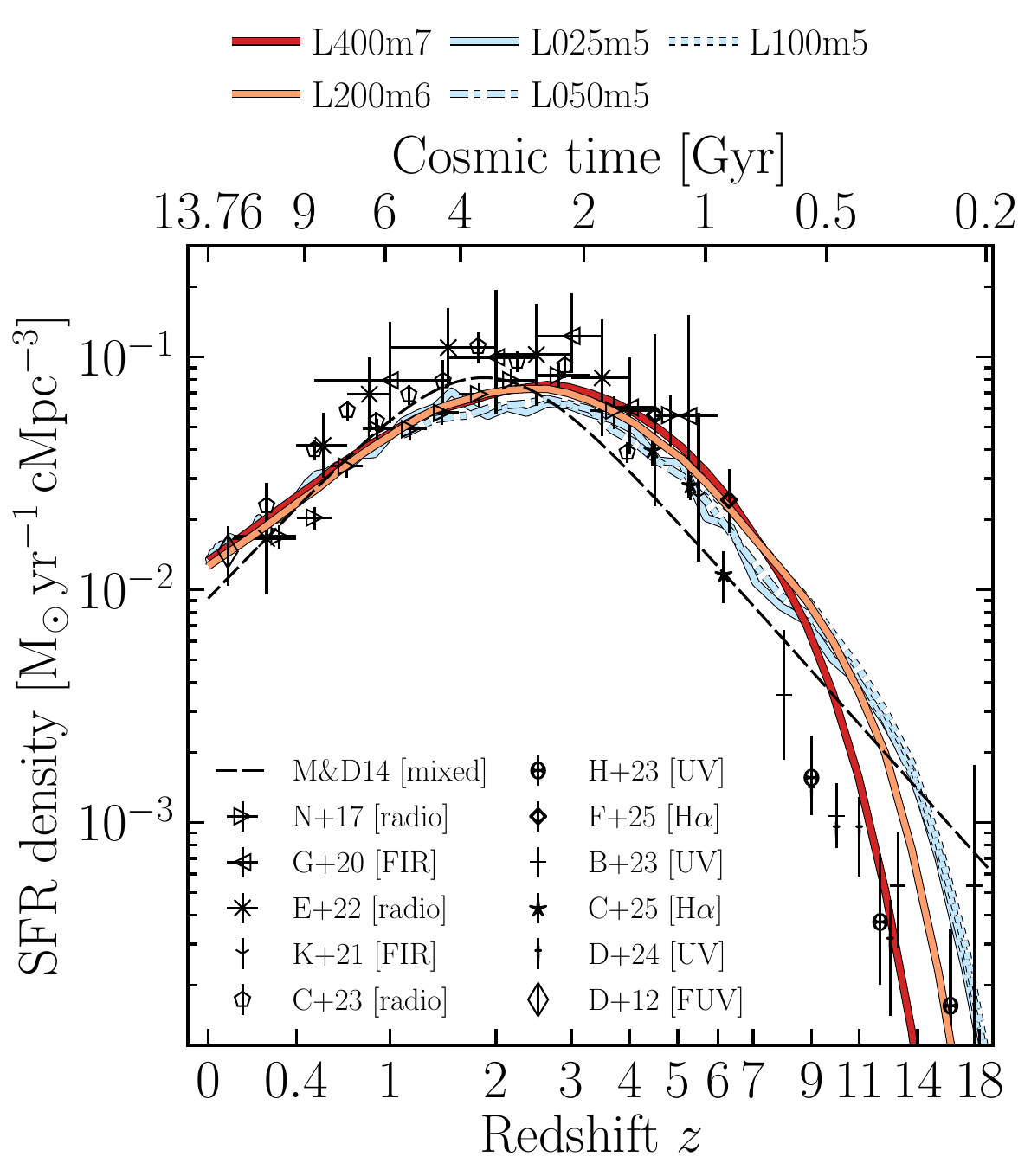}
    \caption{The evolution of the comoving cosmic star formation rate density. Besides L400m7 (solid red), L200m6 (solid orange) and L025m5 (solid blue), we show L050m5 and L100m5 down to their current redshifts ($z=1.1$ and 6, respectively). Data points show recent measurements based on rest-frame ultraviolet \citep{Driver2012, Bouwens2023, Harikane2023, Donnan2024}, H$\alpha$ \citep{Fu2025,Covelo2025}, far infrared \citep{Gruppioni2020, Khusanova2021}, and radio \citep{Novak2017, Enia2022, Cochrane2023} data, as well as the fit to a compilation of older observations from \citet{Madau2014}.  Except at very high redshift ($z>10$), where poorly resolved galaxies dominate and higher resolution thus gives higher star formation rate densities, the convergence between the m6 and m7 resolutions is excellent. The m5 resolution yields slightly lower SFR densities for $2 \lesssim z \lesssim 10$, but only by $\lesssim 0.1$~dex. The agreement with the data is very good. For $z \gtrsim 8$ the simulations overshoot most of the data, but \citet{Chaikin2025smf_evol} show that the agreement is excellent if we account for the UV magnitude limit to which those observational measurements have been normalized.}
    \label{fig:sfh}
\end{figure}

Fig.~\ref{fig:sfh} shows the evolution of the cosmic star formation rate (SFR) density, calculated by summing the instantaneous SFRs of all gas particles. Starting from high redshift, the SFR density initially increases rapidly, reaches a broad peak at $2 \lesssim z \lesssim 4$, and then declines by nearly an order of magnitude. 

At redshift $z > 11$ the SFR density increases systematically with resolution because it is dominated by poorly resolved galaxies. At $z<8$ L200m6 and L400m7 agree very well, but down to $z\approx 2$ the m5 resolution simulations are $\approx 0.1$~dex lower. For $z<2$ all resolutions converge to nearly identical results. The good agreement of L025m5 with L050m5 down to the redshift currently reached by L050m5, $z=1.1$, implies that the L025 volume is sufficiently large to obtain a converged cosmic SFR density at these redshifts. This is in agreement with the results of \citet{Schaye2010} for the OWLS simulations. Looking more closely, we see that while L100m5 agrees very well with the smaller volumes simulated at m5 resolution, for $6 < z < 9$ its SFR density is slightly higher, agreeing nearly perfectly with L200m6, and causing its curve to be hidden by the L200m6 one. However, as can be seen from fig.~A1 of \citet{Chaikin2025smf_evol}, this is due to the L100 initial conditions leading to a slightly higher SFR density at these redshifts than predicted for both smaller and larger volumes. 

The simulations are generally in good agreement with the compilation of SFR densities inferred from observations shown in Fig.~\ref{fig:sfh}. For $7< z < 10$ the simulations appear to overpredict the data, but \citet{Chaikin2025smf_evol} show that this is because of the UV magnitude limit down to which the observations were integrated. If we restrict the cosmic SFR density predicted by the simulations to include only the SFRs of galaxies above this limit, then the agreement with the data is very good also at these redshifts. 
For $1 < z < 3$ several observational studies prefer a slightly higher SFR density, by $\approx 0.2$ dex, though they are mostly consistent with the simulations within the error bars and some observations \citep[e.g.][]{Novak2017} prefer lower values. This tendency of the data at these redshifts to yield slightly higher SFRs than predicted also holds for other simulations and semi-analytic models \citep[e.g.][and references therein]{Leja2022}. This discrepancy was also present for the \eagle\ simulations \citep[e.g.][and Fig.~\ref{fig:eagle}]{Furlong2015} and TNG100 (see Fig.~\ref{fig:tng}), but \citet{Katsianis2020} showed that for \eagle\ it disappears if virtual observations of \eagle\ galaxies are analysed using observational techniques. A similar reduction in the observed SFRs is also required to obtain consistency between the time integral of the SFRs and the evolution of the cosmic stellar mass density obtained by integrating the observed SMF over mass \citep[e.g.][]{Shuntov2025}. 
We conclude that the cosmic star formation history predicted by \colibre\ is consistent with the data.

\subsubsection{The main sequence of star forming galaxies and quenched fractions} \label{sec:sfrs}

\begin{figure}
    \centering
    \includegraphics[width=0.95\linewidth]{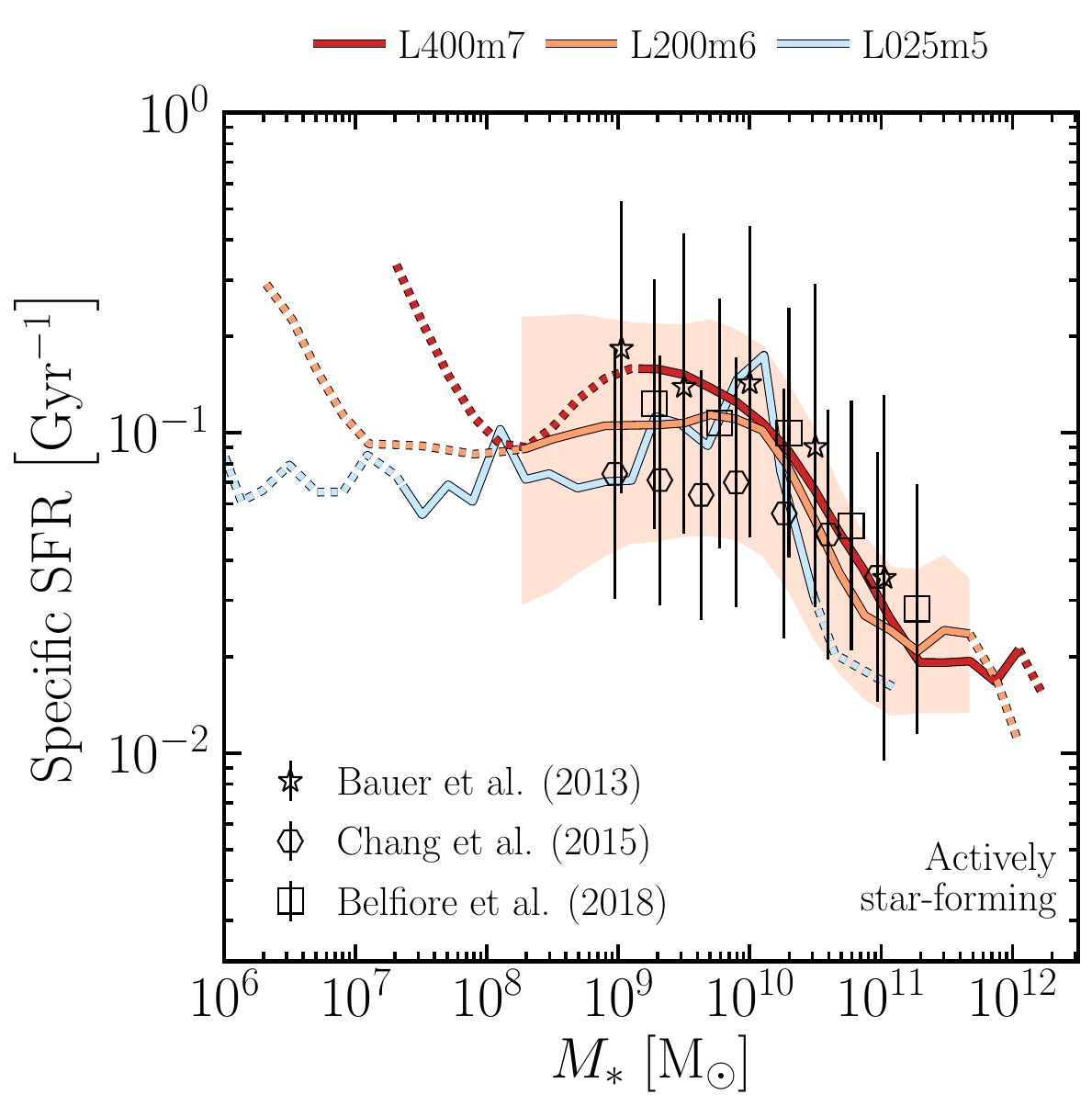} \hspace{0.05\linewidth}
    \caption{Median $z=0$ specific star formation rate (sSFR) as a function of galaxy stellar mass for star-forming galaxies, i.e.\ those with $\text{sSFR} > 10^{-2}\,\text{Gyr}^{-1}$. The meanings of the line colours, line styles, and shading are as in Fig.~\ref{fig:sizes}.  The data points show observations of star-forming galaxies from GAMA \citep{Bauer2013}, SDSS+WISE \citep{Chang2015}, and MANGA \citep{Belfiore2018}, with the error bars indicating the $1\sigma$ scatter. The convergence is excellent for galaxies containing $\gtrsim 10^3$ stellar particles, but even for $\sim 10^2$ particles the differences are $\lesssim 0.2$ dex and smaller than the scatter. The agreement with the data is excellent.}
    \label{fig:ssfr}
\end{figure}

Fig.~\ref{fig:ssfr} shows the $z=0$ relation between the specific star formation rate (sSFR) and stellar mass. Only galaxies that are actively star-forming are included, which we define as $\text{sSFR} > 10^{-2}\,\text{Gyr}^{-1}$. For $M_* \lesssim 10^{10}\,\Msun$ the relation remains relatively flat at $\text{sSFR}\sim 10^{-1}\,\text{Gyr}^{-1}$, which implies $\text{SFR} \sim 1~\Msun\,\yr^{-1} (M_*/10^{10}\,\Msun)$. The track followed by star-forming galaxies over the mass range for which the SFR is roughly proportional to mass is often referred to as the `main sequence'. At very low mass the curves turn up, but comparison of the different simulations shows that this upturn is a resolution effect. The curves bend down above $10^{10}\,\Msun$ and by $10^{11}\,\Msun$ they approach the selection limit of $\text{sSFR} > 10^{-2}\,\text{Gyr}^{-1}$. The convergence is excellent when galaxies are resolved with more than $10^3$ particles. For $10^2$ particles, which is more than sufficient to obtain converged stellar masses (see Fig.~\ref{fig:smhm}), differences of $\approx 0.2$~dex are seen, but these differences are small compared to the scatter, which is indicated for L200m6 by the shaded region. The simulations are compared with observations of star-forming galaxies at $z<0.15$ from MANGA \citep{Belfiore2018}, at $z<0.2$ from SDSS+WISE \citep{Chang2015} and at $0.05 < z < 0.32$ from GAMA \citep{Bauer2013}, selected as main sequence galaxies in the SFR -- $M_*$ diagram, selected in colour-colour space, and selected based on H$\,\alpha$ equivalent width, respectively. The predicted relations are consistent with the data.

\begin{figure}
    \centering
    \includegraphics[width=0.95\linewidth]{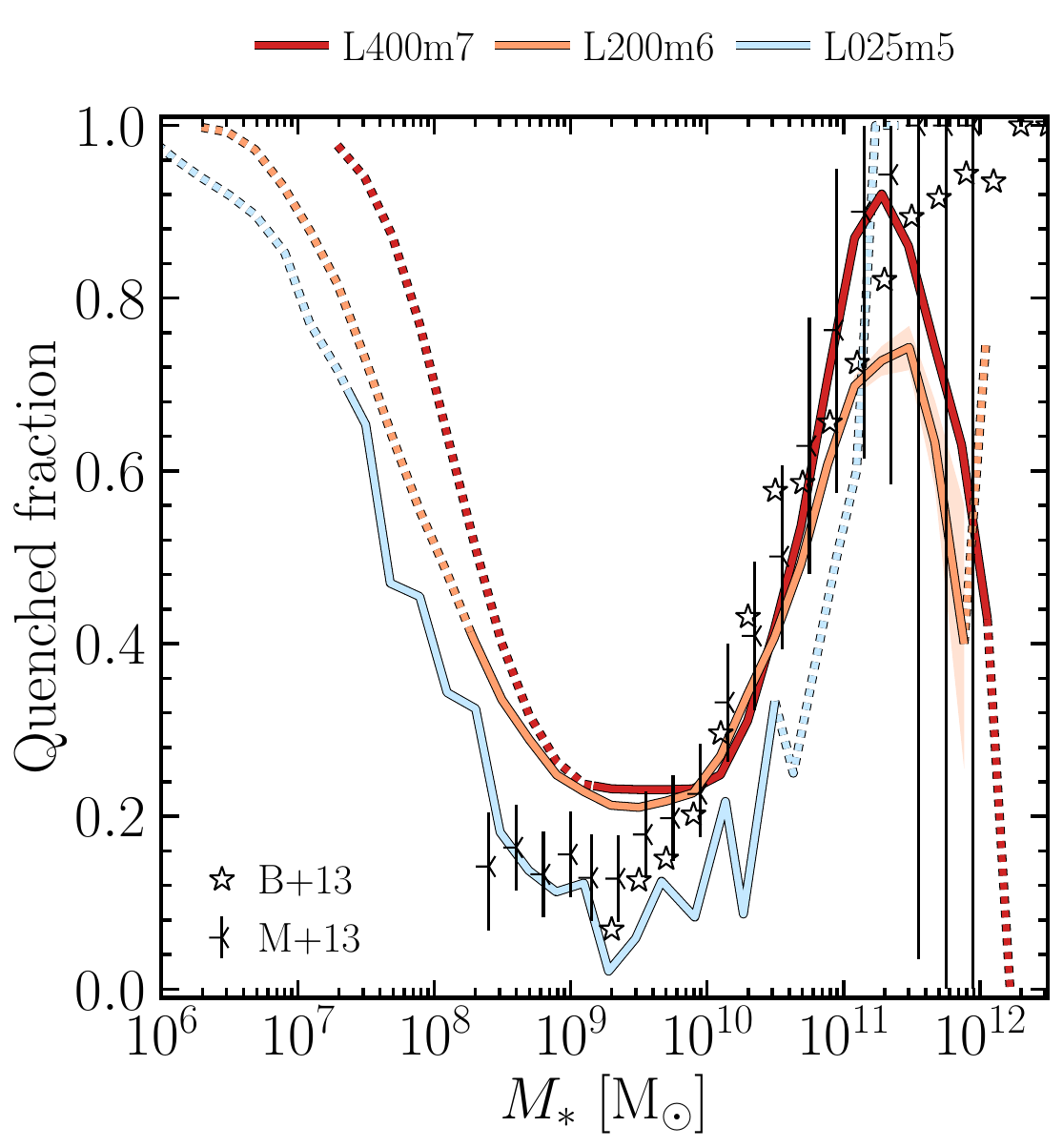}
    \caption{Fraction of $z=0$ galaxies that are quenched, i.e.\ $\text{sSFR} < 10^{-2}\,\text{Gyr}^{-1}$, as a function of galaxy stellar mass. The meanings of the line colours and line styles are as in Fig.~\ref{fig:sizes}. The orange shaded region indicates the Clopper–Pearson 68 per cent confidence interval for L200m6. The data points show observations from GAMA \citep{Bauer2013}, which use the same definition of quenched as applied to the simulations, and from the COSMOS/UltraVISTA Survey \citep{Muzzin2013} at $0.2 < z < 0.5$, for which quenched galaxies are defined based on UVJ colours. The convergence is good in the mass range over which the quenched fraction increases rapidly, particularly between m6 and m7. It is less good for L025m5, but this is mostly a box size effect, which leads to smaller fractions and lower host masses of satellites in L025m5. \citet{Chaikin2025smf_evol} demonstrate that the convergence is excellent for centrals, but that the quenched fraction decreases with resolution for satellites. The agreement with the data is excellent at intermediate masses, but there are hints that the quenched fractions may be too large for $M_* \lesssim 10^9~\Msun$ and too small for $M_* \sim 10^{12}~\Msun$. However, the sSFRs of very high-mass `star-forming' galaxies are only a factor of $\approx 2$ above the critical value (see Fig.~\ref{fig:ssfr}) and the predicted high-mass downturn/dip of the quenched fraction disappears if we measure the star formation rate in an aperture of 10~kpc or if we add 0.3~dex lognormal scatter to the stellar masses.}
    \label{fig:fquenched}
\end{figure}

Fig.~\ref{fig:fquenched} shows the mass dependence of the fraction of galaxies that are excluded from Fig.~\ref{fig:ssfr}, i.e.\ the fraction of $z=0$ galaxies in which star formation is quenched, $\text{sSFR} < 10^{-2}\,\text{Gyr}^{-1}$, and are thus evolving passively. The passive fraction increases rapidly between $M_*\sim 10^{9.5}\,\Msun$ and $10^{11}\,\Msun$. The convergence between L200m6 and L400m7 is excellent in this regime, but L025m5 appears to predict lower quenched fractions below $10^{10}\,\Msun$. \citet{Chaikin2025smf_evol} show that for central galaxies the quenched fractions in L025m5 and L025m6 are converged, but that for satellites with $M_*\lesssim 10^9\,\Msun$ the quenched fraction is lower for L025m5. For $M_* > 10^9\,\Msun$, the difference visible in Fig.~\ref{fig:fquenched} is mostly due to the small simulation volume of L025m5, which biases the satellite fraction and the host masses of satellite galaxies to be low. Because the passive fraction of galaxies with $M_*\ll 10^{11}\,\Msun$ depends mostly on environment \citep[e.g.][]{Peng2010}, this leads to the quenched fraction being biased low. 

Quenched fractions increase with decreasing mass for $M_* <10^{9}\,\Msun$. At the lowest masses, where the sSFR of actively star-forming galaxies turns up in Fig.~\ref{fig:ssfr}, the quenched fractions are boosted by a resolution effect. At such low masses even a single star-forming gas particle can push the galaxies above the main sequence. Because the stellar masses are calibrated to fit the data, these excessively high sSFRs must be compensated for, which implies that most galaxies must have an SFR of zero. The low-mass upturn in the quenched fractions is partly caused by an increase in the satellite fraction towards lower masses, although \citet{Chaikin2025smf_evol} show that there is also an upturn for central galaxies. 

The passive fractions decline towards $M_* \gtrsim 10^{11.5}\,\Msun$, indicating that AGN feedback does not keep the most massive galaxies fully quenched. In Section~\ref{sec:hybrid} we will show that this is also true for the simulations using hybrid AGN feedback. The difference between the m6 and m7 resolutions is greater than at lower masses, with the higher-resolution simulation yielding smaller quenched fractions. This reflects the greater sensitivity to resolution of AGN feedback compared to stellar feedback caused by the increase of the typical density of the gas in which AGN feedback energy is injected with resolution (see \S\ref{sec:birth_properties}).

The quenched fractions are generally in good agreement with the $z\approx 0.1$ observations from the GAMA survey \citep{Bauer2013}, which were converted to our $\text{sSFR} < 10^{-2}\,\text{Gyr}^{-1}$ criterion by \citet{Behroozi2019}, and with the measurements of \citet{Muzzin2013}, who apply UVJ colour selection to $z\approx 0.3$ galaxies from the COSMOS/UltraVISTA survey. There is no sign of an upturn in the data at low masses, $M_* < 10^9\,\Msun$, although this regime is only probed by \citet{Muzzin2013} and selection effects may lead to an underestimate of the passive fraction \citep[e.g.][]{Kaviraj2025}. In contrast to \colibre, at high mass the observed quenched fraction does not turn over. This may suggest that the AGN feedback in the fiducial model is insufficiently effective, though we note that at these masses, $M_* \gtrsim 10^{11.5}\,\Msun$, the star-forming galaxies in the simulation have sSFRs only slightly above the threshold of $\text{sSFR} = 10^{-2}\,\text{Gyr}^{-1}$ (see Fig.~\ref{fig:ssfr}) and that the quenched fraction increases again for $M_*>10^{12}\,\Msun$. Furthermore, the downturn/dip in the sSFR disappears if we measure the SFR in an aperture of $\lesssim 10$~kpc, which may be appropriate for fibre-based measurements, or if we add $\ge 0.3$~dex lognormal scatter to the mass measurements (see Fig.~\ref{fig:edd_bias}), which may be more realistic for BCGs than our fiducial scatter of 0.1~dex.
We conclude that while the agreement with the data is excellent for $10^9\lesssim M_*/\Msun \lesssim 10^{11.5}$, there may be discrepancies outside of this mass range. A more careful comparison with the data, using virtual observations, is necessary to establish whether the quenched fractions are too high for $M_* < 10^9\,\Msun$ and too low for $M_* > 10^{11.5}\,\Msun$.

In Appendix~\ref{app:eagle_tng} we show that the observed main sequence is similarly well reproduced by the \eagle\ and TNG simulations. At m5 resolution quenched fractions are similar in \eagle\ and \colibre, but at m6 resolution \eagle\ fits the data somewhat less well. For $M_* \lesssim 10^{10}\,\Msun$ both the numerical convergence and the level of agreement with the data are similar for TNG and \colibre, but at higher masses both are generally worse for TNG.  

\begin{figure*}
    \centering
    \includegraphics[width=0.45\linewidth]{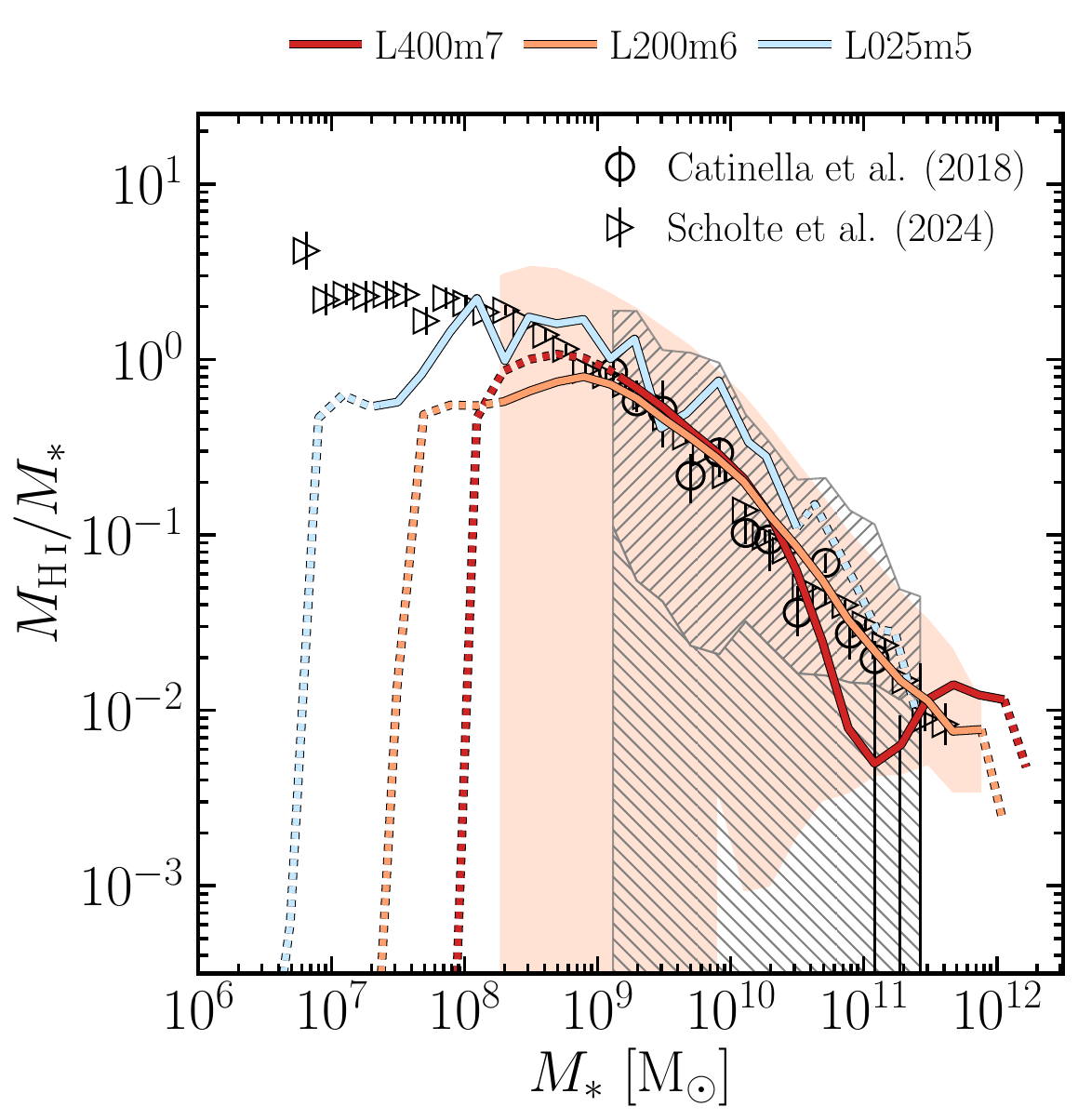}\hspace{0.05\linewidth}
    \includegraphics[width=0.45\linewidth]{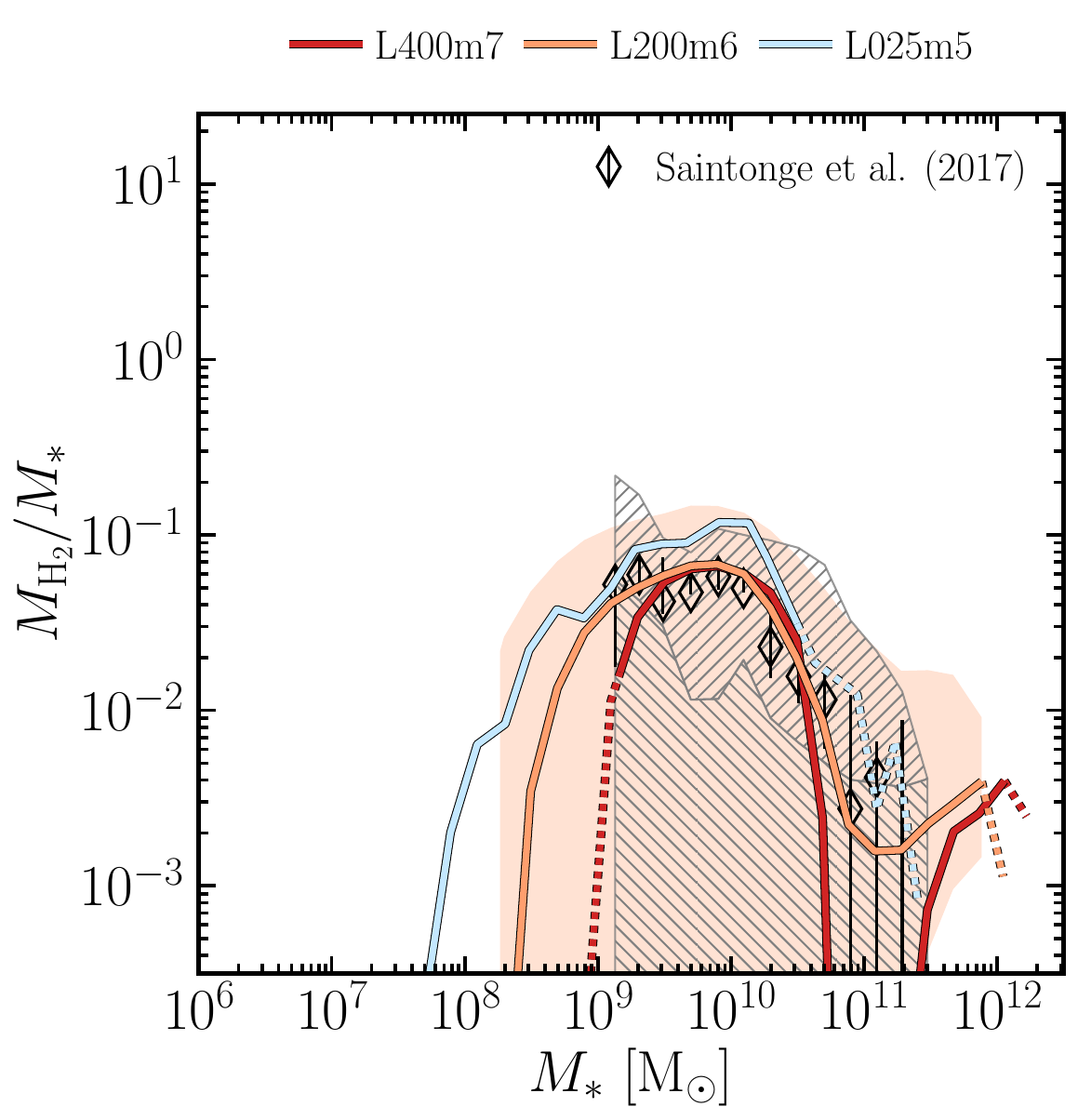}
    \caption{Median ratios of \ion{H}{i} to stellar mass (left panel) and molecular hydrogen to stellar mass (right panel) as a function of galaxy stellar mass at $z=0$. The meanings of the line colours, line styles, and shading are as in Fig.~\ref{fig:sizes}. The data points show $z\approx 0$ observations of \ion{H}{i} from xGASS \citep{catinella2018} and from ALFALFA \citep{Scholte2024}, and of CO from xCOLD GASS, converted to molecular hydrogen masses \citep{Saintonge2017}. These are all stellar-mass selected surveys. The upper grey hatched regions (striking towards the top right) indicate the 16th to 84th percentile scatter in the xGASS and xCOLD GASS surveys with non-detections set to their upper limits. The lower grey hatched regions (striking towards the bottom left) extend the 16th percentiles to their lower limit, obtained by setting non-detections to zero; they therefore indicate the uncertainty on the downward scatter. The convergence is good for intermediate masses, but at the low-mass end the medians turn over at lower masses for higher resolutions. Convergence requires more particles for H$_2$ than for \ion{H}{i}. For $M_* \sim 10^{11}~\Msun$ the differences in the medians between the different resolutions are larger, particularly for H$_2$, but the differences remain small compared to the scatter. The agreement with the data is excellent. The predicted downward scatter (shown only for L200m6) is very large, with the 16th percentile corresponding to negligible gas masses for all but the highest stellar mass bins. This large scatter is fully consistent with the observations.}
    \label{fig:gas}
\end{figure*}

\subsubsection{Atomic and molecular gas} \label{sec:cool_gas}
The fractions of the baryonic mass that are atomic and molecular provide a probe of the warm and cold phases of the ISM. Unlike previous large-volume simulations, such as \eagle, which do not directly model the cold phase, these fractions are directly predicted by \colibre. They therefore do not need to be estimated by post-processing the simulation output using prescriptions whose applicability is doubtful if the gas distribution is affected by the use of a temperature/pressure floor. Furthermore, \colibre\ also predicts the dust content, which is critical given that the formation of molecules occurs predominantly on the surfaces of dust grains and that dust has a strong effect on the thermal balance of the ISM. The direct modelling of dust grains, the coupling between dust physics and chemistry, and the non-equilibrium treatment of hydrogen and helium set \colibre\ apart also from the few cosmological simulations that allow cooling to cold ISM temperatures, such as FIREbox and NewHorizon. 

However, it is important to note that while the resolution of \colibre\ suffices to distinguish warm ($T\sim 10^4\,\K$) and cold ($T\ll 10^4\,\K$) ISM phases, it is insufficient to model the internal structure of molecular clouds. Because dust grain growth occurs predominantly in molecular cloud cores, the predicted dust content would be significantly reduced if the simulation did not boost the rates of gas accretion on to grains and grain coagulation using a subgrid clumping factor (see \S\ref{sec:dust} and \citealt{Trayford2025}). Note that clumping factors are also required by the high-resolution simulations of individual galaxies on which the prescriptions for estimating molecular fractions that are employed by many cosmological simulations are based. For example, \citet{Gnedin2011} boost the growth rate of H$_2$ on dust grains by a factor of 30. Although we do not boost the H$_2$ growth rate, for low-mass galaxies ($M_* < 10^{10}\,\Msun$ at m6 resolution), the galaxy-scale molecular fraction would be much reduced without the use of a clumping factor for dust grain growth. The atomic mass fraction, on the other hand, is insensitive to the use of this clumping factor.

Fig.~\ref{fig:gas} shows the median atomic (left panel) and molecular hydrogen (right panel) mass relative to the stellar mass as a function of stellar mass. The sharp drops at low mass to median fractions of (nearly) zero are clearly a resolution effect because they shift to lower masses for higher resolution. For molecular hydrogen, the different resolutions begin to diverge when the line styles are still solid, indicating that (many) more than 100 stellar particles are required for convergence of the medians. However, given that the scatter (indicated by the shaded region for m6 resolution) is extremely large, which reflects the bimodality of the distribution, the median is more sensitive to resolution than the mean (not shown). At intermediate masses, $10^9 \lesssim M_*/\Msun \lesssim 10^{10.5}$, convergence between m6 and m7 is excellent. The high-resolution simulation predicts slightly higher gas fractions, but the differences are small compared to the $1\sigma$ scatter. At higher masses, where the quenched fraction becomes large (see Fig.~\ref{fig:fquenched}), the differences increase, though they remain smaller than the scatter. In this regime the convergence is much better, and the dips in the gas fractions are much smaller, for the means (not shown) than for the medians, which again reflects the bimodal distributions of the gas fractions. Indeed, for $M_*\sim 10^{11}\,\Msun$, the increase in the gas fractions with increasing resolution mirrors the decrease in the quenched fractions with resolution (see Fig.~\ref{fig:fquenched}). 

In the mass range for which the convergence is good, the gas fractions decrease with the stellar mass. Comparing the two panels, we see that at low mass, $M_* < 10^9\,\Msun$, the molecular fraction is very small compared to the atomic fraction, but that they become comparable at higher masses. The increase in the H$_2$ fraction with stellar mass for $M_*\gtrsim 10^{11.5}\,\Msun$ mirrors the decrease in the quenched fraction (see Fig.~\ref{fig:fquenched}).

The simulation predictions are compared with observations of \ion{H}{i} from the xGASS \citep{catinella2018} and ALFALFA \citep{Scholte2024} surveys, and of CO from the xCOLD GASS survey \citep{Saintonge2017}. Each of these surveys uses a sample of stellar mass-selected galaxies. \citet{Saintonge2017} convert CO luminosity to H$_2$ mass using the multivariate conversion factor of \citet{Accurso2017}. \citet{Saintonge2017} correct for the presence of helium by multiplying the H$_2$ masses by a factor of 1.36, but we have undone that correction. We note that the conversion from CO to H$_2$ introduces a large systematic uncertainty, a factor $\approx 2$ for galaxies such as the Milky Way and a much larger factor for low-metallicity galaxies \citep[e.g.][]{Bolatto2013}. For \citet{catinella2018} and \citet{Saintonge2017}, who published the data for individual galaxies, we have computed the medians (and associated errors using bootstrap resampling), and the 16th to 84th percentile scatter ourselves. Because the fraction of upper limits is substantial (between 16 and 50 per cent for all mass bins), we computed the 16h percentile in two ways: once treating upper limits as detections (16th to 84th percentile scatter indicated by the top grey hatched regions) and once setting them to zero (indicated by the combination of the two hatched regions), in which case the 16h percentiles are zero for all mass bins. The true 16th percentile must be in between these two limiting cases.

For m6 resolution, the predicted median atomic masses are fully consistent with the data from $M_* \sim 10^9\,\Msun$ all the way to the most massive observed galaxies ($M_* \approx 10^{11.8}\,\Msun$). The agreement is similarly good for m7 resolution, except at $M_* \sim 10^{11}\,\Msun$, where the median \ion{H}{i} mass is consistent with \citet{catinella2018} but lower than that measured by \citet{Scholte2024}. At m5 resolution the median atomic masses are $\approx 0.1$~dex too large, but this discrepancy is much smaller than the scatter. For both m6 and m7 resolution the median molecular masses are in excellent agreement with the data over the full observed mass range of $10^9 < M_*/\Msun < 10^{11.3}$. At m5 resolution the H$_2$ masses are slightly higher, typically by $\approx 0.1$~dex, which is again a small difference compared to the scatter. The excellent agreement extends to the scatter, which is fully consistent with both the atomic and molecular gas data over the full observed stellar mass range.

\subsubsection{The mass -- metallicity relation} \label{sec:mzr}

In the publication presenting the \colibre\ model for chemical enrichment, \citet{Correa2025chemo} already compare results from small-volume simulations, L025m6 and L025m7, with a number of observables probing metal abundances at low redshift. Here we will present the $z=0$ relations between stellar mass and both gas and stellar metallicities, finding results consistent with \citet{Correa2025chemo}. We leave evolution, spatially resolved trends, and relative abundances for future work, though we note that \citet{Correa2025chemo} already showed that the small-volume simulations reproduce the observed trend of alpha enhancement with stellar mass for low-$z$ galaxies as well as abundance ratios of individual elements as a function of metallicity observed for stars in the Milky Way Galaxy. 

\begin{figure}
    \centering
    \includegraphics[width=0.95\linewidth]{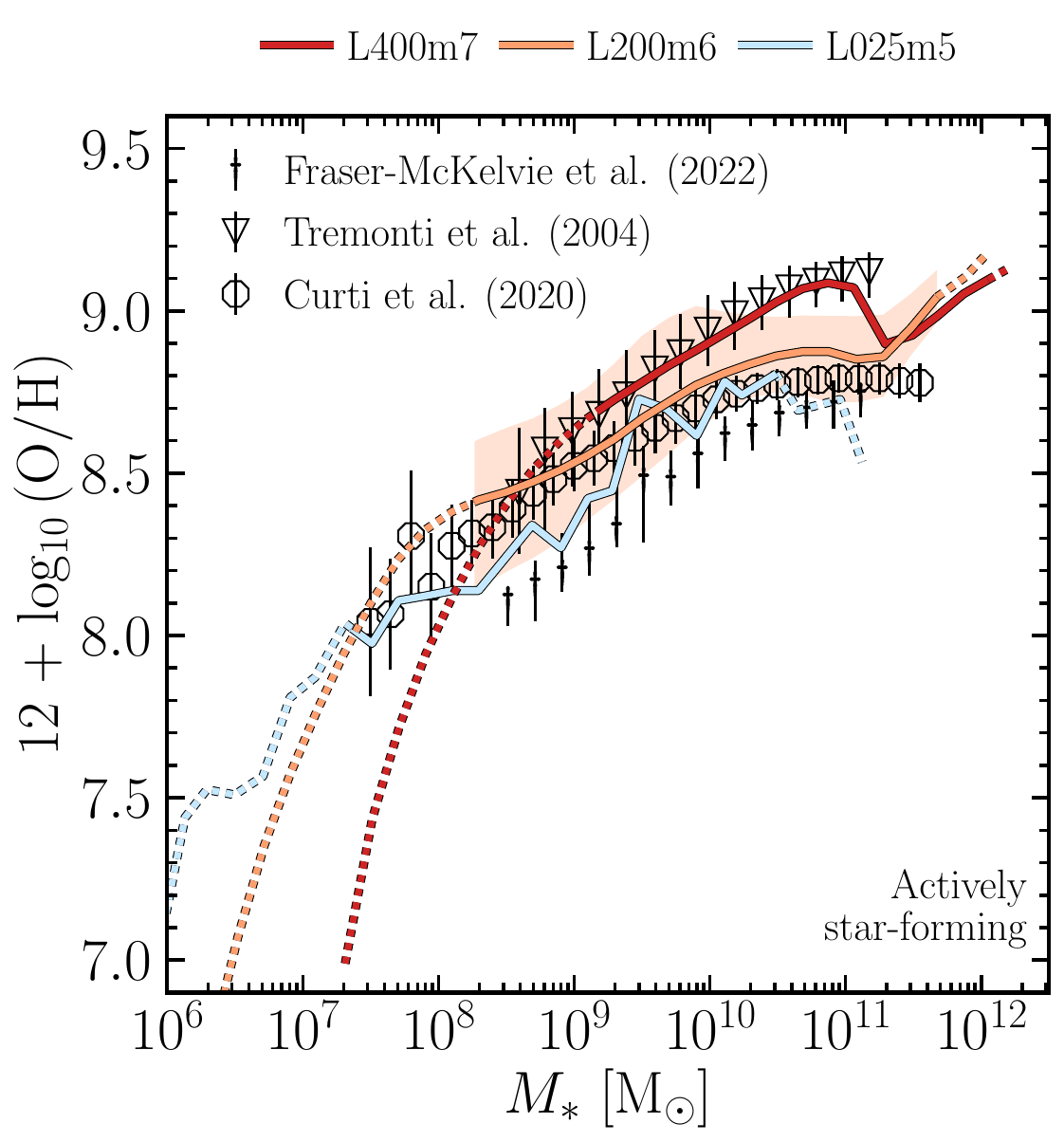}
    \caption{Median $z=0$ gas-phase oxygen abundance as a function of galaxy stellar mass. Only star-forming galaxies, i.e.\ $\text{sSFR} > 10^{-2}~\text{Gyr}^{-1}$, are included. The metallicity is computed as the ratio of the total number of gas-phase oxygen nuclei, i.e.\ excluding oxygen depleted onto dust grains, to the total number of hydrogen nuclei, including only gas with density $n_\text{H} > 0.1~\cm^{-3}$ and temperature $T<10^{4.5}~\K$. The meanings of the line colours, line styles, and shading are as in Fig.~\ref{fig:sizes}. The data points show $z\approx 0.1$ observations from \citet{fraser-mckelvie2022} (based on strong-line diagnostics calibrated on $T_\text{e}$ measurements), \citet{Tremonti2004} (based on strong lines and photoionization models), and \citet{curti2020} (based on strong-line diagnostics calibrated on $T_\text{e}$ measurements). The sharp down-turns at the low-mass end are clearly a resolution effect. For better-sampled galaxies higher resolutions give systematically lower metallicities, with the possible exception of the most massive galaxies. The difference between m7 and m5 is $\lesssim 0.2$~dex, with m6 falling in between. Given the systematic differences between different data sets, all simulations are consistent with the observations.}
    \label{fig:zgas}
\end{figure}

Fig.~\ref{fig:zgas} shows the median oxygen abundance in the ISM as a function of galaxy stellar mass, where the ISM is defined as gas with density $n_\text{H} > 0.1~\cm^{-3}$ and temperature $T<10^{4.5}~\K$. The oxygen abundance, O$/$H, is computed as the ratio of the total number of gas-phase oxygen nuclei, i.e.\ excluding oxygen depleted onto dust grains, to the total number of hydrogen nuclei. Because measurements of gas metallicities are based on observations of \ion{H}{ii} regions, only star-forming galaxies, i.e.\ $\text{sSFR} > 10^{-2}~\text{Gyr}^{-1}$, are included. 

The gas metallicity increases with mass, but the relation becomes less steep with increasing mass. The convergence with resolution does not appear to be as good as for the observables considered so far. While the resolution-dependent sharp drop below stellar masses corresponding to $\sim 10^2$ particles is similar to the resolution effects seen for other observables, differences similar in magnitude to the scatter between objects (indicated by the shaded region for L200m6) remain up to $M_*\sim 10^{11}\,\Msun$, with higher-resolution simulations tending to predict lower metallicities. However, the differences are actually quite small. For $10^9 \lesssim M_*/\Msun \lesssim 10^{11}\,\Msun$ the difference between m7 and m5 is $\approx 0.2$~dex and the difference between m6 and m5 is only $\approx 0.1$~dex. For higher masses, which are not sampled at m5 resolution, convergence between m7 and m6 is very good. We note that \citet{Correa2025chemo} show that the slope of the relation depends somewhat on the assumed value of the turbulent diffusion coefficient, with stronger diffusion resulting in lower gas metallicities in massive galaxies. 

The simulation predictions are compared with three sets of $z\approx 0.1$ observations. \citet{fraser-mckelvie2022} used strong lines calibrated using the electron-temperature ($T_\text{e}$) method to measure the oxygen abundance within one effective radius for 472 star-forming galaxies from the SAMI survey. \citet{Tremonti2004} used comparisons of strong lines with photoionization models to measure the metallicity within the 3~arcsec fibre for $\approx 5\times 10^4$ galaxies from the SDSS survey. \citet{curti2020} used strong-line diagnostics calibrated on $T_\text{e}$ measurements of individual low-metallicity and stacks of high-metallicity galaxies to measure fibre metallicities for $\approx 1.5\times 10^5$ SDSS galaxies. The spread between the different observations reflects well-known systematic uncertainties, with the $T_\text{e}$-calibrated methods giving systematically lower metallicities than the theoretically-calibrated strong line diagnostics. Given these uncertainties, all simulations are consistent with the data. The m7, m6, and m5 resolutions agree particularly well with \citet{Tremonti2004}, \citet{curti2020}, and \citet{fraser-mckelvie2022}, respectively, but each simulation is inconsistent with at least one of these data sets.

\begin{figure*}
    \centering
    \includegraphics[width=0.45\linewidth]{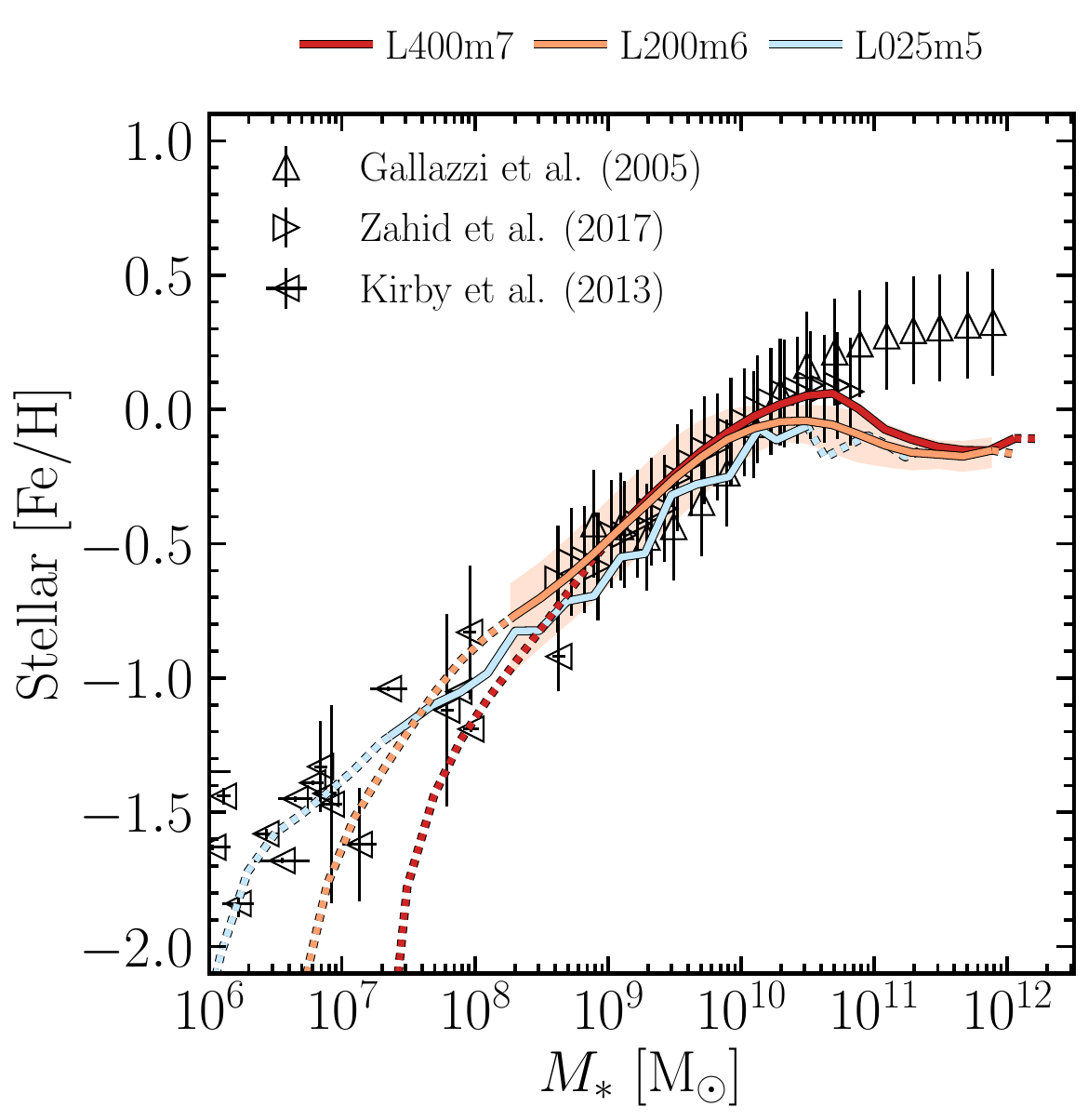}\hspace{0.05\linewidth}
    \includegraphics[width=0.45\linewidth]{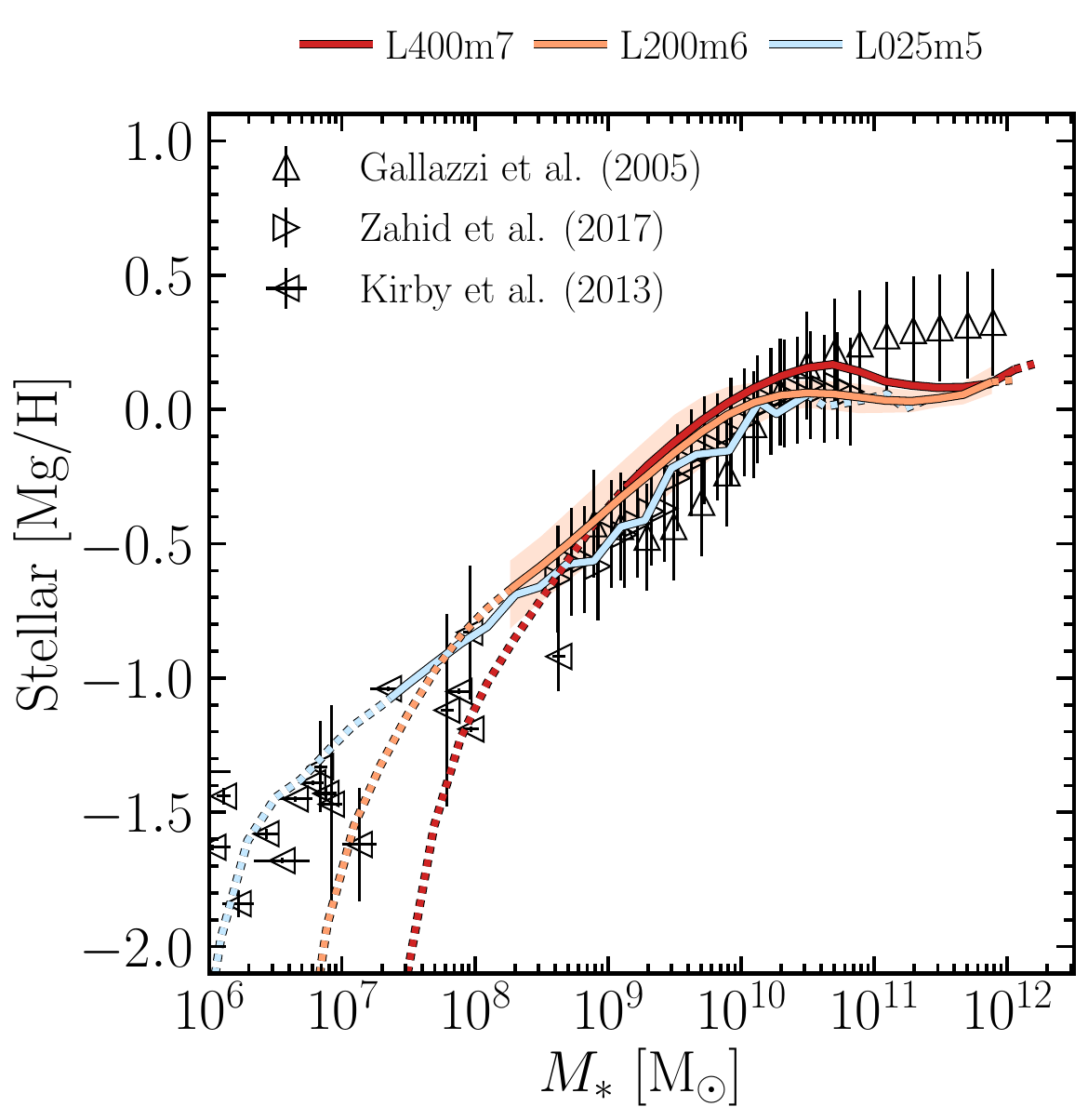}
    \caption{Median $z=0$ stellar iron (left panel) and magnesium (right panel) abundance as a function of galaxy stellar mass. The meanings of the line colours, line styles, and shading are as in Fig.~\ref{fig:sizes}. The data points show observations of \citet{Gallazzi2005} and \citet{Zahid2017}, both for SDSS galaxies measured within the 3~arcsec SDSS fibres,
    and from \citet{Kirby2013} for individual dwarf satellite galaxies of the Milky Way and M31. The observed metallicities in solar units are assumed to be identical for iron and magnesium.
    All abundances are relative to the solar abundances from \citet{Asplund2009}. The downturn below masses corresponding to $\sim 10$ stellar particles is clearly a resolution effect. For larger particle numbers the differences between m7 and m5 are $\lesssim 0.1$~dex. The agreement with the data is generally very good. For massive galaxies, $M_* \gtrsim 10^{11}~\Msun$, the observed metallicity is higher than the predicted iron abundance, but for magnesium the difference is only $\lesssim 0.2$~dex and consistent within the errors.}
    \label{fig:zstars}
\end{figure*}

The relation between stellar metallicity and galaxy stellar mass is shown in Fig.~\ref{fig:zstars}. In contrast to the gas metallicities, these do not depend directly on the dust content. Because of that, and because they are effectively time averages of the metallicity of star-forming gas, predictions for stellar metallicities are more robust than predictions for gas metallicities. As observations typically use diagnostics that are sensitive to iron and magnesium, we show [Fe$/$H] in the left panel and [Mg$/$H] in the right panel, where square brackets indicate $\log_{10}$ of the abundance in solar units ($12+\log_{10}(\text{Fe}/\text{H})_\odot = 7.5$ and $12+\log_{10}(\text{Mg}/\text{H})_\odot = 7.6$; \citealt{Asplund2009}). The Mg metallicity typically exceeds that of Fe, particularly for high-mass galaxies, $M_*\gtrsim 10^{11}\,\Msun$. Such high-mass $\alpha$-enhancement is observed \citep{Thomas2010} and is consistent with the prediction from the \eagle\ simulations, where it is caused by more massive galaxies forming their stars earlier and more rapidly as a result of AGN feedback \citep{Segers2016alpha}. The stellar mass -- metallicity relation resembles that of the gas metallicity (see Fig.~\ref{fig:zgas}), though the decrease in the slope towards higher masses is more pronounced, even becoming negative at $M_*\sim 10^{11}\,\Msun$.

The convergence with numerical resolution is excellent, better than for the gas metallicity, with differences between m7 and m5 $\lesssim 0.1$~dex for galaxies resolved with $\gtrsim 10$ stellar particles. 

We compare with observations from \citet{Gallazzi2005} for $\sim 10^5$ redshift $z\approx 0.1$ SDSS galaxies, measured within the 3~arcsec SDSS fibre, of \citet{Zahid2017} for $2\times 10^5$ redshift $z<0.25$ star-forming SDSS galaxies, also measured within the 3~arcsec SDSS fibre, and from \citet{Kirby2013} for individual dwarf irregular and spheroidal satellite galaxies of the Milky Way and M31. Following \citet{Yates2021}, we assume 0.2~dex metallicity errors for \citet{Zahid2017}. The observed metallicities in solar units are taken to be identical for Fe and Mg. For stellar masses corresponding to $\gtrsim 10$ stellar particles up to $M_*\sim 10^{11}\,\Msun$ the simulations are fully consistent with the data. Higher masses are only probed by \citet{Gallazzi2005}. For $M_*\sim 10^{12}\,\Msun$, they find a typical metallicity that is 0.4~dex higher than the predicted [Fe$/$H], but the predicted [Mg$/$H] is consistent with the data within the error bars. The strong dependence on the chosen element, and the fact that aperture effects and Eddington bias (see Fig.~\ref{fig:edd_bias}) can be important at these very high masses, warrant a like-for-like analysis using virtual observations, which we leave for future work. 

In Appendix~\ref{app:eagle_tng} we show that both \eagle\ and IllustrisTNG predict much shallower mass -- metallicity relations than \colibre, resulting in too high metallicities for $M_* < 10^{10}\,\Msun$, particularly at m5 and m6 resolutions for TNG and \eagle, respectively. However, for $M_* \gtrsim 10^{11.5}\,\Msun$ TNG agrees better with the observations of \citet{Gallazzi2005} than \colibre\ does. 

\begin{figure}
    \centering
    \includegraphics[width=0.95\linewidth]{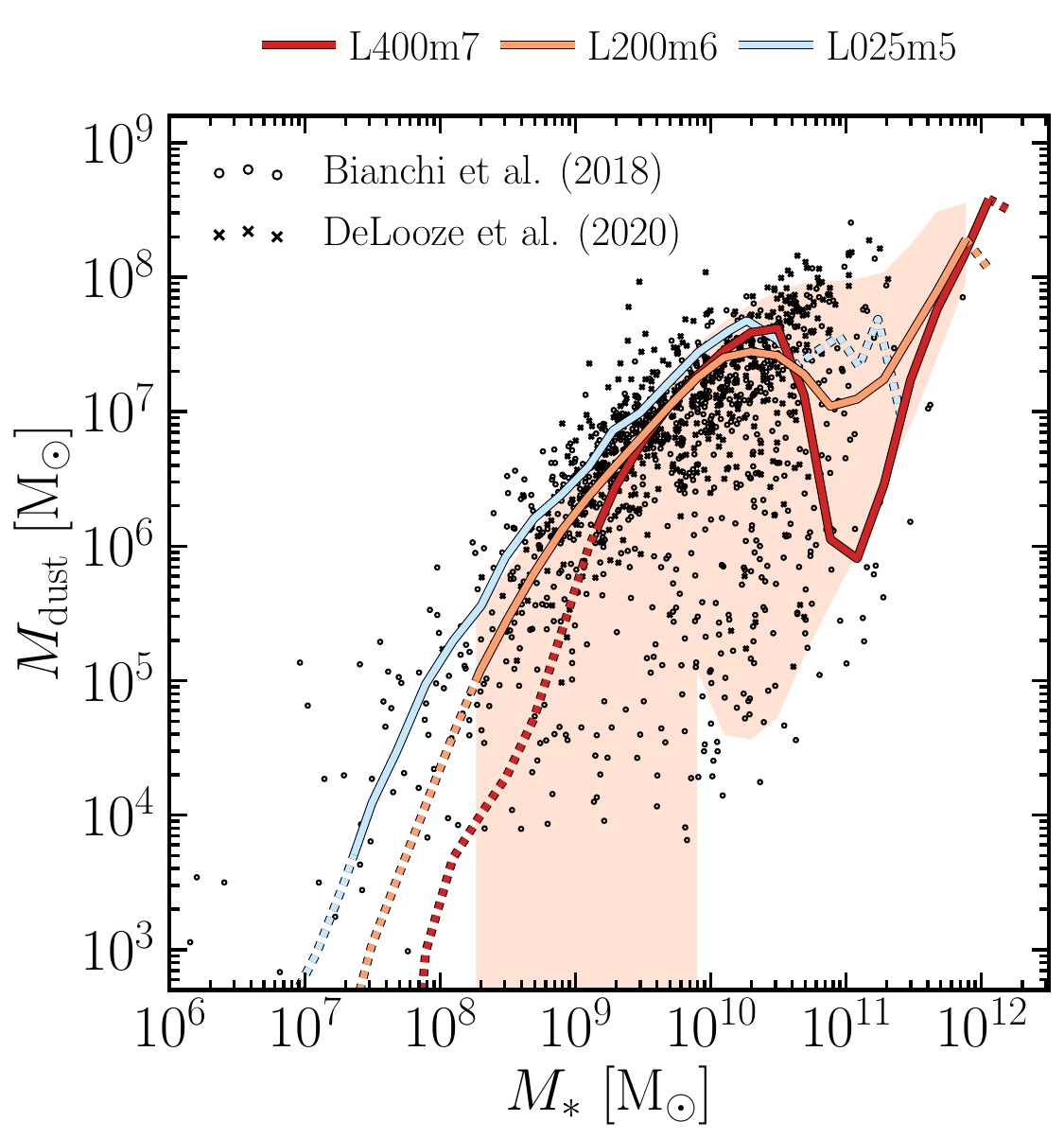}
    \caption{Median $z=0$ dust mass as a function of galaxy stellar mass. The meanings of the line colours, line styles, and shading are as in Fig.~\ref{fig:sizes}. The small open circles indicate observations from \textit{DustPedia} taken from \citet{Bianchi2018} and the small crosses indicate the compilation of survey data from \citet{DeLooze2020}. None of the surveys are mass-selected. At low masses, corresponding to $\ll 10^3$ particles, and at $M_* \sim 10^{11}~\Msun$, higher resolution simulations yield higher dust masses, while the convergence is excellent at other masses. Comparison to Fig.~\ref{fig:gas} shows that the differences between the resolutions reflect differences in the cold gas content. The agreement with the data, including the large downward scatter, is very good.}
    \label{fig:mdust}
\end{figure}

\subsubsection{Dust} \label{sec:dust_masses}
The inclusion of a model for the evolution of dust grains opens up a new class of observables that can be directly predicted. \citet{Trayford2025} already presented predictions from L025m6 simulations with the \colibre\ model, finding good agreement with observations of the evolution of the cosmic mass density in dust, the $z=0$ dust mass function and various dust scaling relations. Here we present the first results for dust from the flagship \colibre\ simulations, which include both lower and higher resolutions than the runs analysed by \citet{Trayford2025} and, for the m6 resolution, have a volume 512 times larger. Future work will present a wider range of dust-related observables as well as their evolution and spatially resolved trends. 

Fig.~\ref{fig:mdust} shows the median dust mass as a function of stellar mass. The convergence with resolution is excellent for stellar masses corresponding to $\sim 10^3$ stellar particles up to $M_*\approx 10^{10.7}\,\Msun$. Towards lower masses systematic differences appear, with higher resolution simulations yielding higher dust masses. At $M_*\sim 10^{11}\,\Msun$ the convergence is poor, though the differences are still within the scatter (indicated by the shaded region for m6). While the m6 and m7 resolutions both show a dip in the median dust content, the dip is an order of magnitude deeper for m7. For higher masses the two resolutions converge, yielding very good agreement for $M_*\gtrsim 10^{11.5}\,\Msun$. The resolution dependence parallels that of the molecular mass fractions (compare with the right panel of Fig.~\ref{fig:gas}). In the mass range that is well converged and for which the quenched fraction is still small (i.e.\ $M_* \lesssim 10^{10}\,\Msun$, see Fig.~\ref{fig:fquenched}), the dust mass is roughly proportional to the stellar mass, which parallels the constant molecular-to-stellar mass ratio in the same mass range (see Fig.~\ref{fig:gas}). As is the case for the molecular mass, there is a very large downward scatter in the dust masses. The close similarity between the dust and molecular mass trends suggests that a large fraction of the dust resides in the cold interstellar gas phase, as can indeed be seen in the bottom left panel of Fig.~\ref{fig:phase_diagrams}. 

As discussed in \citet{Trayford2025} and in Section~\ref{sec:dust}, because \colibre\ does not resolve the molecular cloud cores where most grain growth is expected to occur, the growth rates due to gas accretion and coagulation have to be boosted by a clumping factor. In order to limit the boost to densities typical of the cold ISM, the clumping factor increases monotonically from 1 (i.e.\ no boost) to 100 over the density range $n_\text{H} = 10^{-1}$ to $10^2\,\cm^{-3}$ (see equation~\ref{eq:clumping_factor}). The dust content becomes insensitive to the value of the clumping factor when grain growth is limited by the availability of bottleneck elements. \citet{Trayford2025} show that for m6 resolution this occurs in high-density gas ($n_\text{H}\gg 10~\cm^{-3}$) provided the clumping factor is $\gtrsim 10$. An L025m6 simulation that does not use a clumping factor (not shown) results in median $z=0$ dust masses that are a factor of $\approx 3$ lower for $M_*>10^{10}\,\Msun$. The effect of the clumping factor is maximum for $M_*\sim 10^9\,\Msun$, where it increases the median dust masses by about an order of magnitude. 

The small open circles and crosses in Fig.~\ref{fig:mdust} indicate observational measurements presented by \citet{Bianchi2018} and \citet{DeLooze2020}, respectively. These are based on SED fitting of nearby galaxies from a variety of surveys. Because none of the surveys are mass-selected, it is unclear how representative the trends with mass are. Comparisons with multi-variate trends could reduce this concern, which we leave for future work. The simulation predictions are in good agreement with the data, reproducing both the trend followed by the majority of galaxies and the presence of a population with much smaller dust masses.

\begin{figure}
    \centering
    \includegraphics[width=0.95\linewidth]{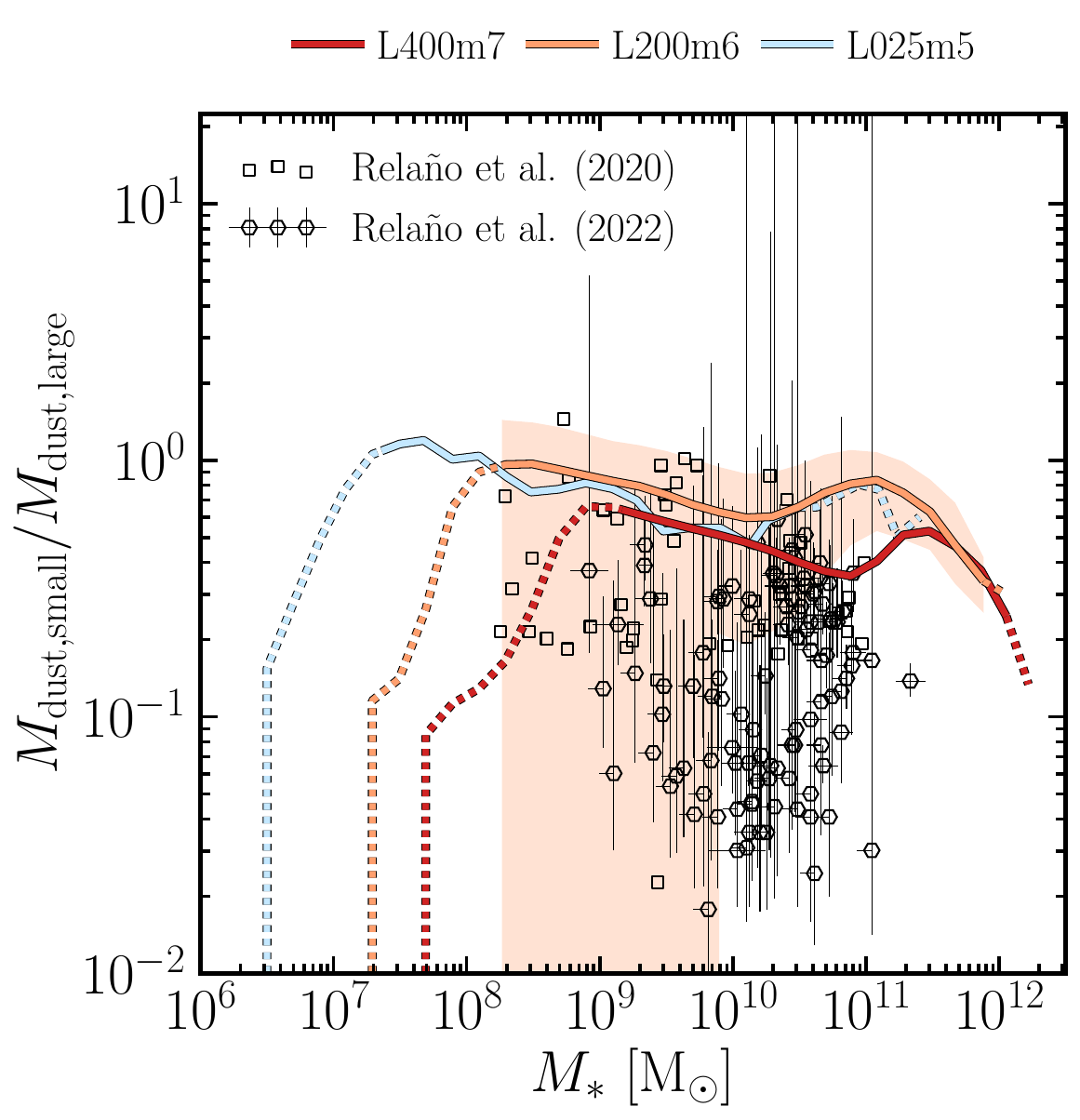}
    \caption{Median $z=0$ ratio of the dust mass in small (0.01 \textmu m) grains and the dust mass in large (0.1~\textmu m) grains as a function of galaxy stellar mass. The meanings of the line colours, line styles, and shading are as in Fig.~\ref{fig:sizes}. The convergence with resolution is good for galaxy masses corresponding to $>10^2$ stellar particles. The downward scatter is large, particularly for low-mass galaxies. The data points show observations of nearby galaxies from \citet{Relano2020,Relano2022}. These measurements are model-dependent and for galaxies from surveys that were not mass-selected. As the observed grain sizes are inferred from SED fitting, they are likely biased to dense gas in star-forming regions. In \colibre\ such regions have lower dust grain size ratios than the medians shown here, which would further improve the agreement with the data.}
    \label{fig:grain_size}
\end{figure}

The \colibre\ dust model includes two grain sizes, with radii of 0.01 and 0.1~\textmu m. As discussed in Section~\ref{sec:dust}, the coagulation rate is boosted by a density-dependent clumping factor. In addition, to improve the convergence with resolution, in the low-resolution m7 simulation the e-folding time for grain coagulation is further reduced by a factor $10^{0.5}$. This decreases the small-to-large grain dust mass ratio. 

Fig.~\ref{fig:grain_size} shows the median ratio of the dust mass in small grains to that in large grains as a function of stellar mass. For stellar masses corresponding to $> 10^2$ particles the convergence with resolution is very good, particularly between m5 and m6. For stellar masses corresponding to $<10^2$ particles the small-to-large ratio drops rapidly towards the value of $10^{-1}$ used for the grain seeds. Above this resolution limit grain sizes tend to increase with galaxy mass, except for a small excursion towards smaller grain sizes at the stellar mass for which the molecular mass fractions and the dust masses are minimum (see Figs.~\ref{fig:gas} and \ref{fig:mdust}, respectively). However, the trend of a decreasing median small-to-large grain ratio with mass is much smaller than the scatter. 

The bottom right panel of Fig.~\ref{fig:phase_diagrams}, and fig.~20 of \citet{Trayford2025}, show that the size distribution depends on the gas density, with higher densities favouring larger grain sizes, as expected due to the stronger density dependence for coagulation than for shattering (see \S\ref{sec:dust}). The density dependence reported by \citet{Trayford2025} is strong, with the size ratio decreasing from $\sim 10^{-0.5}$ for $n_\text{H}\sim 10~\cm^{-3}$ to $\sim 10^{-1.2}$ for $n_\text{H}\sim 10^2~\cm^{-3}$. This suggests that the trend of grain size with mass reflects an increase in the typical density of the dusty ISM with mass. The strong density dependence also implies that grain sizes inferred from virtual observations could be larger, and hence that the size ratio could be smaller, in star-forming regions than the galaxy-wide averages presented here. As the grain sizes of observed galaxies are inferred from SED fitting, the observed grain size ratios may thus be biased low.

The simulation results are compared with a compilation of observations from \citet{Relano2020,Relano2022}. Their measurements are for nearby galaxies from the same surveys as used for the dust mass measurements shown in Fig.~\ref{fig:mdust}. The surveys are not mass-limited and the size ratio, which is determined by SED fitting, is sensitive to the assumed dust model. Despite these limitations, the predictions are in broad agreement with the observations, with most of the low-mass ($M_* < 10^{10}\,\Msun$) data points falling within the $1\sigma$ scatter (indicated by the shaded region for L200m6). A subset of the observed high-mass ($M_*> 10^{10}\,\Msun$) galaxies are inferred to have relatively large grain sizes, with size ratios similar to what \colibre\ predicts for densities of $n_\text{H}\sim 10^2\,\cm^{-3}$ \citep[see fig.~10 of][]{Trayford2025}.

\begin{figure}
    \centering
    \includegraphics[width=0.95\linewidth]{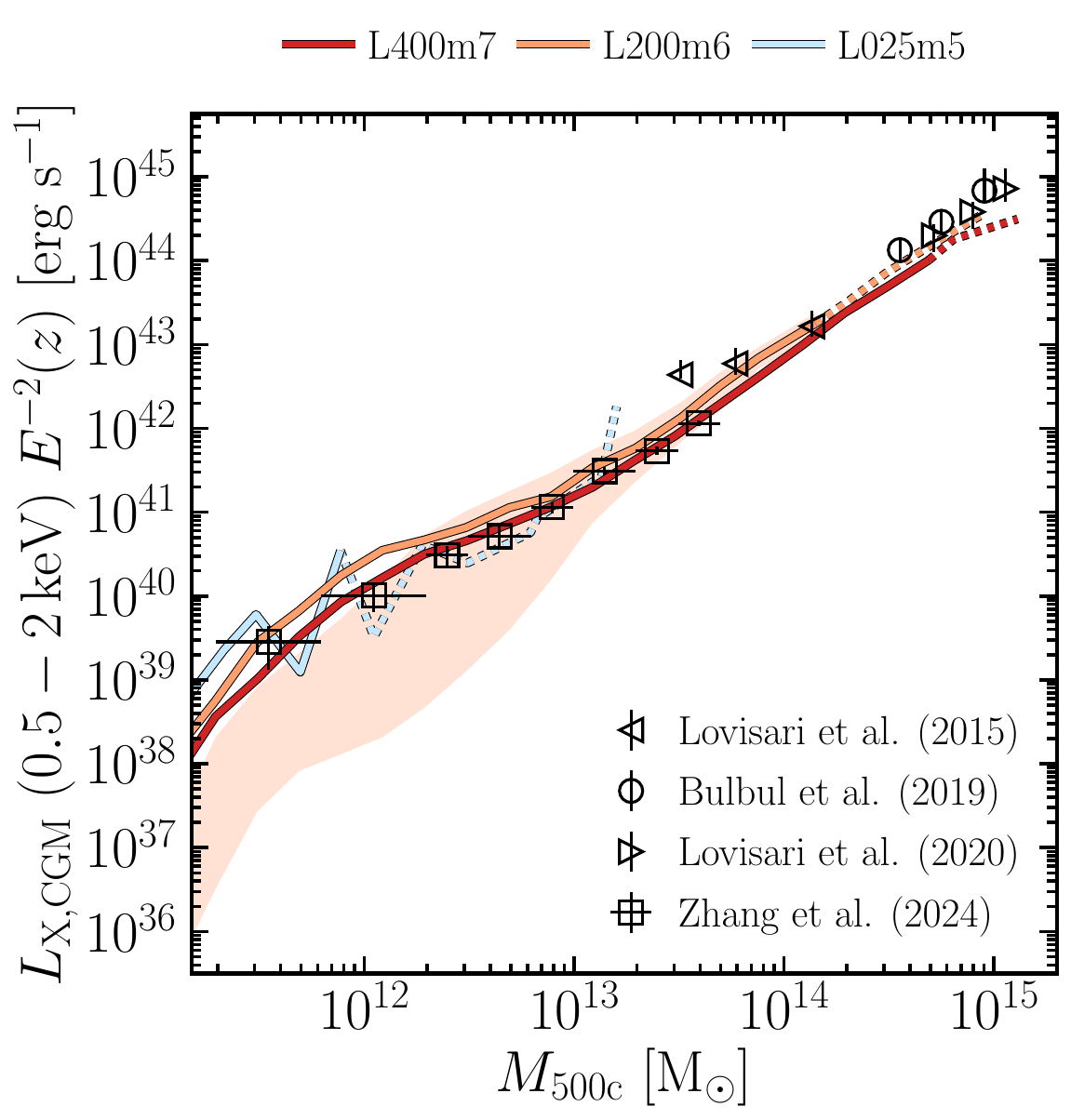}
    \caption{Mean X-ray luminosity from the CGM, i.e.\ integrated over the radial range $(0.15-1)\times R_\text{500c}$, as a function of halo mass, $M_\text{500c}$, at $z=0$. The meanings of the line colours, line styles, and shading are as in Fig.~\ref{fig:sizes}, except that the lines indicate means rather than medians. For masses corresponding to galaxy clusters, $M_\text{500c}>10^{13}\,\Msun$, the medians (not plotted) are very close to the means, but for $M_\text{500c} \lesssim 10^{12}\,\Msun$ the mean exceeds even the 84th percentile, which for L200m6 corresponds to the top of the shaded region. The data points show observations of \citet{Zhang2024LxCGM}, who computed mean CGM luminosities by stacking eROSITA observations of $\sim 10^5$ central $z<0.2$ galaxies, and with the median luminosities for the samples of individually detected clusters (without core excision) from \citet[][$z<0.035$]{Lovisari2015}, \citet[][$0.2<z<0.66$]{Bulbul2019}, and \citet[][$0.059<z<0.55$]{Lovisari2020} compiled by \citet{Braspenning2024}. All observed luminosities have been corrected for the expected self-similar evolution by dividing by $E^2(z) = [H(z)/H_0]^2$. The numerical convergence and the agreement with the data are both very good. Note that at the highest masses the agreement would be further improved if we had matched the radial range (i.e.\ no core excision) and redshifts of the observed clusters.}
    \label{fig:Lx}
\end{figure}

\subsubsection{X-ray luminosity of the circumgalactic medium} \label{sec:xray}
The previous figures compared the predictions for galaxy stellar and ISM properties with observations. Fig.~\ref{fig:Lx} instead shows a property of the CGM that is constrained by observations over a wide range of halo mass. The predicted mean $z=0$ X-ray luminosity of the CGM, $L_\text{X,CGM}$, is plotted as a function of halo mass, $M_\text{500c}$. Here, the CGM X-ray luminosity of an individual object is computed by summing the luminosities of its gas particles in the radial range $(0.15-1)\times R_\text{500c}$, where $R_\text{500c}$ is the radius within which the mean total matter density is 500 times the critical density of the universe, and $M_\text{500c}$ is the mass inside $R_\text{500c}$. Gas particles bound to satellites are excluded, but this has a negligible effect. The particle luminosities are computed in the 0.5--2~keV band using \textsc{cloudy} \citep[version 17.02][]{Ferland2017} with the method described in \citet{Braspenning2024}, but accounting for the non-equilibrium free electron densities from hydrogen and helium. We plot the mean instead of the median luminosity to facilitate comparison with stacks of observed objects, which is how emission is detected statistically in the observations of the CGM of low-mass galaxies from \citet{Zhang2024LxCGM} that we compare with. 

The radial cut $(0.15-1)\times R_\text{500c}$, sometimes referred to as `core excision', is commonly adopted in comparisons of simulations and observations, including by \citet{Zhang2024LxCGM}. Core excision results in the exclusion of emission from the central galaxy's ISM and the inner halo. Such a cut is motivated by the fact that observational analyses attempt to subtract emission from galaxies. Another motivation is that simulations generally do not model several important sources of X-ray emission within galaxies, such as X-ray binaries, AGN and supernova remnants. Furthermore, the X-ray emission from the ISM and inner CGM is sensitive to the precise subgrid implementation of feedback. In \colibre, particles with high gas densities that have recently been directly heated by the large temperature increments associated with SN and AGN feedback, as well as the immediate neighbours of these particles, can temporarily have very high X-ray luminosities. Such high luminosities of individual particles reflect the coarse-grained nature of subgrid models for feedback and are therefore probably artificial. In addition, they would likely be subtracted as point sources in observational analyses. For \colibre, core excision has only a small effect for $M_\text{500c} \gtrsim 10^{13}\,\Msun$, but is very important for $M_\text{500c}\lesssim 10^{12.5}\,\Msun$ ($M_*\lesssim 10^{11}\,\Msun)$. Without core excision, the mean X-ray luminosity would increase by $\approx 0.1-0.2$~dex for $M_\text{500c} \gtrsim 10^{13}\,\Msun$, but by $\approx 1$~dex for $M_\text{500c} \sim 10^{12}\,\Msun$. For low masses the spread in $L_\text{X}$ also increases dramatically if the core region is included.

Fig.~\ref{fig:Lx} shows that the CGM X-ray luminosity increases super-linearly with halo mass. The convergence between the different resolutions is very good. Comparison of the orange line, which shows the mean for L200m6, and the orange shading, which shows the 16th to 84th percentile spread, reveals that for $M_\text{500c}\lesssim 10^{12}\,\Msun$ the mean is higher than the 84th percentile, which is caused by the large and non-Gaussian scatter in $L_\text{X,CGM}$ at low masses. 

For $11.3 < \log_{10} M_\text{500c}/\Msun < 13.7$ the predictions are compared with CGM observations of \citet{Zhang2024LxCGM}, who stacked eROSITA data of $\sim 10^5$ central galaxies with spectroscopic redshifts $z<0.2$ from SDSS and subtracted X-ray emission from the galaxies. The observed luminosities were scaled by $E^{-2}(z) = [H(z)/H_0]^{-2}$ to correct for evolution under the assumption of self-similarity. At higher masses we compare with the median luminosities for the samples of individually detected clusters from \citet[][$z<0.035$]{Lovisari2015}, \citet[][$0.2<z<0.66$]{Bulbul2019}, and \citet[][$0.059<z<0.55$]{Lovisari2020} compiled by \citet{Braspenning2024}. 
As described in \citet{Braspenning2024}, the observed cluster X-ray luminosities have been shifted to the 0.5 - 
2.0~keV band using \textsc{pimms} \citep{Mukai1993} and the observed cluster masses have been divided by 0.743 to correct for the hydrostatic mass bias that affects halo mass measurements based on X-ray observations. 
The cluster luminosities are not core-excised, but for cluster-sized objects core-excision has little impact on the \colibre\ predictions. 

The simulations are in good agreement with the data. At the highest masses, $M_\text{500c} > 10^{14.5}\,\Msun$, the luminosity is slightly underpredicted, but the discrepancy is only 0.3~dex and would be slightly smaller if we had not applied core-excision to the simulations. Matching the redshifts of the observed clusters instead of correcting the observed luminosities to $z=0$ would likely further reduce the small discrepancy because the luminosity evolution is not precisely self-similar \citep{Braspenning2024}. However, we caution that the very good agreement with the low-mass data may be somewhat fortuitous given the large predicted scatter and the importance of subtracting emission from the galaxies, which is done differently for the observations and the simulations.

\begin{figure*}
    \centering
    \includegraphics[width=0.95\linewidth]{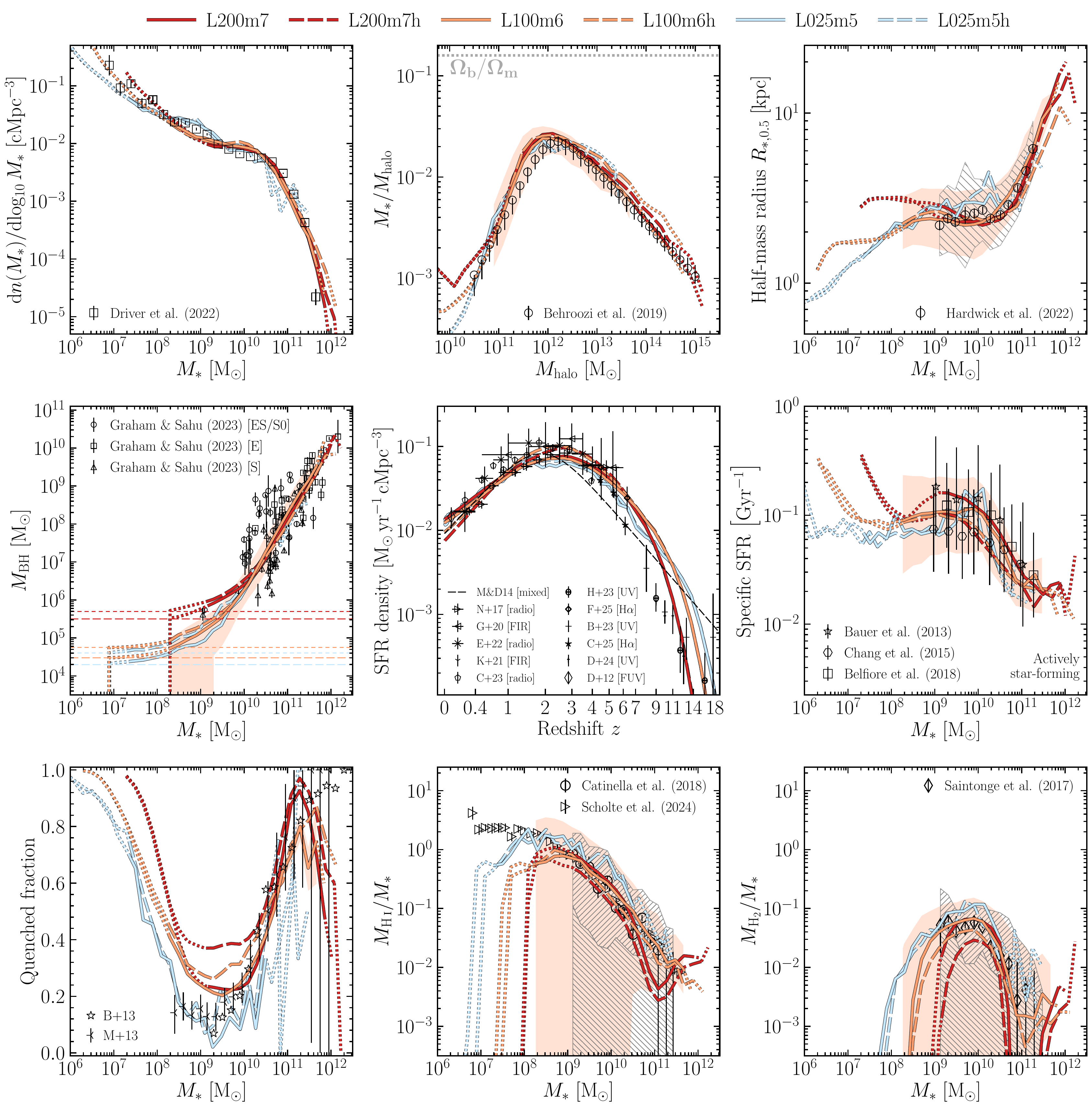}
    \caption{Comparison of the hybrid (dashed lines) and fiducial (solid lines) AGN feedback simulations. The panels show the same observables in the same order as presented for the fiducial simulations in Section~\ref{sec:obs}. The data points show the same observations as in the figures shown there. Red, orange, and blue lines indicate fiducial and hybrid models for L200m7, L100m6, and L025m5, respectively. The volumes of the runs using fiducial AGN feedback are identical to those of the hybrid runs of the same resolution, but up to 8 times smaller than for the fiducial simulations shown in Section~\ref{sec:obs}. The lines become dotted where the stellar mass becomes smaller than 100 times the mean initial baryonic particle mass and at high mass where there are fewer than 10 objects per 0.2 dex mass bin. Where present, the orange shading indicates the 16th to 84th percentile scatter for L100m6, except for the panel showing the quenched fraction, where it indicates the $1\sigma$ confidence interval or L100m6. The differences between the predictions of the two types of AGN feedback models are mostly small and hence their levels of agreement with the data are similarly good.}
    \label{fig:hybrid}
\end{figure*}
\begin{figure*}
    \centering
    \includegraphics[width=0.95\linewidth]{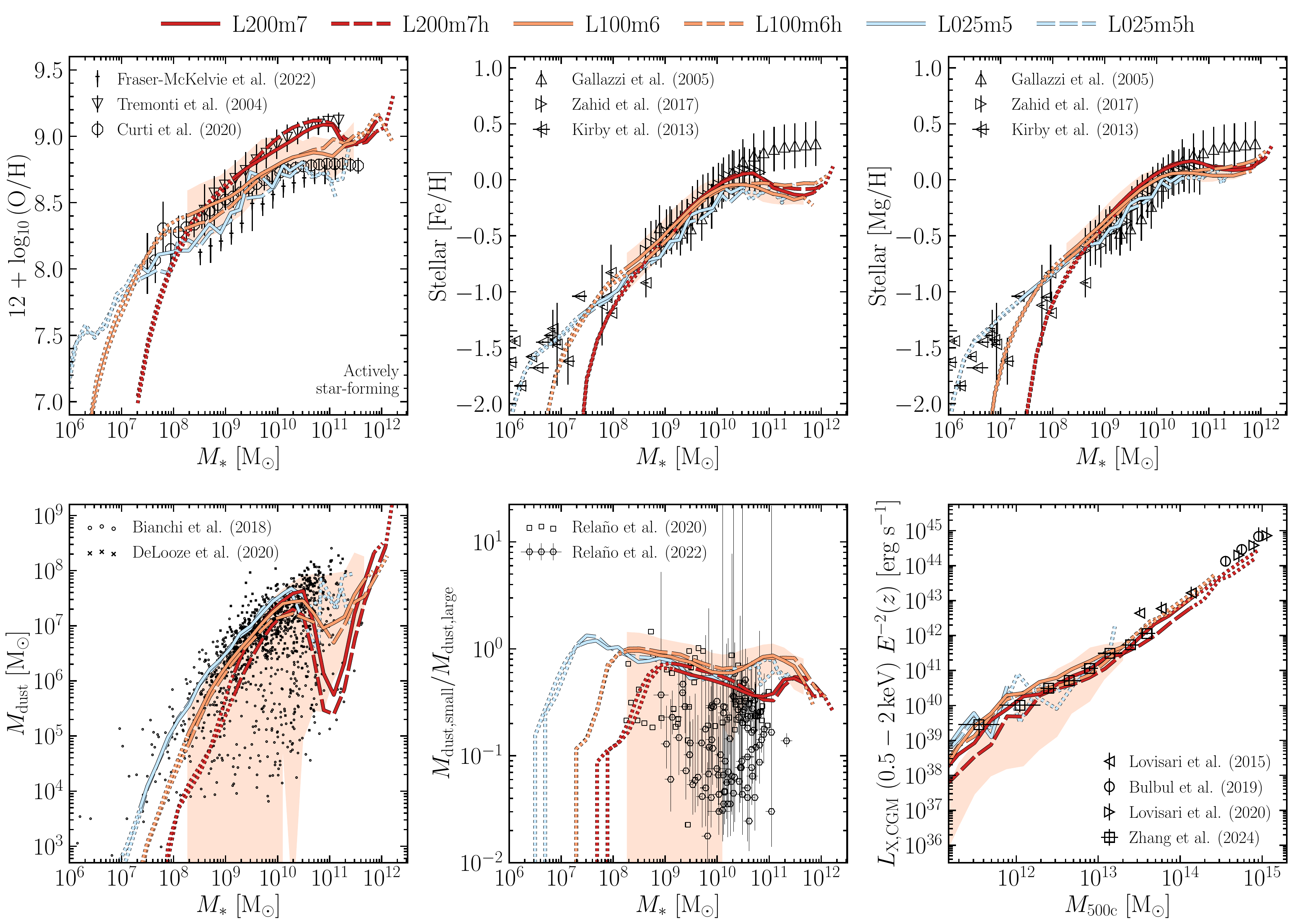}
    \caption{Continuation of Fig.~\ref{fig:hybrid}.}
    \label{fig:hybrid2}
\end{figure*}

\section{Dependence on the implementation of AGN feedback}\label{sec:hybrid}

The comparison between the simulations and observations presented in the previous section was limited to the fiducial \colibre\ simulations, in which the AGN feedback is driven by the injection of thermal energy. Because the nature of AGN feedback is uncertain, and because AGN-powered jets are observed, the \colibre\ suite also includes simulations that use `hybrid AGN feedback', i.e.\ both thermal isotropic and kinetic jet feedback. The model is detailed in \citet{Husko2025method} and summarized in Sections~\ref{sec:bhs} and \ref{sec:agn_fb}. At each resolution, the largest simulation with hybrid AGN feedback was run in a box with a $2^3=8$ times smaller volume than the flagship fiducial simulation (see Table~\ref{tbl:simulations}). 

The hybrid AGN feedback simulations include a sophisticated subgrid model that tracks the evolution of the spin of every BH, which is then used to determine subgrid AGN feedback efficiencies and jet directions. Depending on the Eddington ratio, which is based on the rate of gas accretion onto the outer subgrid BH accretion disc, the accretion disc can be thick ($f_\text{Edd} < 0.01$), thin ($0.01 < f_\text{Edd} < 1$), or slim ($f_\text{Edd} > 1$). Jet feedback occurs in all three disc states, accretion disc winds are present in the thick and slim modes, but radiation pressure-driven winds are limited to the thin disc regime. Injection of thermal energy is used for both types of winds, employing the same implementation as the fiducial simulations. Some of the parameters of the CCSN and AGN feedback subgrid models were adjusted from their values in the fiducial models to maintain agreement with the observables used for the calibration, i.e.\ the $z\approx 0$ galaxy stellar mass function, size -- mass relation, and BH masses at high stellar masses (see \S\ref{sec:calibration}). The most important feedback mechanisms are thermal injections from the thin disc (at all $z$), jets from the thick disc (at $z\lesssim 1$), and jets from the thin disc (at $z\gtrsim 1$). 

Figs.~\ref{fig:hybrid} and \ref{fig:hybrid2} compare the fiducial (solid) and hybrid (dashed) AGN feedback simulations at m5 (blue), m6 (orange), and m7 (red) resolution in volumes of 25, 100, and 200~cMpc on a side, respectively. The panels show the same observables as were presented for the fiducial model in Section~\ref{sec:obs} in the order in which they appeared in that section. The data points show the same observations as were plotted in the figures in Section~\ref{sec:obs}. To isolate the effects of the modelling of BHs and AGN feedback, the hybrid and fiducial AGN feedback simulations shown in these figures were started from the same initial conditions, which means that the latter use volumes smaller than presented in Section~\ref{sec:obs} for the same resolution.  

The differences in the predictions of the two AGN models are typically very small for the observables investigated here. Consequently, the level of agreement with the data is very similar for the two models. For L025m5 the differences are least significant, so we will focus on the larger volumes available at the lower resolutions. In the hybrid model BCG masses and half-mass radii are up to $\approx 0.15$~dex larger and smaller, respectively, but the differences are significantly smaller for m7 than for m6 resolution. The hybrid model predicts up to $\approx 0.2$~dex lower sSFRs for star-forming galaxies, and quenched fractions that are $\approx 20$ percentage points higher for both $M_*\sim 10^{9.5}\,\Msun$ at m7 resolution and $M_*\sim 10^{11.5}\,\Msun$ at m6 resolution, where the differences are largest. Consistent with these differences, the \ion{H}{i}, molecular and dust masses are somewhat smaller for hybrid AGN feedback. For the other $z=0$ observables the differences between the feedback models are smaller than the differences between the simulations of different resolution. The differences between hybrid and fiducial feedback generally have the same sign for both resolutions. 

The $z=0$ cosmic star formation rate density is $\approx 0.25$~dex lower for the hybrid model, consistent with its lower sSFRs and higher quenched fractions. Interestingly, the peak of the cosmic SFR density, which occurs at $z\approx 2-3$, is higher for hybrid AGN feedback, though only by $\approx 0.1-0.2$~dex. This suggests that for other observables the sign of the differences may also change going from low to high redshift. At $z<2$ the cosmic SFR density declines more steeply in the simulation using hybrid AGN feedback. The SFR density drops by a full order of magnitude between its peak and $z=0$ compared to a drop of $\approx 0.75$~dex for the fiducial simulations.   

Although the differences in $z\approx 0$ galaxy properties are relatively small, at least for the spatially integrated quantities investigated here, preliminary work shows that larger differences occur in the IGM and for BHs at high redshift. We conclude that, in terms of the wide variety of observables presented here, the hybrid AGN feedback simulations are similarly realistic as the fiducial models. The good agreement between the two types of simulations suggests that, provided the subgrid models for the feedback are calibrated to reproduce observables that constrain the efficiencies of the feedback on galaxy scales, in our case the $z\approx 0$ stellar and BH masses, self-regulated galaxy and BH growth ensure that differences in other basic observables probing similar scales will also tend to be small. However, this does not preclude the possibility of larger differences on sub- and intergalactic scales and at higher redshifts. \citet{Husko2025method} and future work will provide more in-depth comparisons for a wider range of observables and redshifts.

\section{Summary and discussion} \label{sec:conclusions}

We have introduced the \colibre\ galaxy formation model and a suite of cosmological hydrodynamical simulations that use it. 

Compared to previous simulations run to $z=0$ using similar baryonic resolution, and also compared to most higher-resolution cosmological simulations, \colibre\ breaks new ground in several ways. This includes direct simulation of the multiphase ISM (rather than imposing an equation of state), direct simulation of the evolution of dust grains and their coupling to the chemistry and cooling, using four times more dark matter than baryonic particles in order to suppress spurious transfer of energy from dark matter to stars, repeating simulations using different models for BHs and AGN feedback that have been calibrated to the same data, and the sheer size of the runs ($5\times 3008^3 \approx 1.4\times 10^{11}$ particles; e.g.\ 20 times more than used in the \eagle\ simulation). 

The \colibre\ simulations span a factor of 64 in mass resolution, with particle masses (for both baryons and dark matter) of $\sim 10^5\,\Msun$ (m5), $10^6\,\Msun$ (m6), and $10^7\,\Msun$ (m7) in cubic volumes with comoving side lengths, $L$, of up to 100, 200, and 400~cMpc, respectively (see Table~\ref{tbl:simulations}). The effective resolution is higher than for previous cosmological simulations with similar baryonic particle masses because the equation of state imposed on the ISM in those simulations artificially smooths the gas on sub-kpc scales (i.e.\ scales smaller than the Jeans scale in the warm ISM) even if the quoted force resolution or cell size is much smaller. The effective resolution is also higher because \colibre\ uses more dark matter particles per baryonic particle, by a factor of 4 compared to most previous simulations. The availability of different simulation resolutions and volumes, which were each calibrated to the $z=0$ galaxy stellar mass function, galaxy sizes, and BH masses in massive galaxies, allows testing for numerical convergence. The large number of resolution elements translates into relatively large volumes for a given resolution, enabling the creation of statistically significant samples and the study of environmental effects, and yields a relatively large range of galaxy masses probed for a given volume. The mass range can be further expanded by combining simulations using different resolutions. 

The simulations were run with the \swift\ code using the \sphenix\ implementation of smoothed particle hydrodynamics (SPH). All subgrid prescriptions used by \colibre\ were newly developed for the project. \colibre\ includes models for radiative cooling, dust grains, star formation, stellar mass loss and chemical enrichment, turbulent diffusion, early stellar feedback, supernova feedback, supermassive black holes (BHs) and active galactic nucleus (AGN) feedback. 

In this work, we limit the comparison with observations to $z\approx 0$, except for the cosmic star formation history, which we find to be consistent with the data \citep[Fig.~\ref{fig:sfh}; see also][]{Chaikin2025smf_evol}. We focus on the stellar, gas, metal, and dust content of galaxies, although we do compare with the observed relation between the X-ray luminosity of the CGM and halo mass, finding very good agreement (Fig.~\ref{fig:Lx}).

Radiative cooling rates are computed element-by-element using \textsc{hybrid-chimes} \citep[\S\ref{sec:cooling} and][]{Ploeckinger2025}, which combines non-equilibrium abundances of hydrogen and helium species with tabulated equilibrium abundances for heavy elements. For self-consistency, the tables are computed using the same \textsc{chimes} reaction network \citep{Richings2014a, Richings2014b} used to compute hydrogen and helium rates on-the-fly. A novel aspect is that the radiative cooling rates for heavy elements are computed in `quasi-equilibrium': they assume equilibrium ion/molecular fractions but account for the non-equilibrium free electrons from hydrogen and helium. The rates are computed in the presence of a redshift-dependent metagalactic radiation field and a pressure-dependent local interstellar radiation field (including cosmic rays) and account for local shielding by gas and dust. Another feature that is rare for cosmological simulations is that the reaction network is coupled to the dust grain module. The depletion of heavy elements, dust cooling and photo-electric heating, and the formation of molecules on dust grains all account for the evolving abundances of the different types of dust grains. The predicted $z=0$ \ion{H}{i} and H$_2$ masses as a function of stellar mass agree very well with the data, both for the medians and the scatter (Fig.~\ref{fig:gas}).

The abundances of dust grains of three different chemical compositions (carbon and two types of silicates) and two different grain sizes (0.01 and 0.1~\textmu m) are tracked \citep[\S\ref{sec:dust} and][]{Trayford2025}. Dust masses can increase through seeding by AGB stars and core collapse supernovae (CCSN), and depletion of heavy elements by gas accretion in the ISM. Collisions with hot gas and astration in stars decrease dust masses. Grain sizes evolve as a result of coagulation and shattering. The dust-to-metal ratio and the ratio of large-to-small grains are predicted to increase toward higher gas densities (Fig.~\ref{fig:phase_diagrams}). Visually, within the disc of a typical galaxy, the dust is concentrated towards the centre and in the spiral arms, with a spatial distribution intermediate between that of the atomic and molecular gas (Fig.~\ref{fig:disc}). The relations between $z=0$ dust mass and stellar mass (Fig.~\ref{fig:mdust}), and between grain size and stellar mass (Fig.~\ref{fig:grain_size}) trace the observed trends.  

Gravitationally unstable gas is stochastically converted into collisionless stellar particles according to a Schmidt law with an efficiency per free-fall time of 1 per cent \citep[\S\ref{sec:SF} and ][]{Nobels2024}. Because stability is evaluated at the resolution limit, star formation shifts towards higher densities and lower temperatures as the resolution increases. At all resolutions, most stars form in the cold interstellar gas phase $(T\sim 10^2\,\K)$ (Fig.~\ref{fig:birth_props}), but the warm phase ($T\sim 10^4\,\K$) contributes significantly to star formation in poorly resolved galaxies (Fig.~\ref{fig:birth_props_by_mass}). 

Time- and metallicity-dependent stellar mass loss from AGB stars, massive stars, Type~Ia supernovae, and CCSNe is implemented using up-to-date nucleosynthetic yields for the 11 elements that dominate the cooling rates, as well from AGB stars for the s-process elements Ba and Sr, and from neutron star mergers, common-envelope jets SNe and collapsars for the r-process element Eu \citep[\S\ref{sec:chemo} and ][]{Correa2025chemo}. Diffusion driven by local turbulence is modelled for both gaseous elements and dust grains \citep[\S\ref{sec:diffusion} and ][]{Correa2025chemo}. The predicted $z=0$ relations between stellar mass and both gas and stellar metallicity agree well with the data, with the exception of a predicted turnover of the relations at $M_* \sim 10^{11}\,\Msun$. 

Before the first CCSNe explode, stellar particles start to exert feedback on the surrounding gas through stellar winds and radiation pressure, where the latter depends on the local shielding length and dust content. Furthermore, young stellar particles can heat and ionize gas by creating \ion{H}{ii} regions \citep[\S\ref{sec:early_fb} and][]{BenitezLlambay2025}. However, the most important stellar feedback channel is CCSNe. Most (90\%) of the CCSN energy is used to stochastically heat nearby gas to high temperatures \citep{DallaVecchia2012}, but a fraction (10\%) is injected kinetically in the form of low-velocity ($50~\kms$) kicks \citep{Chaikin2023}. The heating temperature increases with the local gas density in order to meet the competing demands of good sampling, which prefers low temperatures, and overcoming numerical overcooling, which demands higher temperatures at higher densities (\S\ref{sec:sn_fb}). While galactic outflows are driven mainly by the thermal channel, the kinetic channel helps drive turbulence \citep{Chaikin2023}.  The probability of a heating event increases with the thermal gas pressure, which compensates for residual numerical overcooling and may also represent the physical effect of the reduction in thermal losses expected for more strongly clustered CCSNe \citep[\S\ref{sec:calibration} and][]{Chaikin2025calibration}. 

BHs are seeded in haloes identified by a friends-of-friends algorithm and then grow by accreting gas, assuming Bondi-Hoyle accretion limited by turbulence and vorticity, and by merging with other BHs (\S\ref{sec:bhs}). AGN feedback is implemented as thermal energy injections \citep{Booth2009} with a heating temperature that scales with the BH mass to maintain equal sampling of feedback events for BHs with different masses accreting at the same Eddington ratio (\S\ref{sec:agn_fb}). To enable investigations of the sensitivity of predictions to the implementation and type of AGN feedback, a subset of the simulations use `hybrid AGN feedback'. These models track the BH spin. The hybrid model combines thermal AGN feedback, which is implemented as in the fiducial model, with kinetic jets directed along the BH spin axis, where the feedback mode depends on the BH accretion rate, mass, and spin \citep[\S\ref{sec:bhs}, \S\ref{sec:agn_fb} and][]{Husko2025method}. 

The subgrid models for CCSN and AGN feedback are calibrated to the $z=0$ galaxy stellar mass function and the galaxy size-mass relation over the range $9.0 < \log_{10} M_*/\Msun < 11.3$ \citep[\S\ref{sec:calibration} and][]{Chaikin2025calibration}. For the fiducial, low-resolution simulations, this is accomplished by optimizing the seed BH mass and the values of three parameters of the subgrid model for CCSN feedback using Gaussian process emulators trained on modest-volume simulations (50~cMpc on a side; L050m7). In addition, the subgrid AGN feedback efficiency is chosen to yield BH masses consistent with observations of massive $z=0$ galaxies. For higher resolutions and for hybrid AGN feedback, small adjustments are made by hand. The simulations are in excellent agreement with the calibration data (Figs.~\ref{fig:gsmf}, \ref{fig:sizes}, \ref{fig:bhs}). 

The simulations reproduce the observed $z=0$ relation between SFR and stellar mass for actively star-forming galaxies (Fig.~\ref{fig:ssfr}). The fraction of passively evolving galaxies as a function of stellar mass matches the observed relation at intermediate masses (Fig.~\ref{fig:fquenched}), but there are signs that the quenched fractions may be too high for galaxies with $M_*\lesssim 10^9\,\Msun$ and too low for the most massive galaxies, $M_* \sim 10^{12}\,\Msun$. These disagreements either point to subtle shortcomings in the simulations, or they might reflect systematic biases in the comparison to the observations (e.g.\ for BCGs aperture choices or random errors in the stellar masses may be plausible explanations).

To first order, the two AGN feedback implementations yield very similar predictions for all the observables investigated here, and thus yield very similar levels of agreement with the observations. However, there are some small but significant differences. Depending on the resolution, hybrid AGN feedback gives a $0.1-0.2$~dex higher peak cosmic SFR density, while the $z=0$ SFRs are $\approx 0.25$~dex lower. Hence, for hybrid AGN feedback the cosmic SFR density declines more rapidly between $z\approx 2$ and 0. Consistent with the lower present-day SFR density, the $z=0$ specific SFRs and molecular masses are slightly lower, while the quenched fractions are somewhat higher than for the fiducial model. For hybrid AGN feedback BCG masses and half-mass radii are up to $\approx 0.15$~dex larger and smaller, respectively.

Comparison of the predictions of the simulations using different resolutions shows that convergence is generally excellent for objects sampled with more than $\sim 10 - 10^3$ stellar particles, where the limiting particle number depends on the observable. For galaxies of mass $M_* \sim 10^{11}\,\Msun$ convergence is less good for quenched fractions and for median atomic, molecular, and dust masses. The relatively good convergence is notable, especially compared to simulations such as \eagle\ (Fig.~\ref{fig:eagle}) and IllustrisTNG (Fig.~\ref{fig:tng}), whose predictions generally also do not agree as well with the data for the properties investigated here. Like \colibre, \eagle, but not IllustrisTNG, makes small adjustments to some of the subgrid parameters that control the feedback when the resolution is changed. However, unlike \colibre, \eagle\ and IllustrisTNG impose an equation of state on the ISM. This makes the convergence seen in \colibre\ particularly remarkable, since \colibre\ does not resolve the internal structure of molecular clouds. We attribute this to the reduced importance of numerical overcooling in simulations that allow cooling to cold ISM temperatures, and hence enable more of the gas mass to concentrate in spiral arms and high-density clouds. The more inhomogeneous gas distribution results in a lower density for the intercloud medium, where most of the CCSN feedback events occur (Fig.~\ref{fig:feedback_props}), and hence smaller radiative losses. AGN feedback, on the other hand, is concentrated in galactic nuclei, where the gas density reaches very high values (Fig.~\ref{fig:feedback_props}). For that reason, the efficiency of AGN feedback is more sensitive to resolution, which is consistent with the poorer convergence for $M_*\sim 10^{11}\,\Msun$.

Although we consider \colibre\ to be a major step forward compared to previous simulations of representative volumes, it has many known weaknesses (which are not unique to \colibre). Although the resolution suffices to distinguish the cold and warm interstellar gas phases, except in low-mass galaxies sampled with few particles, the simulations do not resolve the structure of molecular clouds. We therefore need to boost the rates of gas accretion onto dust grains and of dust grain coagulation using a subgrid clumping factor. Because of resolution limitations, some star formation occurs in warm gas, particularly in low-mass galaxies. To compensate for numerical overcooling, we need to inject more CCSN energy in gas with a higher thermal pressure, use higher CCSN feedback temperatures in denser gas, and use higher AGN feedback temperatures for more massive BHs. A higher energy per feedback event implies poorer sampling. The parameters controlling stellar and AGN feedback require calibration, which removes some predictive power and which is more difficult to accomplish at higher resolution because of the greater computational expense and longer runtimes.  We cannot resolve dynamical friction for all but the most massive BHs, which necessitates important but ad-hoc corrections to the BH positions. Although we implement prescriptions for self-shielding and a spatially varying interstellar radiation field, radiative transfer is not simulated directly, and we assume flash reionization. We do not vary the gravitational softening with density. There are no magnetic fields, and although cosmic rays are considered when calculating cooling rates and species fractions, they do not exert any pressure. 

To our knowledge, both the level of numerical convergence and the level of agreement with a diverse range of galaxy data that we find for \colibre\ are unprecedented for cosmological hydrodynamical simulations. However, except for the cosmic star formation history, the comparisons presented here are limited to observations of galaxy scaling relations in the low-$z$ Universe, and are performed `in theory space'. In future work, we will compare the simulations with a wider variety of observations, including spatially resolved galaxy properties and environmental dependencies, and extending to high redshift. We also plan to create virtual observations to enable a fairer comparison with the data and to test model-dependent aspects of observational analyses. We plan to complete the L050m5 and L025m5h simulations, to extend the suite to higher resolution and more model variations, and to complement the representative volumes with simulations that zoom in on regions of interest.

\section*{Acknowledgements}
We gratefully acknowledge discussions with researchers working on the analysis of the \colibre\ simulations, as well as discussions with and help from members of the Virgo and \flamingo\ collaborations. We specifically thank Joey Braspenning for help with the X-ray luminosities, Oliver Hahn for help with \monofonIC. We thank Rob Crain and Joel Pfeffer for the addition of a passive subgrid model for globular clusters that we hope they will report on in future work. We are grateful to the late Richard Bower, Tom Theuns, and Bert Vandenbroucke for their contributions and enthusiasm in the early stages of the project. We thank Dylan Nelson and Annalisa Pillepich for comments on Appendix~\ref{app:eagle_tng}. We thank the anonymous reviewer for helpful comments.

We acknowledge the use of the \textsc{SwiftSimIO} open source tools \citep{Borrow2020}. This work used the DiRAC@Durham facility managed by the Institute for Computational Cosmology on behalf of the STFC DiRAC HPC Facility (\url{www.dirac.ac.uk}). The equipment was funded by BEIS capital funding via STFC capital grants ST/K00042X/1, ST/P002293/1, ST/R002371/1 and ST/S002502/1, Durham University and STFC operations grant ST/R000832/1. DiRAC is part of the National e-Infrastructure. This project has received funding from the Netherlands Organization for Scientific Research (NWO) through research programme Athena 184.034.002. SP acknowledges support from the Austrian Science Fund (FWF) through grant number V 982-N. JT acknowledges support from an STFC Early Stage Research and Development grant (ST/X004651/1). ABL acknowledges support by the Italian Ministry for Universities (MUR) program `Dipartimenti di Eccellenza 2023-2027' within the Centro Bicocca di Cosmologia Quantitativa (BiCoQ), and support by UNIMIB’s Fonuk redo Di Ateneo Quota Competitiva (project 2024-ATEQC-0050). CSF acknowledges support from European Research Council (ERC) Advanced Grant DMIDAS (GA 786910). YMB acknowledges support from UK Research and Innovation through a Future Leaders Fellowship (grant agreement MR/X035166/1). This work was supported by the Swiss National Science Foundation (SNSF) under funding reference 200021\_213076. AD is supported by an STFC doctoral studentship. AG acknowledges support from : FWO-Vlaanderen (Fund for Scientific Research Flanders) grant number: FWO.3F0.2021.0030.01. ARJ and CGL acknowledge support from UKRI grants ST/T000244/1 and ST/X001075/1.

\section*{Data Availability}

The data supporting the plots within this article are available on reasonable request to the corresponding author. The \colibre\ simulation data will eventually be made publicly available, although we note that the data volume (several petabytes) may prevent us from simply placing the raw data on a server. In the meantime, people interested in using the simulations are encouraged to contact the corresponding author.
A public version of the \swift\ code \citep{Schaller2024swift} is available at \url{http://www.swiftsim.com}. The \colibre\ modules implemented in \swift\ will be made publicly available after the public release of the simulation data.




\bibliographystyle{mnras}
\bibliography{main} 

@ARTICLE{Abbott2022,
       author = {{Abbott}, T.~M.~C. and {Aguena}, M. and {Alarcon}, A. and {Allam}, S. and {Alves}, O. and {Amon}, A. and {Andrade-Oliveira}, F. and {Annis}, J. and {Avila}, S. and {Bacon}, D. and {Baxter}, E. and {Bechtol}, K. and {Becker}, M.~R. and {Bernstein}, G.~M. and {Bhargava}, S. and {Birrer}, S. and {Blazek}, J. and {Brandao-Souza}, A. and {Bridle}, S.~L. and {Brooks}, D. and {Buckley-Geer}, E. and {Burke}, D.~L. and {Camacho}, H. and {Campos}, A. and {Carnero Rosell}, A. and {Carrasco Kind}, M. and {Carretero}, J. and {Castander}, F.~J. and {Cawthon}, R. and {Chang}, C. and {Chen}, A. and {Chen}, R. and {Choi}, A. and {Conselice}, C. and {Cordero}, J. and {Costanzi}, M. and {Crocce}, M. and {da Costa}, L.~N. and {da Silva Pereira}, M.~E. and {Davis}, C. and {Davis}, T.~M. and {De Vicente}, J. and {DeRose}, J. and {Desai}, S. and {Di Valentino}, E. and {Diehl}, H.~T. and {Dietrich}, J.~P. and {Dodelson}, S. and {Doel}, P. and {Doux}, C. and {Drlica-Wagner}, A. and {Eckert}, K. and {Eifler}, T.~F. and {Elsner}, F. and {Elvin-Poole}, J. and {Everett}, S. and {Evrard}, A.~E. and {Fang}, X. and {Farahi}, A. and {Fernandez}, E. and {Ferrero}, I. and {Fert{\'e}}, A. and {Fosalba}, P. and {Friedrich}, O. and {Frieman}, J. and {Garc{\'\i}a-Bellido}, J. and {Gatti}, M. and {Gaztanaga}, E. and {Gerdes}, D.~W. and {Giannantonio}, T. and {Giannini}, G. and {Gruen}, D. and {Gruendl}, R.~A. and {Gschwend}, J. and {Gutierrez}, G. and {Harrison}, I. and {Hartley}, W.~G. and {Herner}, K. and {Hinton}, S.~R. and {Hollowood}, D.~L. and {Honscheid}, K. and {Hoyle}, B. and {Huff}, E.~M. and {Huterer}, D. and {Jain}, B. and {James}, D.~J. and {Jarvis}, M. and {Jeffrey}, N. and {Jeltema}, T. and {Kovacs}, A. and {Krause}, E. and {Kron}, R. and {Kuehn}, K. and {Kuropatkin}, N. and {Lahav}, O. and {Leget}, P. -F. and {Lemos}, P. and {Liddle}, A.~R. and {Lidman}, C. and {Lima}, M. and {Lin}, H. and {MacCrann}, N. and {Maia}, M.~A.~G. and {Marshall}, J.~L. and {Martini}, P. and {McCullough}, J. and {Melchior}, P. and {Mena-Fern{\'a}ndez}, J. and {Menanteau}, F. and {Miquel}, R. and {Mohr}, J.~J. and {Morgan}, R. and {Muir}, J. and {Myles}, J. and {Nadathur}, S. and {Navarro-Alsina}, A. and {Nichol}, R.~C. and {Ogando}, R.~L.~C. and {Omori}, Y. and {Palmese}, A. and {Pandey}, S. and {Park}, Y. and {Paz-Chinch{\'o}n}, F. and {Petravick}, D. and {Pieres}, A. and {Plazas Malag{\'o}n}, A.~A. and {Porredon}, A. and {Prat}, J. and {Raveri}, M. and {Rodriguez-Monroy}, M. and {Rollins}, R.~P. and {Romer}, A.~K. and {Roodman}, A. and {Rosenfeld}, R. and {Ross}, A.~J. and {Rykoff}, E.~S. and {Samuroff}, S. and {S{\'a}nchez}, C. and {Sanchez}, E. and {Sanchez}, J. and {Sanchez Cid}, D. and {Scarpine}, V. and {Schubnell}, M. and {Scolnic}, D. and {Secco}, L.~F. and {Serrano}, S. and {Sevilla-Noarbe}, I. and {Sheldon}, E. and {Shin}, T. and {Smith}, M. and {Soares-Santos}, M. and {Suchyta}, E. and {Swanson}, M.~E.~C. and {Tabbutt}, M. and {Tarle}, G. and {Thomas}, D. and {To}, C. and {Troja}, A. and {Troxel}, M.~A. and {Tucker}, D.~L. and {Tutusaus}, I. and {Varga}, T.~N. and {Walker}, A.~R. and {Weaverdyck}, N. and {Wechsler}, R. and {Weller}, J. and {Yanny}, B. and {Yin}, B. and {Zhang}, Y. and {Zuntz}, J. and {DES Collaboration}},
        title = "{Dark Energy Survey Year 3 results: Cosmological constraints from galaxy clustering and weak lensing}",
      journal = {\prd},
         year = 2022,
        month = jan,
       volume = {105},
       number = {2},
          eid = {023520},
        pages = {023520},
          doi = {10.1103/PhysRevD.105.023520},
archivePrefix = {arXiv},
       eprint = {2105.13549},
 primaryClass = {astro-ph.CO},
       adsurl = {https://ui.adsabs.harvard.edu/abs/2022PhRvD.105b3520A}
}

@ARTICLE{Accurso2017,
       author = {{Accurso}, G. and {Saintonge}, A. and {Catinella}, B. and {Cortese}, L. and {Dav{\'e}}, R. and {Dunsheath}, S.~H. and {Genzel}, R. and {Gracia-Carpio}, J. and {Heckman}, T.~M. and {Jimmy} and {Kramer}, C. and {Li}, Cheng and {Lutz}, K. and {Schiminovich}, D. and {Schuster}, K. and {Sternberg}, A. and {Sturm}, E. and {Tacconi}, L.~J. and {Tran}, K.~V. and {Wang}, J.},
        title = "{Deriving a multivariate {\ensuremath{\alpha}}$_{CO}$ conversion function using the [C II]/CO (1-0) ratio and its application to molecular gas scaling relations}",
      journal = {\mnras},
         year = 2017,
        month = oct,
       volume = {470},
       number = {4},
        pages = {4750-4766},
          doi = {10.1093/mnras/stx1556},
archivePrefix = {arXiv},
       eprint = {1702.03888},
 primaryClass = {astro-ph.GA},
       adsurl = {https://ui.adsabs.harvard.edu/abs/2017MNRAS.470.4750A}
}

@ARTICLE{Agertz2021,
       author = {{Agertz}, Oscar and {Renaud}, Florent and {Feltzing}, Sofia and {Read}, Justin I. and {Ryde}, Nils and {Andersson}, Eric P. and {Rey}, Martin P. and {Bensby}, Thomas and {Feuillet}, Diane K.},
        title = "{VINTERGATAN - I. The origins of chemically, kinematically, and structurally distinct discs in a simulated Milky Way-mass galaxy}",
      journal = {\mnras},
         year = 2021,
        month = jun,
       volume = {503},
       number = {4},
        pages = {5826-5845},
          doi = {10.1093/mnras/stab322},
archivePrefix = {arXiv},
       eprint = {2006.06008},
 primaryClass = {astro-ph.GA},
       adsurl = {https://ui.adsabs.harvard.edu/abs/2021MNRAS.503.5826A}
}

@ARTICLE{Angulo2016,
       author = {{Angulo}, Raul E. and {Pontzen}, Andrew},
        title = "{Cosmological N-body simulations with suppressed variance}",
      journal = {\mnras},
         year = 2016,
        month = oct,
       volume = {462},
       number = {1},
        pages = {L1-L5},
          doi = {10.1093/mnrasl/slw098},
archivePrefix = {arXiv},
       eprint = {1603.05253},
 primaryClass = {astro-ph.CO},
       adsurl = {https://ui.adsabs.harvard.edu/abs/2016MNRAS.462L...1A}
}

@ARTICLE{Aoyama2017,
       author = {{Aoyama}, Shohei and {Hou}, Kuan-Chou and {Shimizu}, Ikkoh and {Hirashita}, Hiroyuki and {Todoroki}, Keita and {Choi}, Jun-Hwan and {Nagamine}, Kentaro},
        title = "{Galaxy simulation with dust formation and destruction}",
      journal = {\mnras},
         year = 2017,
        month = apr,
       volume = {466},
       number = {1},
        pages = {105-121},
          doi = {10.1093/mnras/stw3061},
archivePrefix = {arXiv},
       eprint = {1609.07547},
 primaryClass = {astro-ph.GA},
       adsurl = {https://ui.adsabs.harvard.edu/abs/2017MNRAS.466..105A}
}

@ARTICLE{Applebaum2021,
       author = {{Applebaum}, Elaad and {Brooks}, Alyson M. and {Christensen}, Charlotte R. and {Munshi}, Ferah and {Quinn}, Thomas R. and {Shen}, Sijing and {Tremmel}, Michael},
        title = "{Ultrafaint Dwarfs in a Milky Way Context: Introducing the Mint Condition DC Justice League Simulations}",
      journal = {\apj},
         year = 2021,
        month = jan,
       volume = {906},
       number = {2},
          eid = {96},
        pages = {96},
          doi = {10.3847/1538-4357/abcafa},
archivePrefix = {arXiv},
       eprint = {2008.11207},
 primaryClass = {astro-ph.GA},
       adsurl = {https://ui.adsabs.harvard.edu/abs/2021ApJ...906...96A}
}

@ARTICLE{Asplund2009,
       author = {{Asplund}, Martin and {Grevesse}, Nicolas and {Sauval}, A. Jacques and {Scott}, Pat},
        title = "{The Chemical Composition of the Sun}",
      journal = {\araa},
         year = 2009,
        month = sep,
       volume = {47},
       number = {1},
        pages = {481-522},
          doi = {10.1146/annurev.astro.46.060407.145222},
archivePrefix = {arXiv},
       eprint = {0909.0948},
 primaryClass = {astro-ph.SR},
       adsurl = {https://ui.adsabs.harvard.edu/abs/2009ARA&A..47..481A}
}

@ARTICLE{Aver2021,
       author = {{Aver}, Erik and {Berg}, Danielle A. and {Olive}, Keith A. and {Pogge}, Richard W. and {Salzer}, John J. and {Skillman}, Evan D.},
        title = "{Improving helium abundance determinations with Leo P as a case study}",
      journal = {\jcap},
         year = 2021,
        month = mar,
       volume = {2021},
       number = {3},
          eid = {027},
        pages = {027},
          doi = {10.1088/1475-7516/2021/03/027},
archivePrefix = {arXiv},
       eprint = {2010.04180},
 primaryClass = {astro-ph.CO},
       adsurl = {https://ui.adsabs.harvard.edu/abs/2021JCAP...03..027A}
}

@ARTICLE{Bagla2003,
   author = {{Bagla}, J.~S. and {Ray}, S.},
    title = "{Performance characteristics of TreePM codes}",
  journal = {\na},
   eprint = {astro-ph/0212129},
     year = 2003,
    month = sep,
   volume = 8,
    pages = {665-677},
      doi = {10.1016/S1384-1076(03)00056-3},
   adsurl = {http://adsabs.harvard.edu/abs/2003NewA....8..665B}
}

@ARTICLE{Bahe2016,
       author = {{Bah{\'e}}, Yannick M. and {Crain}, Robert A. and {Kauffmann}, Guinevere and {Bower}, Richard G. and {Schaye}, Joop and {Furlong}, Michelle and {Lagos}, Claudia and {Schaller}, Matthieu and {Trayford}, James W. and {Dalla Vecchia}, Claudio and {Theuns}, Tom},
        title = "{The distribution of atomic hydrogen in EAGLE galaxies: morphologies, profiles, and H I holes}",
      journal = {\mnras},
         year = 2016,
        month = feb,
       volume = {456},
       number = {1},
        pages = {1115-1136},
          doi = {10.1093/mnras/stv2674},
archivePrefix = {arXiv},
       eprint = {1511.04909},
 primaryClass = {astro-ph.GA},
       adsurl = {https://ui.adsabs.harvard.edu/abs/2016MNRAS.456.1115B}
}

@ARTICLE{Bahe2022,
       author = {{Bah{\'e}}, Yannick M. and {Schaye}, Joop and {Schaller}, Matthieu and {Bower}, Richard G. and {Borrow}, Josh and {Chaikin}, Evgenii and {Kugel}, Roi and {Nobels}, Folkert and {Ploeckinger}, Sylvia},
        title = "{The importance of black hole repositioning for galaxy formation simulations}",
      journal = {\mnras},
         year = 2022,
        month = oct,
       volume = {516},
       number = {1},
        pages = {167-184},
          doi = {10.1093/mnras/stac1339},
archivePrefix = {arXiv},
       eprint = {2109.01489},
 primaryClass = {astro-ph.GA},
       adsurl = {https://ui.adsabs.harvard.edu/abs/2022MNRAS.516..167B}
}

@ARTICLE{Baker2025,
       author = {{Baker}, William M. and {Lim}, Seunghwan and {D'Eugenio}, Francesco and {Maiolino}, Roberto and {Ji}, Zhiyuan and {Arribas}, Santiago and {Bunker}, Andrew J. and {Carniani}, Stefano and {Charlot}, Stephane and {de Graaff}, Anna and {Hainline}, Kevin and {Looser}, Tobias J. and {Lyu}, Jianwei and {Rinaldi}, Pierluigi and {Robertson}, Brant and {Schaller}, Matthieu and {Schaye}, Joop and {Scholtz}, Jan and {{\"U}bler}, Hannah and {Williams}, Christina C. and {Willmer}, Christopher N.~A. and {Willott}, Chris and {Zhu}, Yongda},
        title = "{The abundance and nature of high-redshift quiescent galaxies from JADES spectroscopy and the FLAMINGO simulations}",
      journal = {\mnras},
         year = 2025,
        month = may,
       volume = {539},
       number = {1},
        pages = {557-589},
          doi = {10.1093/mnras/staf475},
archivePrefix = {arXiv},
       eprint = {2410.14773},
 primaryClass = {astro-ph.GA},
       adsurl = {https://ui.adsabs.harvard.edu/abs/2025MNRAS.539..557B}
}

@ARTICLE{Barausse2012,
       author = {{Barausse}, E. and {Morozova}, V. and {Rezzolla}, L.},
        title = "{On the Mass Radiated by Coalescing Black Hole Binaries}",
      journal = {\apj},
         year = 2012,
        month = oct,
       volume = {758},
       number = {1},
          eid = {63},
        pages = {63},
          doi = {10.1088/0004-637X/758/1/63},
archivePrefix = {arXiv},
       eprint = {1206.3803},
 primaryClass = {gr-qc},
       adsurl = {https://ui.adsabs.harvard.edu/abs/2012ApJ...758...63B}
}

@ARTICLE{Barris2006,
       author = {{Barris}, Brian J. and {Tonry}, John L.},
        title = "{The Rate of Type Ia Supernovae at High Redshift}",
      journal = {\apj},
         year = 2006,
        month = jan,
       volume = {637},
       number = {1},
        pages = {427-438},
          doi = {10.1086/498292},
archivePrefix = {arXiv},
       eprint = {astro-ph/0509655},
 primaryClass = {astro-ph},
       adsurl = {https://ui.adsabs.harvard.edu/abs/2006ApJ...637..427B}
}

@ARTICLE{Bate1997,
       author = {{Bate}, Matthew R. and {Burkert}, Andreas},
        title = "{Resolution requirements for smoothed particle hydrodynamics calculations with self-gravity}",
      journal = {\mnras},
         year = 1997,
        month = jul,
       volume = {288},
       number = {4},
        pages = {1060-1072},
          doi = {10.1093/mnras/288.4.1060},
       adsurl = {https://ui.adsabs.harvard.edu/abs/1997MNRAS.288.1060B}
}

@ARTICLE{Bauer2013,
       author = {{Bauer}, Amanda E. and {Hopkins}, Andrew M. and {Gunawardhana}, Madusha and {Taylor}, Edward N. and {Baldry}, Ivan and {Bamford}, Steven P. and {Bland-Hawthorn}, Joss and {Brough}, Sarah and {Brown}, Michael J.~I. and {Cluver}, Michelle E. and {Colless}, Matthew and {Conselice}, Christopher J. and {Croom}, Scott and {Driver}, Simon and {Foster}, Caroline and {Jones}, D. Heath and {Lara-Lopez}, Maritza A. and {Liske}, Jochen and {L{\'o}pez-S{\'a}nchez}, {\'A}ngel R. and {Loveday}, Jon and {Norberg}, Peder and {Owers}, Matt S. and {Pimbblet}, Kevin and {Robotham}, Aaron and {Sansom}, Anne E. and {Sharp}, Rob},
        title = "{Galaxy And Mass Assembly (GAMA): linking star formation histories and stellar mass growth}",
      journal = {\mnras},
         year = 2013,
        month = sep,
       volume = {434},
       number = {1},
        pages = {209-221},
          doi = {10.1093/mnras/stt1011},
archivePrefix = {arXiv},
       eprint = {1306.2424},
 primaryClass = {astro-ph.CO},
       adsurl = {https://ui.adsabs.harvard.edu/abs/2013MNRAS.434..209B}
}

@ARTICLE{Behroozi2019,
       author = {{Behroozi}, Peter and {Wechsler}, Risa H. and {Hearin}, Andrew P. and {Conroy}, Charlie},
        title = "{UNIVERSEMACHINE: The correlation between galaxy growth and dark matter halo assembly from z = 0-10}",
      journal = {\mnras},
         year = 2019,
        month = sep,
       volume = {488},
       number = {3},
        pages = {3143-3194},
          doi = {10.1093/mnras/stz1182},
archivePrefix = {arXiv},
       eprint = {1806.07893},
 primaryClass = {astro-ph.GA},
       adsurl = {https://ui.adsabs.harvard.edu/abs/2019MNRAS.488.3143B}
}

@ARTICLE{Belfiore2018,
       author = {{Belfiore}, Francesco and {Maiolino}, Roberto and {Bundy}, Kevin and {Masters}, Karen and {Bershady}, Matthew and {Oyarz{\'u}n}, Grecco A. and {Lin}, Lihwai and {Cano-Diaz}, Mariana and {Wake}, David and {Spindler}, Ashley and {Thomas}, Daniel and {Brownstein}, Joel R. and {Drory}, Niv and {Yan}, Renbin},
        title = "{SDSS IV MaNGA - sSFR profiles and the slow quenching of discs in green valley galaxies}",
      journal = {\mnras},
         year = 2018,
        month = jul,
       volume = {477},
       number = {3},
        pages = {3014-3029},
          doi = {10.1093/mnras/sty768},
archivePrefix = {arXiv},
       eprint = {1710.05034},
 primaryClass = {astro-ph.GA},
       adsurl = {https://ui.adsabs.harvard.edu/abs/2018MNRAS.477.3014B}
}

@ARTICLE{BenitezLlambay2025,
       author = {{Ben{\'\i}tez-Llambay}, Alejandro and {Ploeckinger}, Sylvia and {Schaye}, Joop and {Richings}, Alexander J. and {Chaikin}, Evgenii and {Schaller}, Matthieu and {Trayford}, James W. and {Frenk}, Carlos S. and {Hu{\v{s}}ko}, Filip and {Correa}, Camila},
        title = "{Non-explosive pre-supernova feedback in the COLIBRE model of galaxy formation}",
      journal = {arXiv e-prints},
         year = 2025,
        month = sep,
          eid = {arXiv:2509.25309},
        pages = {arXiv:2509.25309},
          doi = {10.48550/arXiv.2509.25309},
archivePrefix = {arXiv},
       eprint = {2509.25309},
 primaryClass = {astro-ph.GA},
       adsurl = {https://ui.adsabs.harvard.edu/abs/2025arXiv250925309B}
}

@ARTICLE{Bernardi2010,
       author = {{Bernardi}, M. and {Shankar}, F. and {Hyde}, J.~B. and {Mei}, S. and {Marulli}, F. and {Sheth}, R.~K.},
        title = "{Galaxy luminosities, stellar masses, sizes, velocity dispersions as a function of morphological type}",
      journal = {\mnras},
         year = 2010,
        month = jun,
       volume = {404},
       number = {4},
        pages = {2087-2122},
          doi = {10.1111/j.1365-2966.2010.16425.x},
archivePrefix = {arXiv},
       eprint = {0910.1093},
 primaryClass = {astro-ph.CO},
       adsurl = {https://ui.adsabs.harvard.edu/abs/2010MNRAS.404.2087B}
}

@ARTICLE{Bianchi2018,
       author = {{Bianchi}, S. and {De Vis}, P. and {Viaene}, S. and {Nersesian}, A. and {Mosenkov}, A.~V. and {Xilouris}, E.~M. and {Baes}, M. and {Casasola}, V. and {Cassar{\`a}}, L.~P. and {Clark}, C.~J.~R. and {Davies}, J.~I. and {De Looze}, I. and {Dobbels}, W. and {Galametz}, M. and {Galliano}, F. and {Jones}, A.~P. and {Lianou}, S. and {Madden}, S.~C. and {Tr{\v{c}}ka}, A.},
        title = "{Fraction of bolometric luminosity absorbed by dust in DustPedia galaxies}",
      journal = {\aap},
         year = 2018,
        month = dec,
       volume = {620},
          eid = {A112},
        pages = {A112},
          doi = {10.1051/0004-6361/201833699},
archivePrefix = {arXiv},
       eprint = {1810.01208},
 primaryClass = {astro-ph.GA},
       adsurl = {https://ui.adsabs.harvard.edu/abs/2018A&A...620A.112B}
}

@ARTICLE{Bigwood2025,
       author = {{Bigwood}, Leah and {Bourne}, Martin A. and {Irsic}, Vid and {Amon}, Alexandra and {Sijacki}, Debora},
        title = "{The case for large-scale AGN feedback in galaxy formation simulations: insights from XFABLE}",
      journal = {arXiv e-prints},
         year = 2025,
        month = jan,
          eid = {arXiv:2501.16983},
        pages = {arXiv:2501.16983},
          doi = {10.48550/arXiv.2501.16983},
archivePrefix = {arXiv},
       eprint = {2501.16983},
 primaryClass = {astro-ph.CO},
       adsurl = {https://ui.adsabs.harvard.edu/abs/2025arXiv250116983B}
}

@ARTICLE{Bird2022,
       author = {{Bird}, Simeon and {Ni}, Yueying and {Di Matteo}, Tiziana and {Croft}, Rupert and {Feng}, Yu and {Chen}, Nianyi},
        title = "{The ASTRID simulation: galaxy formation and reionization}",
      journal = {\mnras},
         year = 2022,
        month = may,
       volume = {512},
       number = {3},
        pages = {3703-3716},
          doi = {10.1093/mnras/stac648},
archivePrefix = {arXiv},
       eprint = {2111.01160},
 primaryClass = {astro-ph.GA},
       adsurl = {https://ui.adsabs.harvard.edu/abs/2022MNRAS.512.3703B}
}

@INPROCEEDINGS{Black1987,
       author = {{Black}, John H.},
        title = "{Heating and Cooling of the Interstellar Gas}",
    booktitle = {Interstellar Processes},
         year = 1987,
       editor = {{Hollenbach}, David J. and {Thronson}, Jr., Harley A.},
       volume = {134},
        month = jan,
        pages = {731},
          doi = {10.1007/978-94-009-3861-8_27},
       adsurl = {https://ui.adsabs.harvard.edu/abs/1987ASSL..134..731B},
    series = {}
}

@ARTICLE{Blanc2004,
       author = {{Blanc}, G. and {Afonso}, C. and {Alard}, C. and {Albert}, J.~N. and {Aldering}, G. and {Amadon}, A. and {Andersen}, J. and {Ansari}, R. and {Aubourg}, {\'E}. and {Balland}, C. and {Bareyre}, P. and {Beaulieu}, J.~P. and {Charlot}, X. and {Conley}, A. and {Coutures}, C. and {Dahl{\'e}n}, T. and {Derue}, F. and {Fan}, X. and {Ferlet}, R. and {Folatelli}, G. and {Fouqu{\'e}}, P. and {Garavini}, G. and {Glicenstein}, J.~F. and {Goldman}, B. and {Goobar}, A. and {Gould}, A. and {Graff}, D. and {Gros}, M. and {Haissinski}, J. and {Hamadache}, C. and {Hardin}, D. and {Hook}, I.~M. and {de Kat}, J. and {Kent}, S. and {Kim}, A. and {Lasserre}, T. and {Le Guillou}, L. and {Lesquoy}, {\'E}. and {Loup}, C. and {Magneville}, C. and {Marquette}, J.~B. and {Maurice}, {\'E}. and {Maury}, A. and {Milsztajn}, A. and {Moniez}, M. and {Mouchet}, M. and {Newberg}, H. and {Nobili}, S. and {Palanque-Delabrouille}, N. and {Perdereau}, O. and {Pr{\'e}vot}, L. and {Rahal}, Y.~R. and {Regnault}, N. and {Rich}, J. and {Ruiz-Lapuente}, P. and {Spiro}, M. and {Tisserand}, P. and {Vidal-Madjar}, A. and {Vigroux}, L. and {Walton}, N.~A. and {Zylberajch}, S.},
        title = "{Type Ia supernova rate at a redshift of {\ensuremath{\sim}}0.1}",
      journal = {\aap},
         year = 2004,
        month = sep,
       volume = {423},
        pages = {881-894},
          doi = {10.1051/0004-6361:20035948},
archivePrefix = {arXiv},
       eprint = {astro-ph/0405211},
 primaryClass = {astro-ph},
       adsurl = {https://ui.adsabs.harvard.edu/abs/2004A&A...423..881B}
}

@ARTICLE{Blandford1977,
       author = {{Blandford}, R.~D. and {Znajek}, R.~L.},
        title = "{Electromagnetic extraction of energy from Kerr black holes.}",
      journal = {\mnras},
         year = 1977,
        month = may,
       volume = {179},
        pages = {433-456},
          doi = {10.1093/mnras/179.3.433},
       adsurl = {https://ui.adsabs.harvard.edu/abs/1977MNRAS.179..433B}
}

@ARTICLE{Bolatto2013,
       author = {{Bolatto}, Alberto D. and {Wolfire}, Mark and {Leroy}, Adam K.},
        title = "{The CO-to-H$_{2}$ Conversion Factor}",
      journal = {\araa},
         year = 2013,
        month = aug,
       volume = {51},
       number = {1},
        pages = {207-268},
          doi = {10.1146/annurev-astro-082812-140944},
archivePrefix = {arXiv},
       eprint = {1301.3498},
 primaryClass = {astro-ph.GA},
       adsurl = {https://ui.adsabs.harvard.edu/abs/2013ARA&A..51..207B}
}

@ARTICLE{Bondi1952,
       author = {{Bondi}, H.},
        title = "{On spherically symmetrical accretion}",
      journal = {\mnras},
         year = 1952,
        month = jan,
       volume = {112},
        pages = {195},
          doi = {10.1093/mnras/112.2.195},
       adsurl = {https://ui.adsabs.harvard.edu/abs/1952MNRAS.112..195B}
}

@ARTICLE{Booth2009,
       author = {{Booth}, C.~M. and {Schaye}, Joop},
        title = "{Cosmological simulations of the growth of supermassive black holes and feedback from active galactic nuclei: method and tests}",
      journal = {\mnras},
         year = 2009,
        month = sep,
       volume = {398},
       number = {1},
        pages = {53-74},
          doi = {10.1111/j.1365-2966.2009.15043.x},
archivePrefix = {arXiv},
       eprint = {0904.2572},
 primaryClass = {astro-ph.CO},
       adsurl = {https://ui.adsabs.harvard.edu/abs/2009MNRAS.398...53B}
}

@ARTICLE{Booth2010,
       author = {{Booth}, C.~M. and {Schaye}, Joop},
        title = "{Dark matter haloes determine the masses of supermassive black holes}",
      journal = {\mnras},
         year = 2010,
        month = jun,
       volume = {405},
       number = {1},
        pages = {L1-L5},
          doi = {10.1111/j.1745-3933.2010.00832.x},
archivePrefix = {arXiv},
       eprint = {0911.0935},
 primaryClass = {astro-ph.CO},
       adsurl = {https://ui.adsabs.harvard.edu/abs/2010MNRAS.405L...1B}
}

@ARTICLE{Borrow2020,
       author = {{Borrow}, Josh and {Borrisov}, Alexei},
        title = "{swiftsimio: A Python library for reading SWIFT data}",
      journal = {The Journal of Open Source Software},
         year = 2020,
        month = aug,
       volume = {5},
       number = {52},
          eid = {2430},
        pages = {2430},
          doi = {10.21105/joss.02430},
       adsurl = {https://ui.adsabs.harvard.edu/abs/2020JOSS....5.2430B}
}

@ARTICLE{Borrow2022sphenix,
       author = {{Borrow}, Josh and {Schaller}, Matthieu and {Bower}, Richard G. and {Schaye}, Joop},
        title = "{SPHENIX: smoothed particle hydrodynamics for the next generation of galaxy formation simulations}",
      journal = {\mnras},
         year = 2022,
        month = apr,
       volume = {511},
       number = {2},
        pages = {2367-2389},
          doi = {10.1093/mnras/stab3166},
archivePrefix = {arXiv},
       eprint = {2012.03974},
 primaryClass = {astro-ph.GA},
       adsurl = {https://ui.adsabs.harvard.edu/abs/2022MNRAS.511.2367B}
}

@ARTICLE{Borrow2023,
       author = {{Borrow}, Josh and {Schaller}, Matthieu and {Bah{\'e}}, Yannick M. and {Schaye}, Joop and {Ludlow}, Aaron D. and {Ploeckinger}, Sylvia and {Nobels}, Folkert S.~J. and {Altamura}, Edoardo},
        title = "{The impact of stochastic modelling on the predictive power of galaxy formation simulations}",
      journal = {\mnras},
         year = 2023,
        month = dec,
       volume = {526},
       number = {2},
        pages = {2441-2457},
          doi = {10.1093/mnras/stad2928},
archivePrefix = {arXiv},
       eprint = {2211.08442},
 primaryClass = {astro-ph.GA},
       adsurl = {https://ui.adsabs.harvard.edu/abs/2023MNRAS.526.2441B}
}

@ARTICLE{Botticella2008,
       author = {{Botticella}, M.~T. and {Riello}, M. and {Cappellaro}, E. and {Benetti}, S. and {Altavilla}, G. and {Pastorello}, A. and {Turatto}, M. and {Greggio}, L. and {Patat}, F. and {Valenti}, S. and {Zampieri}, L. and {Harutyunyan}, A. and {Pignata}, G. and {Taubenberger}, S.},
        title = "{Supernova rates from the Southern inTermediate Redshift ESO Supernova Search (STRESS)}",
      journal = {\aap},
         year = 2008,
        month = feb,
       volume = {479},
       number = {1},
        pages = {49-66},
          doi = {10.1051/0004-6361:20078011},
archivePrefix = {arXiv},
       eprint = {0710.3763},
 primaryClass = {astro-ph},
       adsurl = {https://ui.adsabs.harvard.edu/abs/2008A&A...479...49B}
}

@ARTICLE{Bouwens2023,
       author = {{Bouwens}, Rychard and {Illingworth}, Garth and {Oesch}, Pascal and {Stefanon}, Mauro and {Naidu}, Rohan and {van Leeuwen}, Ivana and {Magee}, Dan},
        title = "{UV luminosity density results at z > 8 from the first JWST/NIRCam fields: limitations of early data sets and the need for spectroscopy}",
      journal = {\mnras},
         year = 2023,
        month = jul,
       volume = {523},
       number = {1},
        pages = {1009-1035},
          doi = {10.1093/mnras/stad1014},
archivePrefix = {arXiv},
       eprint = {2212.06683},
 primaryClass = {astro-ph.CO},
       adsurl = {https://ui.adsabs.harvard.edu/abs/2023MNRAS.523.1009B}
}

@ARTICLE{Braspenning2023,
       author = {{Braspenning}, Joey and {Schaye}, Joop and {Borrow}, Josh and {Schaller}, Matthieu},
        title = "{Sensitivity of non-radiative cloud-wind interactions to the hydrodynamic solver}",
      journal = {\mnras},
         year = 2023,
        month = jul,
       volume = {523},
       number = {1},
        pages = {1280-1295},
          doi = {10.1093/mnras/stad1243},
archivePrefix = {arXiv},
       eprint = {2203.13915},
 primaryClass = {astro-ph.GA},
       adsurl = {https://ui.adsabs.harvard.edu/abs/2023MNRAS.523.1280B}
}

@ARTICLE{Braspenning2024,
       author = {{Braspenning}, Joey and {Schaye}, Joop and {Schaller}, Matthieu and {McCarthy}, Ian G. and {Kay}, Scott T. and {Helly}, John C. and {Kugel}, Roi and {Elbers}, Willem and {Frenk}, Carlos S. and {Kwan}, Juliana and {Salcido}, Jaime and {van Daalen}, Marcel P. and {Vandenbroucke}, Bert},
        title = "{The FLAMINGO project: galaxy clusters in comparison to X-ray observations}",
      journal = {\mnras},
         year = 2024,
        month = sep,
       volume = {533},
       number = {3},
        pages = {2656-2676},
          doi = {10.1093/mnras/stae1436},
archivePrefix = {arXiv},
       eprint = {2312.08277},
 primaryClass = {astro-ph.GA},
       adsurl = {https://ui.adsabs.harvard.edu/abs/2024MNRAS.533.2656B}
}

@ARTICLE{Bregman1986,
       author = {{Bregman}, Joel N. and {Harrington}, J. Patrick},
        title = "{Photoionization in the Halo of the Galaxy}",
      journal = {\apj},
         year = 1986,
        month = oct,
       volume = {309},
        pages = {833},
          doi = {10.1086/164652},
       adsurl = {https://ui.adsabs.harvard.edu/abs/1986ApJ...309..833B}
}

@ARTICLE{Bryan1998,
       author = {{Bryan}, Greg L. and {Norman}, Michael L.},
        title = "{Statistical Properties of X-Ray Clusters: Analytic and Numerical Comparisons}",
      journal = {\apj},
         year = 1998,
        month = mar,
       volume = {495},
       number = {1},
        pages = {80-99},
          doi = {10.1086/305262},
archivePrefix = {arXiv},
       eprint = {astro-ph/9710107},
 primaryClass = {astro-ph},
       adsurl = {https://ui.adsabs.harvard.edu/abs/1998ApJ...495...80B}
}

@ARTICLE{Buck2020,
       author = {{Buck}, Tobias and {Obreja}, Aura and {Macci{\`o}}, Andrea V. and {Minchev}, Ivan and {Dutton}, Aaron A. and {Ostriker}, Jeremiah P.},
        title = "{NIHAO-UHD: the properties of MW-like stellar discs in high-resolution cosmological simulations}",
      journal = {\mnras},
         year = 2020,
        month = jan,
       volume = {491},
       number = {3},
        pages = {3461-3478},
          doi = {10.1093/mnras/stz3241},
archivePrefix = {arXiv},
       eprint = {1909.05864},
 primaryClass = {astro-ph.GA},
       adsurl = {https://ui.adsabs.harvard.edu/abs/2020MNRAS.491.3461B}
}

@ARTICLE{Buck2021,
       author = {{Buck}, Tobias and {Rybizki}, Jan and {Buder}, Sven and {Obreja}, Aura and {Macci{\`o}}, Andrea V. and {Pfrommer}, Christoph and {Steinmetz}, Matthias and {Ness}, Melissa},
        title = "{The challenge of simultaneously matching the observed diversity of chemical abundance patterns in cosmological hydrodynamical simulations}",
      journal = {\mnras},
         year = 2021,
        month = dec,
       volume = {508},
       number = {3},
        pages = {3365-3387},
          doi = {10.1093/mnras/stab2736},
archivePrefix = {arXiv},
       eprint = {2103.03884},
 primaryClass = {astro-ph.GA},
       adsurl = {https://ui.adsabs.harvard.edu/abs/2021MNRAS.508.3365B}
}

@ARTICLE{Buehlmann2025,
       author = {{Buehlmann}, Michael and {Winkler}, Lukas and {Hahn}, Oliver and {Helly}, John C. and {Jenkins}, Adrian},
        title = "{cosmICweb: Cosmological Initial Conditions for Zoom-in Simulations in the Cloud}",
      journal = {The Open Journal of Astrophysics},
         year = 2025,
        month = mar,
       volume = {8},
          eid = {32},
        pages = {32},
          doi = {10.33232/001c.133696},
archivePrefix = {arXiv},
       eprint = {2406.02693},
 primaryClass = {astro-ph.CO},
       adsurl = {https://ui.adsabs.harvard.edu/abs/2025OJAp....8E..32B}
}

@ARTICLE{Bulbul2019,
       author = {{Bulbul}, Esra and {Chiu}, I. -Non and {Mohr}, Joseph J. and {McDonald}, Michael and {Benson}, Bradford and {Bautz}, Mark W. and {Bayliss}, Matthew and {Bleem}, Lindsey and {Brodwin}, Mark and {Bocquet}, Sebastian and {Capasso}, Raffaella and {Dietrich}, J{\"o}rg P. and {Forman}, Bill and {Hlavacek-Larrondo}, Julie and {Holzapfel}, W.~L. and {Khullar}, Gourav and {Klein}, Matthias and {Kraft}, Ralph and {Miller}, Eric D. and {Reichardt}, Christian and {Saro}, Alex and {Sharon}, Keren and {Stalder}, Brian and {Schrabback}, Tim and {Stanford}, Adam},
        title = "{X-Ray Properties of SPT-selected Galaxy Clusters at 0.2 < z < 1.5 Observed with XMM-Newton}",
      journal = {\apj},
         year = 2019,
        month = jan,
       volume = {871},
       number = {1},
          eid = {50},
        pages = {50},
          doi = {10.3847/1538-4357/aaf230},
archivePrefix = {arXiv},
       eprint = {1807.02556},
 primaryClass = {astro-ph.CO},
       adsurl = {https://ui.adsabs.harvard.edu/abs/2019ApJ...871...50B}
}

@ARTICLE{Cappellaro1999,
       author = {{Cappellaro}, E. and {Evans}, R. and {Turatto}, M.},
        title = "{A new determination of supernova rates and a comparison with indicators for galactic star formation}",
      journal = {\aap},
         year = 1999,
        month = nov,
       volume = {351},
        pages = {459-466},
          doi = {10.48550/arXiv.astro-ph/9904225},
archivePrefix = {arXiv},
       eprint = {astro-ph/9904225},
 primaryClass = {astro-ph},
       adsurl = {https://ui.adsabs.harvard.edu/abs/1999A&A...351..459C}
}

@ARTICLE{Cappellaro2015,
       author = {{Cappellaro}, E. and {Botticella}, M.~T. and {Pignata}, G. and {Grado}, A. and {Greggio}, L. and {Limatola}, L. and {Vaccari}, M. and {Baruffolo}, A. and {Benetti}, S. and {Bufano}, F. and {Capaccioli}, M. and {Cascone}, E. and {Covone}, G. and {De Cicco}, D. and {Falocco}, S. and {Della Valle}, M. and {Jarvis}, M. and {Marchetti}, L. and {Napolitano}, N.~R. and {Paolillo}, M. and {Pastorello}, A. and {Radovich}, M. and {Schipani}, P. and {Spiro}, S. and {Tomasella}, L. and {Turatto}, M.},
        title = "{Supernova rates from the SUDARE VST-OmegaCAM search. I. Rates per unit volume}",
      journal = {\aap},
         year = 2015,
        month = dec,
       volume = {584},
          eid = {A62},
        pages = {A62},
          doi = {10.1051/0004-6361/201526712},
archivePrefix = {arXiv},
       eprint = {1509.04496},
 primaryClass = {astro-ph.CO},
       adsurl = {https://ui.adsabs.harvard.edu/abs/2015A&A...584A..62C}
}

@ARTICLE{Carnall2024,
       author = {{Carnall}, A.~C. and {Cullen}, F. and {McLure}, R.~J. and {McLeod}, D.~J. and {Begley}, R. and {Donnan}, C.~T. and {Dunlop}, J.~S. and {Shapley}, A.~E. and {Rowlands}, K. and {Almaini}, O. and {Arellano-C{\'o}rdova}, K.~Z. and {Barrufet}, L. and {Cimatti}, A. and {Ellis}, R.~S. and {Grogin}, N.~A. and {Hamadouche}, M.~L. and {Illingworth}, G.~D. and {Koekemoer}, A.~M. and {Leung}, H. -H. and {Lovell}, C.~C. and {P{\'e}rez-Gonz{\'a}lez}, P.~G. and {Santini}, P. and {Stanton}, T.~M. and {Wild}, V.},
        title = "{The JWST EXCELS survey: too much, too young, too fast? Ultra-massive quiescent galaxies at 3 < z < 5}",
      journal = {\mnras},
         year = 2024,
        month = oct,
       volume = {534},
       number = {1},
        pages = {325-348},
          doi = {10.1093/mnras/stae2092},
archivePrefix = {arXiv},
       eprint = {2405.02242},
 primaryClass = {astro-ph.GA},
       adsurl = {https://ui.adsabs.harvard.edu/abs/2024MNRAS.534..325C}
}

@ARTICLE{Chabrier2003,
       author = {{Chabrier}, Gilles},
        title = "{Galactic Stellar and Substellar Initial Mass Function}",
      journal = {\pasp},
         year = 2003,
        month = jul,
       volume = {115},
       number = {809},
        pages = {763-795},
          doi = {10.1086/376392},
archivePrefix = {arXiv},
       eprint = {astro-ph/0304382},
 primaryClass = {astro-ph},
       adsurl = {https://ui.adsabs.harvard.edu/abs/2003PASP..115..763C}
}

@ARTICLE{Chaikin2022,
       author = {{Chaikin}, Evgenii and {Schaye}, Joop and {Schaller}, Matthieu and {Bah{\'e}}, Yannick M. and {Nobels}, Folkert S.~J. and {Ploeckinger}, Sylvia},
        title = "{The importance of the way in which supernova energy is distributed around young stellar populations in simulations of galaxies}",
      journal = {\mnras},
         year = 2022,
        month = jul,
       volume = {514},
       number = {1},
        pages = {249-264},
          doi = {10.1093/mnras/stac1132},
archivePrefix = {arXiv},
       eprint = {2203.07134},
 primaryClass = {astro-ph.GA},
       adsurl = {https://ui.adsabs.harvard.edu/abs/2022MNRAS.514..249C}
}

@ARTICLE{Chaikin2023,
       author = {{Chaikin}, Evgenii and {Schaye}, Joop and {Schaller}, Matthieu and {Ben{\'\i}tez-Llambay}, Alejandro and {Nobels}, Folkert S.~J. and {Ploeckinger}, Sylvia},
        title = "{A thermal-kinetic subgrid model for supernova feedback in simulations of galaxy formation}",
      journal = {\mnras},
         year = 2023,
        month = aug,
       volume = {523},
       number = {3},
        pages = {3709-3731},
          doi = {10.1093/mnras/stad1626},
archivePrefix = {arXiv},
       eprint = {2211.04619},
 primaryClass = {astro-ph.GA},
       adsurl = {https://ui.adsabs.harvard.edu/abs/2023MNRAS.523.3709C}
}

@ARTICLE{Chaikin2025calibration,
       author = {{Chaikin}, Evgenii and {Schaye}, Joop and {Schaller}, Matthieu and {Ploeckinger}, Sylvia and {Bah{\'e}}, Yannick M. and {Ben{\'\i}tez-Llambay}, Alejandro and {Correa}, Camila and {Forouhar Moreno}, Victor J. and {Frenk}, Carlos S. and {Hu{\v{s}}ko}, Filip and {Kugel}, Roi and {McGibbon}, Robert and {Richings}, Alexander J. and {Trayford}, James W. and {Borrow}, Josh and {Crain}, Robert A. and {Helly}, John C. and {Lacey}, Cedric G. and {Ludlow}, Aaron and {Nobels}, Folkert S.~J.},
        title = "{COLIBRE: calibrating subgrid feedback in cosmological simulations that include a cold gas phase}",
      journal = {arXiv e-prints},
         year = 2025,
        month = sep,
          eid = {arXiv:2509.04067},
        pages = {arXiv:2509.04067},
          doi = {10.48550/arXiv.2509.04067},
archivePrefix = {arXiv},
       eprint = {2509.04067},
 primaryClass = {astro-ph.GA},
       adsurl = {https://ui.adsabs.harvard.edu/abs/2025arXiv250904067C}
}

@ARTICLE{Chaikin2025smf_evol,
       author = {{Chaikin}, Evgenii and {Schaye}, Joop and {Schaller}, Matthieu and {Ploeckinger}, Sylvia and {Ben{\'\i}tez-Llambay}, Alejandro and {Frenk}, Carlos S. and {Hu{\v{s}}ko}, Filip and {McGibbon}, Robert and {Richings}, Alexander J. and {Trayford}, James W.},
        title = "{The evolution of the galaxy stellar mass function and star formation rates in the COLIBRE simulations from redshift 17 to 0}",
      journal = {arXiv e-prints},
         year = 2025,
        month = sep,
          eid = {arXiv:2509.07960},
        pages = {arXiv:2509.07960},
          doi = {10.48550/arXiv.2509.07960},
archivePrefix = {arXiv},
       eprint = {2509.07960},
 primaryClass = {astro-ph.GA},
       adsurl = {https://ui.adsabs.harvard.edu/abs/2025arXiv250907960C}
}

@ARTICLE{Chandro-Gomez2025,
       author = {{Chandro-G{\'o}mez}, {\'A}ngel and {Lagos}, Claudia del P. and {Power}, Chris and {Moreno}, Victor J. Forouhar and {Helly}, John C. and {Lacey}, Cedric G. and {McGibbon}, Robert J. and {Schaller}, Matthieu and {Schaye}, Joop},
        title = "{On the accuracy of dark matter halo merger trees and the consequences for semi-analytic models of galaxy formation}",
      journal = {\mnras},
         year = 2025,
        month = may,
       volume = {539},
       number = {2},
        pages = {776-807},
          doi = {10.1093/mnras/staf519},
archivePrefix = {arXiv},
       eprint = {2501.07677},
 primaryClass = {astro-ph.GA},
       adsurl = {https://ui.adsabs.harvard.edu/abs/2025MNRAS.539..776C}
}

@ARTICLE{Chang2015,
       author = {{Chang}, Yu-Yen and {van der Wel}, Arjen and {da Cunha}, Elisabete and {Rix}, Hans-Walter},
        title = "{Stellar Masses and Star Formation Rates for 1M Galaxies from SDSS+WISE}",
      journal = {\apjs},
         year = 2015,
        month = jul,
       volume = {219},
       number = {1},
          eid = {8},
        pages = {8},
          doi = {10.1088/0067-0049/219/1/8},
archivePrefix = {arXiv},
       eprint = {1506.00648},
 primaryClass = {astro-ph.GA},
       adsurl = {https://ui.adsabs.harvard.edu/abs/2015ApJS..219....8C}
}

@article{Cheng1999,
title = "A Fast Adaptive Multipole Algorithm in Three Dimensions",
journal = "Journal of Computational Physics",
volume = "155",
number = "2",
pages = "468 - 498",
year = "1999",
note = "",
issn = "0021-9991",
doi = "http://dx.doi.org/10.1006/jcph.1999.6355",
url = "http://www.sciencedirect.com/science/article/pii/S0021999199963556",
author = "H. Cheng and L. Greengard and V. Rokhlin"
}

@ARTICLE{Chevance2022,
       author = {{Chevance}, M{\'e}lanie and {Kruijssen}, J.~M. Diederik and {Krumholz}, Mark R. and {Groves}, Brent and {Keller}, Benjamin W. and {Hughes}, Annie and {Glover}, Simon C.~O. and {Henshaw}, Jonathan D. and {Herrera}, Cinthya N. and {Kim}, Jaeyeon and {Leroy}, Adam K. and {Pety}, J{\'e}r{\^o}me and {Razza}, Alessandro and {Rosolowsky}, Erik and {Schinnerer}, Eva and {Schruba}, Andreas and {Barnes}, Ashley T. and {Bigiel}, Frank and {Blanc}, Guillermo A. and {Dale}, Daniel A. and {Emsellem}, Eric and {Faesi}, Christopher M. and {Grasha}, Kathryn and {Klessen}, Ralf S. and {Kreckel}, Kathryn and {Liu}, Daizhong and {Longmore}, Steven N. and {Meidt}, Sharon E. and {Querejeta}, Miguel and {Saito}, Toshiki and {Sun}, Jiayi and {Usero}, Antonio},
        title = "{Pre-supernova feedback mechanisms drive the destruction of molecular clouds in nearby star-forming disc galaxies}",
      journal = {\mnras},
         year = 2022,
        month = jan,
       volume = {509},
       number = {1},
        pages = {272-288},
          doi = {10.1093/mnras/stab2938},
archivePrefix = {arXiv},
       eprint = {2010.13788},
 primaryClass = {astro-ph.GA},
       adsurl = {https://ui.adsabs.harvard.edu/abs/2022MNRAS.509..272C}
}

@ARTICLE{Cinquegrana2022,
       author = {{Cinquegrana}, Giulia C. and {Karakas}, Amanda I.},
        title = "{The most metal-rich stars in the universe: chemical contributions of low- and intermediate-mass asymptotic giant branch stars with metallicities within 0.04 {\ensuremath{\leq}} Z {\ensuremath{\leq}} 0.10}",
      journal = {\mnras},
         year = 2022,
        month = feb,
       volume = {510},
       number = {2},
        pages = {1557-1576},
          doi = {10.1093/mnras/stab3379},
archivePrefix = {arXiv},
       eprint = {2111.09527},
 primaryClass = {astro-ph.SR},
       adsurl = {https://ui.adsabs.harvard.edu/abs/2022MNRAS.510.1557C}
}

@ARTICLE{Cochrane2023,
       author = {{Cochrane}, R.~K. and {Kondapally}, R. and {Best}, P.~N. and {Sabater}, J. and {Duncan}, K.~J. and {Smith}, D.~J.~B. and {Hardcastle}, M.~J. and {R{\"o}ttgering}, H.~J.~A. and {Prandoni}, I. and {Haskell}, P. and {G{\"u}rkan}, G. and {Miley}, G.~K.},
        title = "{The LOFAR Two-metre Sky Survey: the radio view of the cosmic star formation history}",
      journal = {\mnras},
         year = 2023,
        month = aug,
       volume = {523},
       number = {4},
        pages = {6082-6102},
          doi = {10.1093/mnras/stad1602},
archivePrefix = {arXiv},
       eprint = {2305.15510},
 primaryClass = {astro-ph.GA},
       adsurl = {https://ui.adsabs.harvard.edu/abs/2023MNRAS.523.6082C}
}

@ARTICLE{Conroy2009,
       author = {{Conroy}, Charlie and {Gunn}, James E. and {White}, Martin},
        title = "{The Propagation of Uncertainties in Stellar Population Synthesis Modeling. I. The Relevance of Uncertain Aspects of Stellar Evolution and the Initial Mass Function to the Derived Physical Properties of Galaxies}",
      journal = {\apj},
         year = 2009,
        month = jul,
       volume = {699},
       number = {1},
        pages = {486-506},
          doi = {10.1088/0004-637X/699/1/486},
archivePrefix = {arXiv},
       eprint = {0809.4261},
 primaryClass = {astro-ph},
       adsurl = {https://ui.adsabs.harvard.edu/abs/2009ApJ...699..486C}
}

@ARTICLE{Conroy2010,
       author = {{Conroy}, Charlie and {Gunn}, James E.},
        title = "{The Propagation of Uncertainties in Stellar Population Synthesis Modeling. III. Model Calibration, Comparison, and Evaluation}",
      journal = {\apj},
         year = 2010,
        month = apr,
       volume = {712},
       number = {2},
        pages = {833-857},
          doi = {10.1088/0004-637X/712/2/833},
archivePrefix = {arXiv},
       eprint = {0911.3151},
 primaryClass = {astro-ph.CO},
       adsurl = {https://ui.adsabs.harvard.edu/abs/2010ApJ...712..833C}
}

@ARTICLE{Correa2025chemo,
       author = {{Correa}, Camila and others},
        title = "{TBD}",
      journal = {\mnras},
         year = 2025,
       volume = {submitted},
}

@ARTICLE{Courant1928,
       author = {{Courant}, R. and {Friedrichs}, K. and {Lewy}, H.},
        title = "{{\"U}ber die partiellen Differenzengleichungen der mathematischen Physik}",
      journal = {Mathematische Annalen},
         year = 1928,
        month = jan,
       volume = {100},
        pages = {32-74},
          doi = {10.1007/BF01448839},
       adsurl = {https://ui.adsabs.harvard.edu/abs/1928MatAn.100...32C}
}

@ARTICLE{Covelo2025,
       author = {{Covelo-Paz}, Alba and {Giovinazzo}, Emma and {Oesch}, Pascal A. and {Meyer}, Romain A. and {Weibel}, Andrea and {Brammer}, Gabriel and {Fudamoto}, Yoshinobu and {Kerutt}, Josephine and {Lin}, Jamie and {Matharu}, Jasleen and {Naidu}, Rohan P. and {Velichko}, Anna and {Bollo}, Victoria and {Bouwens}, Rychard and {Chisholm}, John and {Illingworth}, Garth D. and {Kramarenko}, Ivan and {Magee}, Daniel and {Maseda}, Michael and {Matthee}, Jorryt and {Nelson}, Erica and {Reddy}, Naveen and {Schaerer}, Daniel and {Stefanon}, Mauro and {Xiao}, Mengyuan},
        title = "{An H{\ensuremath{\alpha}} view of galaxy buildup in the first 2 Gyr: Luminosity functions at z {\ensuremath{\sim}} 4‑6.5 from NIRCam/grism spectroscopy}",
      journal = {\aap},
         year = 2025,
        month = feb,
       volume = {694},
          eid = {A178},
        pages = {A178},
          doi = {10.1051/0004-6361/202452363},
archivePrefix = {arXiv},
       eprint = {2409.17241},
 primaryClass = {astro-ph.GA},
       adsurl = {https://ui.adsabs.harvard.edu/abs/2025A&A...694A.178C}
}

@ARTICLE{Crain2009,
       author = {{Crain}, Robert A. and {Theuns}, Tom and {Dalla Vecchia}, Claudio and {Eke}, Vincent R. and {Frenk}, Carlos S. and {Jenkins}, Adrian and {Kay}, Scott T. and {Peacock}, John A. and {Pearce}, Frazer R. and {Schaye}, Joop and {Springel}, Volker and {Thomas}, Peter A. and {White}, Simon D.~M. and {Wiersma}, Robert P.~C.},
        title = "{Galaxies-intergalactic medium interaction calculation - I. Galaxy formation as a function of large-scale environment}",
      journal = {\mnras},
         year = 2009,
        month = nov,
       volume = {399},
       number = {4},
        pages = {1773-1794},
          doi = {10.1111/j.1365-2966.2009.15402.x},
archivePrefix = {arXiv},
       eprint = {0906.4350},
 primaryClass = {astro-ph.CO},
       adsurl = {https://ui.adsabs.harvard.edu/abs/2009MNRAS.399.1773C}
}

@ARTICLE{Crain2015,
       author = {{Crain}, Robert A. and {Schaye}, Joop and {Bower}, Richard G. and {Furlong}, Michelle and {Schaller}, Matthieu and {Theuns}, Tom and {Dalla Vecchia}, Claudio and {Frenk}, Carlos S. and {McCarthy}, Ian G. and {Helly}, John C. and {Jenkins}, Adrian and {Rosas-Guevara}, Yetli M. and {White}, Simon D.~M. and {Trayford}, James W.},
        title = "{The EAGLE simulations of galaxy formation: calibration of subgrid physics and model variations}",
      journal = {\mnras},
         year = 2015,
        month = jun,
       volume = {450},
       number = {2},
        pages = {1937-1961},
          doi = {10.1093/mnras/stv725},
archivePrefix = {arXiv},
       eprint = {1501.01311},
 primaryClass = {astro-ph.GA},
       adsurl = {https://ui.adsabs.harvard.edu/abs/2015MNRAS.450.1937C}
}

@ARTICLE{Crain2023,
       author = {{Crain}, Robert A. and {van de Voort}, Freeke},
        title = "{Hydrodynamical Simulations of the Galaxy Population: Enduring Successes and Outstanding Challenges}",
      journal = {\araa},
         year = 2023,
        month = aug,
       volume = {61},
        pages = {473-515},
          doi = {10.1146/annurev-astro-041923-043618},
archivePrefix = {arXiv},
       eprint = {2309.17075},
 primaryClass = {astro-ph.GA},
       adsurl = {https://ui.adsabs.harvard.edu/abs/2023ARA&A..61..473C}
}

@ARTICLE{Curti2020,
       author = {{Curti}, Mirko and {Mannucci}, Filippo and {Cresci}, Giovanni and {Maiolino}, Roberto},
        title = "{The mass-metallicity and the fundamental metallicity relation revisited on a fully T$_{e}$-based abundance scale for galaxies}",
      journal = {\mnras},
         year = 2020,
        month = jan,
       volume = {491},
       number = {1},
        pages = {944-964},
          doi = {10.1093/mnras/stz2910},
archivePrefix = {arXiv},
       eprint = {1910.00597},
 primaryClass = {astro-ph.GA},
       adsurl = {https://ui.adsabs.harvard.edu/abs/2020MNRAS.491..944C}
}

@ARTICLE{Dahlen2004,
       author = {{Dahlen}, Tomas and {Strolger}, Louis-Gregory and {Riess}, Adam G. and {Mobasher}, Bahram and {Chary}, Ranga-Ram and {Conselice}, Christopher J. and {Ferguson}, Henry C. and {Fruchter}, Andrew S. and {Giavalisco}, Mauro and {Livio}, Mario and {Madau}, Piero and {Panagia}, Nino and {Tonry}, John L.},
        title = "{High-Redshift Supernova Rates}",
      journal = {\apj},
         year = 2004,
        month = sep,
       volume = {613},
       number = {1},
        pages = {189-199},
          doi = {10.1086/422899},
archivePrefix = {arXiv},
       eprint = {astro-ph/0406547},
 primaryClass = {astro-ph},
       adsurl = {https://ui.adsabs.harvard.edu/abs/2004ApJ...613..189D}
}

@ARTICLE{Dahlen2008,
       author = {{Dahlen}, Tomas and {Strolger}, Louis-Gregory and {Riess}, Adam G.},
        title = "{The Extended HST Supernova Survey: The Rate of SNe Ia at z > 1.4 Remains Low}",
      journal = {\apj},
         year = 2008,
        month = jul,
       volume = {681},
       number = {1},
        pages = {462-469},
          doi = {10.1086/587978},
archivePrefix = {arXiv},
       eprint = {0803.1130},
 primaryClass = {astro-ph},
       adsurl = {https://ui.adsabs.harvard.edu/abs/2008ApJ...681..462D}
}

@ARTICLE{DallaVecchia2008,
       author = {{Dalla Vecchia}, Claudio and {Schaye}, Joop},
        title = "{Simulating galactic outflows with kinetic supernova feedback}",
      journal = {\mnras},
         year = 2008,
        month = jul,
       volume = {387},
       number = {4},
        pages = {1431-1444},
          doi = {10.1111/j.1365-2966.2008.13322.x},
archivePrefix = {arXiv},
       eprint = {0801.2770},
 primaryClass = {astro-ph},
       adsurl = {https://ui.adsabs.harvard.edu/abs/2008MNRAS.387.1431D}
}

@ARTICLE{DallaVecchia2012,
       author = {{Dalla Vecchia}, Claudio and {Schaye}, Joop},
        title = "{Simulating galactic outflows with thermal supernova feedback}",
      journal = {\mnras},
         year = 2012,
        month = oct,
       volume = {426},
       number = {1},
        pages = {140-158},
          doi = {10.1111/j.1365-2966.2012.21704.x},
archivePrefix = {arXiv},
       eprint = {1203.5667},
 primaryClass = {astro-ph.GA},
       adsurl = {https://ui.adsabs.harvard.edu/abs/2012MNRAS.426..140D}
}

@ARTICLE{Dave2019,
       author = {{Dav{\'e}}, Romeel and {Angl{\'e}s-Alc{\'a}zar}, Daniel and {Narayanan}, Desika and {Li}, Qi and {Rafieferantsoa}, Mika H. and {Appleby}, Sarah},
        title = "{SIMBA: Cosmological simulations with black hole growth and feedback}",
      journal = {\mnras},
         year = 2019,
        month = jun,
       volume = {486},
       number = {2},
        pages = {2827-2849},
          doi = {10.1093/mnras/stz937},
archivePrefix = {arXiv},
       eprint = {1901.10203},
 primaryClass = {astro-ph.GA},
       adsurl = {https://ui.adsabs.harvard.edu/abs/2019MNRAS.486.2827D}
}

@ARTICLE{DeLooze2020,
       author = {{De Looze}, I. and {Lamperti}, I. and {Saintonge}, A. and {Rela{\~n}o}, M. and {Smith}, M.~W.~L. and {Clark}, C.~J.~R. and {Wilson}, C.~D. and {Decleir}, M. and {Jones}, A.~P. and {Kennicutt}, R.~C. and {Accurso}, G. and {Brinks}, E. and {Bureau}, M. and {Cigan}, P. and {Clements}, D.~L. and {De Vis}, P. and {Fanciullo}, L. and {Gao}, Y. and {Gear}, W.~K. and {Ho}, L.~C. and {Hwang}, H.~S. and {Micha{\l}owski}, M.~J. and {Lee}, J.~C. and {Li}, C. and {Lin}, L. and {Liu}, T. and {Lomaeva}, M. and {Pan}, H. -A. and {Sargent}, M. and {Williams}, T. and {Xiao}, T. and {Zhu}, M.},
        title = "{JINGLE - IV. Dust, H I gas, and metal scaling laws in the local Universe}",
      journal = {\mnras},
         year = 2020,
        month = aug,
       volume = {496},
       number = {3},
        pages = {3668-3687},
          doi = {10.1093/mnras/staa1496},
archivePrefix = {arXiv},
       eprint = {2006.01856},
 primaryClass = {astro-ph.GA},
       adsurl = {https://ui.adsabs.harvard.edu/abs/2020MNRAS.496.3668D}
}

@ARTICLE{Dehnen2014,
       author = {{Dehnen}, Walter},
        title = "{A fast multipole method for stellar dynamics}",
      journal = {Computational Astrophysics and Cosmology},
         year = 2014,
        month = sep,
       volume = {1},
          eid = {1},
        pages = {1},
          doi = {10.1186/s40668-014-0001-7},
archivePrefix = {arXiv},
       eprint = {1405.2255},
 primaryClass = {astro-ph.IM},
       adsurl = {https://ui.adsabs.harvard.edu/abs/2014ComAC...1....1D}
}

@ARTICLE{DellAgli2017,
       author = {{Dell'Agli}, F. and {Garc{\'\i}a-Hern{\'a}ndez}, D.~A. and {Schneider}, R. and {Ventura}, P. and {La Franca}, F. and {Valiante}, R. and {Marini}, E. and {Di Criscienzo}, M.},
        title = "{Asymptotic giant branch and super-asymptotic giant branch stars: modelling dust production at solar metallicity}",
      journal = {\mnras},
         year = 2017,
        month = jun,
       volume = {467},
       number = {4},
        pages = {4431-4440},
          doi = {10.1093/mnras/stx387},
archivePrefix = {arXiv},
       eprint = {1702.03904},
 primaryClass = {astro-ph.SR},
       adsurl = {https://ui.adsabs.harvard.edu/abs/2017MNRAS.467.4431D}
}

@ARTICLE{DiMatteo2008,
       author = {{Di Matteo}, Tiziana and {Colberg}, J{\"o}rg and {Springel}, Volker and {Hernquist}, Lars and {Sijacki}, Debora},
        title = "{Direct Cosmological Simulations of the Growth of Black Holes and Galaxies}",
      journal = {\apj},
         year = 2008,
        month = mar,
       volume = {676},
       number = {1},
        pages = {33-53},
          doi = {10.1086/524921},
archivePrefix = {arXiv},
       eprint = {0705.2269},
 primaryClass = {astro-ph},
       adsurl = {https://ui.adsabs.harvard.edu/abs/2008ApJ...676...33D}
}

@ARTICLE{Dilday2010,
       author = {{Dilday}, Benjamin and {Smith}, Mathew and {Bassett}, Bruce and {Becker}, Andrew and {Bender}, Ralf and {Castander}, Francisco and {Cinabro}, David and {Filippenko}, Alexei V. and {Frieman}, Joshua A. and {Galbany}, Llu{\'\i}s and {Garnavich}, Peter M. and {Goobar}, Ariel and {Hopp}, Ulrich and {Ihara}, Yutaka and {Jha}, Saurabh W. and {Kessler}, Richard and {Lampeitl}, Hubert and {Marriner}, John and {Miquel}, Ramon and {Moll{\'a}}, Mercedes and {Nichol}, Robert C. and {Nordin}, Jakob and {Riess}, Adam G. and {Sako}, Masao and {Schneider}, Donald P. and {Sollerman}, Jesper and {Wheeler}, J. Craig and {{\"O}stman}, Linda and {Bizyaev}, Dmitry and {Brewington}, Howard and {Malanushenko}, Elena and {Malanushenko}, Viktor and {Oravetz}, Dan and {Pan}, Kaike and {Simmons}, Audrey and {Snedden}, Stephanie},
        title = "{Measurements of the Rate of Type Ia Supernovae at Redshift lsim0.3 from the Sloan Digital Sky Survey II Supernova Survey}",
      journal = {\apj},
         year = 2010,
        month = apr,
       volume = {713},
       number = {2},
        pages = {1026-1036},
          doi = {10.1088/0004-637X/713/2/1026},
archivePrefix = {arXiv},
       eprint = {1001.4995},
 primaryClass = {astro-ph.CO},
       adsurl = {https://ui.adsabs.harvard.edu/abs/2010ApJ...713.1026D}
}

@ARTICLE{Doherty2014,
       author = {{Doherty}, Carolyn L. and {Gil-Pons}, Pilar and {Lau}, Herbert H.~B. and {Lattanzio}, John C. and {Siess}, Lionel},
        title = "{Super and massive AGB stars - II. Nucleosynthesis and yields - Z = 0.02, 0.008 and 0.004}",
      journal = {\mnras},
         year = 2014,
        month = jan,
       volume = {437},
       number = {1},
        pages = {195-214},
          doi = {10.1093/mnras/stt1877},
archivePrefix = {arXiv},
       eprint = {1310.2614},
 primaryClass = {astro-ph.SR},
       adsurl = {https://ui.adsabs.harvard.edu/abs/2014MNRAS.437..195D}
}

@ARTICLE{Dolag2025,
       author = {{Dolag}, Klaus and {Remus}, Rhea-Silvia and {Valenzuela}, Lucas M. and {Kimmig}, Lucas C. and {Seidel}, Benjamin and {Fortune}, Silvio and {Stoiber}, Johannes and {Ivleva}, Anna and {Hoffmann}, Tadziu and {Biffi}, Veronica and {Marini}, Ilaria and {Popesso}, Paola and {Vladutescu-Zopp}, Stephan},
        title = "{Encyclopedia Magneticum: Scaling Relations from Cosmic Dawn to Present Day}",
      journal = {arXiv e-prints},
         year = 2025,
        month = apr,
          eid = {arXiv:2504.01061},
        pages = {arXiv:2504.01061},
          doi = {10.48550/arXiv.2504.01061},
archivePrefix = {arXiv},
       eprint = {2504.01061},
 primaryClass = {astro-ph.CO},
       adsurl = {https://ui.adsabs.harvard.edu/abs/2025arXiv250401061D}
}

@ARTICLE{Donnan2024,
       author = {{Donnan}, C.~T. and {McLure}, R.~J. and {Dunlop}, J.~S. and {McLeod}, D.~J. and {Magee}, D. and {Arellano-C{\'o}rdova}, K.~Z. and {Barrufet}, L. and {Begley}, R. and {Bowler}, R.~A.~A. and {Carnall}, A.~C. and {Cullen}, F. and {Ellis}, R.~S. and {Fontana}, A. and {Illingworth}, G.~D. and {Grogin}, N.~A. and {Hamadouche}, M.~L. and {Koekemoer}, A.~M. and {Liu}, F. -Y. and {Mason}, C. and {Santini}, P. and {Stanton}, T.~M.},
        title = "{JWST PRIMER: a new multifield determination of the evolving galaxy UV luminosity function at redshifts z ≃ 9 - 15}",
      journal = {\mnras},
         year = 2024,
        month = sep,
       volume = {533},
       number = {3},
        pages = {3222-3237},
          doi = {10.1093/mnras/stae2037},
archivePrefix = {arXiv},
       eprint = {2403.03171},
 primaryClass = {astro-ph.GA},
       adsurl = {https://ui.adsabs.harvard.edu/abs/2024MNRAS.533.3222D}
}

@ARTICLE{Donnari2021,
       author = {{Donnari}, Martina and {Pillepich}, Annalisa and {Nelson}, Dylan and {Marinacci}, Federico and {Vogelsberger}, Mark and {Hernquist}, Lars},
        title = "{Quenched fractions in the IllustrisTNG simulations: comparison with observations and other theoretical models}",
      journal = {\mnras},
         year = 2021,
        month = oct,
       volume = {506},
       number = {4},
        pages = {4760-4780},
          doi = {10.1093/mnras/stab1950},
archivePrefix = {arXiv},
       eprint = {2008.00004},
 primaryClass = {astro-ph.GA},
       adsurl = {https://ui.adsabs.harvard.edu/abs/2021MNRAS.506.4760D}
}

@ARTICLE{Driver2012,
       author = {{Driver}, S.~P. and {Robotham}, A.~S.~G. and {Kelvin}, L. and {Alpaslan}, M. and {Baldry}, I.~K. and {Bamford}, S.~P. and {Brough}, S. and {Brown}, M. and {Hopkins}, A.~M. and {Liske}, J. and {Loveday}, J. and {Norberg}, P. and {Peacock}, J.~A. and {Andrae}, E. and {Bland-Hawthorn}, J. and {Bourne}, N. and {Cameron}, E. and {Colless}, M. and {Conselice}, C.~J. and {Croom}, S.~M. and {Dunne}, L. and {Frenk}, C.~S. and {Graham}, Alister W. and {Gunawardhana}, M. and {Hill}, D.~T. and {Jones}, D.~H. and {Kuijken}, K. and {Madore}, B. and {Nichol}, R.~C. and {Parkinson}, H.~R. and {Pimbblet}, K.~A. and {Phillipps}, S. and {Popescu}, C.~C. and {Prescott}, M. and {Seibert}, M. and {Sharp}, R.~G. and {Sutherland}, W.~J. and {Taylor}, E.~N. and {Thomas}, D. and {Tuffs}, R.~J. and {van Kampen}, E. and {Wijesinghe}, D. and {Wilkins}, S.},
        title = "{Galaxy And Mass Assembly (GAMA): the 0.013 < z < 0.1 cosmic spectral energy distribution from 0.1 {\ensuremath{\mu}}m to 1 mm}",
      journal = {\mnras},
         year = 2012,
        month = dec,
       volume = {427},
       number = {4},
        pages = {3244-3264},
          doi = {10.1111/j.1365-2966.2012.22036.x},
archivePrefix = {arXiv},
       eprint = {1209.0259},
 primaryClass = {astro-ph.CO},
       adsurl = {https://ui.adsabs.harvard.edu/abs/2012MNRAS.427.3244D}
}

@ARTICLE{Driver2022,
       author = {{Driver}, Simon P. and {Bellstedt}, Sabine and {Robotham}, Aaron S.~G. and {Baldry}, Ivan K. and {Davies}, Luke J. and {Liske}, Jochen and {Obreschkow}, Danail and {Taylor}, Edward N. and {Wright}, Angus H. and {Alpaslan}, Mehmet and {Bamford}, Steven P. and {Bauer}, Amanda E. and {Bland-Hawthorn}, Joss and {Bilicki}, Maciej and {Bravo}, Mat{\'\i}as and {Brough}, Sarah and {Casura}, Sarah and {Cluver}, Michelle E. and {Colless}, Matthew and {Conselice}, Christopher J. and {Croom}, Scott M. and {de Jong}, Jelte and {D'Eugenio}, Franceso and {De Propris}, Roberto and {Dogruel}, Burak and {Drinkwater}, Michael J. and {Dvornik}, Andrej and {Farrow}, Daniel J. and {Frenk}, Carlos S. and {Giblin}, Benjamin and {Graham}, Alister W. and {Grootes}, Meiert W. and {Gunawardhana}, Madusha L.~P. and {Hashemizadeh}, Abdolhosein and {H{\"a}u{\ss}ler}, Boris and {Heymans}, Catherine and {Hildebrandt}, Hendrik and {Holwerda}, Benne W. and {Hopkins}, Andrew M. and {Jarrett}, Tom H. and {Heath Jones}, D. and {Kelvin}, Lee S. and {Koushan}, Soheil and {Kuijken}, Konrad and {Lara-L{\'o}pez}, Maritza A. and {Lange}, Rebecca and {L{\'o}pez-S{\'a}nchez}, {\'A}ngel R. and {Loveday}, Jon and {Mahajan}, Smriti and {Meyer}, Martin and {Moffett}, Amanda J. and {Napolitano}, Nicola R. and {Norberg}, Peder and {Owers}, Matt S. and {Radovich}, Mario and {Raouf}, Mojtaba and {Peacock}, John A. and {Phillipps}, Steven and {Pimbblet}, Kevin A. and {Popescu}, Cristina and {Said}, Khaled and {Sansom}, Anne E. and {Seibert}, Mark and {Sutherland}, Will J. and {Thorne}, Jessica E. and {Tuffs}, Richard J. and {Turner}, Ryan and {van der Wel}, Arjen and {van Kampen}, Eelco and {Wilkins}, Steve M.},
        title = "{Galaxy And Mass Assembly (GAMA): Data Release 4 and the z < 0.1 total and z < 0.08 morphological galaxy stellar mass functions}",
      journal = {\mnras},
         year = 2022,
        month = jun,
       volume = {513},
       number = {1},
        pages = {439-467},
          doi = {10.1093/mnras/stac472},
archivePrefix = {arXiv},
       eprint = {2203.08539},
 primaryClass = {astro-ph.GA},
       adsurl = {https://ui.adsabs.harvard.edu/abs/2022MNRAS.513..439D}
}

@ARTICLE{Dubois2014,
       author = {{Dubois}, Y. and {Pichon}, C. and {Welker}, C. and {Le Borgne}, D. and {Devriendt}, J. and {Laigle}, C. and {Codis}, S. and {Pogosyan}, D. and {Arnouts}, S. and {Benabed}, K. and {Bertin}, E. and {Blaizot}, J. and {Bouchet}, F. and {Cardoso}, J. -F. and {Colombi}, S. and {de Lapparent}, V. and {Desjacques}, V. and {Gavazzi}, R. and {Kassin}, S. and {Kimm}, T. and {McCracken}, H. and {Milliard}, B. and {Peirani}, S. and {Prunet}, S. and {Rouberol}, S. and {Silk}, J. and {Slyz}, A. and {Sousbie}, T. and {Teyssier}, R. and {Tresse}, L. and {Treyer}, M. and {Vibert}, D. and {Volonteri}, M.},
        title = "{Dancing in the dark: galactic properties trace spin swings along the cosmic web}",
      journal = {\mnras},
         year = 2014,
        month = oct,
       volume = {444},
       number = {2},
        pages = {1453-1468},
          doi = {10.1093/mnras/stu1227},
archivePrefix = {arXiv},
       eprint = {1402.1165},
 primaryClass = {astro-ph.CO},
       adsurl = {https://ui.adsabs.harvard.edu/abs/2014MNRAS.444.1453D}
}

@ARTICLE{Dubois2021,
       author = {{Dubois}, Yohan and {Beckmann}, Ricarda and {Bournaud}, Fr{\'e}d{\'e}ric and {Choi}, Hoseung and {Devriendt}, Julien and {Jackson}, Ryan and {Kaviraj}, Sugata and {Kimm}, Taysun and {Kraljic}, Katarina and {Laigle}, Clotilde and {Martin}, Garreth and {Park}, Min-Jung and {Peirani}, S{\'e}bastien and {Pichon}, Christophe and {Volonteri}, Marta and {Yi}, Sukyoung K.},
        title = "{Introducing the NEWHORIZON simulation: Galaxy properties with resolved internal dynamics across cosmic time}",
      journal = {\aap},
         year = 2021,
        month = jul,
       volume = {651},
          eid = {A109},
        pages = {A109},
          doi = {10.1051/0004-6361/202039429},
archivePrefix = {arXiv},
       eprint = {2009.10578},
 primaryClass = {astro-ph.GA},
       adsurl = {https://ui.adsabs.harvard.edu/abs/2021A&A...651A.109D}
}

@ARTICLE{Durier2012,
       author = {{Durier}, Fabrice and {Dalla Vecchia}, Claudio},
        title = "{Implementation of feedback in smoothed particle hydrodynamics: towards concordance of methods}",
      journal = {\mnras},
         year = 2012,
        month = jan,
       volume = {419},
       number = {1},
        pages = {465-478},
          doi = {10.1111/j.1365-2966.2011.19712.x},
archivePrefix = {arXiv},
       eprint = {1105.3729},
 primaryClass = {astro-ph.CO},
       adsurl = {https://ui.adsabs.harvard.edu/abs/2012MNRAS.419..465D}
}

@ARTICLE{Dwek1998,
       author = {{Dwek}, Eli},
        title = "{The Evolution of the Elemental Abundances in the Gas and Dust Phases of the Galaxy}",
      journal = {\apj},
         year = 1998,
        month = jul,
       volume = {501},
        pages = {643},
          doi = {10.1086/305829},
archivePrefix = {arXiv},
       eprint = {astro-ph/9707024},
 primaryClass = {astro-ph},
       adsurl = {https://ui.adsabs.harvard.edu/abs/1998ApJ...501..643D}
}

@ARTICLE{Eke1996,
       author = {{Eke}, Vincent R. and {Cole}, Shaun and {Frenk}, Carlos S.},
        title = "{Cluster evolution as a diagnostic for Omega}",
      journal = {\mnras},
         year = 1996,
        month = sep,
       volume = {282},
        pages = {263-280},
          doi = {10.1093/mnras/282.1.263},
archivePrefix = {arXiv},
       eprint = {astro-ph/9601088},
 primaryClass = {astro-ph},
       adsurl = {https://ui.adsabs.harvard.edu/abs/1996MNRAS.282..263E}
}

@ARTICLE{Elbers2021,
       author = {{Elbers}, Willem and {Frenk}, Carlos S. and {Jenkins}, Adrian and {Li}, Baojiu and {Pascoli}, Silvia},
        title = "{An optimal non-linear method for simulating relic neutrinos}",
      journal = {\mnras},
         year = 2021,
        month = oct,
       volume = {507},
       number = {2},
        pages = {2614-2631},
          doi = {10.1093/mnras/stab2260},
archivePrefix = {arXiv},
       eprint = {2010.07321},
 primaryClass = {astro-ph.CO},
       adsurl = {https://ui.adsabs.harvard.edu/abs/2021MNRAS.507.2614E}
}

@ARTICLE{Eldridge2017,
       author = {{Eldridge}, J.~J. and {Stanway}, E.~R. and {Xiao}, L. and {McClelland}, L.~A.~S. and {Taylor}, G. and {Ng}, M. and {Greis}, S.~M.~L. and {Bray}, J.~C.},
        title = "{Binary Population and Spectral Synthesis Version 2.1: Construction, Observational Verification, and New Results}",
      journal = {\pasa},
         year = 2017,
        month = nov,
       volume = {34},
          eid = {e058},
        pages = {e058},
          doi = {10.1017/pasa.2017.51},
archivePrefix = {arXiv},
       eprint = {1710.02154},
 primaryClass = {astro-ph.SR},
       adsurl = {https://ui.adsabs.harvard.edu/abs/2017PASA...34...58E}
}

@ARTICLE{Enia2022,
       author = {{Enia}, Andrea and {Talia}, Margherita and {Pozzi}, Francesca and {Cimatti}, Andrea and {Delvecchio}, Ivan and {Zamorani}, Gianni and {D'Amato}, Quirino and {Bisigello}, Laura and {Gruppioni}, Carlotta and {Rodighiero}, Giulia and {Calura}, Francesco and {Dallacasa}, Daniele and {Giulietti}, Marika and {Barchiesi}, Luigi and {Behiri}, Meriem and {Romano}, Michael},
        title = "{A New Estimate of the Cosmic Star Formation Density from a Radio-selected Sample, and the Contribution of H-dark Galaxies at z {\ensuremath{\geq}} 3}",
      journal = {\apj},
         year = 2022,
        month = mar,
       volume = {927},
       number = {2},
          eid = {204},
        pages = {204},
          doi = {10.3847/1538-4357/ac51ca},
archivePrefix = {arXiv},
       eprint = {2202.00019},
 primaryClass = {astro-ph.CO},
       adsurl = {https://ui.adsabs.harvard.edu/abs/2022ApJ...927..204E}
}

@ARTICLE{Esteban2020,
       author = {{Esteban}, Ivan and {Gonzalez-Garcia}, M.~C. and {Maltoni}, Michele and {Schwetz}, Thomas and {Zhou}, Albert},
        title = "{The fate of hints: updated global analysis of three-flavor neutrino oscillations}",
      journal = {Journal of High Energy Physics},
         year = 2020,
        month = sep,
       volume = {2020},
       number = {9},
          eid = {178},
        pages = {178},
          doi = {10.1007/JHEP09(2020)178},
archivePrefix = {arXiv},
       eprint = {2007.14792},
 primaryClass = {hep-ph},
       adsurl = {https://ui.adsabs.harvard.edu/abs/2020JHEP...09..178E}
}

@ARTICLE{Engler2021,
       author = {{Engler}, Christoph and {Pillepich}, Annalisa and {Joshi}, Gandhali D. and {Nelson}, Dylan and {Pasquali}, Anna and {Grebel}, Eva K. and {Lisker}, Thorsten and {Zinger}, Elad and {Donnari}, Martina and {Marinacci}, Federico and {Vogelsberger}, Mark and {Hernquist}, Lars},
        title = "{The distinct stellar-to-halo mass relations of satellite and central galaxies: insights from the IllustrisTNG simulations}",
      journal = {\mnras},
         year = 2021,
        month = jan,
       volume = {500},
       number = {3},
        pages = {3957-3975},
          doi = {10.1093/mnras/staa3505},
archivePrefix = {arXiv},
       eprint = {2002.11119},
 primaryClass = {astro-ph.GA},
       adsurl = {https://ui.adsabs.harvard.edu/abs/2021MNRAS.500.3957E}
}

@ARTICLE{Faucher2020,
       author = {{Faucher-Gigu{\`e}re}, Claude-Andr{\'e}},
        title = "{A cosmic UV/X-ray background model update}",
      journal = {\mnras},
         year = 2020,
        month = apr,
       volume = {493},
       number = {2},
        pages = {1614-1632},
          doi = {10.1093/mnras/staa302},
archivePrefix = {arXiv},
       eprint = {1903.08657},
 primaryClass = {astro-ph.CO},
       adsurl = {https://ui.adsabs.harvard.edu/abs/2020MNRAS.493.1614F}
}

@ARTICLE{FaucherGiguere2023,
       author = {{Faucher-Gigu{\`e}re}, Claude-Andr{\'e} and {Oh}, S. Peng},
        title = "{Key Physical Processes in the Circumgalactic Medium}",
      journal = {\araa},
         year = 2023,
        month = aug,
       volume = {61},
        pages = {131-195},
          doi = {10.1146/annurev-astro-052920-125203},
archivePrefix = {arXiv},
       eprint = {2301.10253},
 primaryClass = {astro-ph.GA},
       adsurl = {https://ui.adsabs.harvard.edu/abs/2023ARA&A..61..131F}
}

@ARTICLE{Feldmann2023,
       author = {{Feldmann}, Robert and {Quataert}, Eliot and {Faucher-Gigu{\`e}re}, Claude-Andr{\'e} and {Hopkins}, Philip F. and {{\c{C}}atmabacak}, Onur and {Kere{\v{s}}}, Du{\v{s}}an and {Bassini}, Luigi and {Bernardini}, Mauro and {Bullock}, James S. and {Cenci}, Elia and {Gensior}, Jindra and {Liang}, Lichen and {Moreno}, Jorge and {Wetzel}, Andrew},
        title = "{FIREbox: simulating galaxies at high dynamic range in a cosmological volume}",
      journal = {\mnras},
         year = 2023,
        month = jul,
       volume = {522},
       number = {3},
        pages = {3831-3860},
          doi = {10.1093/mnras/stad1205},
archivePrefix = {arXiv},
       eprint = {2205.15325},
 primaryClass = {astro-ph.GA},
       adsurl = {https://ui.adsabs.harvard.edu/abs/2023MNRAS.522.3831F}
}

@ARTICLE{Feldmann2025,
       author = {{Feldmann}, Robert and {Bieri}, Rebekka},
        title = "{Cosmological Simulations of Galaxies}",
      journal = {arXiv e-prints},
         year = 2025,
        month = jul,
          eid = {arXiv:2507.08925},
        pages = {arXiv:2507.08925},
          doi = {10.48550/arXiv.2507.08925},
archivePrefix = {arXiv},
       eprint = {2507.08925},
 primaryClass = {astro-ph.GA},
       adsurl = {https://ui.adsabs.harvard.edu/abs/2025arXiv250708925F}
}

@ARTICLE{Ferland2017,
       author = {{Ferland}, G.~J. and {Chatzikos}, M. and {Guzm{\'a}n}, F. and {Lykins}, M.~L. and {van Hoof}, P.~A.~M. and {Williams}, R.~J.~R. and {Abel}, N.~P. and {Badnell}, N.~R. and {Keenan}, F.~P. and {Porter}, R.~L. and {Stancil}, P.~C.},
        title = "{The 2017 Release Cloudy}",
      journal = {\rmxaa},
         year = 2017,
        month = oct,
       volume = {53},
        pages = {385-438},
archivePrefix = {arXiv},
       eprint = {1705.10877},
 primaryClass = {astro-ph.GA},
       adsurl = {https://ui.adsabs.harvard.edu/abs/2017RMxAA..53..385F}
}

@ARTICLE{Fishlock2014,
       author = {{Fishlock}, Cherie K. and {Karakas}, Amanda I. and {Lugaro}, Maria and {Yong}, David},
        title = "{Evolution and Nucleosynthesis of Asymptotic Giant Branch Stellar Models of Low Metallicity}",
      journal = {\apj},
         year = 2014,
        month = dec,
       volume = {797},
       number = {1},
          eid = {44},
        pages = {44},
          doi = {10.1088/0004-637X/797/1/44},
archivePrefix = {arXiv},
       eprint = {1410.7457},
 primaryClass = {astro-ph.SR},
       adsurl = {https://ui.adsabs.harvard.edu/abs/2014ApJ...797...44F}
}

@ARTICLE{Font2020,
       author = {{Font}, Andreea S. and {McCarthy}, Ian G. and {Poole-Mckenzie}, Robert and {Stafford}, Sam G. and {Brown}, Shaun T. and {Schaye}, Joop and {Crain}, Robert A. and {Theuns}, Tom and {Schaller}, Matthieu},
        title = "{The ARTEMIS simulations: stellar haloes of Milky Way-mass galaxies}",
      journal = {\mnras},
         year = 2020,
        month = oct,
       volume = {498},
       number = {2},
        pages = {1765-1785},
          doi = {10.1093/mnras/staa2463},
archivePrefix = {arXiv},
       eprint = {2004.01914},
 primaryClass = {astro-ph.GA},
       adsurl = {https://ui.adsabs.harvard.edu/abs/2020MNRAS.498.1765F}
}

@ARTICLE{Forouhar2025Herons,
       author = {{Forouhar Moreno}, Victor J. and {Helly}, John and {McGibbon}, Robert and {Schaye}, Joop and {Schaller}, Matthieu and {Han}, Jiaxin and {Kugel}, Roi and {Bah{\'e}}, Yannick M.},
        title = "{Assessing subhalo finders in cosmological hydrodynamical simulations}",
      journal = {\mnras},
         year = 2025,
        month = oct,
       volume = {543},
       number = {2},
        pages = {1339-1372},
          doi = {10.1093/mnras/staf1478},
archivePrefix = {arXiv},
       eprint = {2502.06932},
 primaryClass = {astro-ph.CO},
       adsurl = {https://ui.adsabs.harvard.edu/abs/2025MNRAS.543.1339F}
}

@ARTICLE{Fraser-mckelvie2022,
       author = {{Fraser-McKelvie}, A. and {Cortese}, L. and {Groves}, B. and {Brough}, S. and {Bryant}, J. and {Catinella}, B. and {Croom}, S. and {D'Eugenio}, F. and {L{\'o}pez-S{\'a}nchez}, {\'A}. R. and {van de Sande}, J. and {Sweet}, S. and {Vaughan}, S. and {Bland-Hawthorn}, J. and {Lawrence}, J. and {Lorente}, N. and {Owers}, M.},
        title = "{The SAMI Galaxy Survey: the drivers of gas and stellar metallicity differences in galaxies}",
      journal = {\mnras},
         year = 2022,
        month = feb,
       volume = {510},
       number = {1},
        pages = {320-333},
          doi = {10.1093/mnras/stab3430},
archivePrefix = {arXiv},
       eprint = {2111.11627},
 primaryClass = {astro-ph.GA},
       adsurl = {https://ui.adsabs.harvard.edu/abs/2022MNRAS.510..320F}
}

@ARTICLE{Frohmaier2019,
       author = {{Frohmaier}, C. and {Sullivan}, M. and {Nugent}, P.~E. and {Smith}, M. and {Dimitriadis}, G. and {Bloom}, J.~S. and {Cenko}, S.~B. and {Kasliwal}, M.~M. and {Kulkarni}, S.~R. and {Maguire}, K. and {Ofek}, E.~O. and {Poznanski}, D. and {Quimby}, R.~M.},
        title = "{The volumetric rate of normal type Ia supernovae in the local Universe discovered by the Palomar Transient Factory}",
      journal = {\mnras},
         year = 2019,
        month = jun,
       volume = {486},
       number = {2},
        pages = {2308-2320},
          doi = {10.1093/mnras/stz807},
archivePrefix = {arXiv},
       eprint = {1903.08580},
 primaryClass = {astro-ph.HE},
       adsurl = {https://ui.adsabs.harvard.edu/abs/2019MNRAS.486.2308F}
}

@ARTICLE{Fu2025,
       author = {{Fu}, Shuqi and {Sun}, Fengwu and {Jiang}, Linhua and {Lin}, Xiaojing and {Diego}, Jose M. and {Furtak}, Lukas J. and {Jauzac}, Mathilde and {Koekemoer}, Anton M. and {Li}, Mingyu and {Oguri}, Masamune and {Patel}, Nency R. and {Willmer}, Christopher N.~A. and {Windhorst}, Rogier A. and {Zitrin}, Adi and {Bauer}, Franz E. and {Chen}, Chian-Chou and {Chen}, Wenlei and {Cheng}, Cheng and {Conselice}, Christopher J. and {Eisenstein}, Daniel J. and {Egami}, Eiichi and {Espada}, Daniel and {Fan}, Xiaohui and {Fujimoto}, Seiji and {Hsiao}, Tiger Yu-Yang and {Jin}, Xiangyu and {Kohno}, Kotaro and {Lagattuta}, David J. and {Li}, Zihao and {Liu}, Weizhe and {Miralda-Escud{\'e}}, Jordi and {Ning}, Yuanhang and {Tacchella}, Sandro and {Tee}, Wei Leong and {Umehata}, Hideki and {Wang}, Feige and {Yan}, Haojing and {Zhu}, Yongda},
        title = "{Medium-band Astrophysics with the Grism of NIRCam In Frontier fields (MAGNIF): Spectroscopic Census of H$\alpha$ Luminosity Functions and Cosmic Star Formation at $z\sim 4.5$ and 6.3}",
      journal = {arXiv e-prints},
         year = 2025,
        month = mar,
          eid = {arXiv:2503.03829},
        pages = {arXiv:2503.03829},
          doi = {10.48550/arXiv.2503.03829},
archivePrefix = {arXiv},
       eprint = {2503.03829},
 primaryClass = {astro-ph.GA},
       adsurl = {https://ui.adsabs.harvard.edu/abs/2025arXiv250303829F}
}

@ARTICLE{Fuente2019,
       author = {{Fuente}, A. and {Navarro}, D.~G. and {Caselli}, P. and {Gerin}, M. and {Kramer}, C. and {Roueff}, E. and {Alonso-Albi}, T. and {Bachiller}, R. and {Cazaux}, S. and {Commercon}, B. and {Friesen}, R. and {Garc{\'\i}a-Burillo}, S. and {Giuliano}, B.~M. and {Goicoechea}, J.~R. and {Gratier}, P. and {Hacar}, A. and {Jim{\'e}nez-Serra}, I. and {Kirk}, J. and {Lattanzi}, V. and {Loison}, J.~C. and {Malinen}, J. and {Marcelino}, N. and {Mart{\'\i}n-Dom{\'e}nech}, R. and {Mu{\~n}oz-Caro}, G. and {Pineda}, J. and {Tafalla}, M. and {Tercero}, B. and {Ward-Thompson}, D. and {Trevi{\~n}o-Morales}, S.~P. and {Rivi{\'e}re-Marichalar}, P. and {Roncero}, O. and {Vidal}, T. and {Ballester}, M.~Y.},
        title = "{Gas phase Elemental abundances in Molecular cloudS (GEMS). I. The prototypical dark cloud TMC 1}",
      journal = {\aap},
         year = 2019,
        month = apr,
       volume = {624},
          eid = {A105},
        pages = {A105},
          doi = {10.1051/0004-6361/201834654},
archivePrefix = {arXiv},
       eprint = {1809.04978},
 primaryClass = {astro-ph.GA},
       adsurl = {https://ui.adsabs.harvard.edu/abs/2019A&A...624A.105F}
}

@ARTICLE{Furlong2015,
       author = {{Furlong}, M. and {Bower}, R.~G. and {Theuns}, T. and {Schaye}, J. and {Crain}, R.~A. and {Schaller}, M. and {Dalla Vecchia}, C. and {Frenk}, C.~S. and {McCarthy}, I.~G. and {Helly}, J. and {Jenkins}, A. and {Rosas-Guevara}, Y.~M.},
        title = "{Evolution of galaxy stellar masses and star formation rates in the EAGLE simulations}",
      journal = {\mnras},
         year = 2015,
        month = jul,
       volume = {450},
       number = {4},
        pages = {4486-4504},
          doi = {10.1093/mnras/stv852},
archivePrefix = {arXiv},
       eprint = {1410.3485},
 primaryClass = {astro-ph.GA},
       adsurl = {https://ui.adsabs.harvard.edu/abs/2015MNRAS.450.4486F}
}

@ARTICLE{Gallazzi2005,
       author = {{Gallazzi}, Anna and {Charlot}, St{\'e}phane and {Brinchmann}, Jarle and {White}, Simon D.~M. and {Tremonti}, Christy A.},
        title = "{The ages and metallicities of galaxies in the local universe}",
      journal = {\mnras},
         year = 2005,
        month = sep,
       volume = {362},
       number = {1},
        pages = {41-58},
          doi = {10.1111/j.1365-2966.2005.09321.x},
archivePrefix = {arXiv},
       eprint = {astro-ph/0506539},
 primaryClass = {astro-ph},
       adsurl = {https://ui.adsabs.harvard.edu/abs/2005MNRAS.362...41G}
}

@ARTICLE{Genina2024,
       author = {{Genina}, Anna and {Springel}, Volker and {Rantala}, Antti},
        title = "{A calibrated model for N-body dynamical friction acting on supermassive black holes}",
      journal = {\mnras},
         year = 2024,
        month = oct,
       volume = {534},
       number = {1},
        pages = {957-977},
          doi = {10.1093/mnras/stae2144},
archivePrefix = {arXiv},
       eprint = {2405.08870},
 primaryClass = {astro-ph.GA},
       adsurl = {https://ui.adsabs.harvard.edu/abs/2024MNRAS.534..957G}
}

@ARTICLE{Gnedin2011,
       author = {{Gnedin}, Nickolay Y. and {Kravtsov}, Andrey V.},
        title = "{Environmental Dependence of the Kennicutt-Schmidt Relation in Galaxies}",
      journal = {\apj},
         year = 2011,
        month = feb,
       volume = {728},
       number = {2},
          eid = {88},
        pages = {88},
          doi = {10.1088/0004-637X/728/2/88},
archivePrefix = {arXiv},
       eprint = {1004.0003},
 primaryClass = {astro-ph.CO},
       adsurl = {https://ui.adsabs.harvard.edu/abs/2011ApJ...728...88G}
}

@ARTICLE{Graham2023,
       author = {{Graham}, Alister W. and {Sahu}, Nandini},
        title = "{Appreciating mergers for understanding the non-linear M$_{bh}$-M$_{*,spheroid}$ and M$_{bh}$-M$_{*, galaxy}$ relations, updated herein, and the implications for the (reduced) role of AGN feedback}",
      journal = {\mnras},
         year = 2023,
        month = jan,
       volume = {518},
       number = {2},
        pages = {2177-2200},
          doi = {10.1093/mnras/stac2019},
archivePrefix = {arXiv},
       eprint = {2209.14526},
 primaryClass = {astro-ph.GA},
       adsurl = {https://ui.adsabs.harvard.edu/abs/2023MNRAS.518.2177G}
}

@ARTICLE{Graham2024erratum,
       author = {{Graham}, Alister W. and {Sahu}, Nandini},
        title = "{Erratum: Appreciating mergers for understanding the non-linear M$_{bh}$-M$_{*,spheroid}$ and M$_{bh}$-M$_{*,galaxy}$ relations, updated herein, and the implications for the (reduced) role of AGN feedback}",
      journal = {\mnras},
         year = 2024,
        month = may,
       volume = {530},
       number = {3},
        pages = {3429-3430},
          doi = {10.1093/mnras/stae1079},
       adsurl = {https://ui.adsabs.harvard.edu/abs/2024MNRAS.530.3429G}
}

@ARTICLE{Granato2021,
       author = {{Granato}, Gian Luigi and {Ragone-Figueroa}, Cinthia and {Taverna}, Antonela and {Silva}, Laura and {Valentini}, Milena and {Borgani}, Stefano and {Monaco}, Pierluigi and {Murante}, Giuseppe and {Tornatore}, Luca},
        title = "{Dust evolution in zoom-in cosmological simulations of galaxy formation}",
      journal = {\mnras},
         year = 2021,
        month = may,
       volume = {503},
       number = {1},
        pages = {511-532},
          doi = {10.1093/mnras/stab362},
archivePrefix = {arXiv},
       eprint = {2010.05919},
 primaryClass = {astro-ph.GA},
       adsurl = {https://ui.adsabs.harvard.edu/abs/2021MNRAS.503..511G}
}

@ARTICLE{Grand2017,
       author = {{Grand}, Robert J.~J. and {G{\'o}mez}, Facundo A. and {Marinacci}, Federico and {Pakmor}, R{\"u}diger and {Springel}, Volker and {Campbell}, David J.~R. and {Frenk}, Carlos S. and {Jenkins}, Adrian and {White}, Simon D.~M.},
        title = "{The Auriga Project: the properties and formation mechanisms of disc galaxies across cosmic time}",
      journal = {\mnras},
         year = 2017,
        month = may,
       volume = {467},
       number = {1},
        pages = {179-207},
          doi = {10.1093/mnras/stx071},
archivePrefix = {arXiv},
       eprint = {1610.01159},
 primaryClass = {astro-ph.GA},
       adsurl = {https://ui.adsabs.harvard.edu/abs/2017MNRAS.467..179G}
}

@ARTICLE{Graur2011,
       author = {{Graur}, O. and {Poznanski}, D. and {Maoz}, D. and {Yasuda}, N. and {Totani}, T. and {Fukugita}, M. and {Filippenko}, A.~V. and {Foley}, R.~J. and {Silverman}, J.~M. and {Gal-Yam}, A. and {Horesh}, A. and {Jannuzi}, B.~T.},
        title = "{Supernovae in the Subaru Deep Field: the rate and delay-time distribution of Type Ia supernovae out to redshift 2}",
      journal = {\mnras},
         year = 2011,
        month = oct,
       volume = {417},
       number = {2},
        pages = {916-940},
          doi = {10.1111/j.1365-2966.2011.19287.x},
archivePrefix = {arXiv},
       eprint = {1102.0005},
 primaryClass = {astro-ph.CO},
       adsurl = {https://ui.adsabs.harvard.edu/abs/2011MNRAS.417..916G}
}

@article{Greengard1987,
title = "A fast algorithm for particle simulations",
journal = "Journal of Computational Physics",
volume = "73",
number = "2",
pages = "325 - 348",
year = "1987",
note = "",
issn = "0021-9991",
doi = "http://dx.doi.org/10.1016/0021-9991(87)90140-9",
url = "http://www.sciencedirect.com/science/article/pii/0021999187901409",
author = "L Greengard and V Rokhlin",
}

@ARTICLE{Greif2009,
       author = {{Greif}, Thomas H. and {Glover}, Simon C.~O. and {Bromm}, Volker and {Klessen}, Ralf S.},
        title = "{Chemical mixing in smoothed particle hydrodynamics simulations}",
      journal = {\mnras},
         year = 2009,
        month = feb,
       volume = {392},
       number = {4},
        pages = {1381-1387},
          doi = {10.1111/j.1365-2966.2008.14169.x},
archivePrefix = {arXiv},
       eprint = {0808.0843},
 primaryClass = {astro-ph},
       adsurl = {https://ui.adsabs.harvard.edu/abs/2009MNRAS.392.1381G}
}

@ARTICLE{Gruppioni2020,
       author = {{Gruppioni}, C. and {B{\'e}thermin}, M. and {Loiacono}, F. and {Le F{\`e}vre}, O. and {Capak}, P. and {Cassata}, P. and {Faisst}, A.~L. and {Schaerer}, D. and {Silverman}, J. and {Yan}, L. and {Bardelli}, S. and {Boquien}, M. and {Carraro}, R. and {Cimatti}, A. and {Dessauges-Zavadsky}, M. and {Ginolfi}, M. and {Fujimoto}, S. and {Hathi}, N.~P. and {Jones}, G.~C. and {Khusanova}, Y. and {Koekemoer}, A.~M. and {Lagache}, G. and {Lemaux}, B.~C. and {Oesch}, P.~A. and {Pozzi}, F. and {Riechers}, D.~A. and {Rodighiero}, G. and {Romano}, M. and {Talia}, M. and {Vallini}, L. and {Vergani}, D. and {Zamorani}, G. and {Zucca}, E.},
        title = "{The ALPINE-ALMA [CII] survey. The nature, luminosity function, and star formation history of dusty galaxies up to z ≃ 6}",
      journal = {\aap},
         year = 2020,
        month = nov,
       volume = {643},
          eid = {A8},
        pages = {A8},
          doi = {10.1051/0004-6361/202038487},
archivePrefix = {arXiv},
       eprint = {2006.04974},
 primaryClass = {astro-ph.GA},
       adsurl = {https://ui.adsabs.harvard.edu/abs/2020A&A...643A...8G}
}

@ARTICLE{Gunawardhana2011,
       author = {{Gunawardhana}, M.~L.~P. and {Hopkins}, A.~M. and {Sharp}, R.~G. and {Brough}, S. and {Taylor}, E. and {Bland-Hawthorn}, J. and {Maraston}, C. and {Tuffs}, R.~J. and {Popescu}, C.~C. and {Wijesinghe}, D. and {Jones}, D.~H. and {Croom}, S. and {Sadler}, E. and {Wilkins}, S. and {Driver}, S.~P. and {Liske}, J. and {Norberg}, P. and {Baldry}, I.~K. and {Bamford}, S.~P. and {Loveday}, J. and {Peacock}, J.~A. and {Robotham}, A.~S.~G. and {Zucker}, D.~B. and {Parker}, Q.~A. and {Conselice}, C.~J. and {Cameron}, E. and {Frenk}, C.~S. and {Hill}, D.~T. and {Kelvin}, L.~S. and {Kuijken}, K. and {Madore}, B.~F. and {Nichol}, B. and {Parkinson}, H.~R. and {Pimbblet}, K.~A. and {Prescott}, M. and {Sutherland}, W.~J. and {Thomas}, D. and {van Kampen}, E.},
        title = "{Galaxy and Mass Assembly (GAMA): the star formation rate dependence of the stellar initial mass function}",
      journal = {\mnras},
         year = 2011,
        month = aug,
       volume = {415},
       number = {2},
        pages = {1647-1662},
          doi = {10.1111/j.1365-2966.2011.18800.x},
archivePrefix = {arXiv},
       eprint = {1104.2379},
 primaryClass = {astro-ph.CO},
       adsurl = {https://ui.adsabs.harvard.edu/abs/2011MNRAS.415.1647G}
}

@ARTICLE{Gutcke2021LYRA,
       author = {{Gutcke}, Thales A. and {Pakmor}, R{\"u}diger and {Naab}, Thorsten and {Springel}, Volker},
        title = "{LYRA - I. Simulating the multiphase ISM of a dwarf galaxy with variable energy supernovae from individual stars}",
      journal = {\mnras},
         year = 2021,
        month = mar,
       volume = {501},
       number = {4},
        pages = {5597-5615},
          doi = {10.1093/mnras/staa3875},
archivePrefix = {arXiv},
       eprint = {2010.07311},
 primaryClass = {astro-ph.GA},
       adsurl = {https://ui.adsabs.harvard.edu/abs/2021MNRAS.501.5597G}
}

@ARTICLE{Gutcke2022,
       author = {{Gutcke}, Thales A. and {Pakmor}, R{\"u}diger and {Naab}, Thorsten and {Springel}, Volker},
        title = "{LYRA - II. Cosmological dwarf galaxy formation with inhomogeneous Population III enrichment}",
      journal = {\mnras},
         year = 2022,
        month = jun,
       volume = {513},
       number = {1},
        pages = {1372-1385},
          doi = {10.1093/mnras/stac867},
archivePrefix = {arXiv},
       eprint = {2110.06233},
 primaryClass = {astro-ph.GA},
       adsurl = {https://ui.adsabs.harvard.edu/abs/2022MNRAS.513.1372G}
}

@ARTICLE{Hahn2011,
       author = {{Hahn}, Oliver and {Abel}, Tom},
        title = "{Multi-scale initial conditions for cosmological simulations}",
      journal = {\mnras},
         year = 2011,
        month = aug,
       volume = {415},
       number = {3},
        pages = {2101-2121},
          doi = {10.1111/j.1365-2966.2011.18820.x},
archivePrefix = {arXiv},
       eprint = {1103.6031},
 primaryClass = {astro-ph.CO},
       adsurl = {https://ui.adsabs.harvard.edu/abs/2011MNRAS.415.2101H}
}

@article{Hahn2020,
  title={MUSIC2-monofonIC: 3LPT initial condition generator},
  author={Hahn, Oliver and Michaux, Micha{\"e}l and Rampf, Cornelius and Uhlemann, Cora and Angulo, Raul E},
  journal={Astrophysics Source Code Library},
  pages={ascl--2008},
  year={2020}
}

@ARTICLE{Hahn2021,
       author = {{Hahn}, Oliver and {Rampf}, Cornelius and {Uhlemann}, Cora},
        title = "{Higher order initial conditions for mixed baryon-CDM simulations}",
      journal = {\mnras},
         year = 2021,
        month = may,
       volume = {503},
       number = {1},
        pages = {426-445},
          doi = {10.1093/mnras/staa3773},
archivePrefix = {arXiv},
       eprint = {2008.09124},
 primaryClass = {astro-ph.CO},
       adsurl = {https://ui.adsabs.harvard.edu/abs/2021MNRAS.503..426H}
}

@ARTICLE{Han2012HBT,
       author = {{Han}, Jiaxin and {Jing}, Y.~P. and {Wang}, Huiyuan and {Wang}, Wenting},
        title = "{Resolving subhaloes' lives with the Hierarchical Bound-Tracing algorithm}",
      journal = {\mnras},
         year = 2012,
        month = dec,
       volume = {427},
       number = {3},
        pages = {2437-2449},
          doi = {10.1111/j.1365-2966.2012.22111.x},
archivePrefix = {arXiv},
       eprint = {1103.2099},
 primaryClass = {astro-ph.CO},
       adsurl = {https://ui.adsabs.harvard.edu/abs/2012MNRAS.427.2437H}
}

@ARTICLE{Han2018HBT+,
       author = {{Han}, Jiaxin and {Cole}, Shaun and {Frenk}, Carlos S. and {Benitez-Llambay}, Alejandro and {Helly}, John},
        title = "{HBT+: an improved code for finding subhaloes and building merger trees in cosmological simulations}",
      journal = {\mnras},
         year = 2018,
        month = feb,
       volume = {474},
       number = {1},
        pages = {604-617},
          doi = {10.1093/mnras/stx2792},
archivePrefix = {arXiv},
       eprint = {1708.03646},
 primaryClass = {astro-ph.CO},
       adsurl = {https://ui.adsabs.harvard.edu/abs/2018MNRAS.474..604H}
}

@ARTICLE{Hardin2000,
       author = {{Hardin}, D. and {Afonso}, C. and {Alard}, C. and {Albert}, J.~N. and {Amadon}, A. and {Andersen}, J. and {Ansari}, R. and {Aubourg}, {\'E}. and {Bareyre}, P. and {Bauer}, F. and {Beaulieu}, J.~P. and {Blanc}, G. and {Bouquet}, A. and {Char}, S. and {Charlot}, X. and {Couchot}, F. and {Coutures}, C. and {Derue}, F. and {Ferlet}, R. and {Glicenstein}, J.~F. and {Goldman}, B. and {Gould}, A. and {Graff}, D. and {Gros}, M. and {Haissinski}, J. and {Hamilton}, J.~C. and {de Kat}, J. and {Kim}, A. and {Lasserre}, T. and {Lesquoy}, {\'E}. and {Loup}, C. and {Magneville}, C. and {Mansoux}, B. and {Marquette}, J.~B. and {Maurice}, {\'E}. and {Milsztajn}, A. and {Moniez}, M. and {Palanque-Delabrouille}, N. and {Perdereau}, O. and {Pr{\'e}vot}, L. and {Regnault}, N. and {Rich}, J. and {Spiro}, M. and {Vidal-Madjar}, A. and {Vigroux}, L. and {Zylberajch}, S. and {EROS Collaboration}},
        title = "{Type Ia supernova rate at z \raisebox{-0.5ex}\textasciitilde 0.1}",
      journal = {\aap},
         year = 2000,
        month = oct,
       volume = {362},
        pages = {419-425},
          doi = {10.48550/arXiv.astro-ph/0006424},
archivePrefix = {arXiv},
       eprint = {astro-ph/0006424},
 primaryClass = {astro-ph},
       adsurl = {https://ui.adsabs.harvard.edu/abs/2000A&A...362..419H}
}

@ARTICLE{Hardwick2022,
       author = {{Hardwick}, Jennifer A. and {Cortese}, Luca and {Obreschkow}, Danail and {Catinella}, Barbara and {Cook}, Robin H.~W.},
        title = "{xGASS: characterizing the slope and scatter of the stellar mass-angular momentum relation for nearby galaxies}",
      journal = {\mnras},
         year = 2022,
        month = jan,
       volume = {509},
       number = {3},
        pages = {3751-3763},
          doi = {10.1093/mnras/stab3261},
archivePrefix = {arXiv},
       eprint = {2111.15048},
 primaryClass = {astro-ph.GA},
       adsurl = {https://ui.adsabs.harvard.edu/abs/2022MNRAS.509.3751H}
}

@ARTICLE{Harikane2023,
       author = {{Harikane}, Yuichi and {Ouchi}, Masami and {Oguri}, Masamune and {Ono}, Yoshiaki and {Nakajima}, Kimihiko and {Isobe}, Yuki and {Umeda}, Hiroya and {Mawatari}, Ken and {Zhang}, Yechi},
        title = "{A Comprehensive Study of Galaxies at z   9-16 Found in the Early JWST Data: Ultraviolet Luminosity Functions and Cosmic Star Formation History at the Pre-reionization Epoch}",
      journal = {\apjs},
         year = 2023,
        month = mar,
       volume = {265},
       number = {1},
          eid = {5},
        pages = {5},
          doi = {10.3847/1538-4365/acaaa9},
archivePrefix = {arXiv},
       eprint = {2208.01612},
 primaryClass = {astro-ph.GA},
       adsurl = {https://ui.adsabs.harvard.edu/abs/2023ApJS..265....5H}
}

@ARTICLE{Hirashita2009,
       author = {{Hirashita}, Hiroyuki and {Yan}, Huirong},
        title = "{Shattering and coagulation of dust grains in interstellar turbulence}",
      journal = {\mnras},
         year = 2009,
        month = apr,
       volume = {394},
       number = {2},
        pages = {1061-1074},
          doi = {10.1111/j.1365-2966.2009.14405.x},
archivePrefix = {arXiv},
       eprint = {0812.3451},
 primaryClass = {astro-ph},
       adsurl = {https://ui.adsabs.harvard.edu/abs/2009MNRAS.394.1061H}
}

@ARTICLE{Hirashita2014,
       author = {{Hirashita}, Hiroyuki and {Voshchinnikov}, Nikolai V.},
        title = "{Effects of grain growth mechanisms on the extinction curve and the metal depletion in the interstellar medium}",
      journal = {\mnras},
         year = 2014,
        month = jan,
       volume = {437},
       number = {2},
        pages = {1636-1645},
          doi = {10.1093/mnras/stt1997},
archivePrefix = {arXiv},
       eprint = {1310.4679},
 primaryClass = {astro-ph.GA},
       adsurl = {https://ui.adsabs.harvard.edu/abs/2014MNRAS.437.1636H}
}

@ARTICLE{Hislop2022,
       author = {{Hislop}, Jessica M. and {Naab}, Thorsten and {Steinwandel}, Ulrich P. and {Lah{\'e}n}, Natalia and {Irodotou}, Dimitrios and {Johansson}, Peter H. and {Walch}, Stefanie},
        title = "{The challenge of simulating the star cluster population of dwarf galaxies with resolved interstellar medium}",
      journal = {\mnras},
         year = 2022,
        month = feb,
       volume = {509},
       number = {4},
        pages = {5938-5954},
          doi = {10.1093/mnras/stab3347},
archivePrefix = {arXiv},
       eprint = {2109.08160},
 primaryClass = {astro-ph.GA},
       adsurl = {https://ui.adsabs.harvard.edu/abs/2022MNRAS.509.5938H}
}

@ARTICLE{Hopkins2015,
       author = {{Hopkins}, Philip F.},
        title = "{A new class of accurate, mesh-free hydrodynamic simulation methods}",
      journal = {\mnras},
         year = 2015,
        month = jun,
       volume = {450},
       number = {1},
        pages = {53-110},
          doi = {10.1093/mnras/stv195},
archivePrefix = {arXiv},
       eprint = {1409.7395},
 primaryClass = {astro-ph.CO},
       adsurl = {https://ui.adsabs.harvard.edu/abs/2015MNRAS.450...53H}
}

@ARTICLE{Hopkins2018FIRE2,
       author = {{Hopkins}, Philip F. and {Wetzel}, Andrew and {Kere{\v{s}}}, Du{\v{s}}an and {Faucher-Gigu{\`e}re}, Claude-Andr{\'e} and {Quataert}, Eliot and {Boylan-Kolchin}, Michael and {Murray}, Norman and {Hayward}, Christopher C. and {Garrison-Kimmel}, Shea and {Hummels}, Cameron and {Feldmann}, Robert and {Torrey}, Paul and {Ma}, Xiangcheng and {Angl{\'e}s-Alc{\'a}zar}, Daniel and {Su}, Kung-Yi and {Orr}, Matthew and {Schmitz}, Denise and {Escala}, Ivanna and {Sanderson}, Robyn and {Grudi{\'c}}, Michael Y. and {Hafen}, Zachary and {Kim}, Ji-Hoon and {Fitts}, Alex and {Bullock}, James S. and {Wheeler}, Coral and {Chan}, T.~K. and {Elbert}, Oliver D. and {Narayanan}, Desika},
        title = "{FIRE-2 simulations: physics versus numerics in galaxy formation}",
      journal = {\mnras},
         year = 2018,
        month = oct,
       volume = {480},
       number = {1},
        pages = {800-863},
          doi = {10.1093/mnras/sty1690},
archivePrefix = {arXiv},
       eprint = {1702.06148},
 primaryClass = {astro-ph.GA},
       adsurl = {https://ui.adsabs.harvard.edu/abs/2018MNRAS.480..800H}
}

@ARTICLE{Hopkins2023FIRE3,
       author = {{Hopkins}, Philip F. and {Wetzel}, Andrew and {Wheeler}, Coral and {Sanderson}, Robyn and {Grudi{\'c}}, Michael Y. and {Sameie}, Omid and {Boylan-Kolchin}, Michael and {Orr}, Matthew and {Ma}, Xiangcheng and {Faucher-Gigu{\`e}re}, Claude-Andr{\'e} and {Kere{\v{s}}}, Du{\v{s}}an and {Quataert}, Eliot and {Su}, Kung-Yi and {Moreno}, Jorge and {Feldmann}, Robert and {Bullock}, James S. and {Loebman}, Sarah R. and {Angl{\'e}s-Alc{\'a}zar}, Daniel and {Stern}, Jonathan and {Necib}, Lina and {Choban}, Caleb R. and {Hayward}, Christopher C.},
        title = "{FIRE-3: updated stellar evolution models, yields, and microphysics and fitting functions for applications in galaxy simulations}",
      journal = {\mnras},
         year = 2023,
        month = feb,
       volume = {519},
       number = {2},
        pages = {3154-3181},
          doi = {10.1093/mnras/stac3489},
archivePrefix = {arXiv},
       eprint = {2203.00040},
 primaryClass = {astro-ph.GA},
       adsurl = {https://ui.adsabs.harvard.edu/abs/2023MNRAS.519.3154H}
}

@ARTICLE{Horesh2008,
       author = {{Horesh}, A. and {Poznanski}, D. and {Ofek}, E.~O. and {Maoz}, D.},
        title = "{The rate of Type Ia supernovae at z \raisebox{-0.5ex}\textasciitilde 0.2 from SDSS-I overlapping fields}",
      journal = {\mnras},
         year = 2008,
        month = oct,
       volume = {389},
       number = {4},
        pages = {1871-1880},
          doi = {10.1111/j.1365-2966.2008.13697.x},
archivePrefix = {arXiv},
       eprint = {0805.1922},
 primaryClass = {astro-ph},
       adsurl = {https://ui.adsabs.harvard.edu/abs/2008MNRAS.389.1871H}
}

@ARTICLE{Hoyle1939,
       author = {{Hoyle}, F. and {Lyttleton}, R.~A.},
        title = "{The effect of interstellar matter on climatic variation}",
      journal = {Proceedings of the Cambridge Philosophical Society},
         year = 1939,
        month = jan,
       volume = {35},
       number = {3},
        pages = {405},
          doi = {10.1017/S0305004100021150},
       adsurl = {https://ui.adsabs.harvard.edu/abs/1939PCPS...35..405H}
}

@ARTICLE{Hu2017,
       author = {{Hu}, Chia-Yu and {Naab}, Thorsten and {Glover}, Simon C.~O. and {Walch}, Stefanie and {Clark}, Paul C.},
        title = "{Variable interstellar radiation fields in simulated dwarf galaxies: supernovae versus photoelectric heating}",
      journal = {\mnras},
         year = 2017,
        month = oct,
       volume = {471},
       number = {2},
        pages = {2151-2173},
          doi = {10.1093/mnras/stx1773},
archivePrefix = {arXiv},
       eprint = {1701.08779},
 primaryClass = {astro-ph.GA},
       adsurl = {https://ui.adsabs.harvard.edu/abs/2017MNRAS.471.2151H}
}

@ARTICLE{Hu2022SF,
       author = {{Hu}, Zipeng and {Krumholz}, Mark R. and {Pokhrel}, Riwaj and {Gutermuth}, Robert A.},
        title = "{High-precision star-formation efficiency measurements in nearby clouds}",
      journal = {\mnras},
         year = 2022,
        month = feb,
       volume = {511},
       number = {1},
        pages = {1431-1438},
          doi = {10.1093/mnras/stac174},
archivePrefix = {arXiv},
       eprint = {2109.04665},
 primaryClass = {astro-ph.GA},
       adsurl = {https://ui.adsabs.harvard.edu/abs/2022MNRAS.511.1431H}
}

@ARTICLE{Hui1997,
       author = {{Hui}, Lam and {Gnedin}, Nickolay Y.},
        title = "{Equation of state of the photoionized intergalactic medium}",
      journal = {\mnras},
         year = 1997,
        month = nov,
       volume = {292},
       number = {1},
        pages = {27-42},
          doi = {10.1093/mnras/292.1.27},
archivePrefix = {arXiv},
       eprint = {astro-ph/9612232},
 primaryClass = {astro-ph},
       adsurl = {https://ui.adsabs.harvard.edu/abs/1997MNRAS.292...27H}
}

@ARTICLE{Husko2022,
       author = {{Hu{\v{s}}ko}, Filip and {Lacey}, Cedric G. and {Schaye}, Joop and {Schaller}, Matthieu and {Nobels}, Folkert S.~J.},
        title = "{Spin-driven jet feedback in idealized simulations of galaxy groups and clusters}",
      journal = {\mnras},
         year = 2022,
        month = nov,
       volume = {516},
       number = {3},
        pages = {3750-3772},
          doi = {10.1093/mnras/stac2278},
archivePrefix = {arXiv},
       eprint = {2206.06402},
 primaryClass = {astro-ph.GA},
       adsurl = {https://ui.adsabs.harvard.edu/abs/2022MNRAS.516.3750H}
}

@ARTICLE{Husko2024,
       author = {{Hu{\v{s}}ko}, Filip and {Lacey}, Cedric G. and {Schaye}, Joop and {Nobels}, Folkert S.~J. and {Schaller}, Matthieu},
        title = "{Winds versus jets: a comparison between black hole feedback modes in simulations of idealized galaxy groups and clusters}",
      journal = {\mnras},
         year = 2024,
        month = jan,
       volume = {527},
       number = {3},
        pages = {5988-6020},
          doi = {10.1093/mnras/stad3548},
archivePrefix = {arXiv},
       eprint = {2307.01409},
 primaryClass = {astro-ph.GA},
       adsurl = {https://ui.adsabs.harvard.edu/abs/2024MNRAS.527.5988H}
}

@ARTICLE{Husko2025SuperEddington,
       author = {{Hu{\v{s}}ko}, Filip and {Lacey}, Cedric G. and {Roper}, William J. and {Schaye}, Joop and {Briggs}, Jemima Mae and {Schaller}, Matthieu},
        title = "{The effects of super-Eddington accretion and feedback on the growth of early supermassive black holes and galaxies}",
      journal = {\mnras},
         year = 2025,
        month = mar,
       volume = {537},
       number = {3},
        pages = {2559-2578},
          doi = {10.1093/mnras/staf146},
archivePrefix = {arXiv},
       eprint = {2410.09450},
 primaryClass = {astro-ph.GA},
       adsurl = {https://ui.adsabs.harvard.edu/abs/2025MNRAS.537.2559H}
}

@ARTICLE{Husko2025coligal,
       author = {{Hu{\v{s}}ko}, Filip and others},
        title = "{TBD}",
      journal = {\mnras},
         year = 2026,
       volume = {in prep.},
}

@ARTICLE{Husko2025method,
       author = {{Hu{\v{s}}ko}, Filip and {Lacey}, Cedric G. and {Schaye}, Joop and {Schaller}, Matthieu and {Chaikin}, Evgenii and {Ploeckinger}, Sylvia and {Ben{\'\i}tez Llambay}, Alejandro and {Richings}, Alexander J. and {Trayford}, James W.},
        title = "{A hybrid active galactic nucleus feedback model with spinning black holes, winds and jets}",
      journal = {arXiv e-prints},
         year = 2025,
        month = sep,
          eid = {arXiv:2509.05179},
        pages = {arXiv:2509.05179},
          doi = {10.48550/arXiv.2509.05179},
archivePrefix = {arXiv},
       eprint = {2509.05179},
 primaryClass = {astro-ph.GA},
       adsurl = {https://ui.adsabs.harvard.edu/abs/2025arXiv250905179H}
}

@ARTICLE{Indriolo2015,
       author = {{Indriolo}, Nick and {Neufeld}, D.~A. and {Gerin}, M. and {Schilke}, P. and {Benz}, A.~O. and {Winkel}, B. and {Menten}, K.~M. and {Chambers}, E.~T. and {Black}, John H. and {Bruderer}, S. and {Falgarone}, E. and {Godard}, B. and {Goicoechea}, J.~R. and {Gupta}, H. and {Lis}, D.~C. and {Ossenkopf}, V. and {Persson}, C.~M. and {Sonnentrucker}, P. and {van der Tak}, F.~F.~S. and {van Dishoeck}, E.~F. and {Wolfire}, Mark G. and {Wyrowski}, F.},
        title = "{Herschel Survey of Galactic OH$^{+}$, H$_{2}$O$^{+}$, and H$_{3}$O$^{+}$: Probing the Molecular Hydrogen Fraction and Cosmic-Ray Ionization Rate}",
      journal = {\apj},
         year = 2015,
        month = feb,
       volume = {800},
       number = {1},
          eid = {40},
        pages = {40},
          doi = {10.1088/0004-637X/800/1/40},
archivePrefix = {arXiv},
       eprint = {1412.1106},
 primaryClass = {astro-ph.GA},
       adsurl = {https://ui.adsabs.harvard.edu/abs/2015ApJ...800...40I}
}

@ARTICLE{Iwamoto1999,
       author = {{Iwamoto}, Koichi and {Brachwitz}, Franziska and {Nomoto}, Ken'ICHI and {Kishimoto}, Nobuhiro and {Umeda}, Hideyuki and {Hix}, W. Raphael and {Thielemann}, Friedrich-Karl},
        title = "{Nucleosynthesis in Chandrasekhar Mass Models for Type IA Supernovae and Constraints on Progenitor Systems and Burning-Front Propagation}",
      journal = {\apjs},
         year = 1999,
        month = dec,
       volume = {125},
       number = {2},
        pages = {439-462},
          doi = {10.1086/313278},
archivePrefix = {arXiv},
       eprint = {astro-ph/0002337},
 primaryClass = {astro-ph},
       adsurl = {https://ui.adsabs.harvard.edu/abs/1999ApJS..125..439I}
}

@ARTICLE{Izquierdo-Villalba2023,
       author = {{Izquierdo-Villalba}, David and {Lupi}, Alessandro and {Regan}, John and {Bonetti}, Matteo and {Franchini}, Alessia},
        title = "{Massive Black Holes in Galactic Nuclei}",
      journal = {arXiv e-prints},
         year = 2023,
        month = nov,
          eid = {arXiv:2311.03152},
        pages = {arXiv:2311.03152},
          doi = {10.48550/arXiv.2311.03152},
archivePrefix = {arXiv},
       eprint = {2311.03152},
 primaryClass = {astro-ph.GA},
       adsurl = {https://ui.adsabs.harvard.edu/abs/2023arXiv231103152I}
}

@ARTICLE{Jenkins2009,
       author = {{Jenkins}, Edward B.},
        title = "{A Unified Representation of Gas-Phase Element Depletions in the Interstellar Medium}",
      journal = {\apj},
         year = 2009,
        month = aug,
       volume = {700},
       number = {2},
        pages = {1299-1348},
          doi = {10.1088/0004-637X/700/2/1299},
archivePrefix = {arXiv},
       eprint = {0905.3173},
 primaryClass = {astro-ph.GA},
       adsurl = {https://ui.adsabs.harvard.edu/abs/2009ApJ...700.1299J}
}

@ARTICLE{Jenkins2013,
       author = {{Jenkins}, Adrian},
        title = "{A new way of setting the phases for cosmological multiscale Gaussian initial conditions}",
      journal = {\mnras},
         year = 2013,
        month = sep,
       volume = {434},
       number = {3},
        pages = {2094-2120},
          doi = {10.1093/mnras/stt1154},
archivePrefix = {arXiv},
       eprint = {1306.5968},
 primaryClass = {astro-ph.CO},
       adsurl = {https://ui.adsabs.harvard.edu/abs/2013MNRAS.434.2094J}
}

@ARTICLE{Jonsson2018,
       author = {{J{\"o}nsson}, Henrik and {Allende Prieto}, Carlos and {Holtzman}, Jon A. and {Feuillet}, Diane K. and {Hawkins}, Keith and {Cunha}, Katia and {M{\'e}sz{\'a}ros}, Szabolcs and {Hasselquist}, Sten and {Fern{\'a}ndez-Trincado}, J.~G. and {Garc{\'\i}a-Hern{\'a}ndez}, D.~A. and {Bizyaev}, Dmitry and {Carrera}, Ricardo and {Majewski}, Steven R. and {Pinsonneault}, Marc H. and {Shetrone}, Matthew and {Smith}, Verne and {Sobeck}, Jennifer and {Souto}, Diogo and {Stringfellow}, Guy S. and {Teske}, Johanna and {Zamora}, Olga},
        title = "{APOGEE Data Releases 13 and 14: Stellar Parameter and Abundance Comparisons with Independent Analyses}",
      journal = {\aj},
         year = 2018,
        month = sep,
       volume = {156},
       number = {3},
          eid = {126},
        pages = {126},
          doi = {10.3847/1538-3881/aad4f5},
archivePrefix = {arXiv},
       eprint = {1807.09784},
 primaryClass = {astro-ph.SR},
       adsurl = {https://ui.adsabs.harvard.edu/abs/2018AJ....156..126J}
}

@ARTICLE{Karakas2010,
       author = {{Karakas}, A.~I.},
        title = "{Updated stellar yields from asymptotic giant branch models}",
      journal = {\mnras},
         year = 2010,
        month = apr,
       volume = {403},
       number = {3},
        pages = {1413-1425},
          doi = {10.1111/j.1365-2966.2009.16198.x},
archivePrefix = {arXiv},
       eprint = {0912.2142},
 primaryClass = {astro-ph.SR},
       adsurl = {https://ui.adsabs.harvard.edu/abs/2010MNRAS.403.1413K}
}

@ARTICLE{Karakas2016,
       author = {{Karakas}, Amanda I. and {Lugaro}, Maria},
        title = "{Stellar Yields from Metal-rich Asymptotic Giant Branch Models}",
      journal = {\apj},
         year = 2016,
        month = jul,
       volume = {825},
       number = {1},
          eid = {26},
        pages = {26},
          doi = {10.3847/0004-637X/825/1/26},
archivePrefix = {arXiv},
       eprint = {1604.02178},
 primaryClass = {astro-ph.SR},
       adsurl = {https://ui.adsabs.harvard.edu/abs/2016ApJ...825...26K}
}

@ARTICLE{Katsianis2020,
       author = {{Katsianis}, A. and {Gonzalez}, V. and {Barrientos}, D. and {Yang}, X. and {Lagos}, C.~D.~P. and {Schaye}, J. and {Camps}, P. and {Tr{\v{c}}ka}, A. and {Baes}, M. and {Stalevski}, M. and {Blanc}, G.~A. and {Theuns}, T.},
        title = "{The high-redshift SFR-M* relation is sensitive to the employed star formation rate and stellar mass indicators: towards addressing the tension between observations and simulations}",
      journal = {\mnras},
         year = 2020,
        month = mar,
       volume = {492},
       number = {4},
        pages = {5592-5606},
          doi = {10.1093/mnras/staa157},
archivePrefix = {arXiv},
       eprint = {2001.06025},
 primaryClass = {astro-ph.GA},
       adsurl = {https://ui.adsabs.harvard.edu/abs/2020MNRAS.492.5592K}
}

@ARTICLE{Katz1996,
       author = {{Katz}, Neal and {Weinberg}, David H. and {Hernquist}, Lars},
        title = "{Cosmological Simulations with TreeSPH}",
      journal = {\apjs},
         year = 1996,
        month = jul,
       volume = {105},
        pages = {19},
          doi = {10.1086/192305},
archivePrefix = {arXiv},
       eprint = {astro-ph/9509107},
 primaryClass = {astro-ph},
       adsurl = {https://ui.adsabs.harvard.edu/abs/1996ApJS..105...19K}
}

@ARTICLE{Katz2022,
       author = {{Katz}, Harley and {Liu}, Shenghua and {Kimm}, Taysun and {Rey}, Martin P. and {Andersson}, Eric P. and {Cameron}, Alex J. and {Rodriguez-Montero}, Francisco and {Agertz}, Oscar and {Devriendt}, Julien and {Slyz}, Adrianne},
        title = "{PRISM: A Non-Equilibrium, Multiphase Interstellar Medium Model for Radiation Hydrodynamics Simulations of Galaxies}",
      journal = {arXiv e-prints},
         year = 2022,
        month = nov,
          eid = {arXiv:2211.04626},
        pages = {arXiv:2211.04626},
          doi = {10.48550/arXiv.2211.04626},
archivePrefix = {arXiv},
       eprint = {2211.04626},
 primaryClass = {astro-ph.GA},
       adsurl = {https://ui.adsabs.harvard.edu/abs/2022arXiv221104626K}
}

@ARTICLE{Kaviraj2025,
       author = {{Kaviraj}, S. and {Lazar}, I. and {Watkins}, A.~E. and {Laigle}, C. and {Martin}, G. and {Jackson}, R.~A.},
        title = "{The quenching of star formation in dwarf galaxies: new perspectives from deep-wide surveys}",
      journal = {\mnras},
         year = 2025,
        month = mar,
       volume = {538},
       number = {1},
        pages = {153-164},
          doi = {10.1093/mnras/staf233},
archivePrefix = {arXiv},
       eprint = {2502.02656},
 primaryClass = {astro-ph.GA},
       adsurl = {https://ui.adsabs.harvard.edu/abs/2025MNRAS.538..153K}
}

@ARTICLE{Keller2014,
       author = {{Keller}, B.~W. and {Wadsley}, J. and {Benincasa}, S.~M. and {Couchman}, H.~M.~P.},
        title = "{A superbubble feedback model for galaxy simulations}",
      journal = {\mnras},
         year = 2014,
        month = aug,
       volume = {442},
       number = {4},
        pages = {3013-3025},
          doi = {10.1093/mnras/stu1058},
archivePrefix = {arXiv},
       eprint = {1405.2625},
 primaryClass = {astro-ph.GA},
       adsurl = {https://ui.adsabs.harvard.edu/abs/2014MNRAS.442.3013K}
}

@ARTICLE{Kennicutt1998,
       author = {{Kennicutt}, Jr., Robert C.},
        title = "{The Global Schmidt Law in Star-forming Galaxies}",
      journal = {\apj},
         year = 1998,
        month = may,
       volume = {498},
       number = {2},
        pages = {541-552},
          doi = {10.1086/305588},
archivePrefix = {arXiv},
       eprint = {astro-ph/9712213},
 primaryClass = {astro-ph},
       adsurl = {https://ui.adsabs.harvard.edu/abs/1998ApJ...498..541K}
}

@ARTICLE{Kennicutt2012,
       author = {{Kennicutt}, Robert C. and {Evans}, Neal J.},
        title = "{Star Formation in the Milky Way and Nearby Galaxies}",
      journal = {\araa},
         year = 2012,
        month = sep,
       volume = {50},
        pages = {531-608},
          doi = {10.1146/annurev-astro-081811-125610},
archivePrefix = {arXiv},
       eprint = {1204.3552},
 primaryClass = {astro-ph.GA},
       adsurl = {https://ui.adsabs.harvard.edu/abs/2012ARA&A..50..531K}
}

@ARTICLE{Khusanova2021,
       author = {{Khusanova}, Y. and {Bethermin}, M. and {Le F{\`e}vre}, O. and {Capak}, P. and {Faisst}, A.~L. and {Schaerer}, D. and {Silverman}, J.~D. and {Cassata}, P. and {Yan}, L. and {Ginolfi}, M. and {Fudamoto}, Y. and {Loiacono}, F. and {Amorin}, R. and {Bardelli}, S. and {Boquien}, M. and {Cimatti}, A. and {Dessauges-Zavadsky}, M. and {Gruppioni}, C. and {Hathi}, N.~P. and {Jones}, G.~C. and {Koekemoer}, A.~M. and {Lagache}, G. and {Maiolino}, R. and {Lemaux}, B.~C. and {Oesch}, P. and {Pozzi}, F. and {Riechers}, D.~A. and {Romano}, M. and {Talia}, M. and {Toft}, S. and {Vergani}, D. and {Zamorani}, G. and {Zucca}, E.},
        title = "{The ALPINE-ALMA [CII] survey. Obscured star formation rate density and main sequence of star-forming galaxies at z > 4}",
      journal = {\aap},
         year = 2021,
        month = may,
       volume = {649},
          eid = {A152},
        pages = {A152},
          doi = {10.1051/0004-6361/202038944},
archivePrefix = {arXiv},
       eprint = {2007.08384},
 primaryClass = {astro-ph.GA},
       adsurl = {https://ui.adsabs.harvard.edu/abs/2021A&A...649A.152K}
}

@ARTICLE{Kim2023Tigress-ncr,
       author = {{Kim}, Chang-Goo and {Kim}, Jeong-Gyu and {Gong}, Munan and {Ostriker}, Eve C.},
        title = "{Introducing TIGRESS-NCR. I. Coregulation of the Multiphase Interstellar Medium and Star Formation Rates}",
      journal = {\apj},
         year = 2023,
        month = mar,
       volume = {946},
       number = {1},
          eid = {3},
        pages = {3},
          doi = {10.3847/1538-4357/acbd3a},
archivePrefix = {arXiv},
       eprint = {2211.13293},
 primaryClass = {astro-ph.GA},
       adsurl = {https://ui.adsabs.harvard.edu/abs/2023ApJ...946....3K}
}

@ARTICLE{Kirby2013,
       author = {{Kirby}, Evan N. and {Cohen}, Judith G. and {Guhathakurta}, Puragra and {Cheng}, Lucy and {Bullock}, James S. and {Gallazzi}, Anna},
        title = "{The Universal Stellar Mass-Stellar Metallicity Relation for Dwarf Galaxies}",
      journal = {\apj},
         year = 2013,
        month = dec,
       volume = {779},
       number = {2},
          eid = {102},
        pages = {102},
          doi = {10.1088/0004-637X/779/2/102},
archivePrefix = {arXiv},
       eprint = {1310.0814},
 primaryClass = {astro-ph.GA},
       adsurl = {https://ui.adsabs.harvard.edu/abs/2013ApJ...779..102K}
}

@ARTICLE{Kobayashi2006,
       author = {{Kobayashi}, Chiaki and {Umeda}, Hideyuki and {Nomoto}, Ken'ichi and {Tominaga}, Nozomu and {Ohkubo}, Takuya},
        title = "{Galactic Chemical Evolution: Carbon through Zinc}",
      journal = {\apj},
         year = 2006,
        month = dec,
       volume = {653},
       number = {2},
        pages = {1145-1171},
          doi = {10.1086/508914},
archivePrefix = {arXiv},
       eprint = {astro-ph/0608688},
 primaryClass = {astro-ph},
       adsurl = {https://ui.adsabs.harvard.edu/abs/2006ApJ...653.1145K}
}

@ARTICLE{Kobayashi2020,
       author = {{Kobayashi}, Chiaki and {Karakas}, Amanda I. and {Lugaro}, Maria},
        title = "{The Origin of Elements from Carbon to Uranium}",
      journal = {\apj},
         year = 2020,
        month = sep,
       volume = {900},
       number = {2},
          eid = {179},
        pages = {179},
          doi = {10.3847/1538-4357/abae65},
archivePrefix = {arXiv},
       eprint = {2008.04660},
 primaryClass = {astro-ph.GA},
       adsurl = {https://ui.adsabs.harvard.edu/abs/2020ApJ...900..179K}
}

@ARTICLE{Kroupa2002,
       author = {{Kroupa}, Pavel},
        title = "{The Initial Mass Function of Stars: Evidence for Uniformity in Variable Systems}",
      journal = {Science},
         year = 2002,
        month = jan,
       volume = {295},
       number = {5552},
        pages = {82-91},
          doi = {10.1126/science.1067524},
archivePrefix = {arXiv},
       eprint = {astro-ph/0201098},
 primaryClass = {astro-ph},
       adsurl = {https://ui.adsabs.harvard.edu/abs/2002Sci...295...82K}
}

@ARTICLE{Krumholz2005,
       author = {{Krumholz}, Mark R. and {McKee}, Christopher F. and {Klein}, Richard I.},
        title = "{Bondi Accretion in the Presence of Vorticity}",
      journal = {\apj},
         year = 2005,
        month = jan,
       volume = {618},
       number = {2},
        pages = {757-768},
          doi = {10.1086/426051},
archivePrefix = {arXiv},
       eprint = {astro-ph/0409454},
 primaryClass = {astro-ph},
       adsurl = {https://ui.adsabs.harvard.edu/abs/2005ApJ...618..757K}
}

@ARTICLE{Krumholz2006,
       author = {{Krumholz}, Mark R. and {McKee}, Christopher F. and {Klein}, Richard I.},
        title = "{Bondi-Hoyle Accretion in a Turbulent Medium}",
      journal = {\apj},
         year = 2006,
        month = feb,
       volume = {638},
       number = {1},
        pages = {369-381},
          doi = {10.1086/498844},
archivePrefix = {arXiv},
       eprint = {astro-ph/0510410},
 primaryClass = {astro-ph},
       adsurl = {https://ui.adsabs.harvard.edu/abs/2006ApJ...638..369K}
}

@ARTICLE{Kugel2023,
       author = {{Kugel}, Roi and {Schaye}, Joop and {Schaller}, Matthieu and {Helly}, John C. and {Braspenning}, Joey and {Elbers}, Willem and {Frenk}, Carlos S. and {McCarthy}, Ian G. and {Kwan}, Juliana and {Salcido}, Jaime and {van Daalen}, Marcel P. and {Vandenbroucke}, Bert and {Bah{\'e}}, Yannick M. and {Borrow}, Josh and {Chaikin}, Evgenii and {Hu{\v{s}}ko}, Filip and {Jenkins}, Adrian and {Lacey}, Cedric G. and {Nobels}, Folkert S.~J. and {Vernon}, Ian},
        title = "{FLAMINGO: calibrating large cosmological hydrodynamical simulations with machine learning}",
      journal = {\mnras},
         year = 2023,
        month = dec,
       volume = {526},
       number = {4},
        pages = {6103-6127},
          doi = {10.1093/mnras/stad2540},
archivePrefix = {arXiv},
       eprint = {2306.05492},
 primaryClass = {astro-ph.CO},
       adsurl = {https://ui.adsabs.harvard.edu/abs/2023MNRAS.526.6103K}
}

@ARTICLE{Kumar2025,
       author = {{Kumar}, Ankit and {Artale}, M. Celeste and {Montero-Dorta}, Antonio D. and {Guaita}, Lucia and {Lee}, Kyoung-Soo and {Pope}, Alexandra and {Schaye}, Joop and {Schaller}, Matthieu and {Gawiser}, Eric and {Hwang}, Ho Seong and {Jeong}, Woong-Seob and {Lee}, Jaehyun and {Padilla}, Nelson and {Park}, Changbom and {Ramakrishnan}, Vandana and {Singh}, Akriti and {Yang}, Yujin},
        title = "{Modeling submillimeter galaxies in cosmological simulations: Contribution to the cosmic star formation density and predictions for future surveys}",
      journal = {arXiv e-prints},
         year = 2025,
        month = jan,
          eid = {arXiv:2501.19327},
        pages = {arXiv:2501.19327},
          doi = {10.48550/arXiv.2501.19327},
archivePrefix = {arXiv},
       eprint = {2501.19327},
 primaryClass = {astro-ph.CO},
       adsurl = {https://ui.adsabs.harvard.edu/abs/2025arXiv250119327K}
}

@ARTICLE{Kuznetsova2008,
       author = {{Kuznetsova}, N. and {Barbary}, K. and {Connolly}, B. and {Kim}, A.~G. and {Pain}, R. and {Roe}, N.~A. and {Aldering}, G. and {Amanullah}, R. and {Dawson}, K. and {Doi}, M. and {Fadeyev}, V. and {Fruchter}, A.~S. and {Gibbons}, R. and {Goldhaber}, G. and {Goobar}, A. and {Gude}, A. and {Knop}, R.~A. and {Kowalski}, M. and {Lidman}, C. and {Morokuma}, T. and {Meyers}, J. and {Perlmutter}, S. and {Rubin}, D. and {Schlegel}, D.~J. and {Spadafora}, A.~L. and {Stanishev}, V. and {Strovink}, M. and {Suzuki}, N. and {Wang}, L. and {Yasuda}, N. and {Cosmology Project}, Supernova},
        title = "{A New Determination of the High-Redshift Type Ia Supernova Rates with the Hubble Space Telescope Advanced Camera for Surveys}",
      journal = {\apj},
         year = 2008,
        month = feb,
       volume = {673},
       number = {2},
        pages = {981-998},
          doi = {10.1086/524881},
archivePrefix = {arXiv},
       eprint = {0710.3120},
 primaryClass = {astro-ph},
       adsurl = {https://ui.adsabs.harvard.edu/abs/2008ApJ...673..981K}
}

@ARTICLE{Lagos2025,
       author = {{Lagos}, Claudia del P. and {Schaye}, Joop and {Schaller}, Matthieu and {Obreschkow}, Danail and {Bahe}, Yannick M. and {Benitez-Llambay}, Alejandro and {Chaikin}, Evgenii and {Correa}, Camila and {Davis}, Timothy A. and {Frenk}, Carlos S. and {Husko}, Filip and {Kaasinen}, Melanie and {McGibbon}, Robert J. and {Oman}, Kyle and {Ploeckinger}, Sylvia and {Richings}, Alexander J. and {Trayford}, James W. and {Wang}, Jing and {Wright}, Ruby J.},
        title = "{Kennicutt-Schmidt relation of galaxies over 13 billion years in the COLIBRE hydrodynamical simulations}",
      journal = {arXiv e-prints},
         year = 2025,
        month = dec,
          eid = {arXiv:2512.11309},
        pages = {arXiv:2512.11309},
          doi = {10.48550/arXiv.2512.11309},
archivePrefix = {arXiv},
       eprint = {2512.11309},
 primaryClass = {astro-ph.GA},
       adsurl = {https://ui.adsabs.harvard.edu/abs/2025arXiv251211309L}
}

@ARTICLE{LeBrun2014,
       author = {{Le Brun}, Amandine M.~C. and {McCarthy}, Ian G. and {Schaye}, Joop and {Ponman}, Trevor J.},
        title = "{Towards a realistic population of simulated galaxy groups and clusters}",
      journal = {\mnras},
         year = 2014,
        month = jun,
       volume = {441},
       number = {2},
        pages = {1270-1290},
          doi = {10.1093/mnras/stu608},
archivePrefix = {arXiv},
       eprint = {1312.5462},
 primaryClass = {astro-ph.CO},
       adsurl = {https://ui.adsabs.harvard.edu/abs/2014MNRAS.441.1270L}
}

@ARTICLE{Leja2022,
       author = {{Leja}, Joel and {Speagle}, Joshua S. and {Ting}, Yuan-Sen and {Johnson}, Benjamin D. and {Conroy}, Charlie and {Whitaker}, Katherine E. and {Nelson}, Erica J. and {van Dokkum}, Pieter and {Franx}, Marijn},
        title = "{A New Census of the 0.2 < z < 3.0 Universe. II. The Star-forming Sequence}",
      journal = {\apj},
         year = 2022,
        month = sep,
       volume = {936},
       number = {2},
          eid = {165},
        pages = {165},
          doi = {10.3847/1538-4357/ac887d},
archivePrefix = {arXiv},
       eprint = {2110.04314},
 primaryClass = {astro-ph.GA},
       adsurl = {https://ui.adsabs.harvard.edu/abs/2022ApJ...936..165L}
}

@ARTICLE{Leroy2025,
       author = {{Leroy}, Adam K. and {Sun}, Jiayi and {Meidt}, Sharon and {Agertz}, Oscar and {Chiang}, I. -Da and {Gensior}, Jindra and {Glover}, Simon C.~O. and {Gnedin}, Oleg Y. and {Hughes}, Annie and {Schinnerer}, Eva and {Barnes}, Ashley T. and {Bigiel}, Frank and {Bolatto}, Alberto D. and {Colombo}, Dario and {den Brok}, Jakob and {Chevance}, M{\'e}lanie and {Chown}, Ryan and {Eibensteiner}, Cosima and {Gleis}, Damian R. and {Grasha}, Kathryn and {Henshaw}, Jonathan D. and {Klessen}, Ralf S. and {Koch}, Eric W. and {Oakes}, Elias K. and {Pan}, Hsi-An and {Querejeta}, Miguel and {Rosolowsky}, Erik and {Saito}, Toshiki and {Sandstrom}, Karin and {Sarbadhicary}, Sumit K. and {Teng}, Yu-Hsuan and {Usero}, Antonio and {Utomo}, Dyas and {Williams}, Thomas G.},
        title = "{Cloud-scale Gas Properties, Depletion Times, and Star Formation Efficiency per Freefall Time in PHANGS{\textendash}ALMA}",
      journal = {\apj},
         year = 2025,
        month = may,
       volume = {985},
       number = {1},
          eid = {14},
        pages = {14},
          doi = {10.3847/1538-4357/adbcab},
archivePrefix = {arXiv},
       eprint = {2502.04481},
 primaryClass = {astro-ph.GA},
       adsurl = {https://ui.adsabs.harvard.edu/abs/2025ApJ...985...14L}
}

@ARTICLE{Leung2018,
       author = {{Leung}, Shing-Chi and {Nomoto}, Ken'ichi},
        title = "{Explosive Nucleosynthesis in Near-Chandrasekhar-mass White Dwarf Models for Type Ia Supernovae: Dependence on Model Parameters}",
      journal = {\apj},
         year = 2018,
        month = jul,
       volume = {861},
       number = {2},
          eid = {143},
        pages = {143},
          doi = {10.3847/1538-4357/aac2df},
archivePrefix = {arXiv},
       eprint = {1710.04254},
 primaryClass = {astro-ph.SR},
       adsurl = {https://ui.adsabs.harvard.edu/abs/2018ApJ...861..143L}
}

@ARTICLE{Li2011,
       author = {{Li}, Weidong and {Chornock}, Ryan and {Leaman}, Jesse and {Filippenko}, Alexei V. and {Poznanski}, Dovi and {Wang}, Xiaofeng and {Ganeshalingam}, Mohan and {Mannucci}, Filippo},
        title = "{Nearby supernova rates from the Lick Observatory Supernova Search - III. The rate-size relation, and the rates as a function of galaxy Hubble type and colour}",
      journal = {\mnras},
         year = 2011,
        month = apr,
       volume = {412},
       number = {3},
        pages = {1473-1507},
          doi = {10.1111/j.1365-2966.2011.18162.x},
archivePrefix = {arXiv},
       eprint = {1006.4613},
 primaryClass = {astro-ph.SR},
       adsurl = {https://ui.adsabs.harvard.edu/abs/2011MNRAS.412.1473L}
}

@ARTICLE{Liang2023,
       author = {{Liang}, Jinning and {Gjergo}, Eda and {Fan}, XiLong},
        title = "{Assessing stellar yields in Galaxy chemical evolution: Observational stellar abundance patterns}",
      journal = {\mnras},
         year = 2023,
        month = jun,
       volume = {522},
       number = {1},
        pages = {863-884},
          doi = {10.1093/mnras/stad1013},
archivePrefix = {arXiv},
       eprint = {2304.00208},
 primaryClass = {astro-ph.GA},
       adsurl = {https://ui.adsabs.harvard.edu/abs/2023MNRAS.522..863L}
}

@ARTICLE{Libeskind2020,
       author = {{Libeskind}, Noam I. and {Carlesi}, Edoardo and {Grand}, Robert J.~J. and {Khalatyan}, Arman and {Knebe}, Alexander and {Pakmor}, Ruediger and {Pilipenko}, Sergey and {Pawlowski}, Marcel S. and {Sparre}, Martin and {Tempel}, Elmo and {Wang}, Peng and {Courtois}, H{\'e}l{\`e}ne M. and {Gottl{\"o}ber}, Stefan and {Hoffman}, Yehuda and {Minchev}, Ivan and {Pfrommer}, Christoph and {Sorce}, Jenny G. and {Springel}, Volker and {Steinmetz}, Matthias and {Tully}, R. Brent and {Vogelsberger}, Mark and {Yepes}, Gustavo},
        title = "{The HESTIA project: simulations of the Local Group}",
      journal = {\mnras},
         year = 2020,
        month = oct,
       volume = {498},
       number = {2},
        pages = {2968-2983},
          doi = {10.1093/mnras/staa2541},
archivePrefix = {arXiv},
       eprint = {2008.04926},
 primaryClass = {astro-ph.GA},
       adsurl = {https://ui.adsabs.harvard.edu/abs/2020MNRAS.498.2968L}
}

@ARTICLE{Lovisari2015,
       author = {{Lovisari}, L. and {Reiprich}, T.~H. and {Schellenberger}, G.},
        title = "{Scaling properties of a complete X-ray selected galaxy group sample}",
      journal = {\aap},
         year = 2015,
        month = jan,
       volume = {573},
          eid = {A118},
        pages = {A118},
          doi = {10.1051/0004-6361/201423954},
archivePrefix = {arXiv},
       eprint = {1409.3845},
 primaryClass = {astro-ph.CO},
       adsurl = {https://ui.adsabs.harvard.edu/abs/2015A&A...573A.118L}
}

@ARTICLE{Lovisari2020,
       author = {{Lovisari}, Lorenzo and {Schellenberger}, Gerrit and {Sereno}, Mauro and {Ettori}, Stefano and {Pratt}, Gabriel W. and {Forman}, William R. and {Jones}, Christine and {Andrade-Santos}, Felipe and {Randall}, Scott and {Kraft}, Ralph},
        title = "{X-Ray Scaling Relations for a Representative Sample of Planck-selected Clusters Observed with XMM-Newton}",
      journal = {\apj},
         year = 2020,
        month = apr,
       volume = {892},
       number = {2},
          eid = {102},
        pages = {102},
          doi = {10.3847/1538-4357/ab7997},
archivePrefix = {arXiv},
       eprint = {2002.11740},
 primaryClass = {astro-ph.CO},
       adsurl = {https://ui.adsabs.harvard.edu/abs/2020ApJ...892..102L}
}

@ARTICLE{Ludlow2019,
       author = {{Ludlow}, Aaron D. and {Schaye}, Joop and {Schaller}, Matthieu and {Richings}, Jack},
        title = "{Energy equipartition between stellar and dark matter particles in cosmological simulations results in spurious growth of galaxy sizes}",
      journal = {\mnras},
         year = 2019,
        month = sep,
       volume = {488},
       number = {1},
        pages = {L123-L128},
          doi = {10.1093/mnrasl/slz110},
archivePrefix = {arXiv},
       eprint = {1903.10110},
 primaryClass = {astro-ph.GA},
       adsurl = {https://ui.adsabs.harvard.edu/abs/2019MNRAS.488L.123L}
}

@ARTICLE{Ludlow2019CDM,
       author = {{Ludlow}, Aaron D. and {Schaye}, Joop and {Bower}, Richard},
        title = "{Numerical convergence of simulations of galaxy formation: the abundance and internal structure of cold dark matter haloes}",
      journal = {\mnras},
         year = 2019,
        month = sep,
       volume = {488},
       number = {3},
        pages = {3663-3684},
          doi = {10.1093/mnras/stz1821},
archivePrefix = {arXiv},
       eprint = {1812.05777},
 primaryClass = {astro-ph.CO},
       adsurl = {https://ui.adsabs.harvard.edu/abs/2019MNRAS.488.3663L}
}

@ARTICLE{Ludlow2020,
       author = {{Ludlow}, Aaron D. and {Schaye}, Joop and {Schaller}, Matthieu and {Bower}, Richard},
        title = "{Numerical convergence of hydrodynamical simulations of galaxy formation: the abundance and internal structure of galaxies and their cold dark matter haloes}",
      journal = {\mnras},
         year = 2020,
        month = apr,
       volume = {493},
       number = {2},
        pages = {2926-2951},
          doi = {10.1093/mnras/staa316},
archivePrefix = {arXiv},
       eprint = {1908.05019},
 primaryClass = {astro-ph.GA},
       adsurl = {https://ui.adsabs.harvard.edu/abs/2020MNRAS.493.2926L}
}

@ARTICLE{Ludlow2021,
       author = {{Ludlow}, Aaron D. and {Fall}, S. Michael and {Schaye}, Joop and {Obreschkow}, Danail},
        title = "{Spurious heating of stellar motions in simulated galactic discs by dark matter halo particles}",
      journal = {\mnras},
         year = 2021,
        month = dec,
       volume = {508},
       number = {4},
        pages = {5114-5137},
          doi = {10.1093/mnras/stab2770},
archivePrefix = {arXiv},
       eprint = {2105.03561},
 primaryClass = {astro-ph.GA},
       adsurl = {https://ui.adsabs.harvard.edu/abs/2021MNRAS.508.5114L}
}

@ARTICLE{Ludlow2023,
       author = {{Ludlow}, Aaron D. and {Fall}, S. Michael and {Wilkinson}, Matthew J. and {Schaye}, Joop and {Obreschkow}, Danail},
        title = "{Spurious heating of stellar motions by dark matter particles in cosmological simulations of galaxy formation}",
      journal = {\mnras},
         year = 2023,
        month = nov,
       volume = {525},
       number = {4},
        pages = {5614-5630},
          doi = {10.1093/mnras/stad2615},
archivePrefix = {arXiv},
       eprint = {2306.05753},
 primaryClass = {astro-ph.GA},
       adsurl = {https://ui.adsabs.harvard.edu/abs/2023MNRAS.525.5614L}
}

@ARTICLE{Ludlow2025,
       author = {{Ludlow}, Aaron D. and others},
        title = "{TBD}",
      journal = {TBD},
         year = 2026,
       volume = {in prep.},
}

@ARTICLE{Lupton2004,
       author = {{Lupton}, Robert and {Blanton}, Michael R. and {Fekete}, George and {Hogg}, David W. and {O'Mullane}, Wil and {Szalay}, Alex and {Wherry}, Nicholas},
        title = "{Preparing Red-Green-Blue Images from CCD Data}",
      journal = {\pasp},
         year = 2004,
        month = feb,
       volume = {116},
       number = {816},
        pages = {133-137},
          doi = {10.1086/382245},
archivePrefix = {arXiv},
       eprint = {astro-ph/0312483},
 primaryClass = {astro-ph},
       adsurl = {https://ui.adsabs.harvard.edu/abs/2004PASP..116..133L}
}

@ARTICLE{Ma2021BHs_not_sinking,
       author = {{Ma}, Linhao and {Hopkins}, Philip F. and {Ma}, Xiangcheng and {Angl{\'e}s-Alc{\'a}zar}, Daniel and {Faucher-Gigu{\`e}re}, Claude-Andr{\'e} and {Kelley}, Luke Zoltan},
        title = "{Seeds don't sink: even massive black hole 'seeds' cannot migrate to galaxy centres efficiently}",
      journal = {\mnras},
         year = 2021,
        month = dec,
       volume = {508},
       number = {2},
        pages = {1973-1985},
          doi = {10.1093/mnras/stab2713},
archivePrefix = {arXiv},
       eprint = {2101.02727},
 primaryClass = {astro-ph.GA},
       adsurl = {https://ui.adsabs.harvard.edu/abs/2021MNRAS.508.1973M}
}

@ARTICLE{Madau2014,
       author = {{Madau}, Piero and {Dickinson}, Mark},
        title = "{Cosmic Star-Formation History}",
      journal = {\araa},
         year = 2014,
        month = aug,
       volume = {52},
        pages = {415-486},
          doi = {10.1146/annurev-astro-081811-125615},
archivePrefix = {arXiv},
       eprint = {1403.0007},
 primaryClass = {astro-ph.CO},
       adsurl = {https://ui.adsabs.harvard.edu/abs/2014ARA&A..52..415M}
}

@ARTICLE{Madgwick2003,
       author = {{Madgwick}, Darren S. and {Hewett}, Paul C. and {Mortlock}, Daniel J. and {Wang}, Lifan},
        title = "{Spectroscopic Detection of Type Ia Supernovae in the Sloan Digital Sky Survey}",
      journal = {\apjl},
         year = 2003,
        month = dec,
       volume = {599},
       number = {1},
        pages = {L33-L36},
          doi = {10.1086/381081},
archivePrefix = {arXiv},
       eprint = {astro-ph/0310887},
 primaryClass = {astro-ph},
       adsurl = {https://ui.adsabs.harvard.edu/abs/2003ApJ...599L..33M}
}

@ARTICLE{Mannucci2005,
       author = {{Mannucci}, F. and {Della Valle}, M. and {Panagia}, N. and {Cappellaro}, E. and {Cresci}, G. and {Maiolino}, R. and {Petrosian}, A. and {Turatto}, M.},
        title = "{The supernova rate per unit mass}",
      journal = {\aap},
         year = 2005,
        month = apr,
       volume = {433},
       number = {3},
        pages = {807-814},
          doi = {10.1051/0004-6361:20041411},
archivePrefix = {arXiv},
       eprint = {astro-ph/0411450},
 primaryClass = {astro-ph},
       adsurl = {https://ui.adsabs.harvard.edu/abs/2005A&A...433..807M}
}

@ARTICLE{Marri2003,
       author = {{Marri}, S. and {White}, S.~D.~M.},
        title = "{Smoothed particle hydrodynamics for galaxy-formation simulations: improved treatments of multiphase gas, of star formation and of supernovae feedback}",
      journal = {\mnras},
         year = 2003,
        month = oct,
       volume = {345},
       number = {2},
        pages = {561-574},
          doi = {10.1046/j.1365-8711.2003.06984.x},
archivePrefix = {arXiv},
       eprint = {astro-ph/0207448},
 primaryClass = {astro-ph},
       adsurl = {https://ui.adsabs.harvard.edu/abs/2003MNRAS.345..561M}
}

@ARTICLE{MartinezSerrano2008,
       author = {{Mart{\'\i}nez-Serrano}, F.~J. and {Serna}, A. and {Dom{\'\i}nguez-Tenreiro}, R. and {Moll{\'a}}, M.},
        title = "{Chemical evolution of galaxies - I. A composition-dependent SPH model for chemical evolution and cooling}",
      journal = {\mnras},
         year = 2008,
        month = jul,
       volume = {388},
       number = {1},
        pages = {39-55},
          doi = {10.1111/j.1365-2966.2008.13383.x},
archivePrefix = {arXiv},
       eprint = {0804.3766},
 primaryClass = {astro-ph},
       adsurl = {https://ui.adsabs.harvard.edu/abs/2008MNRAS.388...39M}
}

@ARTICLE{Mathis1983,
       author = {{Mathis}, J.~S. and {Mezger}, P.~G. and {Panagia}, N.},
        title = "{Interstellar radiation field and dust temperatures in the diffuse interstellar medium and in giant molecular clouds}",
      journal = {\aap},
         year = 1983,
        month = nov,
       volume = {128},
        pages = {212-229},
       adsurl = {https://ui.adsabs.harvard.edu/abs/1983A&A...128..212M}
}

@ARTICLE{Mazzali2014,
       author = {{Mazzali}, P.~A. and {McFadyen}, A.~I. and {Woosley}, S.~E. and {Pian}, E. and {Tanaka}, M.},
        title = "{An upper limit to the energy of gamma-ray bursts indicates that GRBs/SNe are powered by magnetars}",
      journal = {\mnras},
         year = 2014,
        month = sep,
       volume = {443},
       number = {1},
        pages = {67-71},
          doi = {10.1093/mnras/stu1124},
archivePrefix = {arXiv},
       eprint = {1406.1209},
 primaryClass = {astro-ph.HE},
       adsurl = {https://ui.adsabs.harvard.edu/abs/2014MNRAS.443...67M}
}

@ARTICLE{McAlpine2016,
       author = {{McAlpine}, S. and {Helly}, J.~C. and {Schaller}, M. and {Trayford}, J.~W. and {Qu}, Y. and {Furlong}, M. and {Bower}, R.~G. and {Crain}, R.~A. and {Schaye}, J. and {Theuns}, T. and {Dalla Vecchia}, C. and {Frenk}, C.~S. and {McCarthy}, I.~G. and {Jenkins}, A. and {Rosas-Guevara}, Y. and {White}, S.~D.~M. and {Baes}, M. and {Camps}, P. and {Lemson}, G.},
        title = "{The EAGLE simulations of galaxy formation: Public release of halo and galaxy catalogues}",
      journal = {Astronomy and Computing},
         year = 2016,
        month = apr,
       volume = {15},
        pages = {72-89},
          doi = {10.1016/j.ascom.2016.02.004},
archivePrefix = {arXiv},
       eprint = {1510.01320},
 primaryClass = {astro-ph.GA},
       adsurl = {https://ui.adsabs.harvard.edu/abs/2016A&C....15...72M}
}

@ARTICLE{McCarthy2017,
       author = {{McCarthy}, Ian G. and {Schaye}, Joop and {Bird}, Simeon and {Le Brun}, Amandine M.~C.},
        title = "{The BAHAMAS project: calibrated hydrodynamical simulations for large-scale structure cosmology}",
      journal = {\mnras},
         year = 2017,
        month = mar,
       volume = {465},
       number = {3},
        pages = {2936-2965},
          doi = {10.1093/mnras/stw2792},
archivePrefix = {arXiv},
       eprint = {1603.02702},
 primaryClass = {astro-ph.CO},
       adsurl = {https://ui.adsabs.harvard.edu/abs/2017MNRAS.465.2936M}
}

@ARTICLE{McCarthy2018,
       author = {{McCarthy}, Ian G. and {Bird}, Simeon and {Schaye}, Joop and {Harnois-Deraps}, Joachim and {Font}, Andreea S. and {van Waerbeke}, Ludovic},
        title = "{The BAHAMAS project: the CMB-large-scale structure tension and the roles of massive neutrinos and galaxy formation}",
      journal = {\mnras},
         year = 2018,
        month = may,
       volume = {476},
       number = {3},
        pages = {2999-3030},
          doi = {10.1093/mnras/sty377},
archivePrefix = {arXiv},
       eprint = {1712.02411},
 primaryClass = {astro-ph.CO},
       adsurl = {https://ui.adsabs.harvard.edu/abs/2018MNRAS.476.2999M}
}

@ARTICLE{McGibbon2025,
       author = {{McGibbon}, Robert and {Helly}, John and {Schaye}, Joop and {Schaller}, Matthieu and {Vandenbroucke}, Bert},
        title = "{SOAP: A Python Package for Calculating the Properties of Galaxies and Halos Formed in Cosmological Simulations}",
      journal = {The Journal of Open Source Software},
         year = 2025,
        month = jul,
       volume = {10},
       number = {111},
          eid = {8252},
        pages = {8252},
          doi = {10.21105/joss.08252},
archivePrefix = {arXiv},
       eprint = {2507.22669},
 primaryClass = {astro-ph.IM},
       adsurl = {https://ui.adsabs.harvard.edu/abs/2025JOSS...10.8252M}
}

@ARTICLE{McKinnon2016,
       author = {{McKinnon}, Ryan and {Torrey}, Paul and {Vogelsberger}, Mark},
        title = "{Dust formation in Milky Way-like galaxies}",
      journal = {\mnras},
         year = 2016,
        month = apr,
       volume = {457},
       number = {4},
        pages = {3775-3800},
          doi = {10.1093/mnras/stw253},
archivePrefix = {arXiv},
       eprint = {1505.04792},
 primaryClass = {astro-ph.GA},
       adsurl = {https://ui.adsabs.harvard.edu/abs/2016MNRAS.457.3775M}
}

@ARTICLE{Melinder2012,
       author = {{Melinder}, J. and {Dahlen}, T. and {Menc{\'\i}a Trinchant}, L. and {{\"O}stlin}, G. and {Mattila}, S. and {Sollerman}, J. and {Fransson}, C. and {Hayes}, M. and {Kankare}, E. and {Nasoudi-Shoar}, S.},
        title = "{The rate of supernovae at redshift 0.1-1.0. The Stockholm VIMOS Supernova Survey III}",
      journal = {\aap},
         year = 2012,
        month = sep,
       volume = {545},
          eid = {A96},
        pages = {A96},
          doi = {10.1051/0004-6361/201219364},
archivePrefix = {arXiv},
       eprint = {1206.6897},
 primaryClass = {astro-ph.CO},
       adsurl = {https://ui.adsabs.harvard.edu/abs/2012A&A...545A..96M}
}

@ARTICLE{Michaux2021,
       author = {{Michaux}, Micha{\"e}l and {Hahn}, Oliver and {Rampf}, Cornelius and {Angulo}, Raul E.},
        title = "{Accurate initial conditions for cosmological N-body simulations: minimizing truncation and discreteness errors}",
      journal = {\mnras},
         year = 2021,
        month = jan,
       volume = {500},
       number = {1},
        pages = {663-683},
          doi = {10.1093/mnras/staa3149},
archivePrefix = {arXiv},
       eprint = {2008.09588},
 primaryClass = {astro-ph.CO},
       adsurl = {https://ui.adsabs.harvard.edu/abs/2021MNRAS.500..663M}
}

@ARTICLE{Mina2021,
       author = {{Mina}, Mattia and {Shen}, Sijing and {Keller}, Benjamin Walter and {Mayer}, Lucio and {Madau}, Piero and {Wadsley}, James},
        title = "{The baryon cycle of Seven Dwarfs with superbubble feedback}",
      journal = {\aap},
         year = 2021,
        month = nov,
       volume = {655},
          eid = {A22},
        pages = {A22},
          doi = {10.1051/0004-6361/202039420},
archivePrefix = {arXiv},
       eprint = {2009.06646},
 primaryClass = {astro-ph.GA},
       adsurl = {https://ui.adsabs.harvard.edu/abs/2021A&A...655A..22M}
}

@ARTICLE{More2011,
       author = {{More}, Surhud and {Kravtsov}, Andrey V. and {Dalal}, Neal and {Gottl{\"o}ber}, Stefan},
        title = "{The Overdensity and Masses of the Friends-of-friends Halos and Universality of Halo Mass Function}",
      journal = {\apjs},
         year = 2011,
        month = jul,
       volume = {195},
       number = {1},
          eid = {4},
        pages = {4},
          doi = {10.1088/0067-0049/195/1/4},
archivePrefix = {arXiv},
       eprint = {1103.0005},
 primaryClass = {astro-ph.CO},
       adsurl = {https://ui.adsabs.harvard.edu/abs/2011ApJS..195....4M}
}

@ARTICLE{Mukai1993,
       author = {{Mukai}, K.},
        title = "{PIMMS and Viewing: proposal preparation tools}",
      journal = {Legacy},
         year = 1993,
        month = may,
       volume = {3},
        pages = {21-31},
       adsurl = {https://ui.adsabs.harvard.edu/abs/1993Legac...3...21M}
}

@ARTICLE{Muzzin2013,
       author = {{Muzzin}, Adam and {Marchesini}, Danilo and {Stefanon}, Mauro and {Franx}, Marijn and {McCracken}, Henry J. and {Milvang-Jensen}, Bo and {Dunlop}, James S. and {Fynbo}, J.~P.~U. and {Brammer}, Gabriel and {Labb{\'e}}, Ivo and {van Dokkum}, Pieter G.},
        title = "{The Evolution of the Stellar Mass Functions of Star-forming and Quiescent Galaxies to z = 4 from the COSMOS/UltraVISTA Survey}",
      journal = {\apj},
         year = 2013,
        month = nov,
       volume = {777},
       number = {1},
          eid = {18},
        pages = {18},
          doi = {10.1088/0004-637X/777/1/18},
archivePrefix = {arXiv},
       eprint = {1303.4409},
 primaryClass = {astro-ph.CO},
       adsurl = {https://ui.adsabs.harvard.edu/abs/2013ApJ...777...18M}
}

@ARTICLE{Narayan2003,
       author = {{Narayan}, Ramesh and {Igumenshchev}, Igor V. and {Abramowicz}, Marek A.},
        title = "{Magnetically Arrested Disk: an Energetically Efficient Accretion Flow}",
      journal = {\pasj},
         year = 2003,
        month = dec,
       volume = {55},
        pages = {L69-L72},
          doi = {10.1093/pasj/55.6.L69},
archivePrefix = {arXiv},
       eprint = {astro-ph/0305029},
 primaryClass = {astro-ph},
       adsurl = {https://ui.adsabs.harvard.edu/abs/2003PASJ...55L..69N}
}

@ARTICLE{Neill2006,
       author = {{Neill}, J.~D. and {Sullivan}, M. and {Balam}, D. and {Pritchet}, C.~J. and {Howell}, D.~A. and {Perrett}, K. and {Astier}, P. and {Aubourg}, E. and {Basa}, S. and {Carlberg}, R.~G. and {Conley}, A. and {Fabbro}, S. and {Fouchez}, D. and {Guy}, J. and {Hook}, I. and {Pain}, R. and {Palanque-Delabrouille}, N. and {Regnault}, N. and {Rich}, J. and {Taillet}, R. and {Aldering}, G. and {Antilogus}, P. and {Arsenijevic}, V. and {Balland}, C. and {Baumont}, S. and {Bronder}, J. and {Ellis}, R.~S. and {Filiol}, M. and {Gon{\c{c}}alves}, A.~C. and {Hardin}, D. and {Kowalski}, M. and {Lidman}, C. and {Lusset}, V. and {Mouchet}, M. and {Mourao}, A. and {Perlmutter}, S. and {Ripoche}, P. and {Schlegel}, D. and {Tao}, C.},
        title = "{The Type Ia Supernova Rate at z\raisebox{-0.5ex}\textasciitilde0.5 from the Supernova Legacy Survey}",
      journal = {\aj},
         year = 2006,
        month = sep,
       volume = {132},
       number = {3},
        pages = {1126-1145},
          doi = {10.1086/505532},
archivePrefix = {arXiv},
       eprint = {astro-ph/0605148},
 primaryClass = {astro-ph},
       adsurl = {https://ui.adsabs.harvard.edu/abs/2006AJ....132.1126N}
}

@ARTICLE{Nelson2019,
       author = {{Nelson}, Dylan and {Pillepich}, Annalisa and {Springel}, Volker and {Pakmor}, R{\"u}diger and {Weinberger}, Rainer and {Genel}, Shy and {Torrey}, Paul and {Vogelsberger}, Mark and {Marinacci}, Federico and {Hernquist}, Lars},
        title = "{First results from the TNG50 simulation: galactic outflows driven by supernovae and black hole feedback}",
      journal = {\mnras},
         year = 2019,
        month = dec,
       volume = {490},
       number = {3},
        pages = {3234-3261},
          doi = {10.1093/mnras/stz2306},
archivePrefix = {arXiv},
       eprint = {1902.05554},
 primaryClass = {astro-ph.GA},
       adsurl = {https://ui.adsabs.harvard.edu/abs/2019MNRAS.490.3234N}
}

@ARTICLE{Nelson2019TNGrelease,
       author = {{Nelson}, Dylan and {Springel}, Volker and {Pillepich}, Annalisa and {Rodriguez-Gomez}, Vicente and {Torrey}, Paul and {Genel}, Shy and {Vogelsberger}, Mark and {Pakmor}, Ruediger and {Marinacci}, Federico and {Weinberger}, Rainer and {Kelley}, Luke and {Lovell}, Mark and {Diemer}, Benedikt and {Hernquist}, Lars},
        title = "{The IllustrisTNG simulations: public data release}",
      journal = {Computational Astrophysics and Cosmology},
         year = 2019,
        month = may,
       volume = {6},
       number = {1},
          eid = {2},
        pages = {2},
          doi = {10.1186/s40668-019-0028-x},
archivePrefix = {arXiv},
       eprint = {1812.05609},
 primaryClass = {astro-ph.GA},
       adsurl = {https://ui.adsabs.harvard.edu/abs/2019ComAC...6....2N}
}

@ARTICLE{Neumayer2020,
       author = {{Neumayer}, Nadine and {Seth}, Anil and {B{\"o}ker}, Torsten},
        title = "{Nuclear star clusters}",
      journal = {\aapr},
         year = 2020,
        month = jul,
       volume = {28},
       number = {1},
          eid = {4},
        pages = {4},
          doi = {10.1007/s00159-020-00125-0},
archivePrefix = {arXiv},
       eprint = {2001.03626},
 primaryClass = {astro-ph.GA},
       adsurl = {https://ui.adsabs.harvard.edu/abs/2020A&ARv..28....4N}
}

@ARTICLE{Nobels2022,
       author = {{Nobels}, Folkert S.~J. and {Schaye}, Joop and {Schaller}, Matthieu and {Bah{\'e}}, Yannick M. and {Chaikin}, Evgenii},
        title = "{The interplay between AGN feedback and precipitation of the intracluster medium in simulations of galaxy groups and clusters}",
      journal = {\mnras},
         year = 2022,
        month = oct,
       volume = {515},
       number = {4},
        pages = {4838-4859},
          doi = {10.1093/mnras/stac2061},
archivePrefix = {arXiv},
       eprint = {2204.02205},
 primaryClass = {astro-ph.GA},
       adsurl = {https://ui.adsabs.harvard.edu/abs/2022MNRAS.515.4838N}
}

@ARTICLE{Nobels2024,
       author = {{Nobels}, Folkert S.~J. and {Schaye}, Joop and {Schaller}, Matthieu and {Ploeckinger}, Sylvia and {Chaikin}, Evgenii and {Richings}, Alexander J.},
        title = "{Tests of subgrid models for star formation using simulations of isolated disc galaxies}",
      journal = {\mnras},
         year = 2024,
        month = aug,
       volume = {532},
       number = {3},
        pages = {3299-3321},
          doi = {10.1093/mnras/stae1390},
archivePrefix = {arXiv},
       eprint = {2309.13750},
 primaryClass = {astro-ph.GA},
       adsurl = {https://ui.adsabs.harvard.edu/abs/2024MNRAS.532.3299N}
}

@ARTICLE{Nobels2025,
       author = {{Nobels}, Folkert S. J. and others},
        title = "{TBD}",
      journal = {\mnras},
         year = 2026,
       volume = {in prep.},
}

@ARTICLE{Nomoto2013,
       author = {{Nomoto}, Ken'ichi and {Kobayashi}, Chiaki and {Tominaga}, Nozomu},
        title = "{Nucleosynthesis in Stars and the Chemical Enrichment of Galaxies}",
      journal = {\araa},
         year = 2013,
        month = aug,
       volume = {51},
       number = {1},
        pages = {457-509},
          doi = {10.1146/annurev-astro-082812-140956},
       adsurl = {https://ui.adsabs.harvard.edu/abs/2013ARA&A..51..457N}
}

@ARTICLE{Novak2017,
       author = {{Novak}, M. and {Smol{\v{c}}i{\'c}}, V. and {Delhaize}, J. and {Delvecchio}, I. and {Zamorani}, G. and {Baran}, N. and {Bondi}, M. and {Capak}, P. and {Carilli}, C.~L. and {Ciliegi}, P. and {Civano}, F. and {Ilbert}, O. and {Karim}, A. and {Laigle}, C. and {Le F{\`e}vre}, O. and {Marchesi}, S. and {McCracken}, H. and {Miettinen}, O. and {Salvato}, M. and {Sargent}, M. and {Schinnerer}, E. and {Tasca}, L.},
        title = "{The VLA-COSMOS 3 GHz Large Project: Cosmic star formation history since z   5}",
      journal = {\aap},
         year = 2017,
        month = jun,
       volume = {602},
          eid = {A5},
        pages = {A5},
          doi = {10.1051/0004-6361/201629436},
archivePrefix = {arXiv},
       eprint = {1703.09724},
 primaryClass = {astro-ph.GA},
       adsurl = {https://ui.adsabs.harvard.edu/abs/2017A&A...602A...5N}
}

@INPROCEEDINGS{Novikov1973,
       author = {{Novikov}, I.~D. and {Thorne}, K.~S.},
        title = "{Astrophysics of black holes.}",
    booktitle = {Black Holes (Les Astres Occlus)},
         year = 1973,
       editor = {{Dewitt}, C. and {Dewitt}, B.~S.},
        month = jan,
        pages = {343-450},
       adsurl = {https://ui.adsabs.harvard.edu/abs/1973blho.conf..343N}
}

@ARTICLE{Ochsendorf2017,
       author = {{Ochsendorf}, Bram B. and {Meixner}, Margaret and {Roman-Duval}, Julia and {Rahman}, Mubdi and {Evans}, II, Neal J.},
        title = "{What Sets the Massive Star Formation Rates and Efficiencies of Giant Molecular Clouds?}",
      journal = {\apj},
         year = 2017,
        month = jun,
       volume = {841},
       number = {2},
          eid = {109},
        pages = {109},
          doi = {10.3847/1538-4357/aa704a},
archivePrefix = {arXiv},
       eprint = {1704.06965},
 primaryClass = {astro-ph.GA},
       adsurl = {https://ui.adsabs.harvard.edu/abs/2017ApJ...841..109O}
}

@ARTICLE{Oku2024,
       author = {{Oku}, Yuri and {Nagamine}, Kentaro},
        title = "{Osaka Feedback Model. III. Cosmological Simulation CROCODILE}",
      journal = {\apj},
         year = 2024,
        month = nov,
       volume = {975},
       number = {2},
          eid = {183},
        pages = {183},
          doi = {10.3847/1538-4357/ad77d3},
archivePrefix = {arXiv},
       eprint = {2401.06324},
 primaryClass = {astro-ph.GA},
       adsurl = {https://ui.adsabs.harvard.edu/abs/2024ApJ...975..183O}
}

@ARTICLE{Okumura2014,
       author = {{Okumura}, Jun E. and {Ihara}, Yutaka and {Doi}, Mamoru and {Morokuma}, Tomoki and {Pain}, Reynald and {Totani}, Tomonori and {Barbary}, Kyle and {Takanashi}, Naohiro and {Yasuda}, Naoki and {Aldering}, Greg and {Dawson}, Kyle and {Goldhaber}, Gerson and {Hook}, Isobel and {Lidman}, Chris and {Perlmutter}, Saul and {Spadafora}, Anthony and {Suzuki}, Nao and {Wang}, Lifan},
        title = "{The Type Ia supernovae rate with Subaru/XMM-Newton Deep Survey}",
      journal = {\pasj},
         year = 2014,
        month = apr,
       volume = {66},
       number = {2},
          eid = {49},
        pages = {49},
          doi = {10.1093/pasj/psu024},
archivePrefix = {arXiv},
       eprint = {1401.7701},
 primaryClass = {astro-ph.CO},
       adsurl = {https://ui.adsabs.harvard.edu/abs/2014PASJ...66...49O}
}

@ARTICLE{Pain2002,
       author = {{Pain}, R. and {Fabbro}, S. and {Sullivan}, M. and {Ellis}, R.~S. and {Aldering}, G. and {Astier}, P. and {Deustua}, S.~E. and {Fruchter}, A.~S. and {Goldhaber}, G. and {Goobar}, A. and {Groom}, D.~E. and {Hardin}, D. and {Hook}, I.~M. and {Howell}, D.~A. and {Irwin}, M.~J. and {Kim}, A.~G. and {Kim}, M.~Y. and {Knop}, R.~A. and {Lee}, J.~C. and {Lidman}, C. and {McMahon}, R.~G. and {Nugent}, P.~E. and {Panagia}, N. and {Pennypacker}, C.~R. and {Perlmutter}, S. and {Ruiz-Lapuente}, P. and {Schahmaneche}, K. and {Schaefer}, B. and {Walton}, N.~A.},
        title = "{The Distant Type Ia Supernova Rate}",
      journal = {\apj},
         year = 2002,
        month = sep,
       volume = {577},
       number = {1},
        pages = {120-132},
          doi = {10.1086/342129},
archivePrefix = {arXiv},
       eprint = {astro-ph/0205476},
 primaryClass = {astro-ph},
       adsurl = {https://ui.adsabs.harvard.edu/abs/2002ApJ...577..120P}
}

@ARTICLE{Pakmor2023,
       author = {{Pakmor}, R{\"u}diger and {Springel}, Volker and {Coles}, Jonathan P. and {Guillet}, Thomas and {Pfrommer}, Christoph and {Bose}, Sownak and {Barrera}, Monica and {Delgado}, Ana Maria and {Ferlito}, Fulvio and {Frenk}, Carlos and {Hadzhiyska}, Boryana and {Hern{\'a}ndez-Aguayo}, C{\'e}sar and {Hernquist}, Lars and {Kannan}, Rahul and {White}, Simon D.~M.},
        title = "{The MillenniumTNG Project: the hydrodynamical full physics simulation and a first look at its galaxy clusters}",
      journal = {\mnras},
         year = 2023,
        month = sep,
       volume = {524},
       number = {2},
        pages = {2539-2555},
          doi = {10.1093/mnras/stac3620},
archivePrefix = {arXiv},
       eprint = {2210.10060},
 primaryClass = {astro-ph.CO},
       adsurl = {https://ui.adsabs.harvard.edu/abs/2023MNRAS.524.2539P}
}

@ARTICLE{Peng2010,
       author = {{Peng}, Ying-jie and {Lilly}, Simon J. and {Kova{\v{c}}}, Katarina and {Bolzonella}, Micol and {Pozzetti}, Lucia and {Renzini}, Alvio and {Zamorani}, Gianni and {Ilbert}, Olivier and {Knobel}, Christian and {Iovino}, Angela and {Maier}, Christian and {Cucciati}, Olga and {Tasca}, Lidia and {Carollo}, C. Marcella and {Silverman}, John and {Kampczyk}, Pawel and {de Ravel}, Loic and {Sanders}, David and {Scoville}, Nicholas and {Contini}, Thierry and {Mainieri}, Vincenzo and {Scodeggio}, Marco and {Kneib}, Jean-Paul and {Le F{\`e}vre}, Olivier and {Bardelli}, Sandro and {Bongiorno}, Angela and {Caputi}, Karina and {Coppa}, Graziano and {de la Torre}, Sylvain and {Franzetti}, Paolo and {Garilli}, Bianca and {Lamareille}, Fabrice and {Le Borgne}, Jean-Francois and {Le Brun}, Vincent and {Mignoli}, Marco and {Perez Montero}, Enrique and {Pello}, Roser and {Ricciardelli}, Elena and {Tanaka}, Masayuki and {Tresse}, Laurence and {Vergani}, Daniela and {Welikala}, Niraj and {Zucca}, Elena and {Oesch}, Pascal and {Abbas}, Ummi and {Barnes}, Luke and {Bordoloi}, Rongmon and {Bottini}, Dario and {Cappi}, Alberto and {Cassata}, Paolo and {Cimatti}, Andrea and {Fumana}, Marco and {Hasinger}, Gunther and {Koekemoer}, Anton and {Leauthaud}, Alexei and {Maccagni}, Dario and {Marinoni}, Christian and {McCracken}, Henry and {Memeo}, Pierdomenico and {Meneux}, Baptiste and {Nair}, Preethi and {Porciani}, Cristiano and {Presotto}, Valentina and {Scaramella}, Roberto},
        title = "{Mass and Environment as Drivers of Galaxy Evolution in SDSS and zCOSMOS and the Origin of the Schechter Function}",
      journal = {\apj},
         year = 2010,
        month = sep,
       volume = {721},
       number = {1},
        pages = {193-221},
          doi = {10.1088/0004-637X/721/1/193},
archivePrefix = {arXiv},
       eprint = {1003.4747},
 primaryClass = {astro-ph.CO},
       adsurl = {https://ui.adsabs.harvard.edu/abs/2010ApJ...721..193P}
}

@ARTICLE{Perrett2012,
       author = {{Perrett}, K. and {Sullivan}, M. and {Conley}, A. and {Gonz{\'a}lez-Gait{\'a}n}, S. and {Carlberg}, R. and {Fouchez}, D. and {Ripoche}, P. and {Neill}, J.~D. and {Astier}, P. and {Balam}, D. and {Balland}, C. and {Basa}, S. and {Guy}, J. and {Hardin}, D. and {Hook}, I.~M. and {Howell}, D.~A. and {Pain}, R. and {Palanque-Delabrouille}, N. and {Pritchet}, C. and {Regnault}, N. and {Rich}, J. and {Ruhlmann-Kleider}, V. and {Baumont}, S. and {Lidman}, C. and {Perlmutter}, S. and {Walker}, E.~S.},
        title = "{Evolution in the Volumetric Type Ia Supernova Rate from the Supernova Legacy Survey}",
      journal = {\aj},
         year = 2012,
        month = aug,
       volume = {144},
       number = {2},
          eid = {59},
        pages = {59},
          doi = {10.1088/0004-6256/144/2/59},
archivePrefix = {arXiv},
       eprint = {1206.0665},
 primaryClass = {astro-ph.CO},
       adsurl = {https://ui.adsabs.harvard.edu/abs/2012AJ....144...59P}
}

@ARTICLE{Pettini2001,
       author = {{Pettini}, Max and {Bowen}, David V.},
        title = "{A New Measurement of the Primordial Abundance of Deuterium: Toward Convergence with the Baryon Density from the Cosmic Microwave Background?}",
      journal = {\apj},
         year = 2001,
        month = oct,
       volume = {560},
       number = {1},
        pages = {41-48},
          doi = {10.1086/322510},
archivePrefix = {arXiv},
       eprint = {astro-ph/0104474},
 primaryClass = {astro-ph},
       adsurl = {https://ui.adsabs.harvard.edu/abs/2001ApJ...560...41P}
}

@ARTICLE{Pillepich2018TNGmethod,
       author = {{Pillepich}, Annalisa and {Springel}, Volker and {Nelson}, Dylan and {Genel}, Shy and {Naiman}, Jill and {Pakmor}, R{\"u}diger and {Hernquist}, Lars and {Torrey}, Paul and {Vogelsberger}, Mark and {Weinberger}, Rainer and {Marinacci}, Federico},
        title = "{Simulating galaxy formation with the IllustrisTNG model}",
      journal = {\mnras},
         year = 2018,
        month = jan,
       volume = {473},
       number = {3},
        pages = {4077-4106},
          doi = {10.1093/mnras/stx2656},
archivePrefix = {arXiv},
       eprint = {1703.02970},
 primaryClass = {astro-ph.GA},
       adsurl = {https://ui.adsabs.harvard.edu/abs/2018MNRAS.473.4077P}
}

@ARTICLE{Pillepich2018TNG,
       author = {{Pillepich}, Annalisa and {Nelson}, Dylan and {Hernquist}, Lars and {Springel}, Volker and {Pakmor}, R{\"u}diger and {Torrey}, Paul and {Weinberger}, Rainer and {Genel}, Shy and {Naiman}, Jill P. and {Marinacci}, Federico and {Vogelsberger}, Mark},
        title = "{First results from the IllustrisTNG simulations: the stellar mass content of groups and clusters of galaxies}",
      journal = {\mnras},
         year = 2018,
        month = mar,
       volume = {475},
       number = {1},
        pages = {648-675},
          doi = {10.1093/mnras/stx3112},
archivePrefix = {arXiv},
       eprint = {1707.03406},
 primaryClass = {astro-ph.GA},
       adsurl = {https://ui.adsabs.harvard.edu/abs/2018MNRAS.475..648P}
}

@ARTICLE{Ploeckinger2020,
       author = {{Ploeckinger}, Sylvia and {Schaye}, Joop},
        title = "{Radiative cooling rates, ion fractions, molecule abundances, and line emissivities including self-shielding and both local and metagalactic radiation fields}",
      journal = {\mnras},
         year = 2020,
        month = oct,
       volume = {497},
       number = {4},
        pages = {4857-4883},
          doi = {10.1093/mnras/staa2172},
archivePrefix = {arXiv},
       eprint = {2006.14322},
 primaryClass = {astro-ph.GA},
       adsurl = {https://ui.adsabs.harvard.edu/abs/2020MNRAS.497.4857P}
}

@ARTICLE{Ploeckinger2024,
       author = {{Ploeckinger}, Sylvia and {Nobels}, Folkert S.~J. and {Schaller}, Matthieu and {Schaye}, Joop},
        title = "{Resolution criteria to avoid artificial clumping in Lagrangian hydrodynamic simulations with a multiphase interstellar medium}",
      journal = {\mnras},
         year = 2024,
        month = feb,
       volume = {528},
       number = {2},
        pages = {2930-2951},
          doi = {10.1093/mnras/stad3935},
archivePrefix = {arXiv},
       eprint = {2310.10721},
 primaryClass = {astro-ph.GA},
       adsurl = {https://ui.adsabs.harvard.edu/abs/2024MNRAS.528.2930P}
}

@ARTICLE{Ploeckinger2025,
       author = {{Ploeckinger}, Sylvia and {Richings}, Alexander J. and {Schaye}, Joop and {Trayford}, James W. and {Schaller}, Matthieu and {Chaikin}, Evgenii},
        title = "{HYBRID-CHIMES: a model for radiative cooling and the abundances of ions and molecules in simulations of galaxy formation}",
      journal = {\mnras},
         year = 2025,
        month = oct,
       volume = {543},
       number = {2},
        pages = {891-916},
          doi = {10.1093/mnras/staf1402},
archivePrefix = {arXiv},
       eprint = {2506.15773},
 primaryClass = {astro-ph.GA},
       adsurl = {https://ui.adsabs.harvard.edu/abs/2025MNRAS.543..891P}
}

@ARTICLE{Pokhrel2021,
       author = {{Pokhrel}, Riwaj and {Gutermuth}, Robert A. and {Krumholz}, Mark R. and {Federrath}, Christoph and {Heyer}, Mark and {Khullar}, Shivan and {Megeath}, S. Thomas and {Myers}, Philip C. and {Offner}, Stella S.~R. and {Pipher}, Judith L. and {Fischer}, William J. and {Henning}, Thomas and {Hora}, Joseph L.},
        title = "{The Single-cloud Star Formation Relation}",
      journal = {\apjl},
         year = 2021,
        month = may,
       volume = {912},
       number = {1},
          eid = {L19},
        pages = {L19},
          doi = {10.3847/2041-8213/abf564},
archivePrefix = {arXiv},
       eprint = {2104.04551},
 primaryClass = {astro-ph.GA},
       adsurl = {https://ui.adsabs.harvard.edu/abs/2021ApJ...912L..19P}
}

@ARTICLE{Portinari1998,
       author = {{Portinari}, L. and {Chiosi}, C. and {Bressan}, A.},
        title = "{Galactic chemical enrichment with new metallicity dependent stellar yields}",
      journal = {\aap},
         year = 1998,
        month = jun,
       volume = {334},
        pages = {505-539},
archivePrefix = {arXiv},
       eprint = {astro-ph/9711337},
 primaryClass = {astro-ph},
       adsurl = {https://ui.adsabs.harvard.edu/abs/1998A&A...334..505P}
}

@ARTICLE{Poznanski2007,
       author = {{Poznanski}, D. and {Maoz}, D. and {Yasuda}, N. and {Foley}, R.~J. and {Doi}, M. and {Filippenko}, A.~V. and {Fukugita}, M. and {Gal-Yam}, A. and {Jannuzi}, B.~T. and {Morokuma}, T. and {Oda}, T. and {Schweiker}, H. and {Sharon}, K. and {Silverman}, J.~M. and {Totani}, T.},
        title = "{Supernovae in the Subaru Deep Field: an initial sample and Type Ia rate out to redshift 1.6}",
      journal = {\mnras},
         year = 2007,
        month = nov,
       volume = {382},
       number = {3},
        pages = {1169-1186},
          doi = {10.1111/j.1365-2966.2007.12424.x},
archivePrefix = {arXiv},
       eprint = {0707.0393},
 primaryClass = {astro-ph},
       adsurl = {https://ui.adsabs.harvard.edu/abs/2007MNRAS.382.1169P}
}

@ARTICLE{Puchwein2015,
       author = {{Puchwein}, Ewald and {Bolton}, James S. and {Haehnelt}, Martin G. and {Madau}, Piero and {Becker}, George D. and {Haardt}, Francesco},
        title = "{The photoheating of the intergalactic medium in synthesis models of the UV background}",
      journal = {\mnras},
         year = 2015,
        month = jul,
       volume = {450},
       number = {4},
        pages = {4081-4097},
          doi = {10.1093/mnras/stv773},
archivePrefix = {arXiv},
       eprint = {1410.1531},
 primaryClass = {astro-ph.CO},
       adsurl = {https://ui.adsabs.harvard.edu/abs/2015MNRAS.450.4081P}
}

@ARTICLE{Quimby2012,
       author = {{Quimby}, Robert M. and {Yuan}, Fang and {Akerlof}, Carl and {Wheeler}, J. Craig and {Warren}, Michael S.},
        title = "{On the Rates of Type Ia Supernovae in Dwarf and Giant Hosts with ROTSE-IIIb}",
      journal = {\aj},
         year = 2012,
        month = dec,
       volume = {144},
       number = {6},
          eid = {177},
        pages = {177},
          doi = {10.1088/0004-6256/144/6/177},
archivePrefix = {arXiv},
       eprint = {1210.3428},
 primaryClass = {astro-ph.CO},
       adsurl = {https://ui.adsabs.harvard.edu/abs/2012AJ....144..177Q}
}

@ARTICLE{Rahmati2013,
       author = {{Rahmati}, Alireza and {Pawlik}, Andreas H. and {Rai{\v{c}}evi{\'c}}, Milan and {Schaye}, Joop},
        title = "{On the evolution of the H I column density distribution in cosmological simulations}",
      journal = {\mnras},
         year = 2013,
        month = apr,
       volume = {430},
       number = {3},
        pages = {2427-2445},
          doi = {10.1093/mnras/stt066},
archivePrefix = {arXiv},
       eprint = {1210.7808},
 primaryClass = {astro-ph.CO},
       adsurl = {https://ui.adsabs.harvard.edu/abs/2013MNRAS.430.2427R}
}

@ARTICLE{Regan2024,
       author = {{Regan}, John and {Volonteri}, Marta},
        title = "{Massive Black Hole Seeds}",
      journal = {The Open Journal of Astrophysics},
         year = 2024,
        month = sep,
       volume = {7},
          eid = {72},
        pages = {72},
          doi = {10.33232/001c.123239},
archivePrefix = {arXiv},
       eprint = {2405.17975},
 primaryClass = {astro-ph.GA},
       adsurl = {https://ui.adsabs.harvard.edu/abs/2024OJAp....7E..72R}
}

@ARTICLE{Relano2020,
       author = {{Rela{\~n}o}, M. and {Lisenfeld}, U. and {Hou}, K. -C. and {De Looze}, I. and {V{\'\i}lchez}, J.~M. and {Kennicutt}, R.~C.},
        title = "{Evolution of grain size distribution in galactic discs}",
      journal = {\aap},
         year = 2020,
        month = apr,
       volume = {636},
          eid = {A18},
        pages = {A18},
          doi = {10.1051/0004-6361/201937087},
archivePrefix = {arXiv},
       eprint = {2002.01945},
 primaryClass = {astro-ph.GA},
       adsurl = {https://ui.adsabs.harvard.edu/abs/2020A&A...636A..18R}
}

@ARTICLE{Relano2022,
       author = {{Rela{\~n}o}, M. and {De Looze}, I. and {Saintonge}, A. and {Hou}, K. -C. and {Romano}, L.~E.~C. and {Nagamine}, K. and {Hirashita}, H. and {Aoyama}, S. and {Lamperti}, I. and {Lisenfeld}, U. and {Smith}, M.~W.~L. and {Chastenet}, J. and {Xiao}, T. and {Gao}, Y. and {Sargent}, M. and {van der Giessen}, S.~A.},
        title = "{Dust grain size evolution in local galaxies: a comparison between observations and simulations}",
      journal = {\mnras},
         year = 2022,
        month = oct,
       volume = {515},
       number = {4},
        pages = {5306-5334},
          doi = {10.1093/mnras/stac2108},
archivePrefix = {arXiv},
       eprint = {2207.13196},
 primaryClass = {astro-ph.GA},
       adsurl = {https://ui.adsabs.harvard.edu/abs/2022MNRAS.515.5306R}
}

@ARTICLE{Rey2025,
       author = {{Rey}, Martin P. and {Taylor}, Ethan and {Gray}, Emily I. and {Kim}, Stacy Y. and {Andersson}, Eric P. and {Pontzen}, Andrew and {Agertz}, Oscar and {Read}, Justin I. and {Cadiou}, Corentin and {Yates}, Robert M. and {Orkney}, Matthew D.~A. and {Scholte}, Dirk and {Saintonge}, Am{\'e}lie and {Breneman}, Joseph and {McQuinn}, Kristen B.~W. and {Muni}, Claudia and {Das}, Payel},
        title = "{EDGE: The emergence of dwarf galaxy scaling relations from cosmological radiation-hydrodynamics simulations}",
      journal = {arXiv e-prints},
         year = 2025,
        month = mar,
          eid = {arXiv:2503.03813},
        pages = {arXiv:2503.03813},
          doi = {10.48550/arXiv.2503.03813},
archivePrefix = {arXiv},
       eprint = {2503.03813},
 primaryClass = {astro-ph.GA},
       adsurl = {https://ui.adsabs.harvard.edu/abs/2025arXiv250303813R}
}

@ARTICLE{Richings2014a,
       author = {{Richings}, A.~J. and {Schaye}, J. and {Oppenheimer}, B.~D.},
        title = "{Non-equilibrium chemistry and cooling in the diffuse interstellar medium - I. Optically thin regime}",
      journal = {\mnras},
         year = 2014,
        month = jun,
       volume = {440},
       number = {4},
        pages = {3349-3369},
          doi = {10.1093/mnras/stu525},
archivePrefix = {arXiv},
       eprint = {1401.4719},
 primaryClass = {astro-ph.GA},
       adsurl = {https://ui.adsabs.harvard.edu/abs/2014MNRAS.440.3349R}
}

@ARTICLE{Richings2014b,
       author = {{Richings}, A.~J. and {Schaye}, J. and {Oppenheimer}, B.~D.},
        title = "{Non-equilibrium chemistry and cooling in the diffuse interstellar medium - II. Shielded gas}",
      journal = {\mnras},
         year = 2014,
        month = aug,
       volume = {442},
       number = {3},
        pages = {2780-2796},
          doi = {10.1093/mnras/stu1046},
archivePrefix = {arXiv},
       eprint = {1403.6155},
 primaryClass = {astro-ph.GA},
       adsurl = {https://ui.adsabs.harvard.edu/abs/2014MNRAS.442.2780R}
}

@ARTICLE{Richings2022,
       author = {{Richings}, Alexander J. and {Faucher-Gigu{\`e}re}, Claude-Andr{\'e} and {Gurvich}, Alexander B. and {Schaye}, Joop and {Hayward}, Christopher C.},
        title = "{The effects of local stellar radiation and dust depletion on non-equilibrium interstellar chemistry}",
      journal = {\mnras},
         year = 2022,
        month = dec,
       volume = {517},
       number = {2},
        pages = {1557-1583},
          doi = {10.1093/mnras/stac2338},
archivePrefix = {arXiv},
       eprint = {2208.02288},
 primaryClass = {astro-ph.GA},
       adsurl = {https://ui.adsabs.harvard.edu/abs/2022MNRAS.517.1557R}
}

@ARTICLE{Robertson2008,
       author = {{Robertson}, Brant E. and {Kravtsov}, Andrey V.},
        title = "{Molecular Hydrogen and Global Star Formation Relations in Galaxies}",
      journal = {\apj},
         year = 2008,
        month = jun,
       volume = {680},
       number = {2},
        pages = {1083-1111},
          doi = {10.1086/587796},
archivePrefix = {arXiv},
       eprint = {0710.2102},
 primaryClass = {astro-ph},
       adsurl = {https://ui.adsabs.harvard.edu/abs/2008ApJ...680.1083R}
}

@ARTICLE{Rodney2010,
       author = {{Rodney}, Steven A. and {Tonry}, John L.},
        title = "{Revised Supernova Rates from the IfA Deep Survey}",
      journal = {\apj},
         year = 2010,
        month = nov,
       volume = {723},
       number = {1},
        pages = {47-53},
          doi = {10.1088/0004-637X/723/1/47},
archivePrefix = {arXiv},
       eprint = {1008.4573},
 primaryClass = {astro-ph.CO},
       adsurl = {https://ui.adsabs.harvard.edu/abs/2010ApJ...723...47R}
}

@ARTICLE{Rodney2014,
       author = {{Rodney}, Steven A. and {Riess}, Adam G. and {Strolger}, Louis-Gregory and {Dahlen}, Tomas and {Graur}, Or and {Casertano}, Stefano and {Dickinson}, Mark E. and {Ferguson}, Henry C. and {Garnavich}, Peter and {Hayden}, Brian and {Jha}, Saurabh W. and {Jones}, David O. and {Kirshner}, Robert P. and {Koekemoer}, Anton M. and {McCully}, Curtis and {Mobasher}, Bahram and {Patel}, Brandon and {Weiner}, Benjamin J. and {Cenko}, S. Bradley and {Clubb}, Kelsey I. and {Cooper}, Michael and {Filippenko}, Alexei V. and {Frederiksen}, Teddy F. and {Hjorth}, Jens and {Leibundgut}, Bruno and {Matheson}, Thomas and {Nayyeri}, Hooshang and {Penner}, Kyle and {Trump}, Jonathan and {Silverman}, Jeffrey M. and {U}, Vivian and {Azalee Bostroem}, K. and {Challis}, Peter and {Rajan}, Abhijith and {Wolff}, Schuyler and {Faber}, S.~M. and {Grogin}, Norman A. and {Kocevski}, Dale},
        title = "{Type Ia Supernova Rate Measurements to Redshift 2.5 from CANDELS: Searching for Prompt Explosions in the Early Universe}",
      journal = {\aj},
         year = 2014,
        month = jul,
       volume = {148},
       number = {1},
          eid = {13},
        pages = {13},
          doi = {10.1088/0004-6256/148/1/13},
archivePrefix = {arXiv},
       eprint = {1401.7978},
 primaryClass = {astro-ph.CO},
       adsurl = {https://ui.adsabs.harvard.edu/abs/2014AJ....148...13R}
}

@ARTICLE{Romano2010,
       author = {{Romano}, D. and {Karakas}, A.~I. and {Tosi}, M. and {Matteucci}, F.},
        title = "{Quantifying the uncertainties of chemical evolution studies. II. Stellar yields}",
      journal = {\aap},
         year = 2010,
        month = nov,
       volume = {522},
          eid = {A32},
        pages = {A32},
          doi = {10.1051/0004-6361/201014483},
archivePrefix = {arXiv},
       eprint = {1006.5863},
 primaryClass = {astro-ph.GA},
       adsurl = {https://ui.adsabs.harvard.edu/abs/2010A&A...522A..32R}
}

@ARTICLE{Rosas-Guevara2015,
       author = {{Rosas-Guevara}, Y.~M. and {Bower}, R.~G. and {Schaye}, J. and {Furlong}, M. and {Frenk}, C.~S. and {Booth}, C.~M. and {Crain}, R.~A. and {Dalla Vecchia}, C. and {Schaller}, M. and {Theuns}, T.},
        title = "{The impact of angular momentum on black hole accretion rates in simulations of galaxy formation}",
      journal = {\mnras},
         year = 2015,
        month = nov,
       volume = {454},
       number = {1},
        pages = {1038-1057},
          doi = {10.1093/mnras/stv2056},
archivePrefix = {arXiv},
       eprint = {1312.0598},
 primaryClass = {astro-ph.CO},
       adsurl = {https://ui.adsabs.harvard.edu/abs/2015MNRAS.454.1038R}
}

@ARTICLE{Rosdahl2015,
       author = {{Rosdahl}, Joakim and {Schaye}, Joop and {Teyssier}, Romain and {Agertz}, Oscar},
        title = "{Galaxies that shine: radiation-hydrodynamical simulations of disc galaxies}",
      journal = {\mnras},
         year = 2015,
        month = jul,
       volume = {451},
       number = {1},
        pages = {34-58},
          doi = {10.1093/mnras/stv937},
archivePrefix = {arXiv},
       eprint = {1501.04632},
 primaryClass = {astro-ph.GA},
       adsurl = {https://ui.adsabs.harvard.edu/abs/2015MNRAS.451...34R}
}

@ARTICLE{Ruffert1994,
       author = {{Ruffert}, Maximilian and {Arnett}, David},
        title = "{Three-dimensional Hydrodynamic Bondi-Hoyle Accretion. II. Homogeneous Medium at Mach 3 with gamma = 5/3}",
      journal = {\apj},
         year = 1994,
        month = may,
       volume = {427},
        pages = {351},
          doi = {10.1086/174145},
       adsurl = {https://ui.adsabs.harvard.edu/abs/1994ApJ...427..351R}
}

@ARTICLE{Salas2021,
       author = {{\VAN{De}{de}{de} Salas}, P.~F. and {Forero}, D.~V. and {Gariazzo}, S. and {Mart{\'\i}nez-Mirav{\'e}}, P. and {Mena}, O. and {Ternes}, C.~A. and {T{\'o}rtola}, M. and {Valle}, J.~W.~F.},
        title = "{2020 global reassessment of the neutrino oscillation picture}",
      journal = {Journal of High Energy Physics},
         year = 2021,
        month = feb,
       volume = {2021},
       number = {2},
          eid = {71},
        pages = {71},
          doi = {10.1007/JHEP02(2021)071},
archivePrefix = {arXiv},
       eprint = {2006.11237},
 primaryClass = {hep-ph},
       adsurl = {https://ui.adsabs.harvard.edu/abs/2021JHEP...02..071D}
}

@ARTICLE{Sales2010,
       author = {{Sales}, Laura V. and {Navarro}, Julio F. and {Schaye}, Joop and {Dalla Vecchia}, Claudio and {Springel}, Volker and {Booth}, C.~M.},
        title = "{Feedback and the structure of simulated galaxies at redshift z= 2}",
      journal = {\mnras},
         year = 2010,
        month = dec,
       volume = {409},
       number = {4},
        pages = {1541-1556},
          doi = {10.1111/j.1365-2966.2010.17391.x},
archivePrefix = {arXiv},
       eprint = {1004.5386},
 primaryClass = {astro-ph.CO},
       adsurl = {https://ui.adsabs.harvard.edu/abs/2010MNRAS.409.1541S}
}

@ARTICLE{Sawala2016,
       author = {{Sawala}, Till and {Frenk}, Carlos S. and {Fattahi}, Azadeh and {Navarro}, Julio F. and {Bower}, Richard G. and {Crain}, Robert A. and {Dalla Vecchia}, Claudio and {Furlong}, Michelle and {Helly}, John. C. and {Jenkins}, Adrian and {Oman}, Kyle A. and {Schaller}, Matthieu and {Schaye}, Joop and {Theuns}, Tom and {Trayford}, James and {White}, Simon D.~M.},
        title = "{The APOSTLE simulations: solutions to the Local Group's cosmic puzzles}",
      journal = {\mnras},
         year = 2016,
        month = apr,
       volume = {457},
       number = {2},
        pages = {1931-1943},
          doi = {10.1093/mnras/stw145},
archivePrefix = {arXiv},
       eprint = {1511.01098},
 primaryClass = {astro-ph.GA},
       adsurl = {https://ui.adsabs.harvard.edu/abs/2016MNRAS.457.1931S}
}

@ARTICLE{Scannapieco2012,
       author = {{Scannapieco}, C. and {Wadepuhl}, M. and {Parry}, O.~H. and {Navarro}, J.~F. and {Jenkins}, A. and {Springel}, V. and {Teyssier}, R. and {Carlson}, E. and {Couchman}, H.~M.~P. and {Crain}, R.~A. and {Dalla Vecchia}, C. and {Frenk}, C.~S. and {Kobayashi}, C. and {Monaco}, P. and {Murante}, G. and {Okamoto}, T. and {Quinn}, T. and {Schaye}, J. and {Stinson}, G.~S. and {Theuns}, T. and {Wadsley}, J. and {White}, S.~D.~M. and {Woods}, R.},
        title = "{The Aquila comparison project: the effects of feedback and numerical methods on simulations of galaxy formation}",
      journal = {\mnras},
         year = 2012,
        month = jun,
       volume = {423},
       number = {2},
        pages = {1726-1749},
          doi = {10.1111/j.1365-2966.2012.20993.x},
archivePrefix = {arXiv},
       eprint = {1112.0315},
 primaryClass = {astro-ph.GA},
       adsurl = {https://ui.adsabs.harvard.edu/abs/2012MNRAS.423.1726S}
}

@ARTICLE{Schaller2015,
       author = {{Schaller}, Matthieu and {Frenk}, Carlos S. and {Bower}, Richard G. and {Theuns}, Tom and {Jenkins}, Adrian and {Schaye}, Joop and {Crain}, Robert A. and {Furlong}, Michelle and {Dalla Vecchia}, Claudio and {McCarthy}, I.~G.},
        title = "{Baryon effects on the internal structure of {\ensuremath{\Lambda}}CDM haloes in the EAGLE simulations}",
      journal = {\mnras},
         year = 2015,
        month = aug,
       volume = {451},
       number = {2},
        pages = {1247-1267},
          doi = {10.1093/mnras/stv1067},
archivePrefix = {arXiv},
       eprint = {1409.8617},
 primaryClass = {astro-ph.CO},
       adsurl = {https://ui.adsabs.harvard.edu/abs/2015MNRAS.451.1247S}
}

@ARTICLE{Schaller2015hydro,
       author = {{Schaller}, Matthieu and {Dalla Vecchia}, Claudio and {Schaye}, Joop and {Bower}, Richard G. and {Theuns}, Tom and {Crain}, Robert A. and {Furlong}, Michelle and {McCarthy}, Ian G.},
        title = "{The EAGLE simulations of galaxy formation: the importance of the hydrodynamics scheme}",
      journal = {\mnras},
         year = 2015,
        month = dec,
       volume = {454},
       number = {3},
        pages = {2277-2291},
          doi = {10.1093/mnras/stv2169},
archivePrefix = {arXiv},
       eprint = {1509.05056},
 primaryClass = {astro-ph.GA},
       adsurl = {https://ui.adsabs.harvard.edu/abs/2015MNRAS.454.2277S}
}

@ARTICLE{Schaller2024swift,
       author = {{Schaller}, Matthieu and {Borrow}, Josh and {Draper}, Peter W. and {Ivkovic}, Mladen and {McAlpine}, Stuart and {Vandenbroucke}, Bert and {Bah{\'e}}, Yannick and {Chaikin}, Evgenii and {Chalk}, Aidan B.~G. and {Chan}, Tsang Keung and {Correa}, Camila and {van Daalen}, Marcel and {Elbers}, Willem and {Gonnet}, Pedro and {Hausammann}, Lo{\"\i}c and {Helly}, John and {Hu{\v{s}}ko}, Filip and {Kegerreis}, Jacob A. and {Nobels}, Folkert S.~J. and {Ploeckinger}, Sylvia and {Revaz}, Yves and {Roper}, William J. and {Ruiz-Bonilla}, Sergio and {Sandnes}, Thomas D. and {Uyttenhove}, Yolan and {Willis}, James S. and {Xiang}, Zhen},
        title = "{SWIFT: A modern highly-parallel gravity and smoothed particle hydrodynamics solver for astrophysical and cosmological applications}",
      journal = {\mnras},
         year = 2024,
        month = may,
       volume = {530},
       number = {2},
        pages = {2378-2419},
          doi = {10.1093/mnras/stae922},
archivePrefix = {arXiv},
       eprint = {2305.13380},
 primaryClass = {astro-ph.IM},
       adsurl = {https://ui.adsabs.harvard.edu/abs/2024MNRAS.530.2378S}
}

@ARTICLE{Schaye2001maxNHI,
       author = {{Schaye}, Joop},
        title = "{A Physical Upper Limit on the H I Column Density of Gas Clouds}",
      journal = {\apjl},
         year = 2001,
        month = nov,
       volume = {562},
       number = {1},
        pages = {L95-L98},
          doi = {10.1086/338106},
archivePrefix = {arXiv},
       eprint = {astro-ph/0109280},
 primaryClass = {astro-ph},
       adsurl = {https://ui.adsabs.harvard.edu/abs/2001ApJ...562L..95S}
}

@ARTICLE{Schaye2004,
       author = {{Schaye}, Joop},
        title = "{Star Formation Thresholds and Galaxy Edges: Why and Where}",
      journal = {\apj},
         year = 2004,
        month = jul,
       volume = {609},
       number = {2},
        pages = {667-682},
          doi = {10.1086/421232},
archivePrefix = {arXiv},
       eprint = {astro-ph/0205125},
 primaryClass = {astro-ph},
       adsurl = {https://ui.adsabs.harvard.edu/abs/2004ApJ...609..667S}
}

@ARTICLE{Schaye2008,
       author = {{Schaye}, Joop and {Dalla Vecchia}, Claudio},
        title = "{On the relation between the Schmidt and Kennicutt-Schmidt star formation laws and its implications for numerical simulations}",
      journal = {\mnras},
         year = 2008,
        month = jan,
       volume = {383},
       number = {3},
        pages = {1210-1222},
          doi = {10.1111/j.1365-2966.2007.12639.x},
archivePrefix = {arXiv},
       eprint = {0709.0292},
 primaryClass = {astro-ph},
       adsurl = {https://ui.adsabs.harvard.edu/abs/2008MNRAS.383.1210S}
}

@ARTICLE{Schaye2010,
       author = {{Schaye}, Joop and {Dalla Vecchia}, Claudio and {Booth}, C.~M. and {Wiersma}, Robert P.~C. and {Theuns}, Tom and {Haas}, Marcel R. and {Bertone}, Serena and {Duffy}, Alan R. and {McCarthy}, I.~G. and {van de Voort}, Freeke},
        title = "{The physics driving the cosmic star formation history}",
      journal = {\mnras},
         year = 2010,
        month = mar,
       volume = {402},
       number = {3},
        pages = {1536-1560},
          doi = {10.1111/j.1365-2966.2009.16029.x},
archivePrefix = {arXiv},
       eprint = {0909.5196},
 primaryClass = {astro-ph.CO},
       adsurl = {https://ui.adsabs.harvard.edu/abs/2010MNRAS.402.1536S}
}

@ARTICLE{Schaye2015,
       author = {{Schaye}, Joop and {Crain}, Robert A. and {Bower}, Richard G. and {Furlong}, Michelle and {Schaller}, Matthieu and {Theuns}, Tom and {Dalla Vecchia}, Claudio and {Frenk}, Carlos S. and {McCarthy}, I.~G. and {Helly}, John C. and {Jenkins}, Adrian and {Rosas-Guevara}, Y.~M. and {White}, Simon D.~M. and {Baes}, Maarten and {Booth}, C.~M. and {Camps}, Peter and {Navarro}, Julio F. and {Qu}, Yan and {Rahmati}, Alireza and {Sawala}, Till and {Thomas}, Peter A. and {Trayford}, James},
        title = "{The EAGLE project: simulating the evolution and assembly of galaxies and their environments}",
      journal = {\mnras},
         year = 2015,
        month = jan,
       volume = {446},
       number = {1},
        pages = {521-554},
          doi = {10.1093/mnras/stu2058},
archivePrefix = {arXiv},
       eprint = {1407.7040},
 primaryClass = {astro-ph.GA},
       adsurl = {https://ui.adsabs.harvard.edu/abs/2015MNRAS.446..521S}
}

@ARTICLE{Schaye2023,
       author = {{Schaye}, Joop and {Kugel}, Roi and {Schaller}, Matthieu and {Helly}, John C. and {Braspenning}, Joey and {Elbers}, Willem and {McCarthy}, Ian G. and {van Daalen}, Marcel P. and {Vandenbroucke}, Bert and {Frenk}, Carlos S. and {Kwan}, Juliana and {Salcido}, Jaime and {Bah{\'e}}, Yannick M. and {Borrow}, Josh and {Chaikin}, Evgenii and {Hahn}, Oliver and {Hu{\v{s}}ko}, Filip and {Jenkins}, Adrian and {Lacey}, Cedric G. and {Nobels}, Folkert S.~J.},
        title = "{The FLAMINGO project: cosmological hydrodynamical simulations for large-scale structure and galaxy cluster surveys}",
      journal = {\mnras},
         year = 2023,
        month = dec,
       volume = {526},
       number = {4},
        pages = {4978-5020},
          doi = {10.1093/mnras/stad2419},
archivePrefix = {arXiv},
       eprint = {2306.04024},
 primaryClass = {astro-ph.CO},
       adsurl = {https://ui.adsabs.harvard.edu/abs/2023MNRAS.526.4978S}
}

@ARTICLE{Schmidt1959,
       author = {{Schmidt}, Maarten},
        title = "{The Rate of Star Formation.}",
      journal = {\apj},
         year = 1959,
        month = mar,
       volume = {129},
        pages = {243},
          doi = {10.1086/146614},
       adsurl = {https://ui.adsabs.harvard.edu/abs/1959ApJ...129..243S}
}

@ARTICLE{Scholte2024,
       author = {{Scholte}, Dirk and {Saintonge}, Am{\'e}lie and {Moustakas}, John and {Catinella}, Barbara and {Zou}, Hu and {Dey}, Biprateep and {Aguilar}, J. and {Ahlen}, S. and {Anand}, A. and {Blum}, R. and {Brooks}, D. and {Circosta}, C. and {Claybaugh}, T. and {de la Macorra}, A. and {Doel}, P. and {Font-Ribera}, A. and {F{\"o}rster}, P.~U. and {Forero-Romero}, J.~E. and {Gazta{\~n}aga}, E. and {Gontcho A Gontcho}, S. and {Juneau}, S. and {Kehoe}, R. and {Kisner}, T. and {Koposov}, S.~E. and {Kremin}, A. and {Lambert}, A. and {Landriau}, M. and {Maraston}, C. and {Martini}, P. and {Meisner}, A. and {Mighty}, A.~S. and {Miquel}, R. and {Myers}, A.~D. and {Nie}, J. and {Poppett}, C. and {Prada}, F. and {Rezaie}, M. and {Rossi}, G. and {Sanchez}, E. and {Schubnell}, M. and {Silber}, J. and {Sprayberry}, D. and {Siudek}, M. and {Speranza}, F. and {Tarl{\'e}}, G. and {Tojeiro}, R. and {Weaver}, B.~A.},
        title = "{The atomic gas sequence and mass-metallicity relation from dwarfs to massive galaxies}",
      journal = {\mnras},
         year = 2024,
        month = dec,
       volume = {535},
       number = {3},
        pages = {2341-2356},
          doi = {10.1093/mnras/stae2477},
archivePrefix = {arXiv},
       eprint = {2408.03996},
 primaryClass = {astro-ph.GA},
       adsurl = {https://ui.adsabs.harvard.edu/abs/2024MNRAS.535.2341S}
}

@ARTICLE{Segers2016,
       author = {{Segers}, Marijke C. and {Crain}, Robert A. and {Schaye}, Joop and {Bower}, Richard G. and {Furlong}, Michelle and {Schaller}, Matthieu and {Theuns}, Tom},
        title = "{Recycled stellar ejecta as fuel for star formation and implications for the origin of the galaxy mass-metallicity relation}",
      journal = {\mnras},
         year = 2016,
        month = feb,
       volume = {456},
       number = {2},
        pages = {1235-1258},
          doi = {10.1093/mnras/stv2562},
archivePrefix = {arXiv},
       eprint = {1507.08281},
 primaryClass = {astro-ph.GA},
       adsurl = {https://ui.adsabs.harvard.edu/abs/2016MNRAS.456.1235S}
}

@ARTICLE{Segers2016alpha,
       author = {{Segers}, Marijke C. and {Schaye}, Joop and {Bower}, Richard G. and {Crain}, Robert A. and {Schaller}, Matthieu and {Theuns}, Tom},
        title = "{The origin of the {\ensuremath{\alpha}}-enhancement of massive galaxies}",
      journal = {\mnras},
         year = 2016,
        month = sep,
       volume = {461},
       number = {1},
        pages = {L102-L106},
          doi = {10.1093/mnrasl/slw111},
archivePrefix = {arXiv},
       eprint = {1603.05653},
 primaryClass = {astro-ph.GA},
       adsurl = {https://ui.adsabs.harvard.edu/abs/2016MNRAS.461L.102S}
}

@ARTICLE{Sellwood2013,
       author = {{Sellwood}, J.~A.},
        title = "{Relaxation in N-body Simulations of Disk Galaxies}",
      journal = {\apjl},
         year = 2013,
        month = jun,
       volume = {769},
       number = {2},
          eid = {L24},
        pages = {L24},
          doi = {10.1088/2041-8205/769/2/L24},
archivePrefix = {arXiv},
       eprint = {1303.4919},
 primaryClass = {astro-ph.CO},
       adsurl = {https://ui.adsabs.harvard.edu/abs/2013ApJ...769L..24S}
}

@ARTICLE{Sellwood2015,
       author = {{Sellwood}, J.~A.},
        title = "{Relaxation in N-body simulations of spherical systems}",
      journal = {\mnras},
         year = 2015,
        month = nov,
       volume = {453},
       number = {3},
        pages = {2919-2926},
          doi = {10.1093/mnras/stv1846},
archivePrefix = {arXiv},
       eprint = {1504.06500},
 primaryClass = {astro-ph.GA},
       adsurl = {https://ui.adsabs.harvard.edu/abs/2015MNRAS.453.2919S}
}

@ARTICLE{Semenov2017,
       author = {{Semenov}, Vadim A. and {Kravtsov}, Andrey V. and {Gnedin}, Nickolay Y.},
        title = "{The Physical Origin of Long Gas Depletion Times in Galaxies}",
      journal = {\apj},
         year = 2017,
        month = aug,
       volume = {845},
       number = {2},
          eid = {133},
        pages = {133},
          doi = {10.3847/1538-4357/aa8096},
archivePrefix = {arXiv},
       eprint = {1704.04239},
 primaryClass = {astro-ph.GA},
       adsurl = {https://ui.adsabs.harvard.edu/abs/2017ApJ...845..133S}
}

@ARTICLE{Shakura1973,
       author = {{Shakura}, N.~I. and {Sunyaev}, R.~A.},
        title = "{Black holes in binary systems. Observational appearance.}",
      journal = {\aap},
         year = 1973,
        month = jan,
       volume = {24},
        pages = {337-355},
       adsurl = {https://ui.adsabs.harvard.edu/abs/1973A&A....24..337S}
}

@ARTICLE{Sharma2014,
       author = {{Sharma}, Prateek and {Roy}, Arpita and {Nath}, Biman B. and {Shchekinov}, Yuri},
        title = "{In a hot bubble: why does superbubble feedback work, but isolated supernovae do not?}",
      journal = {\mnras},
         year = 2014,
        month = oct,
       volume = {443},
       number = {4},
        pages = {3463-3476},
          doi = {10.1093/mnras/stu1307},
archivePrefix = {arXiv},
       eprint = {1402.6695},
 primaryClass = {astro-ph.GA},
       adsurl = {https://ui.adsabs.harvard.edu/abs/2014MNRAS.443.3463S}
}

@ARTICLE{Shen2010,
       author = {{Shen}, S. and {Wadsley}, J. and {Stinson}, G.},
        title = "{The enrichment of the intergalactic medium with adiabatic feedback - I. Metal cooling and metal diffusion}",
      journal = {\mnras},
         year = 2010,
        month = sep,
       volume = {407},
       number = {3},
        pages = {1581-1596},
          doi = {10.1111/j.1365-2966.2010.17047.x},
archivePrefix = {arXiv},
       eprint = {0910.5956},
 primaryClass = {astro-ph.CO},
       adsurl = {https://ui.adsabs.harvard.edu/abs/2010MNRAS.407.1581S}
}

@ARTICLE{Shuntov2025,
       author = {{Shuntov}, M. and {Ilbert}, O. and {Toft}, S. and {Arango-Toro}, R.~C. and {Akins}, H.~B. and {Casey}, C.~M. and {Franco}, M. and {Harish}, S. and {Kartaltepe}, J.~S. and {Koekemoer}, A.~M. and {McCracken}, H.~J. and {Paquereau}, L. and {Laigle}, C. and {Bethermin}, M. and {Dubois}, Y. and {Drakos}, N.~E. and {Faisst}, A. and {Gozaliasl}, G. and {Gillman}, S. and {Hayward}, C.~C. and {Hirschmann}, M. and {Huertas-Company}, M. and {Jespersen}, C.~K. and {Jin}, S. and {Kokorev}, V. and {Lambrides}, E. and {Le Borgne}, D. and {Liu}, D. and {Magdis}, G. and {Massey}, R. and {McPartland}, C.~J.~R. and {Mercier}, W. and {McCleary}, J.~E. and {McKinney}, J. and {Oesch}, P.~A. and {Renzini}, A. and {Rhodes}, J.~D. and {Rich}, R.~M. and {Robertson}, B.~E. and {Sanders}, D. and {Trebitsch}, M. and {Tresse}, L. and {Valentino}, F. and {Vijayan}, A.~P. and {Weaver}, J.~R. and {Weibel}, A. and {Wilkins}, S.~M. and {Yang}, L.},
        title = "{COSMOS-Web: Stellar mass assembly in relation to dark matter halos across 0.2 < z < 12 of cosmic history}",
      journal = {\aap},
         year = 2025,
        month = mar,
       volume = {695},
          eid = {A20},
        pages = {A20},
          doi = {10.1051/0004-6361/202452570},
archivePrefix = {arXiv},
       eprint = {2410.08290},
 primaryClass = {astro-ph.GA},
       adsurl = {https://ui.adsabs.harvard.edu/abs/2025A&A...695A..20S}
}

@ARTICLE{Smith_Matthew2021,
       author = {{Smith}, Matthew C. and {Bryan}, Greg L. and {Somerville}, Rachel S. and {Hu}, Chia-Yu and {Teyssier}, Romain and {Burkhart}, Blakesley and {Hernquist}, Lars},
        title = "{Efficient early stellar feedback can suppress galactic outflows by reducing supernova clustering}",
      journal = {\mnras},
         year = 2021,
        month = sep,
       volume = {506},
       number = {3},
        pages = {3882-3915},
          doi = {10.1093/mnras/stab1896},
archivePrefix = {arXiv},
       eprint = {2009.11309},
 primaryClass = {astro-ph.GA},
       adsurl = {https://ui.adsabs.harvard.edu/abs/2021MNRAS.506.3882S}
}

@ARTICLE{Springel2003,
       author = {{Springel}, Volker and {Hernquist}, Lars},
        title = "{Cosmological smoothed particle hydrodynamics simulations: a hybrid multiphase model for star formation}",
      journal = {\mnras},
         year = 2003,
        month = feb,
       volume = {339},
       number = {2},
        pages = {289-311},
          doi = {10.1046/j.1365-8711.2003.06206.x},
archivePrefix = {arXiv},
       eprint = {astro-ph/0206393},
 primaryClass = {astro-ph},
       adsurl = {https://ui.adsabs.harvard.edu/abs/2003MNRAS.339..289S}
}

@ARTICLE{Springel2005BHs,
       author = {{Springel}, Volker and {Di Matteo}, Tiziana and {Hernquist}, Lars},
        title = "{Modelling feedback from stars and black holes in galaxy mergers}",
      journal = {\mnras},
         year = 2005,
        month = aug,
       volume = {361},
       number = {3},
        pages = {776-794},
          doi = {10.1111/j.1365-2966.2005.09238.x},
archivePrefix = {arXiv},
       eprint = {astro-ph/0411108},
 primaryClass = {astro-ph},
       adsurl = {https://ui.adsabs.harvard.edu/abs/2005MNRAS.361..776S}
}

@ARTICLE{Stanway2018,
       author = {{Stanway}, E.~R. and {Eldridge}, J.~J.},
        title = "{Re-evaluating old stellar populations}",
      journal = {\mnras},
         year = 2018,
        month = sep,
       volume = {479},
       number = {1},
        pages = {75-93},
          doi = {10.1093/mnras/sty1353},
archivePrefix = {arXiv},
       eprint = {1805.08784},
 primaryClass = {astro-ph.GA},
       adsurl = {https://ui.adsabs.harvard.edu/abs/2018MNRAS.479...75S}
}

@ARTICLE{Steinwandel2023,
       author = {{Steinwandel}, Ulrich P. and {Bryan}, Greg L. and {Somerville}, Rachel S. and {Hayward}, Christopher C. and {Burkhart}, Blakesley},
        title = "{On the impact of runaway stars on dwarf galaxies with resolved interstellar medium}",
      journal = {\mnras},
         year = 2023,
        month = nov,
       volume = {526},
       number = {1},
        pages = {1408-1427},
          doi = {10.1093/mnras/stad2744},
archivePrefix = {arXiv},
       eprint = {2205.09774},
 primaryClass = {astro-ph.GA},
       adsurl = {https://ui.adsabs.harvard.edu/abs/2023MNRAS.526.1408S}
}

@PHDTHESIS{Strolger2003,
       author = {{Strolger}, Louis-Gregory},
        title = "{The Nearby Galaxies Supernova Search project: The rate of supernovae in the local universe}",
       school = {University of Michigan},
         year = 2003,
        month = aug,
       adsurl = {https://ui.adsabs.harvard.edu/abs/2003PhDT........14S}
}

@ARTICLE{Teyssier2002,
       author = {{Teyssier}, R.},
        title = "{Cosmological hydrodynamics with adaptive mesh refinement. A new high resolution code called RAMSES}",
      journal = {\aap},
         year = 2002,
        month = apr,
       volume = {385},
        pages = {337-364},
          doi = {10.1051/0004-6361:20011817},
archivePrefix = {arXiv},
       eprint = {astro-ph/0111367},
 primaryClass = {astro-ph},
       adsurl = {https://ui.adsabs.harvard.edu/abs/2002A&A...385..337T}
}

@ARTICLE{Thacker2000,
       author = {{Thacker}, R.~J. and {Couchman}, H.~M.~P.},
        title = "{Implementing Feedback in Simulations of Galaxy Formation: A Survey of Methods}",
      journal = {\apj},
         year = 2000,
        month = dec,
       volume = {545},
       number = {2},
        pages = {728-752},
          doi = {10.1086/317828},
archivePrefix = {arXiv},
       eprint = {astro-ph/0001276},
 primaryClass = {astro-ph},
       adsurl = {https://ui.adsabs.harvard.edu/abs/2000ApJ...545..728T}
}

@ARTICLE{Thomas2010,
       author = {{Thomas}, Daniel and {Maraston}, Claudia and {Schawinski}, Kevin and {Sarzi}, Marc and {Silk}, Joseph},
        title = "{Environment and self-regulation in galaxy formation}",
      journal = {\mnras},
         year = 2010,
        month = jun,
       volume = {404},
       number = {4},
        pages = {1775-1789},
          doi = {10.1111/j.1365-2966.2010.16427.x},
archivePrefix = {arXiv},
       eprint = {0912.0259},
 primaryClass = {astro-ph.CO},
       adsurl = {https://ui.adsabs.harvard.edu/abs/2010MNRAS.404.1775T}
}

@ARTICLE{Thorne2021,
       author = {{Thorne}, Jessica E. and {Robotham}, Aaron S.~G. and {Davies}, Luke J.~M. and {Bellstedt}, Sabine and {Driver}, Simon P. and {Bravo}, Mat{\'\i}as and {Bremer}, Malcolm N. and {Holwerda}, Benne W. and {Hopkins}, Andrew M. and {Lagos}, Claudia del P. and {Phillipps}, Steven and {Siudek}, Malgorzata and {Taylor}, Edward N. and {Wright}, Angus H.},
        title = "{Deep Extragalactic VIsible Legacy Survey (DEVILS): SED fitting in the D10-COSMOS field and the evolution of the stellar mass function and SFR-M$_{{\ensuremath{\star}}}$ relation}",
      journal = {\mnras},
         year = 2021,
        month = jul,
       volume = {505},
       number = {1},
        pages = {540-567},
          doi = {10.1093/mnras/stab1294},
archivePrefix = {arXiv},
       eprint = {2011.13605},
 primaryClass = {astro-ph.GA},
       adsurl = {https://ui.adsabs.harvard.edu/abs/2021MNRAS.505..540T}
}

@ARTICLE{Tonry2003,
       author = {{Tonry}, John L. and {Schmidt}, Brian P. and {Barris}, Brian and {Candia}, Pablo and {Challis}, Peter and {Clocchiatti}, Alejandro and {Coil}, Alison L. and {Filippenko}, Alexei V. and {Garnavich}, Peter and {Hogan}, Craig and {Holland}, Stephen T. and {Jha}, Saurabh and {Kirshner}, Robert P. and {Krisciunas}, Kevin and {Leibundgut}, Bruno and {Li}, Weidong and {Matheson}, Thomas and {Phillips}, Mark M. and {Riess}, Adam G. and {Schommer}, Robert and {Smith}, R. Chris and {Sollerman}, Jesper and {Spyromilio}, Jason and {Stubbs}, Christopher W. and {Suntzeff}, Nicholas B.},
        title = "{Cosmological Results from High-z Supernovae}",
      journal = {\apj},
         year = 2003,
        month = sep,
       volume = {594},
       number = {1},
        pages = {1-24},
          doi = {10.1086/376865},
archivePrefix = {arXiv},
       eprint = {astro-ph/0305008},
 primaryClass = {astro-ph},
       adsurl = {https://ui.adsabs.harvard.edu/abs/2003ApJ...594....1T}
}

@ARTICLE{Trayford2025,
       author = {{Trayford}, James W. and {Schaye}, Joop and {Correa}, Camila and {Ploeckinger}, Sylvia and {Richings}, Alexander J. and {Chaikin}, Evgenii and {Schaller}, Matthieu and {Benitez-Llambay}, Alejandro and {Frenk}, Carlos and {Husko}, Filip},
        title = "{Modelling the evolution and influence of dust in cosmological simulations that include the cold phase of the interstellar medium}",
      journal = {arXiv e-prints},
         year = 2025,
        month = may,
          eid = {arXiv:2505.13056},
        pages = {arXiv:2505.13056},
          doi = {10.48550/arXiv.2505.13056},
archivePrefix = {arXiv},
       eprint = {2505.13056},
 primaryClass = {astro-ph.GA},
       adsurl = {https://ui.adsabs.harvard.edu/abs/2025arXiv250513056T}
}

@ARTICLE{Tremmel2015,
       author = {{Tremmel}, M. and {Governato}, F. and {Volonteri}, M. and {Quinn}, T.~R.},
        title = "{Off the beaten path: a new approach to realistically model the orbital decay of supermassive black holes in galaxy formation simulations}",
      journal = {\mnras},
         year = 2015,
        month = aug,
       volume = {451},
       number = {2},
        pages = {1868-1874},
          doi = {10.1093/mnras/stv1060},
archivePrefix = {arXiv},
       eprint = {1501.07609},
 primaryClass = {astro-ph.GA},
       adsurl = {https://ui.adsabs.harvard.edu/abs/2015MNRAS.451.1868T}
}




\appendix

\section{Possible suppression of gravitational instability due to gravitational softening} \label{app:softening}

\begin{figure*}
    \centering
    \includegraphics[width=0.95\linewidth]{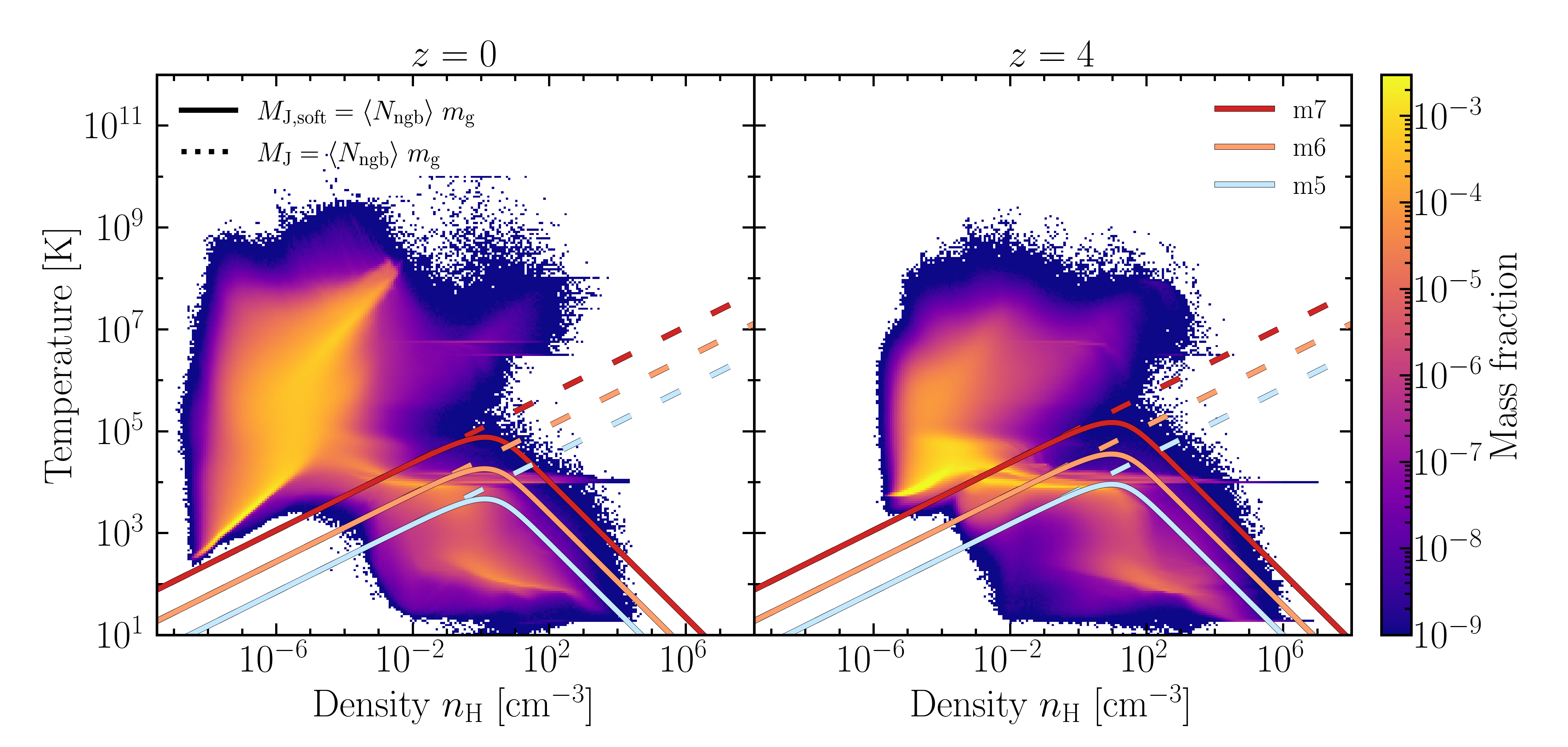}
    \caption{Temperature -- density plots for the gas in the L200m6 simulation at $z=0$ (left panel) and $z=4$ (right panel). The colour encodes the fraction of the gas mass per pixel. Gas below the dotted (solid) contours is gravitationally unstable at the resolution limit in Newtonian (softened) gravity, if the turbulent velocity dispersion is negligible. The line colour indicates the numerical resolution of the simulation, with orange corresponding to the resolution of the simulation whose density-temperature distribution is plotted (m6). Gas clouds with mass equal to the kernel mass that fall in between the solid and dotted contours of the same colour are stabilized by softening, but this is not necessarily true for higher mass clouds.}  
    \label{fig:phase_diagrams_softening}
\end{figure*}

For all \colibre\ resolutions, gravity is softened for baryonic densities $n_\text{H} > n_{\text{H},\epsilon} \equiv m_\text{g}X/(m_\text{H}\epsilon^3) \approx 0.2~\cm^{-3}$ at $z<1.57$ and at densities a factor $[(1+z)/2.57]^3$ higher than $n_{\text{H},\epsilon}$ at $z\ge1.57$. As discussed and demonstrated by \citet{Ploeckinger2024} and \citet{Nobels2024}, for these high densities, softening does not necessarily eliminate gravitational instability on scales greater than the SPH kernel, because the gravitational force is not zero below the softening scale and because such dense gas tends to be cold ($T\ll 10^4\,\K$), which facilitates gravitational instability. 

Assuming that turbulence is subsonic, gas clouds of mass equal to the SPH kernel mass are physically gravitationally unstable (i.e.\ the kernel mass exceeds the thermal Jeans mass) below the dotted contours in the temperature -- density plots shown in Fig.~\ref{fig:phase_diagrams_softening}. If there were turbulence, then the contours would truncate below the density for which the thermal velocity dispersion drops below the turbulent value (e.g.\ fig.~1 of \citealt{Nobels2024}). If the turbulence is supersonic, which is commonly the case for the cold ISM ($T \ll 10^4~\K$), then there is thus a minimum density for gravitational instability at the mass resolution limit. The different line colours show different numerical resolutions, where orange corresponds to m6, which is the resolution of the simulation whose mass distribution is indicated by the colour image. The left and right panels show redshifts $z=0$ and 4, respectively. Because the kernel mass decreases with increasing resolution, higher densities and/or lower temperatures are required for gravitational instability at higher resolution.  Above the solid contours (kernel mass equals the softened Jeans mass; see \citealt{Ploeckinger2024}), which bend down above $n_{\text{H},\epsilon}$, gas clouds with mass equal to the kernel mass are gravitationally stable in the simulation. Gas clouds in the region in between the solid and dotted contours should be unstable but are stabilized by softening, at least for a cloud mass equal to the kernel mass. Higher-mass clouds can still be unstable. 

Fig.~\ref{fig:phase_diagrams_softening} shows that at both redshifts there is little gas in the potentially problematic region between the solid and dotted contours. Note that gas in this region is eligible to form stars because the star formation criterion does not account for softening. Hence, star formation in this region is only slowed down indirectly by the suppression of gravitational instability in low-mass clouds. Even if the gas is unstable in the simulation, i.e.\ below the solid contours, if $n_\text{H} > n_{\text{H},\epsilon}$, then softening may slow gravitational collapse in the regime where the dynamical time is small compared to the sound-crossing time. Here, the time-scale for gravitational collapse is the softened free-fall time, instead of the classical Newtonian free-fall time \citep{Ploeckinger2024}. If the weight of overlying gas layers pushes the gas to sufficiently high densities, then gas clouds that are stabilized by softening may still become unstable, but such layers will only be resolved if the galaxy contains a sufficient number of particles. The problem of suppression of gravitational instability due to softening is thus more likely to occur in lower-mass galaxies because they are resolved with fewer particles. Another reason is that lower-mass galaxies also tend to have gas with lower metallicities, which tends to be hotter. 

If we increase the resolution, then the contours in Fig.~\ref{fig:phase_diagrams_softening} shift down, and higher gas densities are required for star formation (see equation~\ref{eq:SFcrit}). Hence, for resolutions much higher than those considered here, the slow-down of gravitational collapse by softening could become a bottleneck. This may result in excessively long gas consumption time-scales and hence too much mass in the ISM. If softening were to cause problems, then we would expect to see signs first in the lowest mass galaxies in the highest resolution runs. Resolving such issues would require switching to adaptive (i.e.\ density-dependent) gravitational softening\footnote{Note that adaptive softening can only help if the softening length is allowed to become sufficiently small.}, which, however, would require recalibration of subgrid models and would, if it results in smaller softening lengths for collisionless particles, exacerbate spurious energy transfer due to discreteness effects. Future work, particularly with simulations using higher resolution than used for the \colibre\ simulations presented here, could benefit from a switch to adaptive softening, at least for gas particles.

\section{The linear phases for the initial conditions}\label{app:panphasia}
The Gaussian phases for the initial conditions of all simulation volumes use an updated version of the Panphasia hierarchical Gaussian White Noise field introduced by \citet{Jenkins2013}. We used the same version as the \flamingo\ simulations \citep[see appendix~B of][]{Schaye2023}. 
The text descriptors in Table~\ref{tbl:panphasia} specify the initial Gaussian phases for the different simulation volumes presented in this work. The \monofonIC code \citep{Hahn2020,Michaux2021} can be used to create uniform mass resolution initial conditions using these text descriptors. The same descriptors can be read by the \textsc{music2} code \citep{Hahn2011,Buehlmann2025} to generate zoom initial conditions for regions or objects within the \colibre\ volumes.

\begin{table*}
    \centering
    \caption{PANPHASIA descriptors for different box sizes. Simulation variations in the same volume use the same descriptor.} 
    \label{tbl:panphasia}
    \begin{tabular}{rl}
	\hline
        $L$ (cMpc) & Descriptor \\
        \hline
        25 & [Panph6,L20,(967096,215599,663706),S1,KK1025,CH3222183807,COLIBRE025] \\
        50 & [Panph6,L20,(235287,445214,422255),S1,KK1025,CH2001508634,COLIBRE050] \\
        100 & [Panph6,L19,(266507,351694,328156),S1,KK1025,CH1356507509,COLIBRE100] \\
        200 & [Panph6,L19,(218575,323145,460872),S1,KK1025,CH2306922479,COLIBRE200] \\
        400 & [Panph6,L18,(200557,163876,161484),S1,KK1025,CH3518244376,COLIBRE400] \\
        \hline
    \end{tabular}
\end{table*}

\section{Stellar birth properties in galaxy stellar mass bins} \label{app:birth_by_mass}
\begin{figure}
    \centering
    \includegraphics[width=0.95\linewidth]{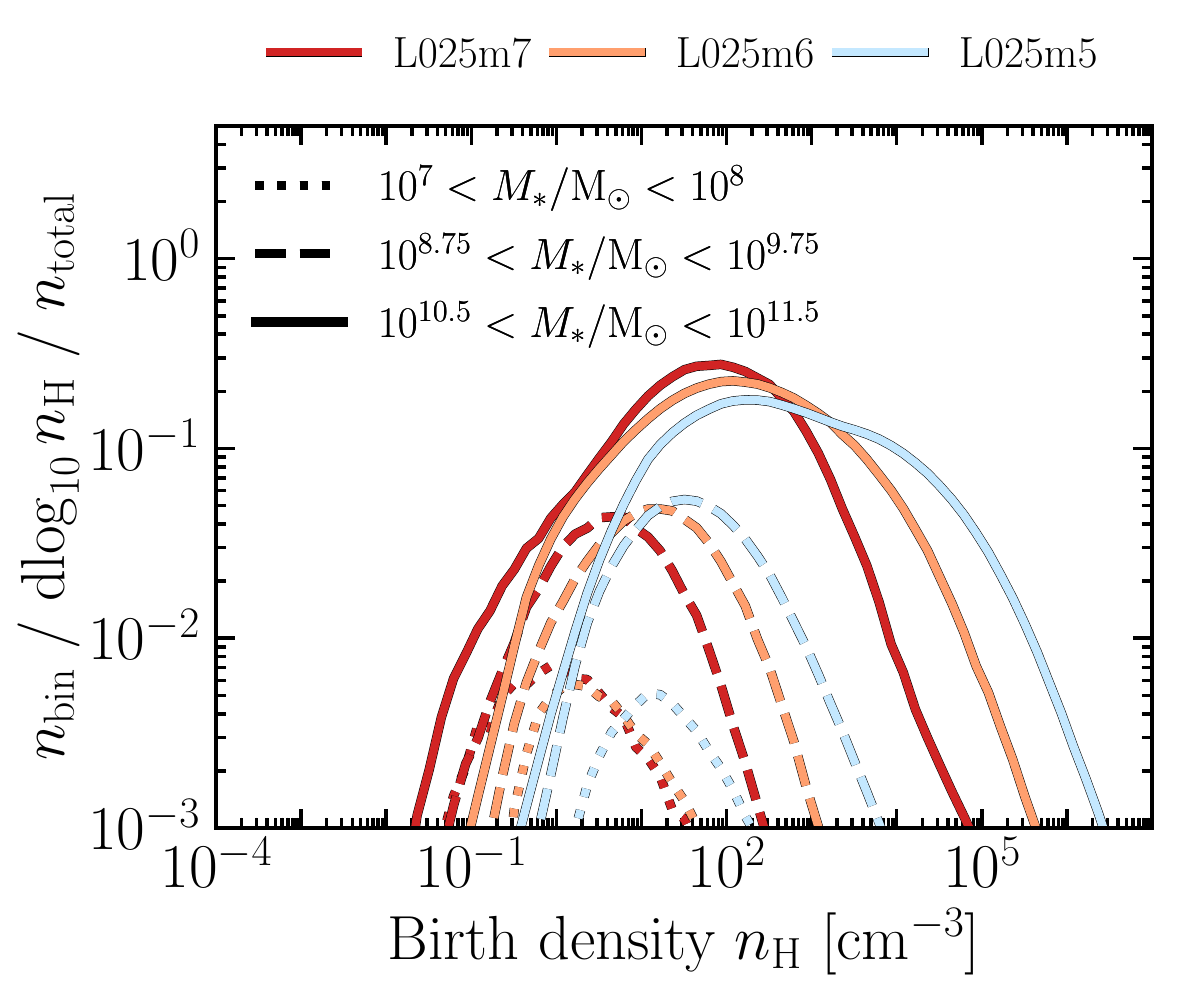}\\
    \includegraphics[width=0.95\linewidth]{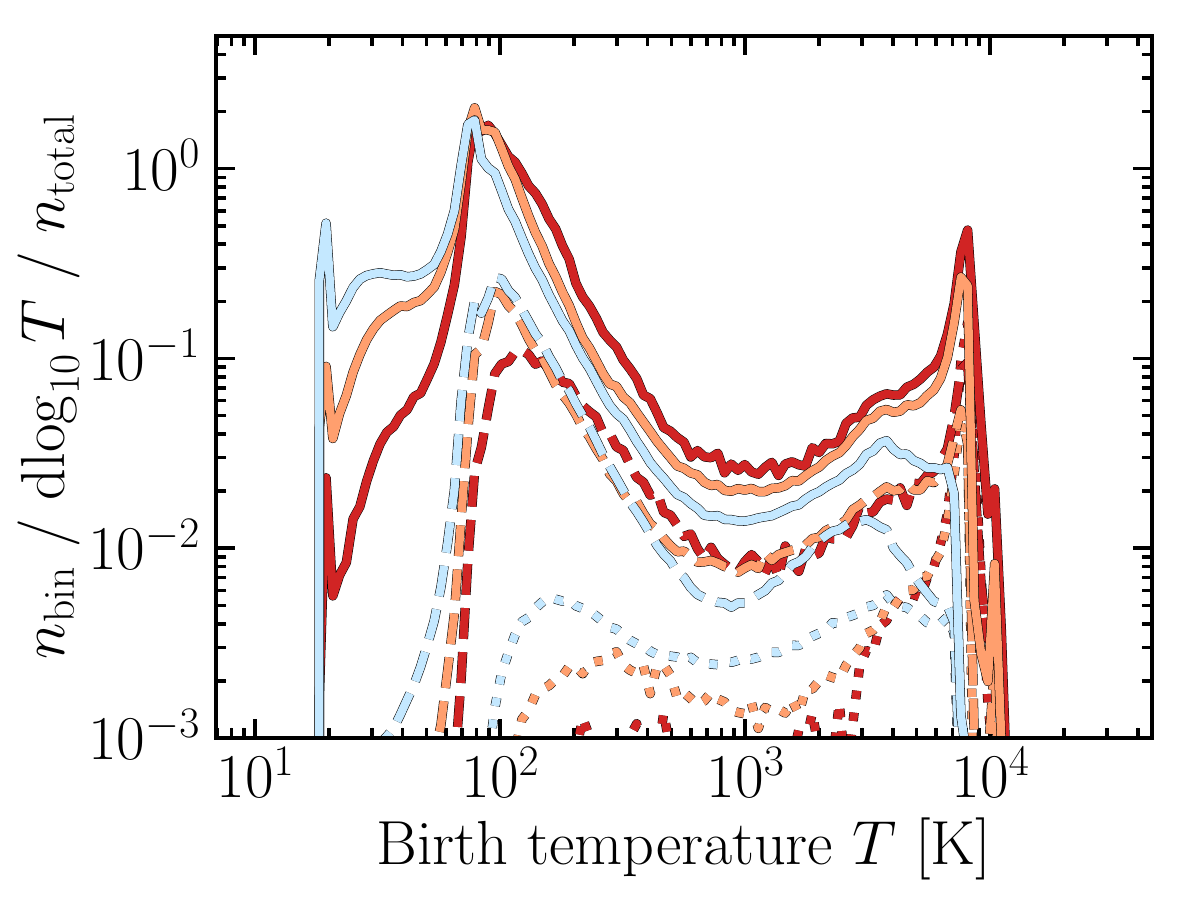}\\
    \includegraphics[width=0.95\linewidth]{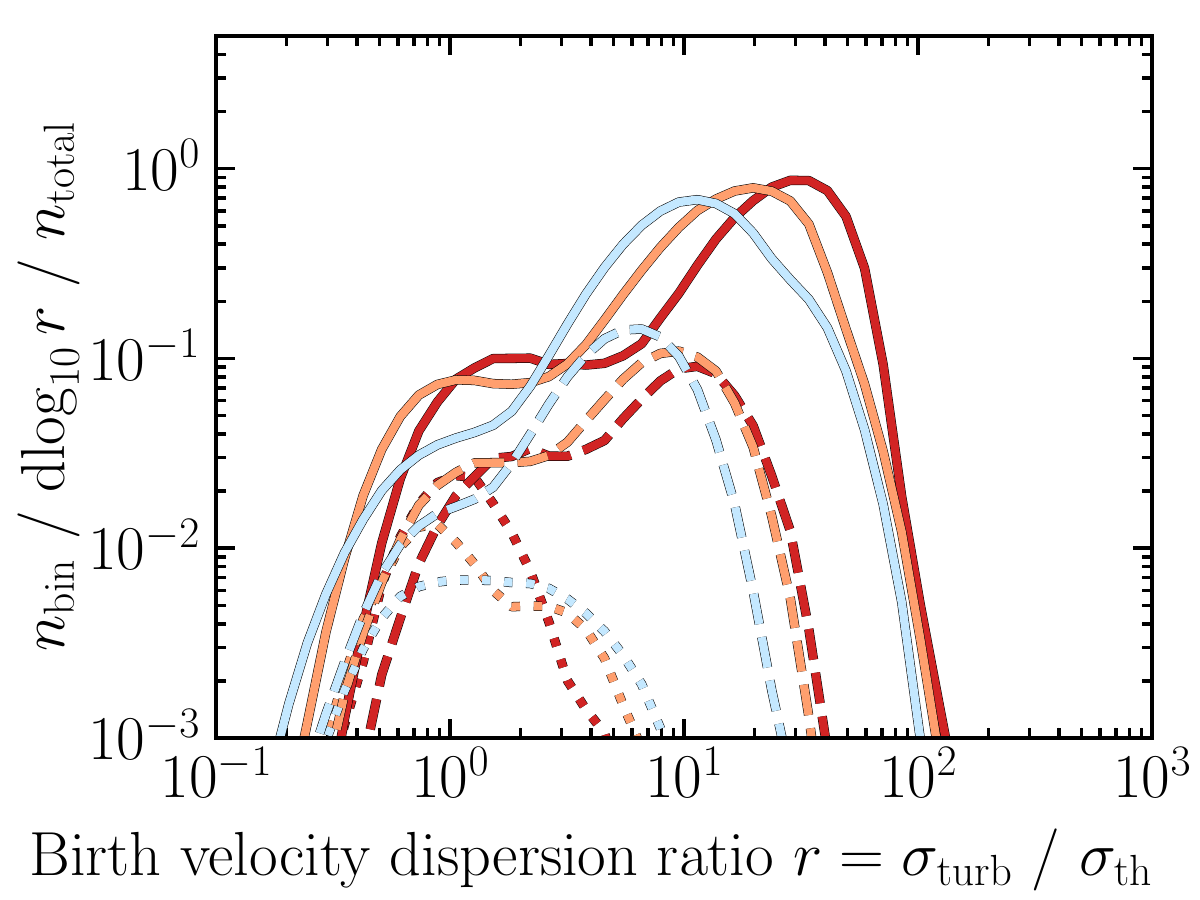}
    \caption{As Fig.~\ref{fig:birth_props}, but for three different non-contiguous bins of $z=0$ galaxy stellar mass (different line styles). In more massive galaxies stars tend to form in gas that is denser, has a lower temperature, and in which turbulence is more supersonic.}
    \label{fig:birth_props_by_mass}
\end{figure}

Fig.~\ref{fig:birth_props_by_mass} is analogous to Fig.~\ref{fig:birth_props}, but compares the stellar birth properties in three different bins of $z=0$ galaxy stellar mass (different line styles). We see that the physical conditions of the gas in which the stellar particles were born depend strongly on galaxy mass. In more massive galaxies stars tend to form from gas that is denser, colder and where turbulence is more supersonic. Because the masses are $z=0$ values, the low-density, high-temperature, low-turbulence parts of the distributions for massive galaxies largely reflect stars born at higher redshifts in lower-mass progenitors. 

\section{The effect of random errors in stellar masses} \label{app:edd_bias}
\begin{figure*}
    \centering
    \includegraphics[width=0.93\linewidth]{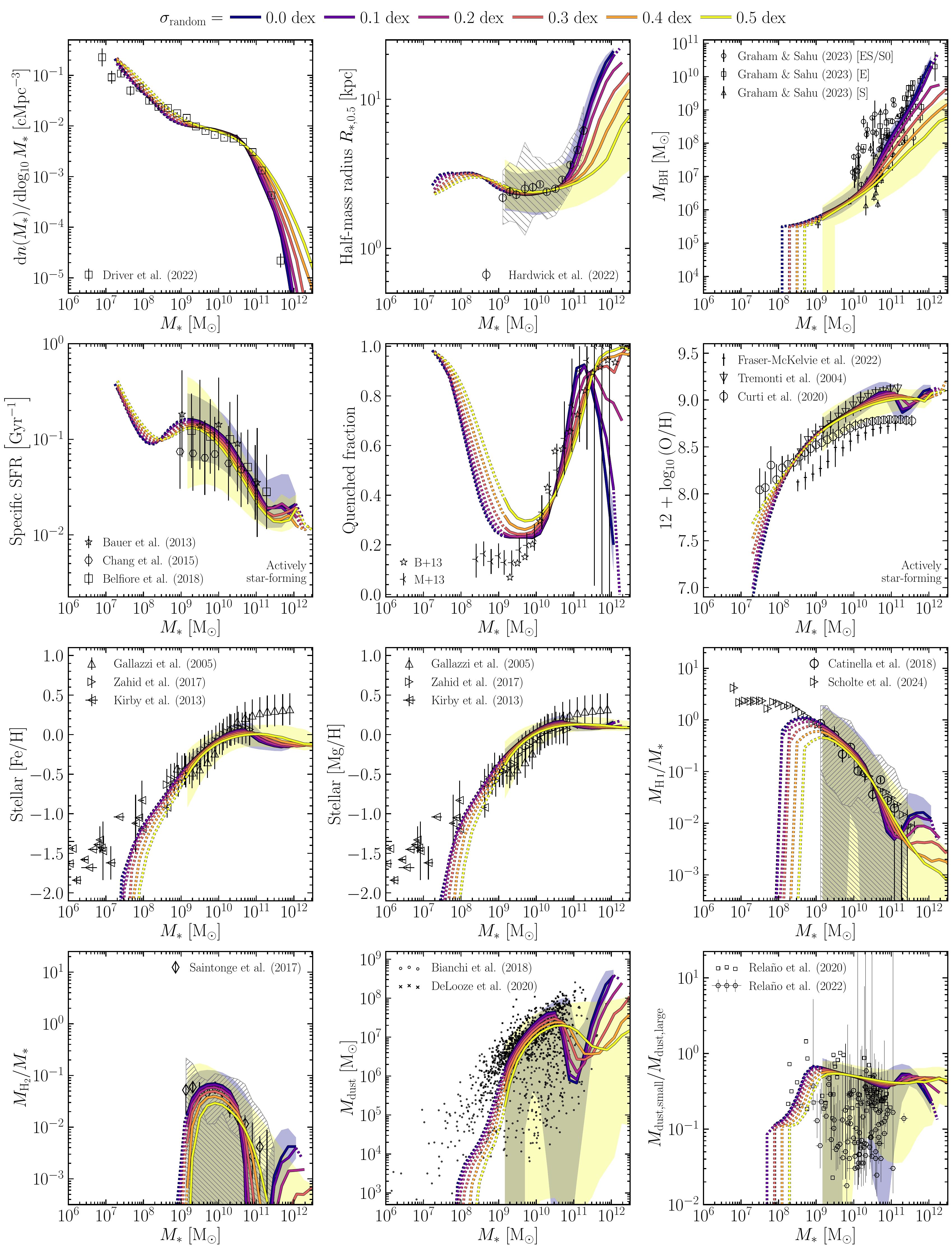}
    \caption{The effect of imposing random lognormal scatter on galaxy stellar mass, with zero mean and standard deviation $\sigma_\text{random}$ on $\log_{10}M_*$, on $z=0$ galaxy scaling relations in the L400m7 simulation. Different colour lines correspond to different amounts of mass scatter, with $\sigma_\text{random}$ varying between 0 (i.e.\, no errors) to 0.5~dex. Dark blue and yellow shaded regions, which are present in all panels except for the one showing quenched fractions, show the 16th to 84th percentile scatter for $\sigma_\text{random} = 0$ and 0.5~dex, respectively. The different panels repeat all the plots with stellar mass on the $x$-axis shown in Section~\ref{sec:obs}, where we assumed $\sigma_\text{random} = 0.1$~dex. The Eddington bias resulting from random mass errors tends to smooth out mass trends and increase the scatter. It can be important near masses at which galaxy properties change rapidly with mass and for massive galaxies, $M_*\gtrsim 10^{11}\,\Msun$, where the SMF drops exponentially.}
    \label{fig:edd_bias}
\end{figure*}

As discussed in Section~\ref{sec:obs}, when comparing to observations, we mimic the effects of random errors on the observed galaxy stellar mass by adding random lognormal scatter to the predicted masses with zero mean and standard deviation $\sigma_\text{random}(\log_{10}M_*)$, where $\sigma_\text{random}=0.1$~dex at $z=0$ (equation~\ref{eq:random_scatter}). Because it is unclear exactly how large the random errors should be, Fig.~\ref{fig:edd_bias} shows the effect of varying $\sigma_\text{random}$ between 0 and 0.5~dex for all plots shown in Section~\ref{sec:obs} that have stellar mass along the $x$-axis. For clarity, we only show the results for L400m7, which provides the best statistics at the massive end. We do not add scatter due to random errors on quantities other than galaxy mass, which may, however, also be important.

While the effect of our fiducial scatter of $0.1$~dex is small, larger stellar mass errors can have a significant impact on galaxy scaling relations. They increase the scatter at fixed mass and can substantially affect trends with mass. Scatter in the stellar masses can be particularly important for galaxy scaling relations near masses where galaxy properties change rapidly with mass, as is the case for BH mass at the stellar mass corresponding to the halo mass used for BH seeding, and for the quenched fraction near the low-mass upturn. In the L400m7 simulation these transitions are sharp due to numerical effects. The effect of mass scatter on the high-mass end of the scaling relations, $M_*\gtrsim 10^{11}\,\Msun$, is physically more interesting. Here random scatter results in Eddington bias, i.e.\ the fact that there is significantly more upscatter than downscatter if the mass function declines steeply, which smooths out both the SMF itself and trends of galaxy properties with mass.   

\section{The relations between stellar and halo mass} \label{app:smhm}
\begin{figure*}
    \centering
    \includegraphics[width=0.45\linewidth]{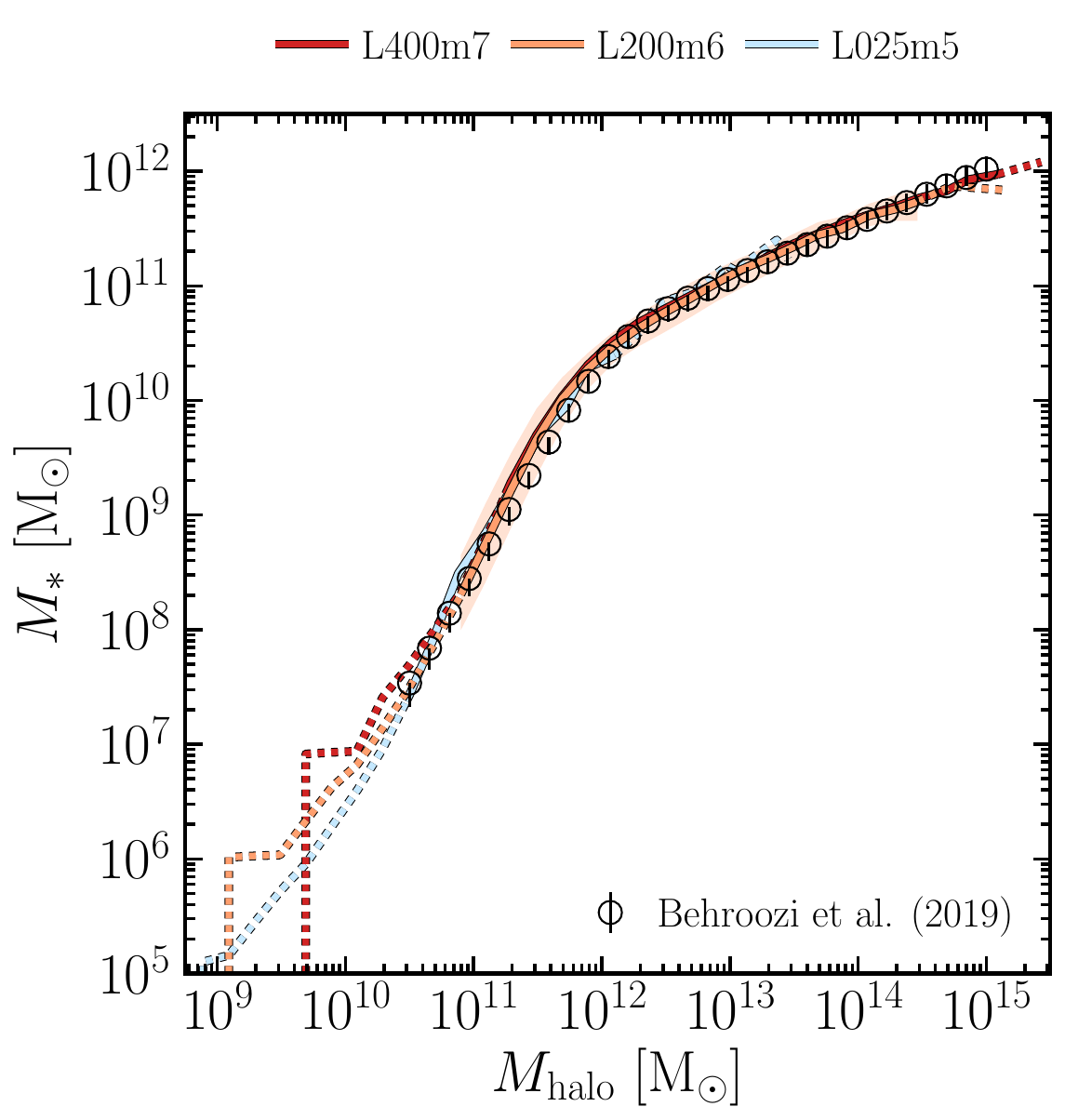} \hspace{0.05\linewidth}
    \includegraphics[width=0.45\linewidth]{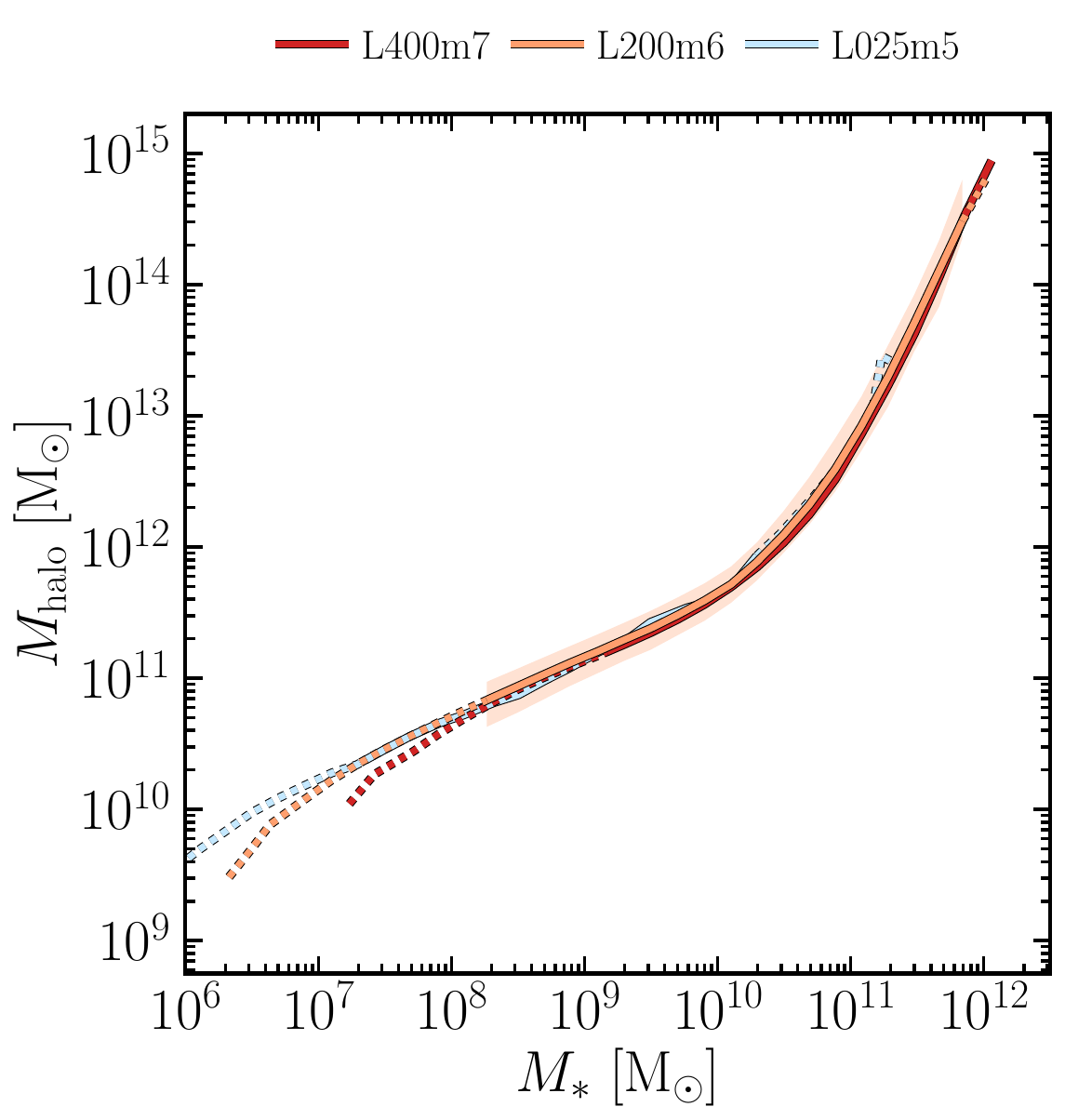}
    \caption{As Fig.~\ref{fig:smhm}, but showing stellar mass as a function of halo mass in the left panel and vice versa in the right panel. Halo mass is defined as the mass within the radius for which the mean internal density is equal to the critical overdensity provided by \citet{Bryan1998}.}
    \label{fig:smhm_2panels}
\end{figure*}

To facilitate easy conversion between the stellar and halo masses of central galaxies, Fig.~\ref{fig:smhm_2panels} shows the median stellar mass as a function of halo mass (left panel) and the median halo mass as a function of stellar mass (right panel).

\section{Comparison with the \eagle\ and IllustrisTNG simulations} \label{app:eagle_tng}
In this appendix we compare the \colibre\ predictions with those from the two suites of cosmological hydrodynamical simulations that are currently the most widely used, \eagle\ \citep{Schaye2015,Crain2015,McAlpine2016} and IllustrisTNG \citep{Pillepich2018TNGmethod,Pillepich2018TNG,Nelson2019TNGrelease}. Like \colibre, both simulations span multiple resolutions. \eagle\ Recal\_L025N752 and Ref\_L100N1504 have baryonic particle masses of $2.3\times 10^5$ and $1.8\times 10^6\,\Msun$ in box sizes of 25 and 100~Mpc, respectively. These baryonic particle masses are similar to \colibre's m5 and m6 resolutions, although for dark matter the \eagle\ resolutions are instead similar to \colibre's m6 and m7 resolutions. \eagle\ does not include a model with a baryonic mass resolution similar to m7. TNG50, TNG100, and TNG300 use box sizes of 51.7, 110.7, and 302.6~Mpc and have gas cell/stellar particle masses of $8.5\times 10^4$, $1.4\times 10^6$, and $1.1\times 10^7\,\Msun$, respectively. These baryonic mass resolutions are similar to \colibre's m5, m6, and m7 resolutions, although for dark matter they are instead similar to m6, m7, and m8 using our notation. 

Like \colibre, both sets of simulations include radiative cooling, star formation, chemical evolution, CCSN and SNIa feedback, BHs and AGN feedback. In contrast to \colibre, neither \eagle\ nor TNG allows the ISM to cool to temperatures below $\approx 10^4\,\K$, and neither models the evolution of dust grains. As for \colibre, the stellar and AGN feedback in \eagle\ are calibrated to the present-day SMF, galaxy sizes and BH masses in massive galaxies. TNG100 is calibrated to these observables and in addition to the stellar-to-halo-mass relation, the mass-metallicity relation, the mass scale above which star formation is quenched, the halo gas fraction in galaxy groups, and the cosmic star formation history. Unlike \colibre\ and \eagle, the TNG subgrid feedback parameters are kept fixed when the resolution is changed.

\begin{figure*}
    \centering
    \includegraphics[width=\linewidth]{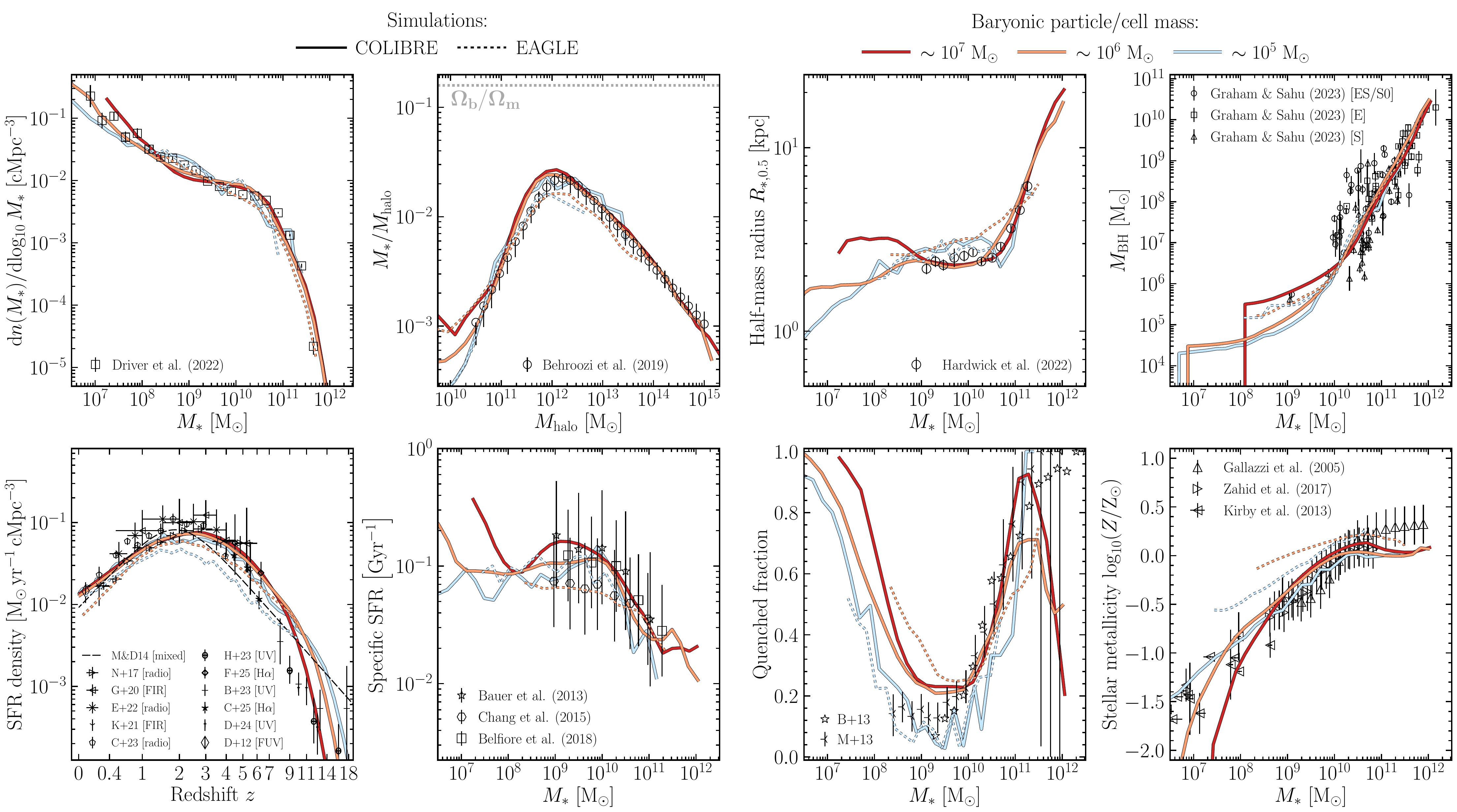}
    \caption{Comparison of \colibre\ (solid) and \eagle\ (dashed) for galaxy observables that are direct results for both sets of simulations. From the top left to the bottom right the different panels show the galaxy stellar mass function, the stellar-to-halo mass ratio for central galaxies, galaxy sizes (i.e.\ projected half-mass radii), BH masses, the cosmic star formation history, specific star formation rates of star-forming galaxies, quenched fractions, and stellar metallicities. Except for the star formation history, all results are for $z=0$. Line colours indicate the approximate baryonic particle mass, which is similar to \colibre's L200m6 for \eagle\ Ref\_L100N1504 (orange) and similar to \colibre's L025m5 for \eagle\ Recal\_L025N752 (blue). There is no \eagle\ simulation with m7 resolution. Note that for \eagle\ the corresponding dark matter mass resolution is a factor of four times worse than for the baryons, whereas they are approximately the same for \colibre. For \eagle\ all quantities are computed using 30~kpc apertures, which is the fiducial aperture choice for \eagle, whereas we use 50~kpc for \colibre. The data points show the same observations as shown in Section~\ref{sec:obs}. Halo mass is consistently defined as the mass within the radius for which the mean internal density is equal to the critical overdensity provided by \citet{Bryan1998}. Different from the plots shown in that section, none of the predicted stellar masses were perturbed to mimic the effect of random errors on the observed stellar masses.}
    \label{fig:eagle}
\end{figure*}

\begin{figure*} 
    \centering
    \includegraphics[width=\linewidth]{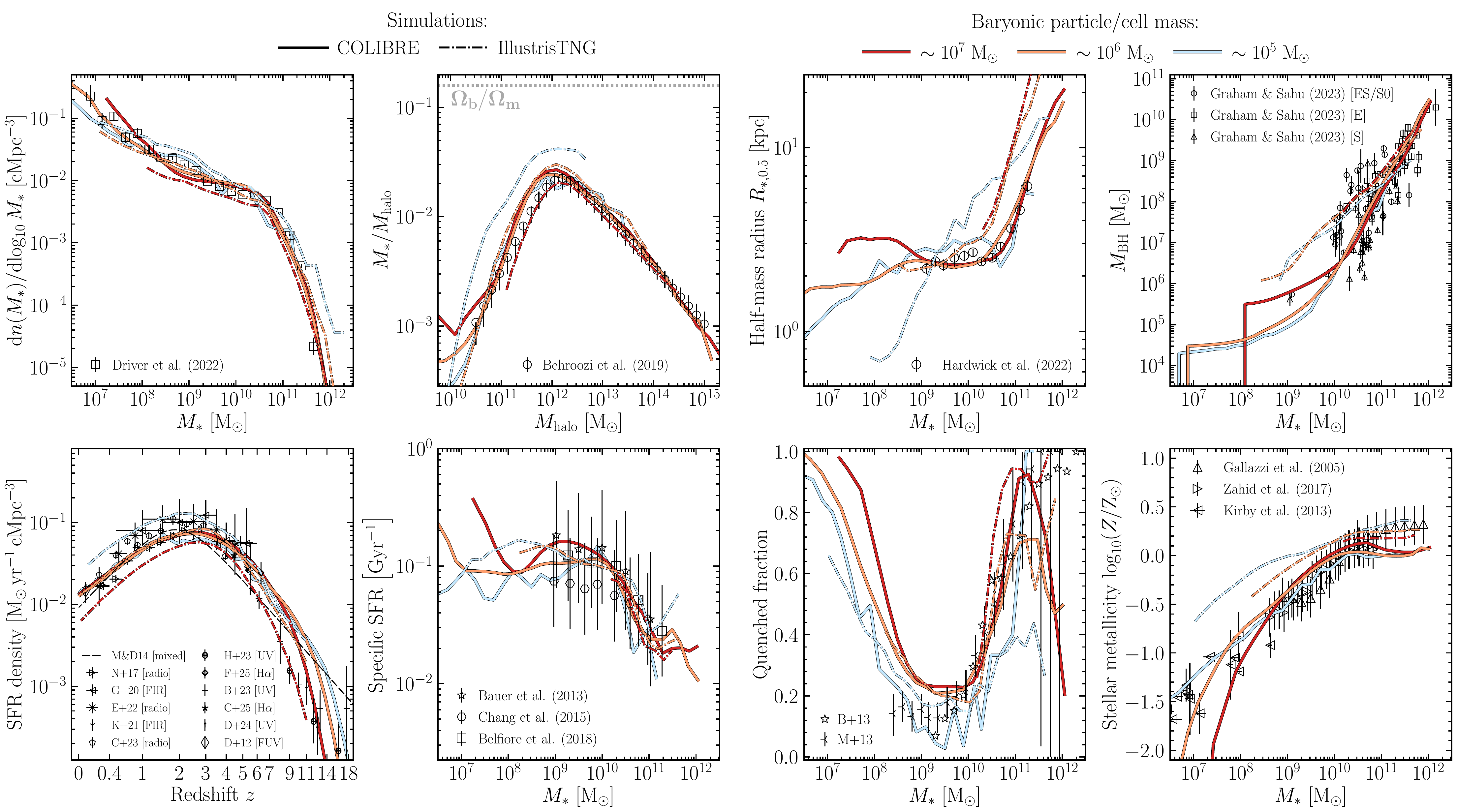}
    \caption{As Fig.~\ref{fig:eagle}, but comparing to IllustrisTNG instead of \eagle. Line colours indicate the approximate baryonic particle/cell mass, which is similar to our L400m7 for TNG300 (red), to L200m6 for TNG100 (orange), and to L025m5 for TNG50 (blue). Note that for TNG the corresponding dark matter mass resolution is a factor of four times worse than for the baryons, whereas they are approximately the same for \colibre. For TNG galaxy properties are computed using 30~kpc apertures, the fiducial TNG value for e.g.\ the mass function, whereas we use 50~kpc for \colibre. The exception is the stellar mass in the stellar mass -- halo mass relation, which for TNG300 we have taken from \citet{Pakmor2023} because they computed the relation down to lower masses than available in the public catalogue, but who use all bound mass for this relation. The TNG300 SMF, which has also been taken from \citet{Pakmor2023}, does however use 30~kpc apertures. For \colibre\ and the data from \citet{Behroozi2019} halo mass is defined as the mass within the radius for which the mean internal density is equal to the critical overdensity provided by \citet{Bryan1998}, while for TNG it is equal to 200 times the critical density.}
    \label{fig:tng}
\end{figure*}

Fig.~\ref{fig:eagle} compares the SMF, stellar-to-halo mass ratio, galaxy sizes, BH masses, cosmic star formation history, sSFRs of star-forming galaxies, quenched fractions (i.e.\ fraction of galaxies with sSFR$<10^{-11}\,\yr^{-1}$), and stellar metallicities for \colibre\ (solid) to \eagle\ (dashed) and to the same observational data as shown in the corresponding figures from Section~\ref{sec:obs}. Except for the star formation history, all predictions are for $z=0$ and were taken from the public catalogues rather than computed from snapshot data. Fig.~\ref{fig:tng} shows a similar comparison with IllustrisTNG. We do not show \ion{H}{i}, H$_2$, and dust masses (and dust grain sizes) because these properties are not directly predicted by \eagle\ and IllustrisTNG. We show total stellar metallicities instead of iron and magnesium abundances. We do not compare gas metallicities because only \colibre\ accounts for the depletion of metals onto dust grains. We also do not compare the X-ray luminosity of the CGM, which is not part of the public catalogues. While we again use apertures with radii of 50~kpc for \colibre, we use 30~kpc for \eagle\ and TNG, which are their fiducial values\footnote{For IllustrisTNG \citet{Pillepich2018TNGmethod} use apertures of twice the half-mass radius. We use 30~kpc where possible for consistency, noting also that twice the half-mass radius can be very small, e.g.\ only $\approx 5$~kpc for galaxies with masses similar to that of the Milky Way. For the stellar mass - halo mass relation, which for TNG300 we take from \citet{Pakmor2023} because they compute it down to lower masses than available in the public catalogue, we follow those authors and use all bound stellar mass.}. However, the two different apertures give nearly identical results for $M_*\lesssim 10^{11}\,\Msun$ \cite[e.g][]{Schaye2015,Chaikin2025smf_evol}. As for \colibre, galaxy SFRs are instantaneous and computed from the gas particles. Different from the plots in the main text, we do not impose random stellar mass errors on any of the predicted masses because such errors were not considered in the calibration of \eagle\ and TNG. However, as shown in Appendix~\ref{app:edd_bias}, the effect of the 0.1~dex scatter that we included in the plots shown in the main text is very small. The TNG300 stellar masses are the original predictions as opposed to the masses corrected for resolution effects.

Fig.~\ref{fig:eagle} shows that for the SMF and BH masses the numerical convergence and agreement with the data are similarly good for \eagle\ and \colibre, although at m6 resolution \eagle\ slightly undershoots the knee of the SMF. Unlike \colibre, \eagle\ m6, though not m5, predicts slightly too large galaxy sizes for $M_* < 10^{11}\,\Msun$, but the difference is likely smaller than the systematics affecting the comparison of \eagle\ with observations \citep{deGraaff2022}. The cosmic star formation history is better converged and reproduced by \colibre, with \eagle\ underestimating the peak, particularly for m5 resolution. At $M_*\sim 10^{10}\,\Msun$ the main sequence of star-forming galaxies is a factor $\approx 2$ lower for \eagle\ than for \colibre, but at m5 resolution the two are in good agreement. For the quenched fractions the numerical convergence is worse for \eagle, and for m6 resolution the agreement with the data is worse than for \colibre. However, at m5 resolution \eagle\ and \colibre\ predict similar quenched fractions. Finally, the stellar mass -- metallicity relation is much shallower for \eagle, and both the agreement with the data and the numerical convergence are clearly poorer than for \colibre.

Fig.~\ref{fig:tng} shows that the numerical convergence tends to be substantially worse for TNG than for \colibre, as might be expected from the fact that TNG was only calibrated at m6 resolution (TNG100). Galaxy stellar masses are a factor $\approx 2-4$ too high for TNG50, while they are too low for TNG300 (see e.g.\ \citealt{Pillepich2018TNG} and \citealt{Engler2021} for a discussion of the effects of resolution on galaxy stellar masses in IllustrisTNG). While galaxy sizes agree well with \colibre\ and the data for $M_*\lesssim 10^{10}\,\Msun$ at m6 resolution, at higher masses they are too large for all TNG resolutions. TNG50 sizes only agree with \colibre\ and the observations for $M_*\approx 10^9\,\Msun$. While BH masses are similar for \colibre\ and TNG at the high mass end, $M_*> 10^{11}\,\Msun$, TNG predicts much higher BH masses in low-mass galaxies. For $M_*\lesssim 10^{10}\,\Msun$ the BH masses in TNG appear too high compared to the data. For m6 resolution, the cosmic star formation histories of TNG and \colibre\ are in good agreement, but at m7 resolution the SFR in TNG is significantly below that of \colibre\ and the observations, while at m5 resolution it is significantly above both \colibre\ and the data for $z<3$. For $M_* > 10^9\,\Msun$ the sSFRs and quenched fractions are in broad agreement for each resolution, although TNG50 predicts lower quenched fractions at the highest stellar masses (see \citealt{Donnari2021} for a discussion of the impact of resolution and statistical sampling). Finally, like \eagle, TNG predicts a substantially shallower mass -- metallicity relation than \colibre, resulting in too high metallicities for $M_*\lesssim 10^{10}\,\Msun$.


\bsp	
\label{lastpage}
\end{document}